\documentclass[a4paper,11pt]{article}
\pdfoutput=1 

\usepackage{jheppub} 
\usepackage[vcentermath]{youngtab}
\usepackage{url,amsmath,amssymb,amsfonts,amsxtra, mathrsfs,graphics,graphicx,amsthm,epsfig, youngtab,bm,longtable,float,tikz,caption}
\usepackage[T1]{fontenc} 
\usepackage[margin=0.5cm]{caption}
\usepackage{jheppub}
\usepackage{float}
\restylefloat{table}
\usepackage{comment}
\usepackage{subfig}

\addtocontents{toc}{\setcounter{tocdepth}{2}}

\title{Exact relativistic Toda chain eigenfunctions \\
from Separation of Variables and gauge theory}

\author[a]{Antonio Sciarappa}

\affiliation[a]{School of Physics, Korea Institute for Advanced Study \\
85 Hoegiro, Dongdaemun-gu, Seoul 130-722, Republic of Korea}

\emailAdd{asciara@kias.re.kr}

\abstract{We provide a proposal, motivated by Separation of Variables and gauge theory arguments, for constructing exact solutions to the quantum Baxter equation associated to the $N$-particle relativistic Toda chain and test our proposal against numerical results.
\linebreak Quantum Mechanical non-perturbative corrections, essential in order to obtain a sensible solution, are taken into account in our gauge theory approach by considering codimension two defects on curved backgrounds (squashed $S^5$ and degenerate limits) rather than flat space; 
this setting also naturally incorporates exact quantization conditions and energy spectrum of the relativistic Toda chain as well as its modular dual structure.}

\preprint{KIAS-P17042}

\begin{document}

\maketitle



\section{Introduction}

Apart from very special cases like the harmonic oscillator, it is often quite hard to solve a generic Quantum Mechanical system (i.e. computing eigenvalues and eigenfunctions in the appropriate Hilbert space), even if we just limit ourselves to consider systems with discrete spectrum: sometimes the solution can be found but cannot be expressed in a closed form, while some other time we simply don't know how to solve the problem with the standard tools of Quantum Mechanics. It would therefore be very interesting to develop new, more powerful tools to approach these problems.

In recent years it has been realized that supersymmetric gauge theories 
may be thought of as a new framework to study (at least some class of) Quantum Mechanical problems \cite{2010maph.conf..265N}. The original proposal of \cite{2010maph.conf..265N} consists in considering four-dimensional gauge theories with \linebreak
$\mathcal{N} = 2$ supersymmetry living on $\mathbb{R}^2_{\epsilon_1} \times \mathbb{R}^2_{\epsilon_2}$, with $\epsilon_1$, $\epsilon_2$ Omega background parameters regularizing the infinite volume of $\mathbb{R}^4$, in the so-called NS limit $\epsilon_1 = i \hbar$, $\epsilon_2 \rightarrow 0$: in this limit the remaining parameter $\hbar \in \mathbb{R}_+$ should be thought of as the Planck constant of some Quantum Mechanical system, while gauge theory observables will admit an interpretation in terms of various quantities of interest in the Quantum Mechanical problem such as eigenvalues or quantization conditions. 
Different Quantum Mechanical systems will be associated to different supersymmetric gauge theories, and this association can often be understood by matching spectral curves with Seiberg-Witten curves.
It is then clear what the advantage of the gauge theory approach is: since we know how to compute gauge theory observables via supersymmetric localization, we can provide an explicit solution to the Quantum Mechanical system (although often not in closed form) once the dictionary between Quantum Mechanical quantities and gauge theories observables has been established.

Although it may be hard to prove it rigorously, the proposal by \cite{2010maph.conf..265N} can at least be checked against analytical results in those cases for which the solution to the corresponding Quantum Mechanical system is known, or against numerical results otherwise. A notable example is the $N$-particle Toda chain, which is in correspondence with four-dimensional pure $\mathcal{N} = 2$ $SU(N)$ Yang-Mills. This system has been solved via a technique known as Separation of Variables in \cite{GUTZWILLER1980347,GUTZWILLER1981304,Kharchev:1999bh,Kharchev:2000yj,An2009}; it was later shown in \cite{Kozlowski:2010tv} that the Separation of Variables solution coincides with the gauge theory one proposed by \cite{2010maph.conf..265N}. \\

Ordinary Quantum Mechanical systems are often associated to differential operators: in fact  many of them can be obtained from a classical Hamiltonian $H(x,p)$ with polynomial dependence (usually quadratic) on the momentum $p$, therefore canonical quantization $p \rightarrow -i \hbar \partial_x$ promotes the Hamiltonian to a differential operator $\widehat{H}(x, -i\hbar \partial_x)$.
There may however be well-defined Quantum Mechanical systems, which we call ``relativistic'', arising from a classical Hamiltonian $\mathcal{H}(e^x, e^p)$ with \textit{exponential} dependence on the momentum $p$ (which therefore is a periodic variable); these lead to \textit{finite-difference} operators $\widehat{\mathcal{H}}(e^{x}, e^{-i\hbar \partial_x})$ after canonical quantization.
An interesting example is the $N$-particle ``relativistic'' Toda chain, which simply is the finite-difference version of the Toda chain mentioned above.
Relativistic Quantum Mechanical systems are typically harder to solve than usual ones; in particular, differently from its ``non-relativistic'' (differential) counterpart, the general solution to the relativistic Toda chain is not known.

Always according to the proposal by \cite{2010maph.conf..265N}, relativistic Quantum Mechanical systems may also be studied in the framework of supersymmetric gauge theory: the dictionary between Quantum Mechanical quantities and gauge theories observables should remain the same, and the only difference is that this time the gauge theory will have to be five-dimensional $\mathcal{N} = 1$ and living on $\mathbb{R}^2_{\epsilon_1} \times \mathbb{R}^2_{\epsilon_2} \times S^1_R$ (always in the NS limit), where $R$ is related to the periodicity of $p$ in an appropriate normalization. 
Since relativistic Quantum Mechanical systems are harder to solve than usual ones,  
one could hope to rely on the gauge theory approach when dealing with relativistic problems.

Unfortunately, this turns out not to be the case. In fact even if the analytic solution is not known, we can always find a numerical solution for the relativistic system of interest and compare it against the proposal by \cite{2010maph.conf..265N}: this was done in \cite{Grassi:2014zfa,2015arXiv151102860H} for the relativistic Toda chain, and it was found that the energy spectrum predicted by \cite{2010maph.conf..265N} does not match with the numerical one. 
The mismatch is due to the fact that the gauge theory approach seems to give the exact WKB answer, but does not take into account possible non-perturbative corrections in $\hbar$, which turn out to be present and very important for the cases at hand; one should therefore understand how to refine the gauge theory proposal in such a way to include Quantum Mechanical instantons.

A more refined proposal, which seems to correctly include these non-perturbative corrections since it matches with numerical results, was put forward in \cite{Grassi:2014zfa} for the 2-particle relativistic Toda chain (as well as for other relativistic systems, see also subsequent works \cite{Codesido:2015dia,Gu:2015pda}), at least for what quantization conditions and energy spectrum are concerned. Differently from the original suggestion by \cite{2010maph.conf..265N}, this refined proposal implies that we shouldn't simply consider observables of five-dimensional gauge theories on flat space but we should work with a full, non-perturbatively complete theory of topological strings; quite remarkably, the prescription of \cite{Grassi:2014zfa} also seems to agree with results in resurgence theory \cite{Couso-Santamaria:2016vwq}. 

Another proposal which seems to properly take into account these non-perturbative corrections for the 2-particle case, again at the level of quantization conditions and energy spectrum, was given in \cite{Wang:2015wdy}. The basic idea of \cite{Wang:2015wdy} is to add an appropriate term to the ``naive'' quantization conditions one would get from \cite{2010maph.conf..265N} in such a way that the total, exact quantization conditions will be invariant under the exchange $\hbar \leftrightarrow 4\pi^2/\hbar$ (sometimes referred to as ``$S$-duality''); this symmetry is particularly nice since it naturally fits with the \textit{modular duality} property of relativistic systems. 
This second proposal was later extended to the general $N$-particle relativistic Toda chain in \cite{2015arXiv151102860H}, where it was also checked that it gives an energy spectrum matching the numerical one (at least for the 3-particle case), and to other cluster integrable systems in \cite{Franco:2015rnr}.
Remarkably, although very different in spirit (and in formulae), the two proposals by \cite{Grassi:2014zfa} and \cite{Wang:2015wdy},\cite{2015arXiv151102860H} are compatible and can be related to each other as shown in \cite{Sun:2016obh,Grassi:2016nnt}, at least for a large class of relativistic Quantum Mechanical systems. 
\\

Despite having two precise (and related) proposals for determining exact quantization conditions and energy spectrum of many relativistic Quantum Mechanical systems, and in particular of the relativistic Toda chain, a complete solution would also require knowing  the eigenfunctions for the system. 
Not much is known about these eigenfunctions. For what the relativistic Toda chain is concerned, the general analysis of \cite{Kharchev:2001rs} in the Separation of Variables formalism reduces the problem of finding the relativistic Toda eigenfunctions to the problem of finding an appropriate solution to the Baxter equation associated to the system; however, it is not known how to find this solution to the Baxter equation. 
More recent works studying the eigenfunction problem for the 2-particle relativistic Toda chain (at least in some limit) are \cite{Marino:2016rsq,Kashaev:2017zmv}; their approach is quite different from the one of \cite{Kharchev:2001rs}, and in particular in \cite{Marino:2016rsq} it is proposed to construct the eigenfunction in terms of a non-perturbatively complete theory of open topological strings, much in the spirit of \cite{Grassi:2014zfa}
(as such, the proposal by \cite{Marino:2016rsq} is actually more general and can be used to construct eigenfunctions of quantum mirror curves for all toric geometries).
Nevertheless, a general construction for the relativistic Toda eigenfunctions is still lacking, either in the standard Quantum Mechanical approach or in a (refined) gauge theory/string theory approach.

In this work we will provide a gauge theory proposal for constructing the solution to the Baxter equation associated to the relativistic Toda chain, out of which the relativistic Toda eigenfunctions can be obtained via Separation of Variables. The idea is to start from the  ``naive'' Baxter solution one would get from \cite{2010maph.conf..265N} and modify it by an appropriate term chosen is such a way that the final Baxter solution will be invariant under ``$S$-duality'' (i.e. $\hbar \leftrightarrow 4\pi^2/\hbar$), much like what done in \cite{Wang:2015wdy} for the quantization conditions and as natural from the modular duality property of the relativistic Toda chain. A similar idea was considered in \citep{Kashani-Poor:2016edc,Sciarappa:2016ctj}, but the results there are incomplete; here instead we will check our proposal against numerical results and find good agreement.

Our proposal, partially motivated by the work \cite{2012arXiv1210.5909L} and subsequent observations in \cite{2015arXiv150704799H}, can be thought of as a refinement of the original one by \cite{2010maph.conf..265N} in which, instead of considering the NS limit of five-dimensional $\mathcal{N} = 1$ gauge theories on the flat background $\mathbb{R}^2_{\epsilon_1} \times \mathbb{R}^2_{\epsilon_2} \times S^1_R$, we study the ``NS limit'' of the same theories but on the curved, squashed five-sphere background $S^5_{\omega_1, \omega_2, \omega_3}$ with $\omega_i$ squashing parameters (or some other background with the same ``NS limit''). Exact quantization conditions, exact spectrum, exact eigenfunctions, ``$S$-duality'' symmetry, and the modular double structure of the relativistic Toda chain seem to be naturally unified in this framework; however it is not clear to us why this should be so, therefore at the moment considering gauge theories on $S^5_{\omega_1, \omega_2, \omega_3}$ should mostly be regarded as a computational tool.
\\

This paper is organized as follows. Section \ref{sec2} is dedicated to the (``non-relativistic'') $N$-particle Toda chain, whose solution (eigenvalues and eigenfunctions) is known: after reminding the definition of the system (Section \ref{sectoda}), we will review its solution both in the Separation of Variables framework (Section \ref{sec2.2}) and in the gauge theory framework of \cite{2010maph.conf..265N} (Section \ref{sec2.3}). Additional comments are collected in Section \ref{sec2.4}. Although pretty much everything is already known, the concepts and ideas reviewed in Section \ref{sec2} will be useful in order to understand how to approach the relativistic case.

In Section \ref{sec3} instead we will analyse the relativistic $N$-particle Toda chain. We will first recall its definition (Section \ref{sec3.1}) and study its solution via numerical methods (Section \ref{sec3.2}). We will then move to discuss how an analytic solution for this system can be constructed by considering gauge theories on $S^5_{\omega_1, \omega_2, \omega_3}$, especially for the 2-particle case: after recovering the exact quantization conditions of \cite{Wang:2015wdy} and energy spectrum in our framework (Section \ref{sec3.3}), we will state our proposal for the solution to the Baxter equation and compare it against numerical results (Section \ref{sec3.4}). A number of remarks are discussed in Section \ref{seccomments5d}. We conclude Section \ref{sec3} with a discussion on the general $N$-particle case, where we provide some additional explicit check against numerical results for $N = 3$ (Section \ref{sec3.5}).

Section \ref{sec4} contains our conclusions and a discussion on open problems; in Appendices \ref{appA} and \ref{appdoublesine} instead we collected the relevant formulae used to compute the gauge theory observable of interest.



\newpage

\section{Toda integrable systems} \label{sec2}

In order to understand how quantum mechanical problems can be solved in terms of gauge theory quantities, we briefly revisit in this Section the case of the $N$-particle Toda chain. This system has extensively been studied and the solution (eigenfunctions and spectrum) is explicitly known in the integrable system literature as reviewed in Section \ref{sec2.2}. The gauge theory approach to solving the Toda chain has also been studied in detail and was found to coincide with the integrable system community result; this will be reviewed in Section \ref{sec2.3}. 
The main purpose of this Section is to introduce the relevant concepts and conventions that will be used in the rest of the paper, as well as to provide a guideline for how to proceed in finding the solution to the relativistic Toda chain that will be discussed in Section \ref{sec3};
as such there will be no new results here, apart possibly from a few comments in Section \ref{sec2.4}.

\subsection{Quantum Toda chain (open and closed)} \label{sectoda}

The quantum $N$-particle Toda chain is a quantum mechanical system describing $N$ particles on a line interacting via the Hamiltonian 
\begin{equation}
\widehat{H}_2 = 
\sum_{m<n}^N p_m p_n - \sum_{m=1}^N e^{x_m - x_{m+1}}
. \label{Htoda}
\end{equation}
Here $x_m$, $p_m$ are position and momentum of the $m$-th particle\footnote{In this work we will only consider $x_m \in \mathbb{R}$ since this choice leads to quantum mechanical operators with discretized energy levels (at least for the closed Toda chain). Choosing $x_m$ imaginary leads to operators with continuous spectrum and will not be discussed here.}, satisfying the commutation relations 
\begin{equation}
[p_l, x_m] = -i \hbar \delta_{l,m}, \;\;\; \hbar \in \mathbb{R}_+,
\end{equation}
and we imposed the boundary condition
\begin{equation}
x_{N+1} = x_1 - \ln Q \,,\;\;\; Q \in \mathbb{R}_+. \label{bc}
\end{equation}
The value of the parameter $Q$ allows us to distinguish between two very different cases:
\begin{itemize}
\item $Q = 0$: open Toda chain (non-confining potential, continuous spectrum); 
\item $Q > 0$: closed Toda chain (confining potential, discrete spectrum). 
\end{itemize}
Independently on the value of $Q$, both open and closed Toda chains are actually integrable models, and as such admit a total of $N$ commuting Hamiltonians $\widehat{H}_m$ ($m = 1,\ldots, N$) including \eqref{Htoda}; schematically we have

\begin{eqnarray}
\widehat{H}_1 &=& \sum_{m=1}^N p_m, \nonumber \\
\widehat{H}_2 &=& \sum_{m<n}^N p_m p_n - \sum_{m=1}^N e^{x_m - x_{m+1}}, \nonumber \\ 
\widehat{H}_3 &=& \sum_{m<n<r}^N p_m p_n p_r + \ldots \nonumber, \\
&\vdots & \nonumber \\
\widehat{H}_N &=& \ldots. \label{hams}
\end{eqnarray}
Decoupling the center of mass of the system is equivalent to impose $\widehat{H}_1 = 0$. 
Sometimes it will be more convenient to consider particular combinations $\widehat{H}'_m$ of the Hamiltonians $\widehat{H}_m$ given by
\begin{eqnarray}
\widehat{H}'_1 &=& \sum_{m=1}^N p_m, \nonumber \\
\widehat{H}'_2 &=& \dfrac{1}{2}\sum_{m = 1}^N p_m^2 + \sum_{m=1}^N e^{x_m - x_{m+1}}, \nonumber \\ 
&\vdots & \nonumber \\
\widehat{H}'_N &=& \dfrac{1}{N!} \sum_{m=1}^N p_m^N + \ldots. \label{hamsprime}
\end{eqnarray}
The spectral problem we want to solve consists of finding the common eigenfunctions $\Psi_{\vec{E}}(\vec{x}, \hbar)$ of the Toda Hamiltonians $\widehat{H}_m$, 
\begin{equation}
\widehat{H}_m \Psi_{\vec{E}}(x_1, \ldots, x_N, \hbar) = E_m \Psi_{\vec{E}}(x_1, \ldots, x_N, \hbar) \,,\;\;\;\; m = 1, \ldots, N, \label{sptoda}
\end{equation}
with $\vec{E} = (E_1, \ldots, E_N)$ set of eigenvalues (continuous or discrete), satisfying the appropriate boundary conditions (after decoupling the center of mass) imposed by the form of the potential: 
\begin{itemize}
\item for the open Toda chain, $\Psi_{\vec{E}}(\vec{x}, \hbar)$ should vanish fast enough as $x_k - x_{k+1} \rightarrow \infty$;
\item for the closed Toda chain, normalizability requires $\Psi_{\vec{E}}(\vec{x}, \hbar) \in L^2(\mathbb{R}^{N-1})$. 
\end{itemize}
In a slightly more compact notation, by defining the generating function for the Toda Hamiltonians
\begin{equation}
\widehat{t}(\sigma) = \sum_{m=0}^N (-1)^m \sigma^{N-m}\widehat{H}_m,  \;\;\;\;\;\;  \widehat{H}_0 = 1
\end{equation}
which satisfies the commutation relation 
\begin{equation}
[\widehat{t}(\sigma), \widehat{t}(\sigma')] = 0,
\end{equation}
the spectral problem \eqref{sptoda} is equivalent to requiring 
\begin{equation}
\widehat{t}(\sigma)\Psi_{\vec{E}}(\vec{x}, \hbar) = t(\sigma, \vec{E})\Psi_{\vec{E}}(\vec{x}, \hbar)
\end{equation}
with the appropriate boundary conditions imposed, where 
\begin{equation}
t(\sigma, \vec{E}) = \sum_{m=0}^N (-1)^m \sigma^{N-m} E_m, \;\;\;\;\;\; E_0 = 1
\end{equation}
is the generating function of the Toda eigenvalues.
The polynomial $t(\sigma, \vec{E})$ also enters in the definition of the spectral curve of the classical Toda chain: this is a Riemann surface embedded in $(\sigma, y) \in \mathbb{C} \times \mathbb{C}^{\times}$ given by 
\begin{equation}
y + Q y^{-1} = t(\sigma, \vec{E}), \label{spcurve4d}
 \end{equation}  
out of which one can compute the action-angle variables of the classical Toda system. Canonical quantization of the $(\sigma, y)$ variables turns equation \eqref{spcurve4d} into a quantum operator known as Baxter equation (after a small redefinition of parameters); as we will see, solutions to the Baxter equation play a key role in constructing the eigenfunctions of the Toda chain. 

Since it will turn out to be more natural from the point of view of gauge theory, let us mention here that one can also re-express the polynomial $t(\sigma, \vec{E})$ in terms of an auxiliary set of variables $\vec{a} = (a_1, \ldots, a_N)$ or $\vec{\tau} = (\tau_1, \ldots, \tau_N)$ for the open and closed case respectively as
\renewcommand\arraystretch{1.5}
\begin{equation}
t(\sigma, \vec{E}) = \left\{ \begin{array}{ll}
\prod_{m=1}^N (\sigma - a_m)\;\;\;\;\Longrightarrow & \;\;\;\; E_m(\vec{a}) = e_m(\vec{a}) \;\;\;\; \text{(open chain)},\\
\prod_{m=1}^N (\sigma - \tau_m)\;\;\;\;\Longrightarrow & \;\;\;\; E_m(\vec{\tau}) = e_m(\vec{\tau}) \;\;\;\; \text{(closed chain)},
\end{array}
\right.  \label{tpol}
\end{equation}
\renewcommand\arraystretch{1}
\hspace{-0.1 cm}where $e_m(\vec{x})$ are elementary symmetric polynomials in the variables $\vec{x}$ and $a_1 + \ldots + a_N = \tau_1 + \ldots + \tau_N$. Since the $\tau_m$ variables should reduce to the $a_m$ ones at $Q=0$ we expect that $\tau_m = a_m + o(Q)$; more in general we will often think of $\tau_m = \tau_m(\vec{a})$ as functions of the $a_m$, so that also the closed Toda eigenvalues $E_m = e_m(\vec{\tau}(\vec{a}))$ will indirectly be functions of the $a_m$'s.

\subsection{Solution via Separation of Variables} \label{sec2.2}

Explicit solutions to both the open and closed Toda chain spectral problems \eqref{sptoda} are well-known in the integrable system literature. 
Historically the first step was done by Gutzwiller in \cite{GUTZWILLER1980347,GUTZWILLER1981304}, where he provided expressions for the $N = 2, 3, 4$-particles closed Toda chain eigenfunctions in terms of linear combinations of the open Toda chain ones and also obtained quantization conditions and spectrum for these systems. His works already contained the basic idea underlying what is today known as quantum Separation of Variables method, later developed in more detail by Sklyanin \cite{Sklyanin:1984sb}, see also \cite{0305-4470-25-20-007}.
The Separation of Variables method reduces the problem of finding eigenfunctions of the $N$-particle Toda chain to the problem of finding solutions of a one-dimensional Baxter equation; Gutzwiller's quantization conditions follow from certain analytic requirements on the solution to the Baxter problem. 
Separation of Variables has later been used in \cite{Kharchev:1999bh,Kharchev:2000yj} to provide a solution to the $N$-particles Toda problem (both open and closed) at any $N$; in this framework the Toda eigenfunctions admit an explicit expression in terms of multiple contour integrals. Finally, additional technical problems were rigorously solved in \cite{An2009}. In this Section we will quickly review the general $N$ solution by \cite{Kharchev:1999bh,Kharchev:2000yj}; we will later see in Section \ref{sec2.3} how this solution can be recovered in terms of gauge theory quantities. \\


For what the open Toda chain is concerned, it is shown in \cite{Kharchev:1999bh,Kharchev:2000yj} that, in terms of the auxiliary variables $\vec{a}$, the eigenfunction $\Psi_{\vec{a}}(\vec{x},\hbar)$ for the $N$-particle open Toda chain with the appropriate boundary conditions can be expressed in a recursive way as 
\begin{equation}
\begin{split}
& \Psi_{a_1, \ldots, a_N}(x_1, \ldots, x_N, \hbar) =  \\
& = \int_{\mathcal{C}} 
\left[\prod_{j=1}^{N-1} d\sigma_j \right] \mu(\vec{\sigma}, \hbar)\, \mathcal{Q}(\vec{\sigma}, \vec{a}, \hbar) \\
&\;\;\;\;\;\; \Psi_{\sigma_1, \ldots, \sigma_{N-1}}(x_1, \ldots, x_{N-1}, \hbar)  e^{\frac{i}{\hbar}x_N \left( \sum_{m=1}^N a_m - \sum_{j=1}^{N-1}\sigma_j \right)} \;\;\;\;\;\; \text{(open chain)}, \label{eigopen}
\end{split}
\end{equation}
starting from the eigenfunction of the 1-particle problem $\Psi_{a_1}(x_1) = e^{i a_1 x_1/\hbar}$, for an appropriate contour of integration $\mathcal{C}$; it is also possible to show that \eqref{eigopen} is an entire functions of the $\vec{a}$ parameters. Here $\mu(\vec{\sigma}, \hbar)$ is an integration measure,
\begin{equation}
\mu(\vec{\sigma}, \hbar) = \prod_{j < k}^{N-1} \dfrac{(\sigma_j - \sigma_k)}{\hbar} \sinh\left( \pi\frac{\sigma_j - \sigma_k}{\hbar} \right),
\end{equation}
while
\begin{equation}
\mathcal{Q}(\vec{\sigma}, \vec{a}, \hbar) = \prod_{j=1}^{N-1} q(\sigma_j, \vec{a}, \hbar)= \prod_{j=1}^{N-1} \prod_{m=1}^{N} 
\hbar^{-i\frac{\sigma_j - a_m}{\hbar}} \Gamma\left(-i\frac{\sigma_j - a_m}{\hbar}\right) \label{Qtildeopen}
\end{equation}
is sometimes referred to as the Harish-Chandra function for the open Toda chain, or wave-function in separated variables; the function $q(\sigma, \vec{a}, \hbar)$, and consequently $\mathcal{Q}(\vec{\sigma}, \vec{a}, \hbar)$, formally satisfies the one-dimensional \textit{finite-difference} Baxter equation for the open Toda chain  
\begin{equation}
(i)^N q(\sigma + i \hbar, \vec{a}, \hbar) = t(\sigma, \vec{E}(\vec{a})) q(\sigma, \vec{a}, \hbar), \label{baxopen}
\end{equation}
with $t(\sigma, \vec{E}(\vec{a})))$ as in the first line of \eqref{tpol}. \\

Moving to the $N$-particle closed Toda chain, its eigenfunctions $\Psi_{\vec{a}}(\vec{x},\hbar, Q)$ can be obtained from the eigenfunctions of the $(N-1)$-particle \textit{open} Toda chain via an expression similar to \eqref{eigopen}, the only difference being in the function $\mathcal{Q}(\vec{\sigma}, \vec{a}, \hbar, Q)$; more precisely we have (modulo normalizations)\footnote{Here and in the following we will often distinguish $\Psi_{\vec{a}}(\vec{x},\hbar, Q)$, $\mathcal{Q}(\vec{\sigma}, \vec{a}, \hbar, Q)$ associated to the closed Toda chain from $\Psi_{\vec{a}}(\vec{x},\hbar)$, $\mathcal{Q}(\vec{\sigma}, \vec{a}, \hbar)$ associated to the open Toda chain by explicitly indicating the $Q$-dependence in the former case.}

\begin{equation}
\begin{split}
& \Psi_{a_1, \ldots, a_N}(x_1, \ldots, x_N, \hbar, Q) =  \\
& = \int_{\mathcal{C}} 
\left[\prod_{j=1}^{N-1} d\sigma_j \right] \mu(\vec{\sigma}, \hbar) \, \mathcal{Q}(\vec{\sigma}, \vec{\tau}(\vec{a}), \hbar, Q) \\
&\;\;\;\;\;\; \Psi_{\sigma_1, \ldots, \sigma_{N-1}}(x_1, \ldots, x_{N-1}, \hbar)  e^{\frac{i}{\hbar}x_N \left( \sum_{m=1}^N a_m - \sum_{j=1}^{N-1}\sigma_j \right)} \;\;\;\;\;\; \text{(closed chain)}, \label{eigclosed}
\end{split}
\end{equation}
where this time
\begin{equation}
\mathcal{Q}(\vec{\sigma}, \vec{\tau}(\vec{a}), \hbar, Q) = \prod_{j=1}^{N-1} q(\sigma_j, \vec{\tau}(\vec{a}), \hbar, Q) \label{Qsep}
\end{equation}
and $q(\sigma, \vec{\tau}, \hbar, Q)$ satisfies the one-dimensional \textit{finite-difference} Baxter equation for the closed Toda chain (suppressing the implicit $\vec{a}$ dependence)
\begin{equation}
(i)^N q(\sigma + i \hbar, \vec{\tau}, \hbar, Q) + Q (i)^{-N} q(\sigma - i \hbar, \vec{\tau}, \hbar, Q) = t(\sigma, \vec{E}(\vec{\tau})) q(\sigma, \vec{\tau}, \hbar, Q), \label{baxclosed}
\end{equation}
with $t(\sigma, \vec{E}(\vec{\tau}))$ as in the second line of \eqref{tpol}. It is important to remark that this equation is nothing else than the quantization of the classical spectral curve \eqref{spcurve4d} if we identify $y = (i)^N e^{- p_{\sigma}} = (i)^N e^{i \hbar \partial_{\sigma}}$. 

To sum up, Separation of Variables reduces the problem of finding eigenfunctions $\Psi_{\vec{a}}(\vec{x},\hbar,Q)$ to the closed Toda chain \eqref{sptoda} to the problem of finding solutions $q(\sigma, \vec{\tau}, \hbar, Q)$ to the Baxter equation \eqref{baxclosed}; this is a great simplification since the Baxter equation is a 1-dimensional problem. However, $q(\sigma, \vec{\tau}, \hbar, Q)$ cannot just be any function formally satisfying \eqref{baxclosed}: in fact as discussed in \cite{Kharchev:1999bh,Kharchev:2000yj}, our expression \eqref{eigclosed} is a solution to the closed Toda spectral problem only if the solution $q(\sigma, \vec{\tau}, \hbar, Q)$ to \eqref{baxclosed} is \textit{entire} in $\sigma$ and goes to zero fast enough as $\vert \sigma \vert \rightarrow \infty$; among other things, these additional conditions help us fixing the quasi-constant ambiguity affecting $q(\sigma, \vec{\tau}, \hbar, Q)$ (being a finite-difference equation, any formal solution to \eqref{baxclosed} is only defined modulo $i\hbar$-periodic functions). It turns out that these conditions are only satisfied when the auxiliary parameters $a_m$ (or equivalently $\tau_m$) have certain special values, from which quantization of the energies $E_m(\vec{\tau})$ follows. 

More in detail, in Gutzwiller's approach to the study of entire solutions to \eqref{baxclosed}
one first considers the half-infinite determinants $K^{(\mp)}(\sigma, \vec{\tau}, \hbar, Q)$ given by
\begin{equation}
K^{(-)}(\sigma, \vec{\tau}, \hbar, Q) = \left\vert \begin{array}{cccc}
\ldots & \ldots & \ldots & \ldots \\
\ldots & 1 & \dfrac{1}{t(\sigma - 3i\hbar, \vec{E}(\vec{\tau}))} & 0 \\
0 & \dfrac{Q}{t(\sigma - 2i\hbar, \vec{E}(\vec{\tau}))} & 1 & \dfrac{1}{t(\sigma - 2i\hbar, \vec{E}(\vec{\tau}))} \\
\ldots & 0 & \dfrac{Q}{t(\sigma - i\hbar, \vec{E}(\vec{\tau}))} & 1
\end{array} \right\vert,
\end{equation}
\begin{equation}
K^{(+)}(\sigma, \vec{\tau}, \hbar, Q) = \left\vert \begin{array}{cccc}
1 & \dfrac{1}{t(\sigma + i\hbar, \vec{E}(\vec{\tau}))} & 0 & \ldots \\ 
\dfrac{Q}{t(\sigma + 2i\hbar, \vec{E}(\vec{\tau}))} & 1 & \dfrac{1}{t(\sigma + 2i\hbar, \vec{E}(\vec{\tau}))} & 0 \\
0 & \dfrac{Q}{t(\sigma + 3i\hbar, \vec{E}(\vec{\tau}))} & 1 & \ldots \\
\ldots & \ldots & \ldots & \ldots
\end{array} \right\vert,
\end{equation}
from which two linearly independent entire solutions $q^{(\mp)}_0(\sigma, \vec{\tau}, \hbar, Q)$ can be constructed:
\begin{equation}
q^{(-)}_0(\sigma, \vec{\tau}, \hbar, Q) =
\dfrac{e^{-\frac{\pi N \sigma}{\hbar}} K^{(-)}(\sigma, \vec{\tau}, \hbar, Q)}{\prod_{m=1}^N \hbar^{\frac{i \sigma}{\hbar}} \Gamma (1 + i\frac{\sigma - \tau_m}{\hbar})},
\end{equation}
\begin{equation}
\;\;\;\;\;\;q^{(+)}_0(\sigma, \vec{\tau}, \hbar, Q) =
\dfrac{Q^{-\frac{i \sigma}{\hbar}}e^{-\frac{\pi N \sigma}{\hbar}} K^{(+)}(\sigma, \vec{\tau}, \hbar, Q)}{\prod_{m=1}^N \hbar^{- \frac{i \sigma}{\hbar}} \Gamma (1- i\frac{\sigma - \tau_m}{\hbar})}.
\end{equation}
Despite being entire, these solutions do not go to zero fast enough at both $\sigma = \pm \infty$.  To obtain the desired asymptotic behaviour, we use the fact that solutions to \eqref{baxclosed} are only defined up to $i\hbar$-periodic functions; we can then multiply $q^{(\mp)}_0(\sigma, \vec{\tau}, \hbar, Q)$ by an $i\hbar$-periodic factor which modifies the asymptotics as needed. The simplest choice is to consider
\begin{equation}
q^{(\mp)}(\sigma, \vec{\tau}, \hbar, Q) = \dfrac{q^{(\mp)}_0(\sigma, \vec{\tau}, \hbar, Q)}{\prod_{m=1}^N e^{-\frac{\pi \sigma}{\hbar}} \sinh\left( \pi \frac{\sigma - a_m}{\hbar} \right)}, \label{tempx}
\end{equation}
where $a_m$ are the zeroes of the ($i\hbar$-periodic) Hill determinant
\begin{equation}
\mathcal{H}(\sigma, \vec{\tau}, \hbar, Q) = \left\vert \begin{array}{cccccc}
\ldots & \ldots & \ldots & \ldots & \ldots & \ldots \\
\ldots & \dfrac{Q}{t(\sigma - i\hbar, \vec{E}(\vec{\tau}))} & 1 & \dfrac{1}{t(\sigma - i\hbar, \vec{E}(\vec{\tau}))} & 0 & \ldots \\
\ldots & 0 & \dfrac{Q}{t(\sigma, \vec{E}(\vec{\tau}))} & 1 & \dfrac{1}{t(\sigma, \vec{E}(\vec{\tau}))} & \ldots \\
\ldots & \ldots & \ldots & \ldots & \ldots & \ldots
\end{array} \right\vert,
\end{equation}
which can also be written as
\begin{equation}
\mathcal{H}(\sigma, \vec{\tau}, \hbar, Q) = \prod_{m=1}^N \dfrac{\sinh\left(\pi \frac{\sigma - a_m}{\hbar} \right)}{\sinh\left(\pi \frac{\sigma - \tau_m}{\hbar} \right)}. \label{hilldet}
\end{equation}
By the analysis of \cite{Kozlowski:2010tv}, the zeroes of $\mathcal{H}(\sigma)$ coincide with the parameters $a_m$ we introduced earlier\footnote{The definition $\mathcal{H}(a_m) = 0$, $m = 1,\ldots, N$ implicitly relates the $a_m$ variables to the $\tau_m$ ones, taking into account the constraint $\sum_{m=1}^N a_m = \sum_{m=1}^N \tau_m$. Let us also remark that because of $i\hbar$-periodicity, $\mathcal{H}(a_m + i \hbar n_m) = 0$ for any $n_m \in \mathbb{Z}$.}. 
However step \eqref{tempx} does not come without a price: although our new functions $q^{(\mp)}(\sigma)$ satisfy \eqref{baxclosed} and go to zero fast enough at both $\sigma = \pm \infty$, they are no longer entire but meromorphic since they present poles at $\sigma = a_m + i \hbar n_m$, $n_m \in \mathbb{Z}$. To solve this problem and construct a solution $q(\sigma)$ which is both entire and with the desired asymptotic behaviour, we may consider a linear combination
\begin{equation}
\begin{split}
q(\sigma, \vec{a}, \hbar, Q) &\;=\; q^{(+)}(\sigma, \vec{a}, \hbar, Q) - \xi q^{(-)}(\sigma, \vec{a}, \hbar, Q) \;= \\
& \;=\; \dfrac{q^{(+)}_0(\sigma, \vec{a}, \hbar, Q) - \xi q^{(-)}_0(\sigma, \vec{a}, \hbar, Q)}{\prod_{m=1}^N e^{-\frac{\pi \sigma}{\hbar}} \sinh\left( \pi \frac{\sigma - a_m}{\hbar} \right)} , \label{qsigma}
\end{split}
\end{equation}
with $\xi$ some constant chosen in such a way to cancel the poles, that is
\begin{equation}
q^{(+)}_0(a_m + i \hbar n_m, \vec{a}, \hbar, Q) = \xi q^{(-)}_0(a_m + i \hbar n_m, \vec{a}, \hbar, Q). \label{equality}
\end{equation}
A necessary condition for this to be valid is that the Wronskian
\begin{equation}
W(\sigma) = q_0^{(+)}(\sigma)q_0^{(-)}(\sigma + i \hbar) - q_0^{(+)}(\sigma + i \hbar)q_0^{(-)}(\sigma)
\end{equation}
vanishes at $\sigma = a_m + i \hbar n_m$; this is however satisfied since computation of the Wronskian shows that
\begin{equation}
W(\sigma) \;\propto \; \mathcal{H}(\sigma)\prod_{m=1}^N \sinh\left(\pi \frac{\sigma - \tau_m}{\hbar} \right)
\end{equation}
and $\sigma = a_m + i \hbar n_m$ are exactly the zeroes of the Hill determinant. To sum up, the solution to the Baxter equation \eqref{baxclosed} of interest to us is the linear combination \eqref{qsigma}, which is entire and with the appropriate asymptotic behaviour provided that the constant $\xi$ is chosen in such a way that
\begin{equation}
\xi = \dfrac{q_0^{(+)}(a_1)}{q_0^{(-)}(a_1)} = \ldots = \dfrac{q_0^{(+)}(a_N)}{q_0^{(-)}(a_N)}. \label{quantcond}
\end{equation}
It turns out that these conditions are only satisfied for particular values of $\vec{a} = (a_1, \ldots, a_N)$; as such, these are interpreted as quantization conditions for the spectrum of the closed Toda chain and were first obtained in \cite{GUTZWILLER1980347,GUTZWILLER1981304}. 

Having solved the Baxter equation, from $q(\sigma, \vec{\tau}(\vec{a}), \hbar, Q)$ we can construct $\mathcal{Q}(\vec{\sigma},\vec{\tau}(\vec{a}), \hbar, Q)$ according to \eqref{Qsep}, and with this we have all the ingredients entering in the expression for the $N$-particle closed Toda chain eigenfunctions \eqref{eigclosed}; the only problem left at this point consists in performing the integrations. This can be easily done by evaluation of residues. To give an idea of the procedure, let us first consider integrating over $\sigma_1$; denoting 
$\Psi'_{\vec{\sigma}}(\vec{x},\hbar) = \Psi_{\sigma_1, \ldots, \sigma_{N-1}}(x_1, \ldots, x_{N-1},\hbar)  e^{\frac{i}{\hbar}x_N \left( \sum_{m=1}^N a_m - \sum_{j=1}^{N-1}\sigma_j \right)}$ for shortness, we can split the integral as
\begin{equation}
\begin{split}
& \Psi_{\vec{a}}(\vec{x},\hbar,Q) = \\
& = \int_{-\infty}^{\infty} d\sigma_1 \left[\prod_{j=2}^{N-1} d\sigma_j \right]
\dfrac{q^{(+)}_0(\sigma_1) - \xi q^{(-)}_0(\sigma_1)}{\prod_{m=1}^N e^{-\frac{\pi \sigma_1}{\hbar}} \sinh\left( \pi \frac{\sigma_1 - a_m}{\hbar} \right)} \mu(\vec{\sigma}) \Psi'_{\vec{\sigma}}(\vec{x},\hbar) 
 \prod_{j=2}^{N-1} q(\sigma_j) = \\
& = \;\;\int_{\mathcal{C}_+} d\sigma_1 \left[\prod_{j=2}^{N-1} d\sigma_j \right]
\dfrac{q^{(+)}_0(\sigma_1)}{\prod_{m=1}^N e^{-\frac{\pi \sigma_1}{\hbar}} \sinh\left( \pi \frac{\sigma_1 - a_m}{\hbar} \right)} 
\mu(\vec{\sigma}) \Psi'_{\vec{\sigma}}(\vec{x},\hbar)  \prod_{j=2}^{N-1} q(\sigma_j) \\
& + \xi \int_{\mathcal{C}_-} d\sigma_1 \left[\prod_{j=2}^{N-1} d\sigma_j \right]
\dfrac{q^{(-)}_0(\sigma_1)}{\prod_{m=1}^N e^{-\frac{\pi \sigma_1}{\hbar}} \sinh\left( \pi \frac{\sigma_1 - a_m}{\hbar} \right)} 
\mu(\vec{\sigma}) \Psi'_{\vec{\sigma}}(\vec{x},\hbar)  \prod_{j=2}^{N-1} q(\sigma_j),
\end{split}
\end{equation}
where the contour $\mathcal{C}_+$ contains the poles $a_m + i\hbar n_m$ for $n_m \geqslant 0$, $m = 1, \ldots, N$ while $\mathcal{C}_-$ contains the poles $a_m + i\hbar n_m$ for $n_m < 0$. The two integrals can be computed by residues, and by using \eqref{equality} the result can be expressed in terms of the function $q_0^{(+)}$ only. 
Integration over other variables is performed along the same lines; the final result has the form of an infinite linear combination of $(N-1)$-particle \textit{open} Toda chain eigenfunctions:
\begin{equation}
\Psi_{\vec{a}}(\vec{x},\hbar,Q) \;\propto\; \sum_{m=1}^N (-1)^{N-m} \sum_{\vec{n}^{(m)} \in \mathbb{Z}^{N-1}} \Delta(\vec{a}^{(m)} + i \hbar \vec{n}^{(m)}) \Psi'_{\vec{a}^{(m)} + i \hbar \vec{n}^{(m)}}(\vec{x},\hbar) \prod_{j=1}^{N-1} q_0^{(+)}(\vec{a}^{(m)} + i \hbar \vec{n}^{(m)}). \label{rompi}
\end{equation}
Here $\Delta(\vec{\sigma}) = \prod_{j<k} (\sigma_j - \sigma_k)$ is the usual Vandermonde determinant and we introduced the vectors $\vec{a}^{(m)} = (a_1, \ldots, a_{m-1}, a_{m+1}, \ldots, a_N)$ and $\vec{n}^{(m)} = (n_1, \ldots, n_{m-1}, n_{m+1}, \ldots, n_N)$.
More precise formulae and additional technical details on the computation and properties of the closed Toda chain eigenfunctions in the Separation of Variables formalism can be found in \cite{Kharchev:1999bh,Kharchev:2000yj}.

\subsection{Solution via gauge theory} \label{sec2.3}

As mentioned in the Introduction, starting with the seminal work \cite{2010maph.conf..265N} it has been gradually understood that supersymmetric four-dimensional $\mathcal{N} = 2$ gauge theories in flat space in the presence of non-trivial Omega background parameters $\epsilon_1$, $\epsilon_2$ can provide an alternative framework for solving stationary quantum mechanical problems with discrete spectrum such as Toda chains. More precisely, such quantum mechanical problems are related to supersymmetric gauge theories when we consider the so-called NS limit $\epsilon_1 = i \hbar$, $\epsilon_2 = 0$: in this limit the second Omega background parameter is sent to zero, while the first one plays the role of the Planck constant (here and in the following we will be assuming $\hbar \in \mathbb{R}_+$). 

Different quantum mechanical systems are associated to different gauge theories: for example, for $N$-particle Toda chains the relevant gauge theory is pure $\mathcal{N}=2$ $SU(N)$ Yang-Mills, where the instanton counting parameter $Q_{\text{4d}} = e^{-8\pi^2/g^2_{\text{4d}}}$ is identified with the parameter $Q$ introduced in \eqref{bc} (open Toda chains therefore appear when the Yang-Mills coupling constant $g_{\text{4d}}$ is sent to zero). 
A partial solution (i.e. quantization conditions and spectrum) to the Toda chain in terms of gauge theory quantities was provided in \cite{2010maph.conf..265N},
while eigenfunctions were later discussed in \cite{Kozlowski:2010tv} where it was also shown that gauge theory results coincide with the known solution we reviewed in Section \ref{sec2.2}.
Here we will recall the main results of \cite{2010maph.conf..265N,Kozlowski:2010tv} for the case of interest, slightly reinterpreting them in a way more suitable for generalizations.

\subsubsection{Quantization conditions and energy spectrum} \label{2.3.1}

Let us start by considering quantization conditions and spectrum. When applied to pure $\mathcal{N}=2$ $SU(N)$ Yang-Mills, the proposal of \cite{2010maph.conf..265N} provides new explicit expressions for quantization conditions and energy spectrum of the $N$-particle closed Toda chain; these may look very different from the ones we already know from Section \ref{sec2.2}, but it is actually possible to show that they are in fact equivalent \cite{Kozlowski:2010tv}. 

More in detail, for what quantization conditions are concerned the idea is to start by considering the pure $\mathcal{N} = 2$ $SU(N)$ gauge theory partition function on $\mathbb{R}^2_{\epsilon_1} \times \mathbb{R}^2_{\epsilon_2}$ \cite{2002hep.th....6161N,2003hep.th....6238N}

\begin{equation}
Z_{\text{4d}}(\vec{a}, \epsilon_1, \epsilon_2, Q_{\text{4d}}) = 
Z_{\text{4d}}^{\text{pert}}(\vec{a}, \epsilon_1, \epsilon_2, Q_{\text{4d}}) 
Z_{\text{4d}}^{\text{inst}}(\vec{a}, \epsilon_1, \epsilon_2, Q_{\text{4d}}). \label{Z4d}
\end{equation}
Here we divided the partition function into its perturbative part (classical + 1-loop) and instanton part, while $\vec{a} = (a_1, \ldots, a_N)$ are the vacuum expectation values of the adjoint scalar field $\phi$ in the $\mathcal{N} = 2$ vector multiplet; on the Toda system side, these will be identified with the auxiliary parameters $\vec{a}$ we introduced in \eqref{tpol} (i.e. with the zeroes of the Hill determinant \eqref{hilldet}). The NS limit of \eqref{Z4d} is divergent, but the \textit{effective twisted superpotential} $\mathcal{W}_{\text{4d}}(\vec{a}, \epsilon_1, Q_{\text{4d}})$ defined as
\begin{equation}
\dfrac{1}{\epsilon_1} \mathcal{W}_{\text{4d}}(\vec{a}, \epsilon_1, Q_{\text{4d}}) = \lim_{\epsilon_2 \rightarrow 0}\left[- \epsilon_2 \log Z_{\text{4d}}(\vec{a}, \epsilon_1, \epsilon_2, Q_{\text{4d}})\right] 
\end{equation}
is instead finite; this can also be separated into its perturbative and instanton part as
\begin{equation}
\mathcal{W}_{\text{4d}}(\vec{a}, \epsilon_1, Q_{\text{4d}}) = 
\mathcal{W}_{\text{4d}}^{\text{pert}}(\vec{a}, \epsilon_1, Q_{\text{4d}}) +
\mathcal{W}_{\text{4d}}^{\text{inst}}(\vec{a}, \epsilon_1, Q_{\text{4d}}). \label{twes4d}
\end{equation}
Replacing $\epsilon_1 = i\hbar$, the proposal by \cite{2010maph.conf..265N} states that the quantization conditions for the Toda chain are equivalent to the set of supersymmetric vacua equations
\begin{equation}
\exp\left( -\dfrac{1}{i \hbar} \dfrac{\partial}{\partial a_m}  \mathcal{W}_{\text{4d}}(\vec{a}, i\hbar, Q_{\text{4d}}) \right) = 1 \,, \;\;\;\; m = 1, \ldots, N,
\end{equation}
or alternatively
\begin{equation}
\dfrac{\partial}{\partial a_m} \mathcal{W}_{\text{4d}}(\vec{a}, i\hbar, Q_{\text{4d}}) = - 2\pi \hbar n_m \,,\;\;\;\; n_m \in \mathbb{Z} , \;\;\;\; m = 1, \ldots, N. \label{quant4d}
\end{equation}
The equivalence between gauge theory quantization conditions \eqref{quant4d} and Gutzwiller's quantization conditions \eqref{quantcond} was later shown to be true in \cite{Kozlowski:2010tv}. 

Moving to the spectrum of the Toda system, \cite{2010maph.conf..265N} suggests that the energies $E_m$ should correspond to the NS limit of the vacuum expectation value of gauge-invariant operators built out of $\phi$, that is
\begin{equation}
E_m^{(\vec{n})} = \dfrac{1}{m!} \langle \text{Tr}\, \phi^m \rangle_{\text{NS}}^{(\vec{n})}, \label{gaugeinv4d}
\end{equation}
evaluated at those particular values of the $a_m$ parameters which solve the quantization conditions \eqref{quant4d} for a given set of integers $(\vec{n})$. The energies $E_1$, $E_2$ of the first two Toda Hamiltonians are particulary simple to evaluate; in fact $E_1$ is just the total momentum
\begin{equation}
E_1(\vec{a}) = \sum_{m=1}^N a_m
\end{equation}
(independent of $Q_{\text{4d}}$) which is often set to zero, while $E_2$ can be obtained from $\mathcal{W}_{\text{4d}}(\vec{a}, i \hbar, Q_{\text{4d}})$ via the Matone relation
\begin{equation}
E_2(\vec{a}, \hbar, Q_{\text{4d}})  = Q_{\text{4d}} \dfrac{d}{dQ_{\text{4d}}} \mathcal{W}_{\text{4d}}(\vec{a}, i \hbar, Q_{\text{4d}}). \label{matone}
\end{equation}
The proposal of \cite{2010maph.conf..265N} therefore provides a detailed prescription on how to compute the energy spectrum of the closed Toda chain by means of gauge theory quantities; as discussed in \cite{Kozlowski:2010tv}, this prescription is equivalent to Gutzwiller's one. Conjecturally the gauge theory prescription should also work for other quantum mechanical system, which will be associated to different gauge theories. 
\\

\renewcommand\arraystretch{1.5}
\begin{table}[]
\begin{center}
\begin{tabular}{cccc}
\hline 
\rule[-1ex]{0pt}{2.5ex} & $E_2^{(0)}$ & $E_2^{(1)}$ & $E_2^{(2)}$ \\ 
\hline 
\rule[-1ex]{0pt}{2.5ex} $Q = 1$, $\hbar = 1.5$ & 3.63012093325678\ldots & 7.11913797277331\ldots & 11.03476737232167\ldots \\ 
\rule[-1ex]{0pt}{2.5ex} $Q = \frac{1}{\sqrt{2}}$, $\hbar = \sqrt{3}$ & 3.44076329369006\ldots & 7.25213834512333 \ldots & 11.60628188091683 \ldots \\ 
\hline 
\end{tabular} \caption{Numerical eigenvalues $E_2^{(n)}$ of $\widehat{H}'_2$ at level $n = 0,1,2$ for different $Q$, $\hbar$.} \label{numenergy4d}
\end{center} 
\end{table}
\renewcommand\arraystretch{1}

\subsubsection{Numerical study} \label{2.3.2}

\begin{figure}[]
  \centering
  \subfloat[Level $n = 0, 1, 2$ closed 2-particle Toda numerical eigenfunctions for $Q = 1$, $\hbar = 1.5$.]{\includegraphics[width=0.9\textwidth]{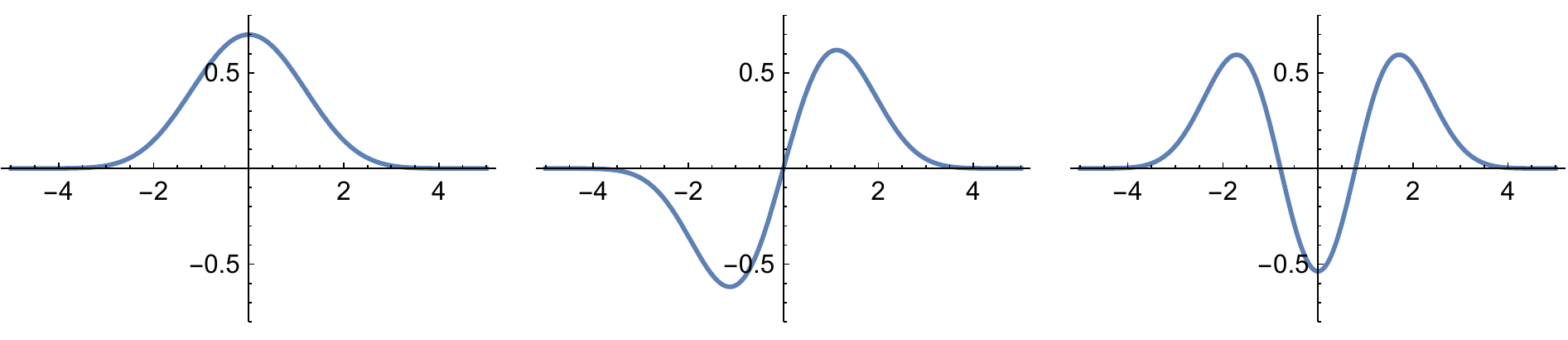}\label{fig:numeig4d}}
\newline
  \subfloat[Level $n = 0, 1, 2$ closed 2-particle Toda numerical eigenfunctions for $Q = \frac{1}{\sqrt{2}}$, $\hbar = \sqrt{3}$.]{\includegraphics[width=0.9\textwidth]{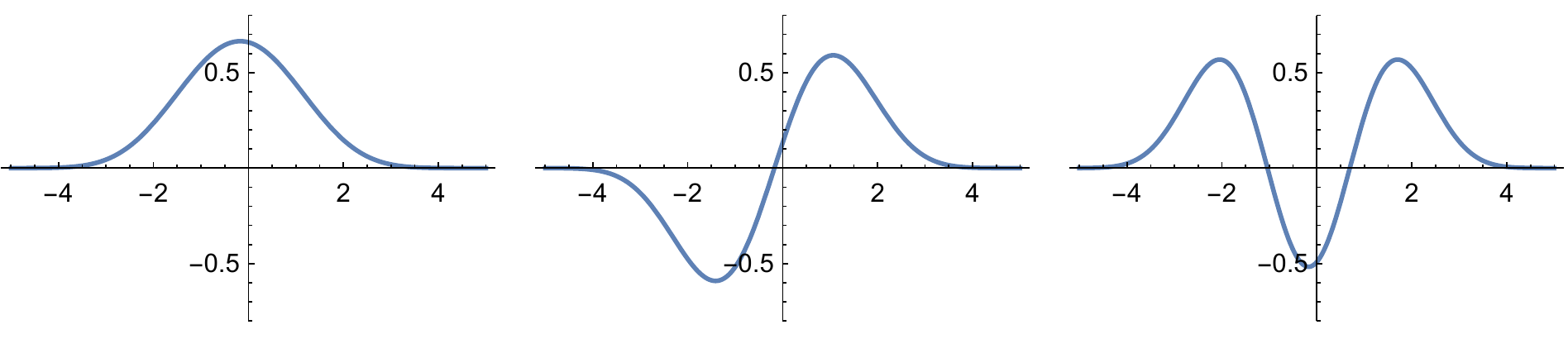}\label{numeig4dmod}}
 \caption{2-particle closed Toda numerical eigenfunctions.}
 \label{firstfigure}
\end{figure}

The gauge theory solution proposed by \cite{2010maph.conf..265N} can also be explicitly tested against numerical results, as done for example in \cite{2015arXiv151102860H} for the 2-particle Toda chain. In order to give a better idea on how the prescription of \cite{2010maph.conf..265N} works in practice, and also because we will need to do something similar later in Section \ref{sec3}, let us quickly review the numerical tests performed by \cite{2015arXiv151102860H}. Let us consider the 2-particle closed Toda chain and decouple the center of mass for simplicity; then $E_1 = 0$, that is $a_1 = - a_2 = a$. We can use the Hamiltonian $\widehat{H}'_2$ of \eqref{hamsprime} to define the quantum mechanical problem; when the center of mass is decoupled this reduces to
\begin{equation}
\widehat{H}'_2 \;=\; p^2 + e^x + Q e^{-x} \label{hamex}
\end{equation}
with $[p,x] = -i\hbar$. This is a problem defined on $x \in \mathbb{R}$, and given the form of the potential we expect a discrete energy spectrum and $L^2(\mathbb{R})$-normalizable eigenfunctions; we can then try to diagonalize \eqref{hamex} in terms of an orthonormal basis on $L^2(\mathbb{R})$ such as the harmonic oscillator one, given by
\begin{equation}
\psi_k(x) = \dfrac{1}{\sqrt{2^k k!}} \left( \dfrac{m \omega}{\pi \hbar} \right)^{\frac{1}{4}} e^{-\frac{m \omega x^2}{2\hbar}} H_k \left( \sqrt{\frac{m \omega}{\hbar}} x \right) \,, \;\;\;\; k \geqslant 0 \label{harmonicbasis}
\end{equation}
for a harmonic oscillator of mass $m$ and frequency $\omega$\footnote{Here we are considering a harmonic oscillator with Hamiltonian $\widehat{H} = \frac{p^2}{2m} + \frac{1}{2}m \omega^2 x^2$.}, where $H_k(x)$ are Hermite polynomials. Although for  diagonalization we should in principle consider an $\infty \times \infty$ matrix, from a practical point of view one computes the matrix elements $\langle \psi_{k_1} \vert \widehat{H}'_2  \vert \psi_{k_2} \rangle$ up to a certain order and  evaluates numerically the eigenvalues of this finite matrix; the parameters $m$, $\omega$ can be fixed by expanding \eqref{hamex} at small $x$. The expectation is that the numerical eigenvalues thus obtained should approach the exact ones by increasing the size of the matrix. In the same way, one can construct numerical eigenfunctions (normalized in such a way to have norm 1) by considering eigenvectors of this finite matrix. Examples of numerical results for the energy of the ground state and the first two excited states at various values of $Q$ and $\hbar$ can be found in Table \ref{numenergy4d}, while Figure \ref{firstfigure} shows the corresponding eigenfunctions (symmetric with respect to $x = \log Q^{1/2}$); for numerical computations we considered 400 $\times$ 400 matrices.

Let us now try to reproduce the numerical results in Table \ref{numenergy4d} from the proposal by \cite{2010maph.conf..265N}. In order to do this we first need the twisted effective superpotential \eqref{twes4d}. This can be easily computed from the formulas collected in Appendix \ref{formula4d}; we find that the instanton part is given by a series expansion in $Q_{\text{4d}}$ starting as
\begin{equation}
\mathcal{W}^{\text{inst}}_{\text{4d}}(a, i\hbar, Q_{\text{4d}})  = Q_{\text{4d}} \frac{2}{4a^2+\hbar ^2} + Q_{\text{4d}}^2 \frac{20 a^2 - 7 \hbar ^2}{4 \left(4a^2+\hbar ^2\right){}^3 \left(a^2+ \hbar ^2\right)} + o(Q_{\text{4d}}^3),
\end{equation}
while the $a$-derivative of the perturbative part reads
\begin{equation}
\dfrac{\partial}{\partial a} \mathcal{W}^{\text{pert}}_{\text{4d}}(a, i\hbar, Q_{\text{4d}}) 
= 4a \log \left( \dfrac{Q_{\text{4d}}^{1/4}}{\hbar} \right)^2 + 2i\hbar \log \dfrac{\Gamma(1 + \frac{2ia}{\hbar})}{\Gamma(1 - \frac{2ia}{\hbar})} + 2 \pi \hbar.
\end{equation}
We can then look for solutions to the quantization condition \eqref{quant4d}, that is
\begin{equation}
\dfrac{\partial}{\partial a} \left[ \mathcal{W}^{\text{pert}}_{\text{4d}}(a, i\hbar, Q_{\text{4d}}) + \mathcal{W}^{\text{inst}}_{\text{4d}}(a, i\hbar, Q_{\text{4d}}) \right] = - 2\pi \hbar n , \;\;\;\; n \in \mathbb{N}. \label{museum}
\end{equation}
At fixed $Q_{\text{4d}}$, $\hbar$ and energy level $n \geqslant 0$ this equation will be solved for a particular positive real value of $a$ which we call $a^{(n)}$; for the examples considered in Table \ref{numenergy4d}, a 12-instanton computation produces the values listed in Table \ref{a4d}. Having determined the $a^{(n)}$, all we have to do is to substitute them into the expression for the energy \eqref{matone}; this will again be a series in $Q_{\text{4d}}$ starting as\footnote{The parameters $\tau_1$, $\tau_2$ entering in \eqref{tpol} in this case are $\tau_1= -\tau_2 = \sqrt{E_2}$.}
\begin{equation}
E_2(a, \hbar, Q_{\text{4d}}) = a^2 + Q_{\text{4d}} \frac{2}{4a^2+\hbar ^2} + Q_{\text{4d}}^2 \frac{20 a^2 - 7 \hbar ^2}{2 \left(4a^2+\hbar ^2\right){}^3 \left(a^2+ \hbar ^2\right)} + o(Q_{\text{4d}}^3). \label{E2}
\end{equation}
Adding more instanton corrections to \eqref{E2} the gauge theory results approach the numerical ones better and better, and already a 12-instanton computation reproduces the numerical results of Table \ref{numenergy4d}, thus providing some evidence for the validity of the proposal by \cite{2010maph.conf..265N}. \\





\renewcommand\arraystretch{1.5}
\begin{table}[h]
\begin{center}
\begin{tabular}{cccc}
\hline 
\rule[-1ex]{0pt}{2.5ex} & $a^{(0)}$ & $a^{(1)}$ & $a^{(2)}$ \\ 
\hline 
\rule[-1ex]{0pt}{2.5ex} $Q_{\text{4d}} = 1$, $\hbar = 1.5$ & 1.87246705538171 \ldots & 2.65579461654756 \ldots & 3.31533009851180 \ldots \\ 
\rule[-1ex]{0pt}{2.5ex} $Q_{\text{4d}} = \frac{1}{\sqrt{2}}$, $\hbar = \sqrt{3}$ & 1.83145069816166 \ldots & 2.68470017820023 \ldots & 3.40258311045043 \ldots \\ 
\hline 
\end{tabular} \caption{Quantized values of $a^{(n)}$ at level $n = 0,1,2$ for different $Q_{\text{4d}}$, $\hbar$.} \label{a4d}
\end{center}
\end{table}
\renewcommand\arraystretch{1}

\subsubsection{Eigenfunctions} \label{2.3.3}

\begin{figure}[]
  \centering
  \subfloat[]{\includegraphics[width=0.9\textwidth]{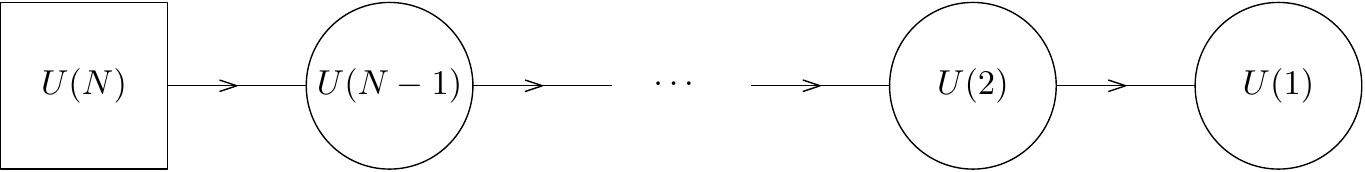}\label{TSUN}}
\newline
  \subfloat[]{\includegraphics[width=0.107 \textwidth]{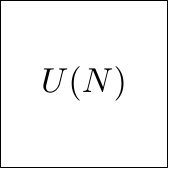}\label{freechiral}}
  \hspace{3 cm}
  \subfloat[]{\includegraphics[width=0.3\textwidth]{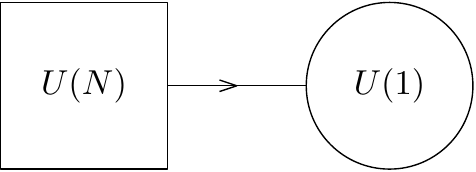} \label{chiralgauge}}
 \caption{Two-dimensional $\mathcal{N} = (2,2)$ defects of type I (a), II (b) and III (c).}
 \label{secondfigure}
\end{figure}

Having discussed quantization conditions and energy spectrum, it now remains to understand how eigenfunctions of the Toda chain can be realized in gauge theory. Although this point was not discussed in \cite{2010maph.conf..265N}, subsequent works \cite{Alday:2010vg,Kanno:2011fw,Gaiotto:2014ina} and \cite{Kozlowski:2010tv} showed that these eigenfunctions should correspond to the partition function of four-dimensional pure $\mathcal{N} = 2$ $SU(N)$ Yang-Mills 
in the presence of particular codimension two defects,
again in the NS limit $\epsilon_2 \rightarrow 0$, $\epsilon_1 \rightarrow i \hbar$ (from this point of view, local operators \eqref{gaugeinv4d} associated to the eigenvalues can be thought as codimension four defects). It turns out that there are many different codimension two defects one can consider, and most of them can be realized in different ways; here we will only discuss defects which admit a realization as two-dimensional $\mathcal{N} = (2,2)$ theories coupled to our four-dimensional $\mathcal{N} = 2$ $SU(N)$ Yang-Mills theory. Three two-dimensional  $\mathcal{N} = (2,2)$ theories are of particular interest to us:
\begin{enumerate}
\item[I)] The quiver theory of Figure \ref{TSUN}, associated to the common eigenfunctions $\Psi_{\vec{a}}(\vec{x},\hbar,Q)$ of the Toda chain spectral problem \eqref{sptoda}
\begin{equation}
\widehat{H}_m \Psi_{\vec{a}}(\vec{x}, \hbar, Q) = E_m(\vec{a}, \hbar, Q) \Psi_{\vec{a}}(\vec{x}, \hbar, Q), \;\;\;\; m = 1, \ldots, N,
\end{equation}
which can equivalently be formulated in terms of the Hamiltonians $\widehat{H}'_m$ in \eqref{hamsprime};
\item[II)] The theory of $N$ free chiral (or antichiral) multiplets of Figure \ref{freechiral}, associated to the eigenfunction $q(\sigma)$ of the Baxter equation \eqref{baxclosed}
\begin{equation}
(i)^N q(\sigma + i \hbar) + Q (i)^{-N} q(\sigma - i \hbar) = t(\sigma) q(\sigma),
\end{equation}
which as we already mentioned can be thought of as a quantized version of the classical spectral curve \eqref{spcurve4d} if $\sigma$ is chosen as coordinate and $y = (i)^N e^{-p_{\sigma}}$ as conjugate momentum, that is
\begin{equation}
(i)^N e^{-p_{\sigma}} + Q (i)^{-N} e^{p_{\sigma}} = t(\sigma) ; \label{temp}
\end{equation}
\item[III)] The theory of $N$ chiral (or antichiral) multiplets coupled to a $U(1)$ gauge multiplet of Figure \ref{chiralgauge}, associated to the function $q_{\text{FT}}(\sigma)$ which can be roughly though of as a ``Fourier-transformed'' version of $q(\sigma)$; that is, $q_{\text{FT}}(\sigma)$ will be an eigenfunction of the quantized version of the curve
\begin{equation}
(i)^N e^{\sigma} + Q (i)^{-N} e^{-\sigma} = t(p_{\sigma}),
\end{equation}
which is obtained from the spectral curve \eqref{temp} via a canonical change of variables $\sigma \rightarrow p_{\sigma}$, $p_{\sigma} \rightarrow -\sigma$. For the special case $N = 2$, this coincides with the type I defect.
\end{enumerate}
In our following discussion we will mostly focus on defects of type I and II living on a disc $D_{\epsilon_1}$ of radius $\hbar^{-1} = i \epsilon_1^{-1}$ coupled to our four-dimensional theory on $D_{\epsilon_1} \times \mathbb{R}^2_{\epsilon_2}$, and review how the NS limit of the partition function in the presence of these defects provide expressions for the Baxter and Toda eigenfunctions equivalent to the ones discussed in Section \ref{sec2.2}.\footnote{Considering all possible vacua in $\mathbb{R}^2_{\epsilon_1} \times \mathbb{R}^2_{\epsilon_2}$ is equivalent to considering $D_{\epsilon_1} \times \mathbb{R}^2_{\epsilon_2}$, while a single vacuum in $\mathbb{R}^2_{\epsilon_1} \times \mathbb{R}^2_{\epsilon_2}$  is only a formal eigenfunction since it does not satisfy the correct asymptotic behaviour (see comments near \eqref{2todaopen})\label{fn1}.} \\

\noindent \textit{Open Toda chains} \\

\noindent Let us proceed step by step and start by discussing the eigenfunctions for the open Toda chain and its Baxter equation. Since this case corresponds to $Q = Q_{\text{4d}} = 0$, on the gauge theory side it means that we are decoupling the four-dimensional gauge interaction, so that we only remain with the two-dimensional theory on the defect. The partition function on the disc for type II defects, that is $N$ free chiral/antichiral multiplets, is simply \cite{Hori:2013ika,Honda:2013uca,Sugishita:2013jca}
\begin{subequations} \label{pappa}
\begin{align}
q^{(\text{c})}(\sigma, \vec{a}, \hbar) &= \prod_{m=1}^N \hbar^{- i \frac{\sigma - a_m}{\hbar}} \Gamma\left(-i\frac{\sigma - a_m}{\hbar}\right) \;\;\;\;\;\;\;\; (\text{chiral}), \label{pappaA} \\
q^{(\text{ac})}(\sigma, \vec{a}, \hbar) &= \prod_{m=1}^N \hbar^{i \frac{\sigma - a_m}{\hbar}} \Gamma\left(i\frac{\sigma - a_m}{\hbar}\right) \;\;\;\;\;\;\;\; (\text{antichiral}),  \label{pappaB}
\end{align}
\end{subequations}
where $\hbar$ is interpreted as the inverse radius of the disc while $\sigma$ and $a_m$ are twisted masses associated to the Cartan of the flavour symmetry $U(1) \times U(N)$; going to $SU(N)$ by decoupling an overall $U(1)$ factor sets the constraint $\sum_{m=1}^N a_m = 0$ and corresponds to decoupling the center of mass from the point of view of the open Toda chain. As we can notice, \eqref{pappaA} is nothing else than the separated variables wavefunction $q(\sigma, \vec{a}, \hbar)$ which appeared in \eqref{Qtildeopen} and therefore formally satisfies the open Toda Baxter equation
\begin{equation}
\begin{split}
(i)^N q^{(\text{c})}(\sigma + i \hbar, \vec{a}, \hbar) \;=\; t(\sigma, \vec{E}(\vec{a})) \,q^{(\text{c})}(\sigma, \vec{a}, \hbar);
\end{split} 
\end{equation}
similarly, \eqref{pappaB} formally satisfies
\begin{equation}
\begin{split}
(i)^{-N} q^{(\text{ac})}(\sigma - i \hbar, \vec{a}, \hbar) \;=\; t(\sigma, \vec{E}(\vec{a})) \,q^{(\text{ac})}(\sigma, \vec{a}, \hbar).
\end{split} 
\end{equation}
As a comment let us remark that, as in Section \ref{sec2.2}, these are just formal solutions since they have poles; proper solutions $q^{(\text{c})}_0(\sigma, \vec{a}, \hbar)$, $q^{(\text{ac})}_0(\sigma, \vec{a}, \hbar)$ can be obtained by removing an appropriate $i\hbar$-periodic factor according to (decoupling the center of mass)
\begin{equation}
\begin{split}
q^{(\text{c})}(\sigma, \vec{a}, \hbar) = \dfrac{q^{(\text{c})}_0(\sigma, \vec{a}, \hbar)}{\prod_{m=1}^N e^{-\frac{\pi \sigma}{\hbar}} \sinh \left( \pi\frac{\sigma - a_m}{\hbar} \right)}, \\
q^{(\text{ac})}(\sigma, \vec{a}, \hbar) = \dfrac{q^{(\text{ac})}_0(\sigma, \vec{a}, \hbar)}{\prod_{m=1}^N e^{-\frac{\pi \sigma}{\hbar}} \sinh \left( \pi\frac{\sigma - a_m}{\hbar} \right)}. \label{cacchio1}
\end{split}
\end{equation}
For example, using the properties of the Gamma function we obtain
\begin{equation}
q^{(\text{c})}_0(\sigma, \vec{a}, \hbar) \;\;\propto \;\; \prod_{m=1}^N \dfrac{e^{-\frac{\pi \sigma}{\hbar}}}{\hbar^{\frac{i\sigma}{\hbar}}\Gamma \left(1 +  i\frac{\sigma - a_m}{\hbar} \right)}, \label{cacchio2}
\end{equation}
which is free of poles being an entire function.

Moving to type I defects, their disc partition function can also be evaluated explicitly and for the quiver theory of Figure \ref{TSUN} consisting of chiral multiplets it is given by \cite{Hori:2013ika,Honda:2013uca,Sugishita:2013jca}
\begin{equation}
\begin{split}
& \Psi^{(\text{c})}_{a_1, \ldots, a_N}(x_1, \ldots, x_N, \hbar) = \\
& = e^{i \frac{x_N}{\hbar}\sum_{m=1}^N a_m} \int_{\mathcal{C}} \prod_{s=1}^{N-1} \prod_{j=1}^s d\sigma^{(s)}_j
\prod_{s=1}^{N-1}\prod_{j<k}^s(\sigma_j^{(s)} - \sigma_k^{(s)}) \sinh \left[ \pi \frac{\sigma_j^{(s)} - \sigma_k^{(s)}}{\hbar} \right] \\
& \;\;\;\;\;\;
\prod_{s=1}^{N-1}\prod_{j=1}^{s}\prod_{k=1}^{s+1} \hbar^{-i\frac{\sigma_j^{(s)} - \sigma_k^{(s+1)}}{\hbar}} \Gamma\left( -i \frac{\sigma_j^{(s)} - \sigma_k^{(s+1)}}{\hbar} \right)
\prod_{s=1}^{N-1} e^{i\frac{x_s - x_{s+1}}{\hbar}\sum_{j=1}^s \sigma_j^{(s)}}. \label{thisopen}
\end{split}
\end{equation}
A similar expression can be obtained considering antichiral multiplets instead of chiral ones. Here $\sigma^{(s)}_j$ are vevs of vectormultiplet scalars of the $U(s)$ gauge group ($s = 1, \ldots, N-1$), while $\sigma^{(N)}_m = a_m$ are the twisted masses of the $U(N)$ flavour group and $x_s - x_{s+1}$ is the Fayet-Iliopoulos parameter for the $s$-th gauge group. The integration contour $\mathcal{C}$ goes along the lines Im$\,\sigma^{(s)}_j >$ max$_k \{$Im$\,\sigma^{(s+1)}_k \}$. It is easy to see that \eqref{thisopen} can be re-expressed in a recursive form equivalent to \eqref{eigopen}; recursion in gauge theory language simply means that the length $N$ quiver theory can be obtained from the length $N-1$ one by gauging the $U(N-1)$ flavour symmetry and coupling it to a set of $N$ chiral multiplets. 

\begin{figure}[]
  \centering
\includegraphics[width=0.9\textwidth]{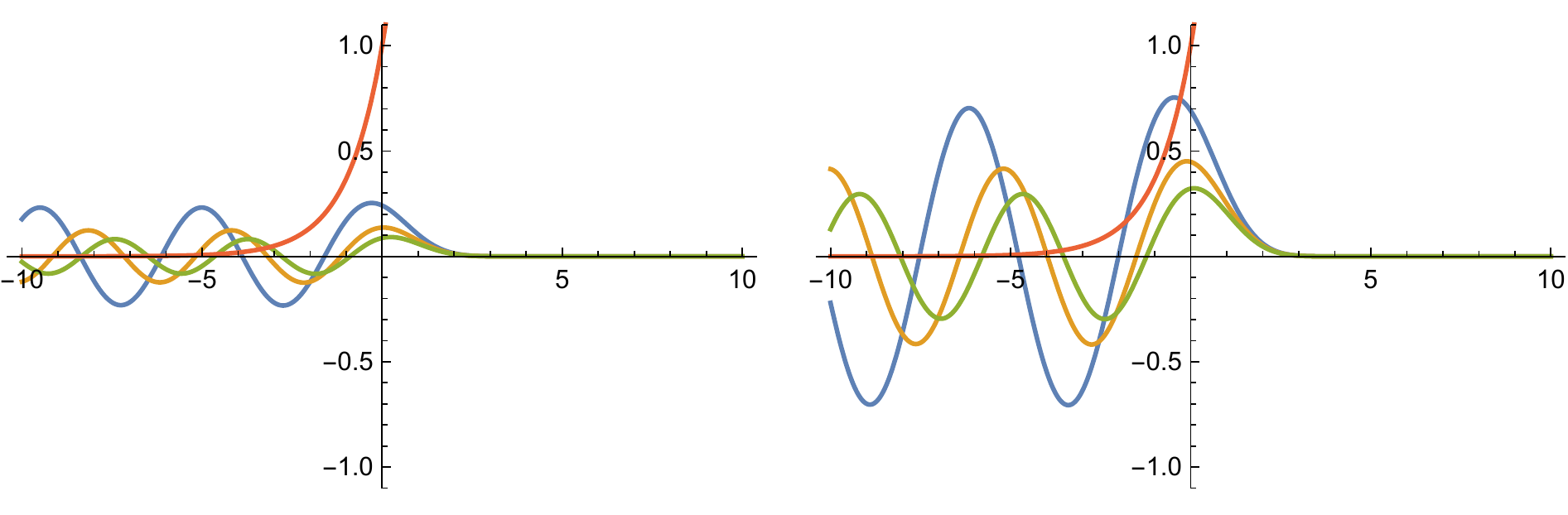}
  \caption{2-particle open Toda eigenfunction $\Psi^{(\text{c})}_a(x, \hbar)$ at $\hbar = 1$ (left), $\hbar = \sqrt{1.5}$ (right) and fixed $a = 1.4, \sqrt{2.5}, 1.702$ (blue, orange, green); the potential $e^x$ is in red.}
  \label{besselk}
\end{figure}

We therefore conclude that, at least as far as open Toda chain are concerned, gauge theory and Separation of Variables give the same result; then by \cite{Kharchev:1999bh,Kharchev:2000yj} we know that \eqref{thisopen} is the correct eigenfunction of the open Toda chain with the proper asymptotic behaviour. To consider an example, the eigenfunction of the 2-particle open Toda system (with center of mass decoupled)
\begin{equation}
\Psi^{(\text{c})}_a(x, \hbar) = \int_{-\infty}^{\infty} \hbar^{-\frac{2 i \sigma}{\hbar}} \Gamma\left(-i\frac{\sigma - a}{\hbar}\right) \Gamma\left(-i\frac{\sigma + a}{\hbar}\right) e^{\frac{i \sigma x}{\hbar}} d\sigma = 4\pi \hbar \, K_{\frac{2ia}{\hbar}}\left(\text{\footnotesize{$\frac{2}{\hbar}$}} e^{\frac{x}{2}}\right) \label{katekyo}
\end{equation}
satisfying
\begin{equation}
\left( -\hbar^2 \partial_x^2 + e^{x} \right) \Psi^{(\text{c})}_a(x, \hbar)  = E \Psi^{(\text{c})}_a(x, \hbar)  = a^2 \Psi^{(\text{c})}_a(x, \hbar) \label{bababa}
\end{equation}
decreases very fast at $x \rightarrow \infty$ while it oscillates at $x<0$ as we can see from Figure \ref{besselk}. Considering antichiral multiplets instead of chiral, we would have obtained 
\begin{equation}
\Psi_a^{(\text{ac})}(x, \hbar) = \int_{-\infty}^{\infty} \hbar^{\frac{2 i \sigma}{\hbar}} \Gamma\left(i\frac{\sigma - a}{\hbar}\right) \Gamma\left(i\frac{\sigma + a}{\hbar}\right) e^{\frac{i \sigma x}{\hbar}} d\sigma = 4\pi \hbar \, K_{\frac{2ia}{\hbar}}\left(\text{\footnotesize{$\frac{2}{\hbar}$}} e^{-\frac{x}{2}}\right),
\end{equation}
which is an eigenfunction of
\begin{equation}
\left(-\hbar^2 \partial_x^2 + e^{-x} \right) \Psi^{(\text{ac})}_a(x, \hbar) = E \, \Psi^{(\text{ac})}_a(x, \hbar) = a^2 \Psi^{(\text{ac})}_a(x, \hbar) 
\end{equation}
with behaviour at $x = \pm \infty$ opposite to the one of \eqref{katekyo}. Let us remark that $\Psi^{(\text{c})}_{\vec{a}}(\vec{x}, \hbar)$ (respectively $\Psi^{(\text{ac})}_{\vec{a}}(\vec{x}, \hbar)$) are related to $q^{(\text{c})}(\sigma, \vec{a}, \hbar)$ (or $q^{(\text{ac})}(\sigma, \vec{a}, \hbar)$) in \eqref{pappa} by Separation of Variables as discussed in Section \ref{sec2.2}. Notice also that \eqref{katekyo} (and similarly its antichiral analogue), when interpreted as a partition function on the disc $D_{\epsilon_1}$, 
is equivalent to the sum over all vacua of the vortex partition function (i.e. partition function on $\mathbb{R}^2_{\epsilon_1}$) of the two-dimensional theory as
\begin{equation}
\begin{split}
\Psi^{(\text{c})}_a(x, \hbar) &\;=\; 2\pi \hbar \left( \dfrac{e^{x/2}}{\hbar}\right)^{\frac{2ia}{\hbar}} \Gamma\left( -\frac{2 i a}{\hbar} \right) \sum_{n \geqslant 0} \dfrac{1}{n!(1+\frac{2ia}{\hbar})_n} \left( \dfrac{e^{x}}{\hbar^2}\right)^{n} \\
& \;+\; 2\pi \hbar \left( \dfrac{e^{x/2}}{\hbar}\right)^{-\frac{2ia}{\hbar}} \Gamma\left( \frac{2 i a}{\hbar} \right) \sum_{n \geqslant 0} \dfrac{1}{n!(1-\frac{2ia}{\hbar})_n} \left( \dfrac{e^{x}}{\hbar^2}\right)^{n}, \label{2todaopen}
\end{split}
\end{equation}
as can be seen by performing the integration; both terms in this sum formally satisfy \eqref{bababa}, but separately they do not have the correct asymptotic behaviour and only their combination is a proper eigenfunction (see also footnote \ref{fn1}).
\\

\noindent \textit{Closed Toda chains} \\

\noindent We can now proceed to discuss the eigenfunctions for the closed Toda chain. In this case $Q = Q_{\text{4d}} > 0$; our two-dimensional theories are then coupled to the four-dimensional one and we therefore expect instanton corrections to the previous expressions given by series in powers of $Q_{\text{4d}}$. 

Let us focus here on defects of type II, i.e. $N$ free chiral/antichiral multiplets; the partition function of this 2d-4d coupled system will be
\begin{equation}
\begin{split}
q^{(\text{c})}(\sigma, \vec{a}, \hbar, Q_{\text{4d}}) &= q^{(\text{c})}(\sigma, \vec{a}, \hbar) q_{\text{inst}}^{(\text{c}), \text{NS}}(\sigma, \vec{a}, \hbar, Q_{\text{4d}}),  \\
q^{(\text{ac})}(\sigma, \vec{a}, \hbar, Q_{\text{4d}}) &= Q_{\text{4d}}^{-\frac{i \sigma}{\hbar}} q^{(\text{ac})}(\sigma, \vec{a}, \hbar) q_{\text{inst}}^{(\text{ac}), \text{NS}}(\sigma, \vec{a}, \hbar, Q_{\text{4d}}), 
\end{split} \label{pappaQ}
\end{equation}
where $q^{(\text{c})}(\sigma, \vec{a}, \hbar)$ and $q^{(\text{ac})}(\sigma, \vec{a}, \hbar)$ are the ones in \eqref{pappa}, the factor $Q_{\text{4d}}^{-\frac{i \sigma}{\hbar}}$ is introduced for later convenience, are $q_{\text{inst}}^{(\text{c}), \text{NS}}(\sigma, \vec{a}, \hbar, Q_{\text{4d}})$ and $q_{\text{inst}}^{(\text{ac}), \text{NS}}(\sigma, \vec{a}, \hbar, Q_{\text{4d}})$ are the NS limit of 
the instanton corrections coming from coupling to the four-dimensional gauge theory.
These corrections can be computed by using the contour integral formulae \eqref{baxfirst}, \eqref{baxsecond} in Appendix \ref{formula4d} according to
\begin{equation}
\begin{split}
q_{\text{inst}}^{(\text{c}), \text{NS}}(\sigma, \vec{a}, \hbar, Q_{\text{4d}}) \;&=\; \lim_{\substack{\epsilon_2 \rightarrow 0}} \dfrac{Z_{\text{2d/4d}}^{\text{(c),inst}}(\sigma, \vec{a}, \epsilon_1, \epsilon_2, Q_{\text{4d}})}{Z_{\text{4d}}^{\text{inst}}(\vec{a}, \epsilon_1, \epsilon_2, Q_{\text{4d}})} \Bigg{\vert}_{\epsilon_1 = i \hbar}, \\
q_{\text{inst}}^{(\text{ac}), \text{NS}}(\sigma, \vec{a}, \hbar, Q_{\text{4d}}) \;&=\; \lim_{\substack{\epsilon_2 \rightarrow 0}} \dfrac{Z_{\text{2d/4d}}^{\text{(ac),inst}}(\sigma, \vec{a}, \epsilon_1, \epsilon_2, Q_{\text{4d}})}{Z_{\text{4d}}^{\text{inst}}(\vec{a}, \epsilon_1, \epsilon_2, Q_{\text{4d}})} \Bigg{\vert}_{\epsilon_1 = i \hbar} \\
\;&=\; q_{\text{inst}}^{(\text{c}), \text{NS}}(\sigma, \vec{a}, -\hbar, Q_{\text{4d}}). \label{2d4dinst}
\end{split}
\end{equation} 
Alternatively, they can be computed via the TBA formulae \eqref{A22} in Appendix \ref{formula4d}. 
The functions \eqref{pappaQ} with the instanton corrections \eqref{2d4dinst} were used in \cite{Kozlowski:2010tv} to construct two linearly independent meromorphic solutions to the Baxter equation \eqref{baxclosed} with the correct asymptotic behaviour at $\sigma \rightarrow \pm \infty$, and were also shown to be equivalent to \eqref{tempx} obtained in the context of Separation of Variables. The $i\hbar$-periodic factor added by hand at the denominator of \eqref{tempx}, source of meromorphicity for the function, appears instead naturally from the gauge theory expression \eqref{pappaQ} if we use the properties of the Gamma functions appearing in \eqref{pappa} as done in \eqref{cacchio1}, \eqref{cacchio2}; it is then very natural from the gauge theory point of view to have poles at $\sigma = a_m + i \hbar n_m$. From the short review of Section \ref{sec2.2} we already know that $L^2(\mathbb{R})$ entire eigenfunctions $q(\sigma, \vec{a}, \hbar, Q_{\text{4d}})$ of the Baxter equation
\begin{equation}
(i)^N q(\sigma + i \hbar, \vec{a}, \hbar, Q_{\text{4d}}) + Q_{\text{4d}} (i)^{-N} q(\sigma - i \hbar, \vec{a}, \hbar, Q_{\text{4d}}) = t(\sigma, \vec{E}(\vec{a}, Q_{\text{4d}})) q(\sigma, \vec{a}, \hbar, Q_{\text{4d}})
\end{equation}
will be given by linear combinations 
\begin{equation}
q(\sigma, \vec{a}, \hbar, Q_{\text{4d}}) = q^{(\text{ac})}(\sigma, \vec{a}, \hbar, Q_{\text{4d}})  - \xi q^{(\text{c})}(\sigma, \vec{a}, \hbar, Q_{\text{4d}}) 
\end{equation}
only for particular values of the $a_m$ parameters determined by \eqref{quantcond}, or equivalently by \eqref{quant4d} in gauge theory language \cite{Kozlowski:2010tv}. We have therefore recovered the known solution of the Baxter equation for the closed Toda chain in terms of gauge theory quantities.\footnote{Our gauge theory solution will in general not be normalized to 1 since we were not able to find the correct normalization factor from gauge theory arguments.} \\

\begin{figure}[]
  \centering
\includegraphics[width=1\textwidth]{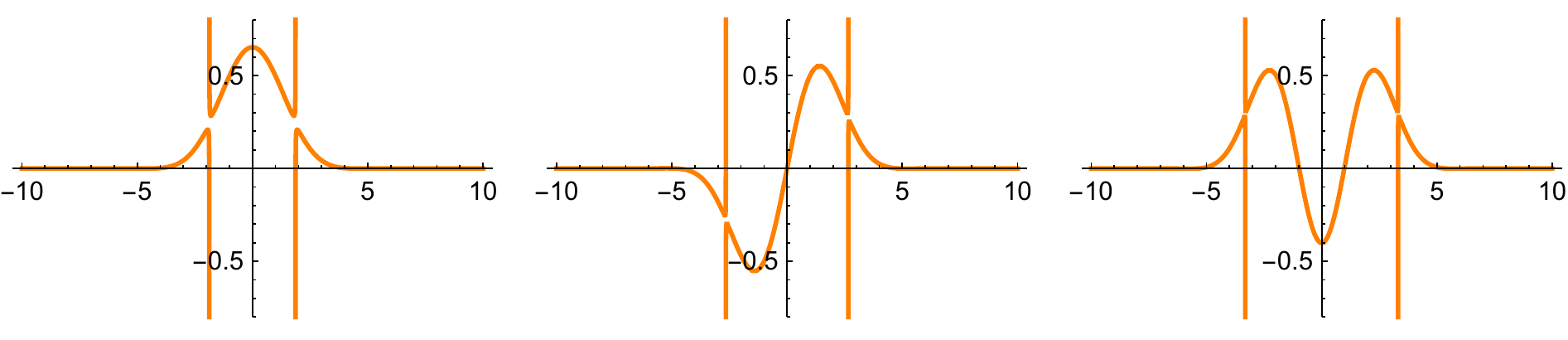}
 \caption{Gauge theory Baxter solution $\overline{q}(\sigma)$ for $Q = 1$, $\hbar = 1.5$ at $a$ not satisfying quantization conditions. Left: $a= 1.87$; center: $a = 2.655$; right: $a = 3.315$. }
  \label{bax4d1wrong}
\end{figure}

Let us study in detail an example and consider the 2-particle closed Toda chain with center of mass decoupled (i.e. $a_1 = -a_2 = a$). In this case the Baxter equation reads
\begin{equation}
- q(\sigma + i \hbar, a, \hbar, Q_{\text{4d}}) - Q_{\text{4d}} q(\sigma - i \hbar, \vec{a}, \hbar, Q_{\text{4d}}) = (\sigma^2 - E_2(a, \hbar, Q_{\text{4d}})) q(\sigma, a, \hbar, Q_{\text{4d}}),
\end{equation}
with $E_2(a, \hbar, Q_{\text{4d}})$ as computed in \eqref{E2}. This Baxter equation can also be written as
\begin{equation}
\left(e^{i\hbar\partial_{\sigma}} + Q_{\text{4d}} e^{-i\hbar\partial_{\sigma}} + \sigma^2\right) q(\sigma, a, \hbar, Q_{\text{4d}}) = E_2(a, \hbar, Q_{\text{4d}}) q(\sigma, a, \hbar, Q_{\text{4d}}) \label{cafe}
\end{equation}
and can be thought as the quantized version of the classical Hamiltonian
\begin{equation}
e^{-p} + Q_{\text{4d}} e^{p} + \sigma^2 = E_2,
\end{equation}
which is the Fourier-transform of the 2-particle Toda Hamiltonian $\widehat{H}_2'$
\begin{equation}
p^2 + e^{\sigma} + Q_{\text{4d}} e^{-\sigma} = E_2.
\end{equation}
This is only true for the $2$-particle Toda chain, since as we mentioned earlier in this case type III defects coincide with type I ones. The (not normalized) solution to the Baxter equation \eqref{cafe} will be a linear combination 
\begin{equation}
q(\sigma, a, \hbar, Q_{\text{4d}}) = q^{(\text{ac})}(\sigma, a, \hbar, Q_{\text{4d}})  - \xi q^{(\text{c})}(\sigma, a, \hbar, Q_{\text{4d}}) \label{tot}
\end{equation}
where, from the formulae in Appendix \ref{formula4d},
\begin{equation}
\begin{split}
q^{(\text{c})}(\sigma, a, \hbar, Q_{\text{4d}}) \;=\; & \hbar^{-\frac{2 i \sigma}{\hbar}} \Gamma\left(-i\frac{\sigma - a}{\hbar}\right) \Gamma\left(-i\frac{\sigma + a}{\hbar}\right) \\
& \left( 1 + Q_{\text{4d}} \dfrac{2\sigma - 3 i \hbar}{i \hbar (4a^2 + \hbar^2)(\sigma - a - i \hbar)(\sigma + a - i \hbar)} + o(Q_{\text{4d}}^2) \right), \\
q^{(\text{ac})}(\sigma, a, \hbar, Q_{\text{4d}})  \;=\; & Q_{\text{4d}}^{-\frac{i \sigma}{\hbar}} \hbar^{\frac{2 i \sigma}{\hbar}} \Gamma\left(i\frac{\sigma - a}{\hbar}\right) \Gamma\left(i\frac{\sigma + a}{\hbar}\right) \\
& \left( 1 - Q_{\text{4d}} \dfrac{2\sigma + 3 i \hbar}{i \hbar (4a^2 + \hbar^2)(\sigma - a + i \hbar)(\sigma + a + i \hbar)}   + o(Q_{\text{4d}}^2) \right).
\end{split}
\end{equation}
This linear combination will be an entire function of $\sigma$ only for $a$ satisfying the quantization condition \eqref{museum} and $\xi$ determined by \eqref{quantcond}; for the values of $a^{(n)}$ in Table \ref{a4d} we find $\xi = -(-1)^n$. On the other hand, singularities do not disappear when $a \neq a^{(n)}$, as shown for example in Figure \ref{bax4d1wrong}.

\begin{figure}[]
\begin{center}
  \subfloat[Numerical results]{\includegraphics[width=1\textwidth]{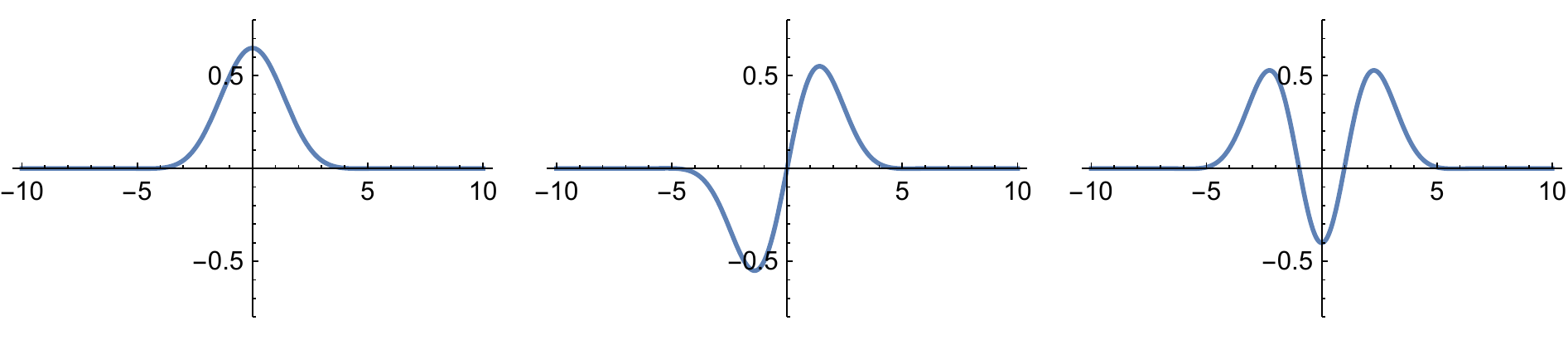}\label{bax4d1num}}
\newline
  \subfloat[Gauge theory results]{\includegraphics[width=1\textwidth]{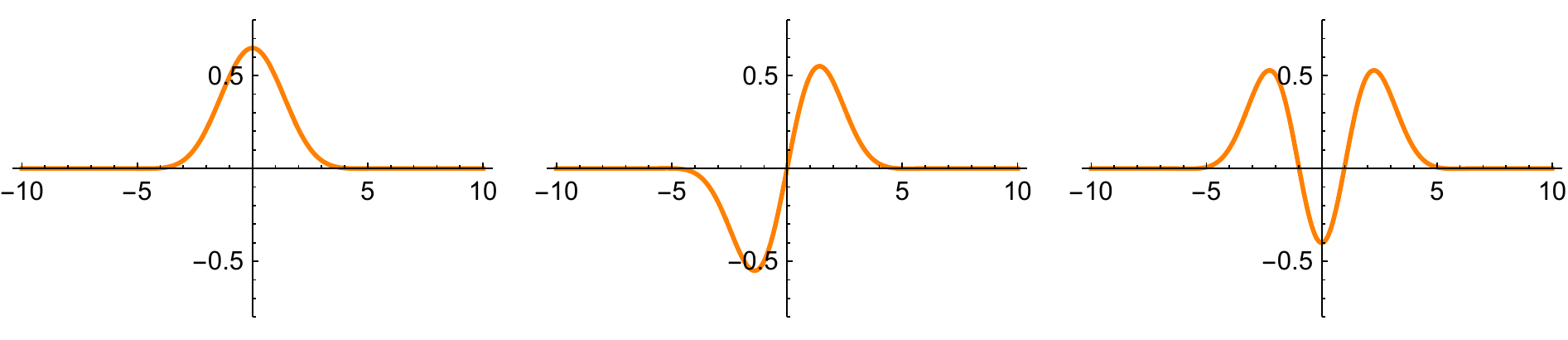}\label{bax4d1analnorm}}
\newline
\subfloat[Difference]{\includegraphics[width=1\textwidth]{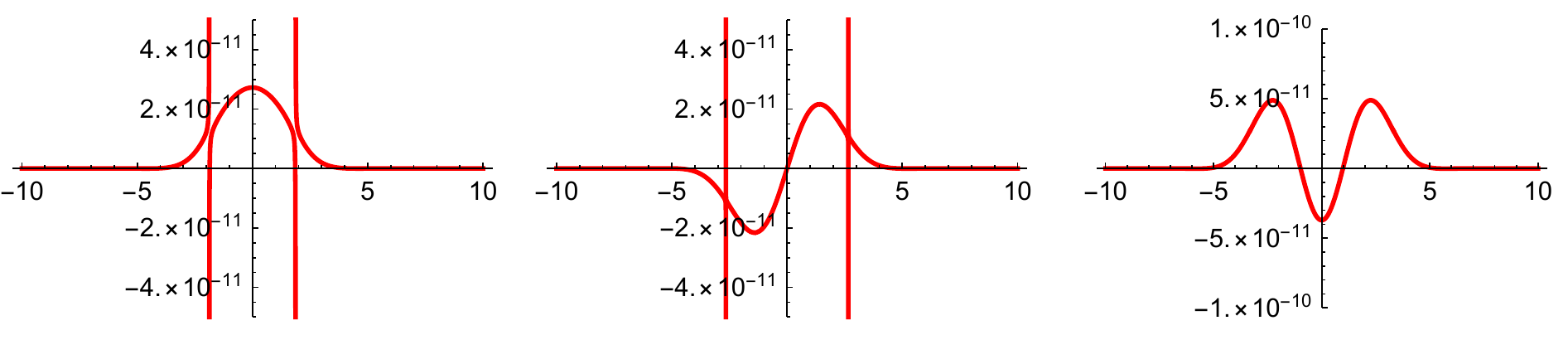}\label{bax4d1diff}}
 \caption{Level $n = 0, 1, 2$ Baxter solution $\overline{q}(\sigma)$ for $Q = 1$, $\hbar = 1.5$.}
\end{center}
\end{figure}

\begin{figure}[h]
  \centering
  \subfloat[Numerical results]{\includegraphics[width=1\textwidth]{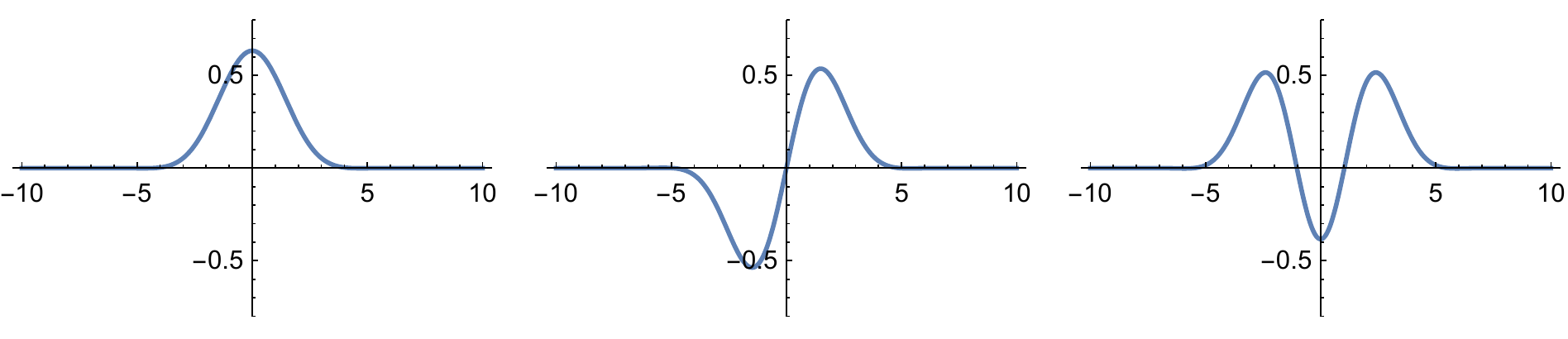}\label{bax4dmodnum}}
\newline
  \subfloat[Gauge theory results]{\includegraphics[width=1\textwidth]{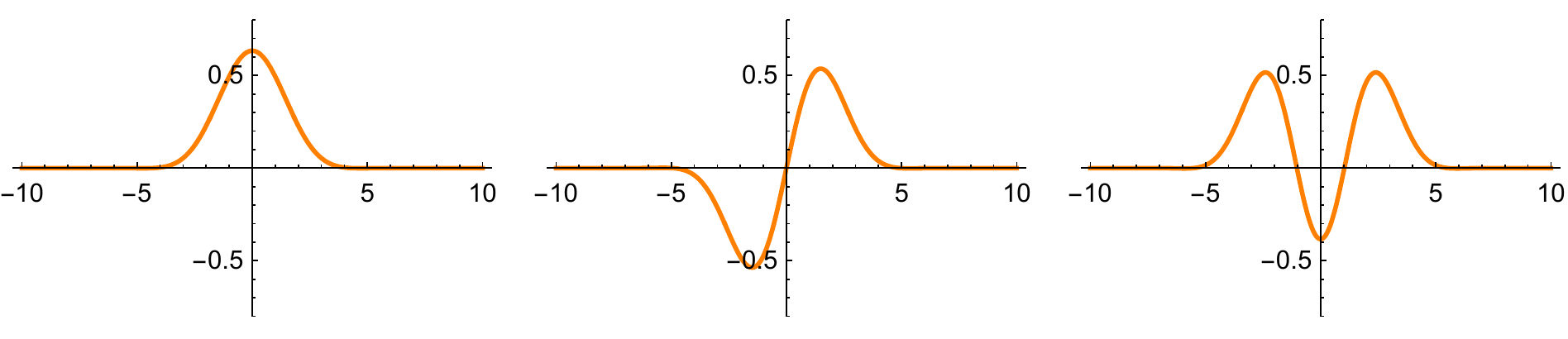}\label{bax4dmodanalnorm}}
  \newline
  \subfloat[Difference]{\includegraphics[width=1\textwidth]{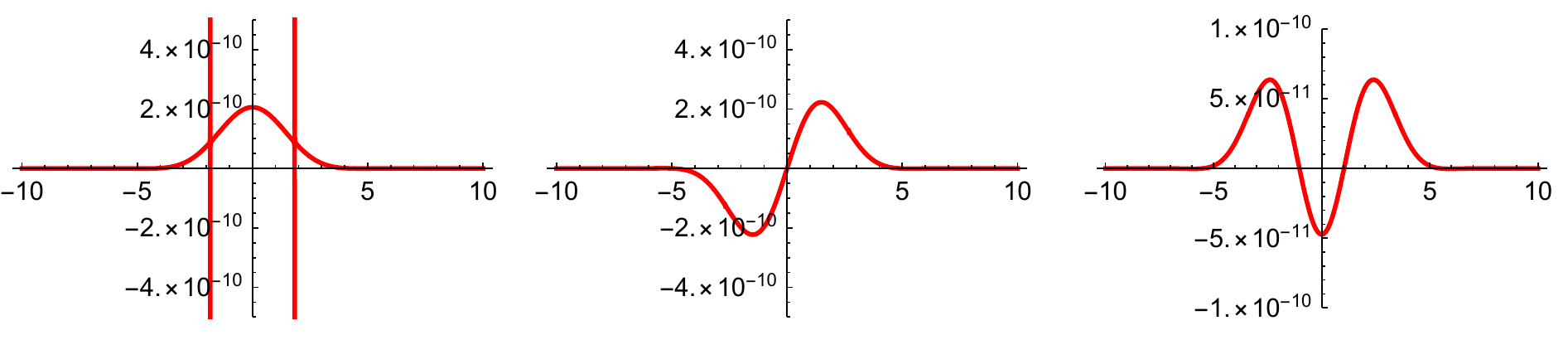}\label{bax4dmoddiff}}
 \caption{Level $n = 0, 1, 2$ Baxter solution $\overline{q}(\sigma)$ for $Q = \frac{1}{\sqrt{2}}$, $\hbar = \sqrt{3}$.}
\end{figure}

In order to check these claims (which were however proven in \cite{Kozlowski:2010tv}), we can compare the gauge theory solution with numerical results; to do this it is actually more convenient to consider the rescaled function
\begin{equation}
\overline{q}(\sigma, a, \hbar, Q_{\text{4d}}) = \left(Q_{\text{4d}}\right)^{\frac{i \sigma}{2 \hbar}}q(\sigma, a, \hbar, Q_{\text{4d}}), \label{rescaled}
\end{equation}
which satisfies the more symmetric problem
\begin{equation}
\left(Q_{\text{4d}}^{1/2} e^{i\hbar\partial_{\sigma}} + Q_{\text{4d}}^{1/2} e^{-i\hbar\partial_{\sigma}} + \sigma^2 \right) \overline{q}(\sigma, a, \hbar, Q_{\text{4d}}) = E_2(a, \hbar, Q_{\text{4d}}) \overline{q}(\sigma, a, \hbar, Q_{\text{4d}}). \label{cafe2}
\end{equation}
Numerical eigenvalues and eigenfunctions for the quantum problem \eqref{cafe2} can be determined with the procedure described in Section \ref{2.3.2}. Numerical eigenvalues are the same as the $\widehat{H}'_2$ ones (Table \ref{numenergy4d}), as expected since the two problems are related by a Fourier transformation; numerical eigenfunctions (normalized to 1) are however different and are shown in Figure \ref{bax4d1num} and \ref{bax4dmodnum} (blue). The gauge theory solution \eqref{rescaled} to the Baxter equation, computed up to 6-instantons and evaluated at the values of $a^{(n)}$ in Table \ref{a4d}, is instead shown in Figure \ref{bax4d1analnorm} and \ref{bax4dmodanalnorm} (orange) where we plotted Re$\left[\widetilde{q}(\sigma)\right]$, Im$\left[\widetilde{q}(\sigma)\right]$, Re$\left[\widetilde{q}(\sigma)\right]$ for
$n = 0, 1, 2$ respectively, the other imaginary/real/imaginary component being identically zero.\footnote{Remember that the gauge theory expression for \eqref{rescaled} is not normalized to 1; what we show in Figure \ref{bax4d1analnorm} and \ref{bax4dmodanalnorm} is the gauge theory result divided by its norm.} 
As we can see, numerical and gauge theory results seem to agree well and are hard to distinguish by the naked eye; moreover, the singularities at $\sigma = \pm a^{(n)}$ of the gauge theory results seem 

\newpage

\vspace{0.75 cm}

\begin{figure}[h]
  \centering
\includegraphics[width=1.05\textwidth]{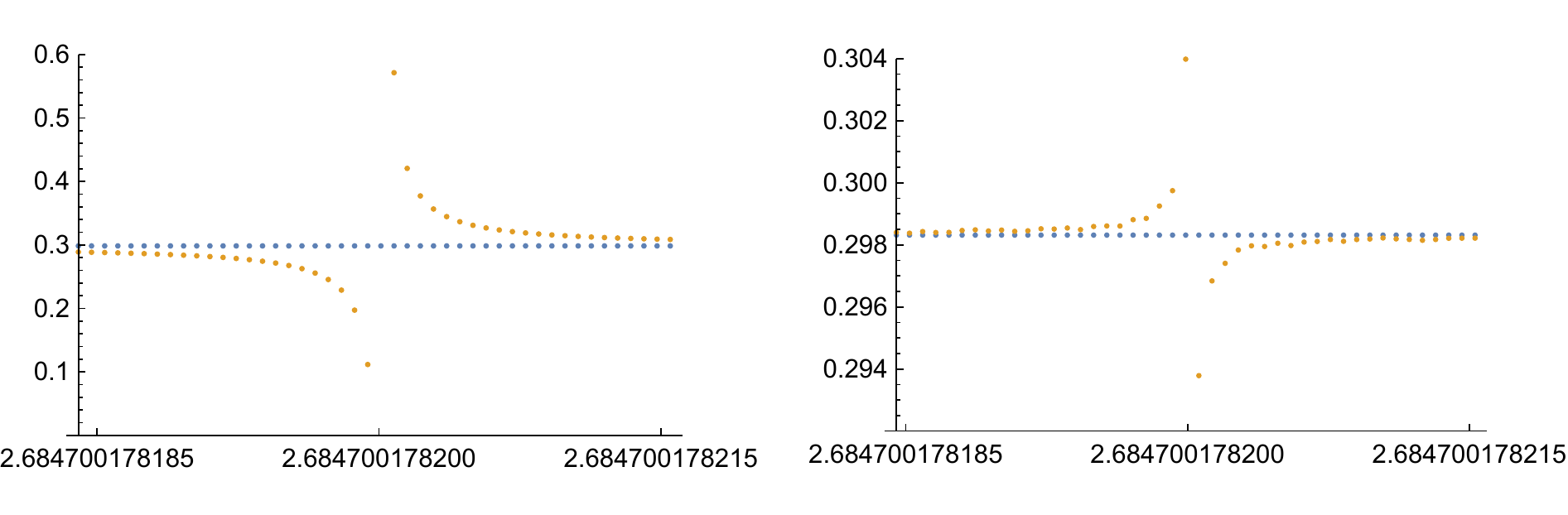}
 \caption{Level $n = 1$ numerical (blue dots) and gauge theory (orange dots) 
 Baxter solution $\overline{q}(\sigma)$ for $Q = \frac{1}{\sqrt{2}}$, $\hbar = \sqrt{3}$ near $\sigma = a^{(1)}$. Left: 4-instanton results; right: 6-instanton results.}
  \label{4dpoints}
\end{figure}

\newpage
\noindent to disappear already at 6-instantons. This is however an artifact, since singularities are only expected to disappear when considering \textit{all} instanton contributions: in fact some of them are visible in Figure \ref{bax4d1diff} and \ref{bax4dmoddiff} (red) which show the difference between numerical and gauge theory results, while when they are not visible we just need to ``zoom'' more near the singular point. What really happens is that divergences tend to close when we add more and more instanton corrections: an example of this behaviour for the level $n = 1$ solution at $Q = \frac{1}{\sqrt{2}}$, $\hbar = \sqrt{3}$ is shown in Figure \ref{4dpoints}, where we can see how the singularity becomes less pronounced when moving from 4-instanton formulae (left) to 6-instanton formulae (right).  \\





\begin{figure}[]
  \centering
\includegraphics[width=1\textwidth]{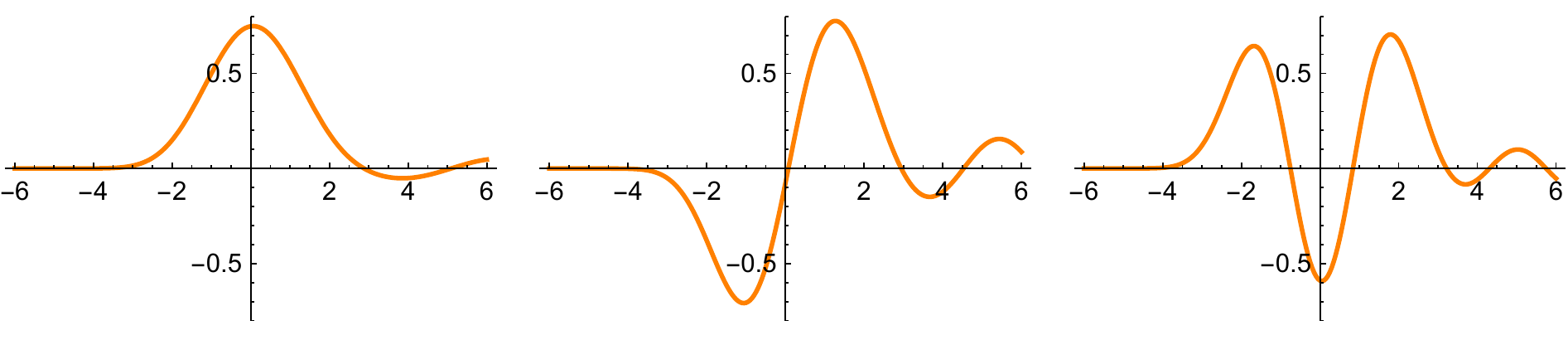}
 \caption{Gauge theory 2-particle Toda solution $\Psi_a(x)$ for $Q = 1$, $\hbar = 1.5$ at $a$ not solving the quantization conditions. Left: $a= 1.85$; center: $a = 2.60$; right: $a = 3.28$. }
  \label{toda4d1wrong}
\end{figure}

\begin{figure}[]
  \subfloat[Numerical results]{\includegraphics[width=1\textwidth]{4dothernum.pdf}\label{toda4d1numnum}}
\newline
  \subfloat[Gauge theory results]{\includegraphics[width=1\textwidth]{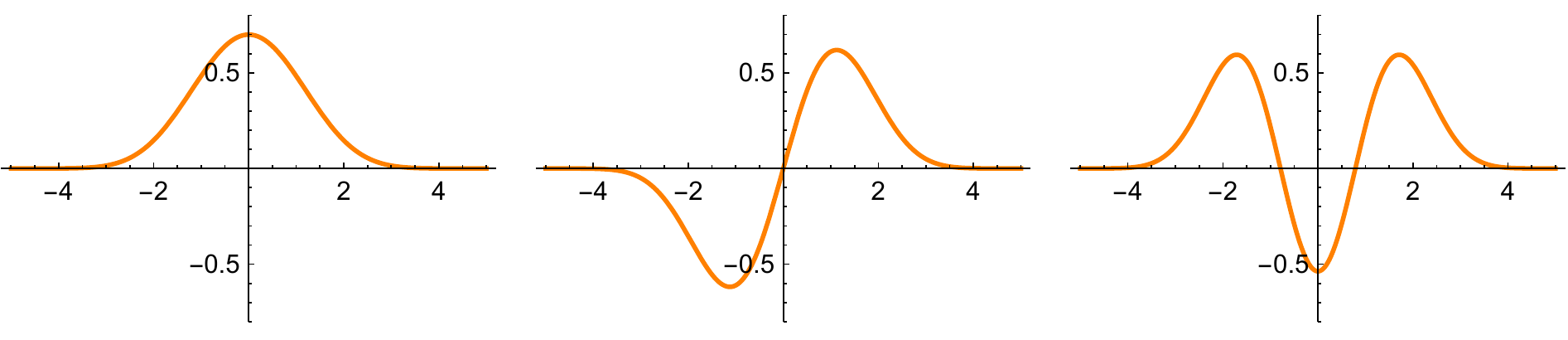}\label{toda4d1analnorm}}
\newline
 \subfloat[Difference]{\includegraphics[width=1\textwidth]{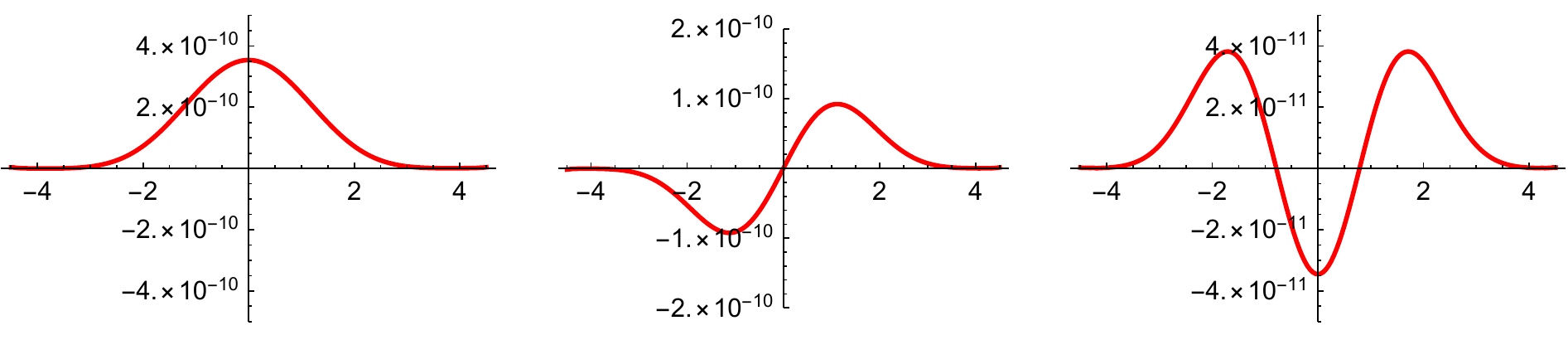}\label{toda4d1diff}}
\caption{Level $n = 0, 1, 2$ 2-particle Toda eigenfunction $\Psi_a(x)$ for $Q = 1$, $\hbar = 1.5$.}
\end{figure}

\begin{figure}[]
  \subfloat[Numerical results]{\includegraphics[width=1\textwidth]{4dmodnum.pdf}\label{toda4dmodnumnum}}
\newline
  \subfloat[Gauge theory results]{\includegraphics[width=1\textwidth]{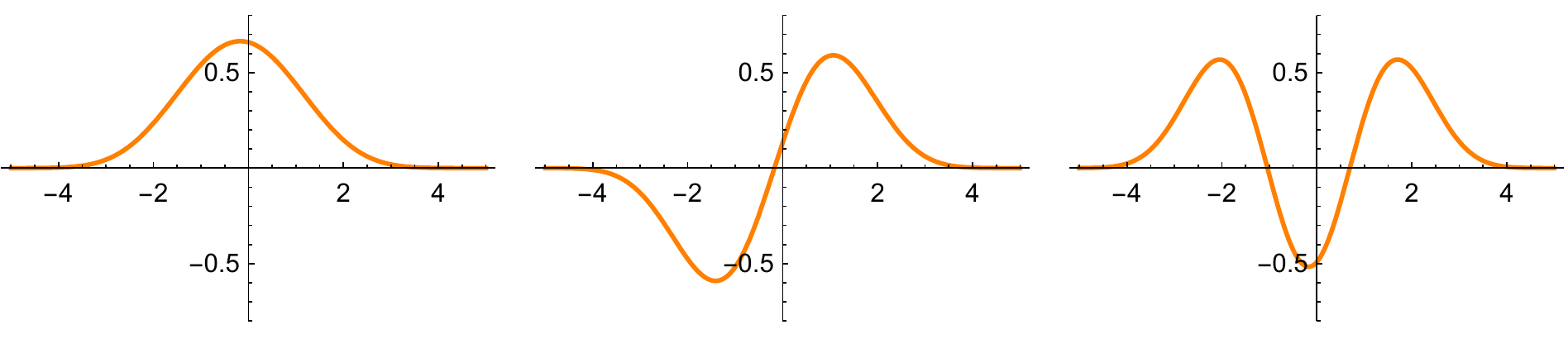}\label{toda4dmodanalnorm}}
\newline
 \subfloat[Difference]{\includegraphics[width=1\textwidth]{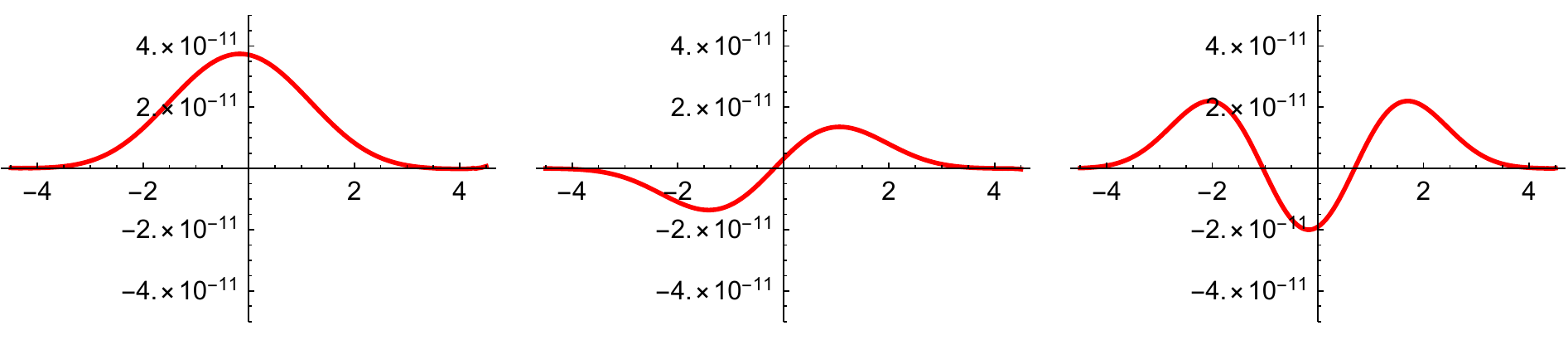}\label{toda4dmoddiff}}
\caption{Level $n = 0, 1, 2$ 2-particle Toda eigenfunction $\Psi_a(x)$ at $Q = \frac{1}{\sqrt{2}}$, $\hbar = \sqrt{3}$.}
\end{figure}

Having determined the solution to the $N$-particle Baxter equation, we can now compute the (not normalized) common eigenfunctions $\Psi_{\vec{a}}(\vec{x}, \hbar,  Q_{\text{4d}})$ of the $N$-particle closed Toda Hamiltonians in terms of the eigenfunctions of the $(N-1)$-particle open Toda chain by using formula \eqref{eigclosed} obtained from Separation of Variables, as reviewed in Section \ref{sec2.2}.
 For the 2-particle case \eqref{eigclosed} reduces to
\begin{equation}
\Psi_a(x,\hbar,Q_{\text{4d}}) = \int_{\mathbb{R}} d\sigma \,q(\sigma, a, \hbar, Q_{\text{4d}}) e^{\frac{i x \sigma}{\hbar}}, \label{ven}
\end{equation}
where $e^{\frac{i x \sigma}{\hbar}}$ is the 1-particle open Toda eigenfunction and  $q(\sigma, a, \hbar, Q_{\text{4d}})$ is as given in \eqref{tot}; when $a = a^{(n)}$ satisfies the quantization conditions \eqref{quant4d} (and therefore $q(\sigma, a, \hbar, Q_{\text{4d}})$ is entire) this satisfies 
\begin{equation}
\left( -\hbar^2 \partial_x^2 + e^x + Q_{\text{4d}}e^{-x} \right) \Psi_a(x,\hbar,Q_{\text{4d}}) = E_2(a, \hbar, Q_{\text{4d}}) \Psi_a(x,\hbar,Q_{\text{4d}}), \label{sat}
\end{equation}
always with $E_2(a, \hbar, Q_{\text{4d}}) $ as in \eqref{E2}. The integral can be explicitly evaluated as explained in Section \ref{sec2.2}. As shown in Figure \ref{toda4d1wrong}, at generic $a \neq a^{(n)}$ this expression does not correspond to a true eigenfunction since it does not go to zero at both $x \rightarrow \pm \infty$ but oscillates in one direction. On the other hand, for the values of $a = a^{(n)}$ in Table \ref{a4d} we obtain the eigenfunctions shown in Figure \ref{toda4d1analnorm} and \ref{toda4dmodanalnorm}, where again we divided gauge theory results by their norms in order to perform comparison. The difference between numerical and gauge theory results is shown in Figure \ref{toda4d1diff} and \ref{toda4dmoddiff}; as we can see the difference is very small already for gauge theory expressions evaluated up to 6-instantons. 

\subsection{Comments} \label{sec2.4}

In Section \ref{sec2.3} we reviewed how the NS limit of various quantities one can compute in the four-dimensional pure $\mathcal{N} = 2$ $SU(N)$ gauge theory admit an interpretation in terms of the $N$-particle quantum Toda chain, and we performed some check against numerical results for the $N = 2$ case. In particular we discussed how the disc partition function of the two-dimensional $\mathcal{N} = (2,2)$ theories corresponding to defects of type II and I precisely coincide with the eigenfunctions of the open Toda chain Baxter equation or with the eigenfunctions of the open Toda chain Hamiltonians as computed in \cite{Kharchev:1999bh,Kharchev:2000yj} via Separation of Variables. Similarly, the partition function of the two-dimensional type II theory coupled to the four-dimensional $SU(N)$ gauge theory provides a solution to the closed Toda chain Baxter equation (that is, the operator arising from quantizing the spectral/Seiberg-Witten curve) which is the same as the one found in \cite{GUTZWILLER1980347,GUTZWILLER1981304,Kharchev:1999bh,Kharchev:2000yj} as already discussed in \cite{Kozlowski:2010tv}; from this solution to the Baxter equation we can then construct eigenfunctions for the closed Toda chain Hamiltonians via Separation of Variables.

There are a few comments we would like to make before closing this Section.

\begin{itemize}

\item The first comment regards the eigenfunction $\Psi_{\vec{a}}(\vec{x}, \hbar,  Q_{\text{4d}})$ of the closed Toda chain Hamiltonians $\widehat{H}_m$. In Section \ref{sec2.2} and \ref{sec2.3} we computed this eigenfunction via Separation of Variables, but this is actually not the most natural way to proceed in gauge theory: the gauge theory prescription would rather involve the computation of the NS limit $Z^{ \text{(I),NS}}_{\text{2d/4d}}(\vec{x}, \vec{a}, \hbar, Q_{\text{4d}}) $ of the partition function of the two-dimensional type I theory coupled to the four-dimensional $SU(N)$ gauge theory. Contour integral formulae for the instanton corrections to the partition function of this 2d-4d coupled system are given by \eqref{4d2dI}, \eqref{4d2dIbis} in Appendix \ref{formula4d}; the perturbative part is instead given by \eqref{katekyo}. In the 2-particle Toda case, this would lead to
\begin{equation}
\begin{split}
\Psi_a(x, \hbar,Q_{\text{4d}}) &\;\propto\; \left( \dfrac{e^{x/2}}{\hbar}\right)^{\frac{2ia}{\hbar}} \Gamma\left( -\frac{2 i a}{\hbar} \right)
Z^{ \text{(I),NS}}_{\text{2d/4d}}(x, a, \hbar, Q_{\text{4d}}) \\
& \;+\; \left( \dfrac{e^{x/2}}{\hbar}\right)^{-\frac{2ia}{\hbar}} \Gamma\left( \frac{2 i a}{\hbar} \right) Z^{ \text{(I),NS}}_{\text{2d/4d}}(x, -a, \hbar, Q_{\text{4d}}). \label{unused}
\end{split}
\end{equation} 
This expression is nothing else but the 2-particle open Toda chain eigenfunction \eqref{2todaopen} in which we replaced the vortex partition function by the NS limit of the instanton-vortex partition function of the 2d type I theory coupled to the 4d theory \eqref{4d2dI}. 
Both the first and second line of \eqref{unused} independently satisfy \eqref{sat} formally for any $a$, in constrast to the expression \eqref{ven} arising from Separation of Variables which is a solution only for $a = a^{(n)}$.  Nevertheless it is expected that only when we consider first and second line together, and only when evaluated at the values of $a= a^{(n)}$ satisfying the quantization condition \eqref{museum}, they provide a proper eigenfunction of the Toda Hamiltonian with the correct $L^2(\mathbb{R})$ asymptotic behaviour. Therefore we expect that the two expressions \eqref{ven} and \eqref{unused}, although different for generic values of $a$, will coincide for $a = a^{(n)}$ satisfying the quantization condition (mainly due to \eqref{equality}). Unfortunately we were not able to show this equality even at the numerical level, mainly because of the difficulty in calculating a sufficiently large number of terms for the series $Z^{ \text{(I),NS}}_{\text{2d/4d}}(\vec{x}, \vec{a}, \hbar, Q_{\text{4d}})$; however it should be possible to prove it analytically by using the results of \cite{Gorsky:2017hro}. 

\item The second comment concerns the quantization condition \eqref{quant4d}, and in particular the one for the $SU(2)$ theory \eqref{museum}. Instead of the NS limit, let us now consider the \textit{unrefined} limit $\epsilon_1 = - \epsilon_2 = i \hbar$ of the pure $\mathcal{N} = 2$ $SU(2)$ partition function:
\begin{equation}
Z_{\text{4d}}(a, \epsilon_1, \epsilon_2, Q_{\text{4d}}) \;\;\Longrightarrow \;\; 
Z_{\text{4d}}^{\text{unref}}(a, \hbar, Q_{\text{4d}}) = Z_{\text{4d}}(a, i\hbar, -i\hbar, Q_{\text{4d}}).
\end{equation}
In this limit the perturbative part reduces to a product of exponential terms and Barnes G-functions, so we can rewrite $Z_{\text{4d}}^{\text{unref}}(a, \hbar, Q_{\text{4d}})$ as
\begin{equation}
Z_{\text{4d}}^{\text{unref}}(a, \hbar, Q_{\text{4d}}) = \dfrac{\left(\hbar^{-4}  Q_{\text{4d}} \right)^{-\frac{a^2}{\hbar^2}}}{G(1+\frac{2ia}{\hbar})G(1-\frac{2ia}{\hbar})} Z_{\text{4d}}^{\text{inst}}(a, i\hbar, -i\hbar, Q_{\text{4d}}).
\end{equation}
We can now consider the Zak transform
of $Z_{\text{4d}}^{\text{unref}}(a, \hbar, Q_{\text{4d}})$, also known as dual partition function \cite{2003hep.th....6238N}:
\begin{equation}
\tau(a, \hbar, Q_{\text{4d}}, \eta) = \sum_{n \in \mathbb{Z}} e^{\frac{i \eta}{\hbar}} Z_{\text{4d}}^{\text{unref}}(a + i n \hbar, \hbar, Q_{\text{4d}}). \label{zak}
\end{equation}
This object has received much attention recently due to its relation to the theory of Painlev\'{e} equations: in fact it has been shown in \cite{Gamayun:2013auu,Its:2014lga,Bershtein:2014yia,Bershtein:2016uov} that $\tau(a, \hbar, Q_{\text{4d}}, \eta)$ is the short-distance (i.e. small $Q_{\text{4d}}$) expansion of the $\tau$-function associated to the Painlev\'{e} III$_{3}$ equation. This means that \eqref{zak} is a solution to a non-linear ordinary differential equation in $Q_{\text{4d}}$, the $\tau$-PIII$_3$ equation; the parameters $a$ and $\eta$ are two integration constants/initial values for this equation, while $\hbar$ can be considered as an overall scale which can be reabsorbed in the definition of the other parameters. From the point of view of gauge theory the interpretation of $\eta$ is not completely clear; however based on \cite{2003hep.th....6238N} and on the analysis of the similar PVI case (associated to four-dimensional $\mathcal{N} = 2$ $SU(2)$ $N_F = 4$ super-QCD) performed in \cite{Gamayun:2012ma,Iorgov:2013uoa,Iorgov:2014vla} we expect $\eta$ to coincide with the $a$-derivative of the twisted effective superpotential $\mathcal{W}_{\text{4d}}(a, i\hbar, Q_{\text{4d}})$, which is a quantity computed in the NS limit instead of the unrefined one:
\begin{equation}
\eta = \dfrac{\partial}{\partial a}\mathcal{W}_{\text{4d}}(a, i\hbar, Q_{\text{4d}}). \label{eta}
\end{equation}
If this is the case then $\eta = 2\pi \hbar n$ when the quantization condition \eqref{museum} is satisfied; due to periodicity, this is equivalent to setting $\eta = 0$. What is relevant to the present discussion is that \eqref{zak} admits another interpretation when $\eta = 0$ (mod $2\pi \hbar$): as shown in \cite{Bonelli:2016idi} based on \cite{Zamolodchikov:1994uw}, $\tau(a, \hbar, Q_{\text{4d}}, 0)$ coincides with the spectral/Fredholm determinant of an ideal Fermi gas whose partition function is given by an $O(2)$ (or polymer) matrix model; this spectral determinant can also be obtained by taking a particular 4d limit of the spectral determinant associated to the local $\mathbb{F}_0$ toric Calabi-Yau geometry which was introduced and analyzed in \cite{Grassi:2014zfa}. As such, the zeroes of $\tau(a, \hbar, Q_{\text{4d}}, 0)$ should contain information about the spectrum of the Fermi gas. The procedure to determine this spectrum was fully explained in \cite{Bonelli:2016idi}: by defining $a = \widetilde{a} + i \frac{\hbar}{2}$, the equation 
\begin{equation}
\tau\left(\widetilde{a} + i \text{\footnotesize{$\frac{\hbar}{2}$}} , \hbar, Q_{\text{4d}}, 0\right) = 0  \label{tauzero}
\end{equation}
admits solutions for real $\widetilde{a}^{(n)}$ at fixed value of $\hbar$, $Q_{\text{4d}}$, where $n$ labels the $n$-th zero; 
the energy of the $n$-th energy level of the Fermi gas is then given by
\begin{equation}
e^{E_{\text{gas}}(\widetilde{a}, \hbar)} = \dfrac{1}{4\pi}\left( e^{\frac{2\pi \widetilde{a}}{\hbar}} + e^{-\frac{2\pi \widetilde{a}}{\hbar}} \right) \label{wilson}
\end{equation}
evaluated at $\widetilde{a}^{(n)}$. What does this have to do with the Toda system? 
Equation \eqref{tauzero} defines quantization conditions for $\widetilde{a}$; it turns out that the values $\widetilde{a}^{(n)}$ thus determined coincide with the ones for $a^{(n)}$ listed in Table \ref{a4d} obtained from \eqref{museum} at same fixed $\hbar$, $Q_{\text{4d}}$. There exist therefore two different ways to express the same quantization conditions for the closed Toda chain from gauge theory: 
\begin{itemize}
\item[A)] the first one is given in terms of $\mathcal{W}_{\text{4d}}(a, i\hbar, Q_{\text{4d}})$, computed in the \textit{NS limit};
\item[B)] the second one is given in terms of $\tau(a, \hbar, Q_{\text{4d}}, 0)$ computed in the \textit{unrefined limit}.
\end{itemize}
This is reminescent of a similar situation occurring in five dimensions: the condition of vanishing of the spectral determinant associated to the local $\mathbb{F}_0$ geometry (i.e. pure $\mathcal{N} = 1$ $SU(2)$ Yang-Mills) studied in \cite{Grassi:2014zfa}, which involves the unrefined limit of the 5d partition function (+ non-perturbative corrections in $\hbar$), seems to be equivalent to extremizing the twisted effective superpotential which is obtained from the NS limit of the 5d $\mathcal{N} = 1$ $SU(2)$ partition function (+ non-perturbative corrections in $\hbar$) \cite{Wang:2015wdy}. This equivalence in five dimensions has been proven in \cite{Grassi:2016nnt} by making use of the 5d blow-up equations of \cite{Nakajima:2005fg,Gottsche:2006bm}; it is then natural to expect that by considering an appropriate 4d limit of \cite{Grassi:2016nnt}, or by using the 4d blow-up equations of \cite{Nakajima:2003pg,Nakajima:2003uh}, the equivalence between four-dimensional NS and unrefined quantization conditions should follow.

Regardless of how we express them, these quantization conditions determine the discrete energy levels of two different quantum systems: 
\begin{itemize}
\item the 2-particle closed Toda chain via \eqref{E2}; 
\item a Fermi gas associated to the $O(2)$ matrix model via \eqref{wilson}.
\end{itemize} 
Some relation between the ``off-shell'' closed Toda chain energy \eqref{E2} and the ``off-shell'' Fermi gas energy \eqref{wilson} already appeared in the mathematical literature in the context of wild nonabelian Hodge correspondence and integrable systems associated to Hitchin moduli spaces in different complex structures ($I$ and $J$ respectively), see \cite{2013arXiv1309.7202B}, \cite{2014arXiv1411.3692W} and references therein. In that context however \eqref{wilson} is usually interpreted not as the energy of a Fermi gas, but as the energy of the relativistic open Toda chain (the energies being identical modulo overall factors), a quantum mechanical system with \textit{continuous} spectrum which will be discussed in Section \ref{sec3}; consequently $\hbar = R^{-1}$ is not thought of as the Planck constant but as the ``speed of light'' of the relativistic system. Moreover, from the point of view of $\mathcal{N} = 2$ gauge theories in four dimension, \eqref{wilson} is interpreted as a Wilson loop wrapping $S^1_R$ for a 4d gauge theory placed on $\mathbb{R}^3 \times S^1_R$ (the most natural setting for studying Hitchin systems from a gauge theory point of view \cite{Gaiotto:2009hg}). 
It might be worth to investigate if the usual mathematical interpretation of \eqref{wilson} being related to the relativistic open Toda chain should be kept or replaced by an interpretation in terms of Fermi gases with discrete spectrum as suggested from the results of \cite{Bonelli:2016idi}.

As a final comment, let us mention that the relation between \eqref{E2} and \eqref{wilson} might be clearer if we start from considering the relativistic closed Toda chain (or five-dimensional gauge theories) and take an appropriate limit, as already suggested in \cite{2014arXiv1411.3692W}; we will come back to this point in Section \ref{seccomments5d} after having studied in some detail this relativistic system.
\end{itemize}

\section{Relativistic Toda integrable systems} \label{sec3}

In Section \ref{sec2} we reviewed in some detail how the $N$-particle closed Toda chain can be solved, both via the Separation of Variables technique and via gauge theory, and we provided some numerical check of the solution. 
In this Section instead we will study the solution of the ``relativistic'' generalization of the Toda system: that is, we will consider a version of the Toda chain in which the \textit{differential} operators appearing in the (``non-relativistic'') Toda Hamiltonians $\widehat{H}_m$ \eqref{hams} (arising from quantizing polynomials in the momenta) get replaced by appropriate \textit{finite-difference} operators, which arise from quantizing exponentials of the momenta. While exact quantization conditions and spectrum for this ``relativistic'' system have been discussed in detail in \cite{Grassi:2014zfa,Wang:2015wdy,2015arXiv151102860H}, it is not yet clear how to construct proper eigenfunctions for the relativistic closed Toda chain or its Baxter equation in full generality, although various particular cases were considered in \cite{Marino:2016rsq,Kashaev:2017zmv} (eigenfunctions for the relativistic open Toda chain have instead been constructed in \cite{Kharchev:2001rs}). 
Here we propose a solution to this problem via gauge theoretical arguments similar to the ones used in Section \ref{sec2.3}, appropriately modified to take into account the novel features appearing in the problem at hand; we will mainly focus on constructing solutions to the relativistic Toda Baxter equation based on analogy with the non-relativistic case, and we will check our proposed solution against numerical results.

\subsection{Quantum relativistic Toda chain (open and closed)} \label{sec3.1}

The ``relativistic'' generalization of the quantum $N$-particle Toda chain is a quantum mechanical system of $N$ particles on a line interacting via the Hamiltonian (in conventions similar to \cite{Kharchev:2001rs})\footnote{We will use $\widehat{\mathcal{H}}_m$ and $\mathcal{E}_m$ to denote Hamiltonians and energies of the relativistic Toda system in order to distinguish them from the ones of the Toda chain.}
\begin{equation}
\widehat{\mathcal{H}}_1 = \sum_{m=1}^N \left[ 1 + q^{-1/2} 
e^{\frac{2\pi}{\omega_2}(x_m - x_{m+1})} \right] e^{\omega_1 p_m}. \label{Htodarel}
\end{equation}
Here $x_m$, $p_m$ are position and momentum of the $m$-th particle (rescaled with respect to the ones we used in Section \ref{sectoda}) and satisfy the commutation relations 
\begin{equation}
[p_l, x_m] = -i \delta_{l,m}.
\end{equation}
This Hamiltonian is self-adjoint on $L^2(\mathbb{R}^{N-1})$ when the parameters $\omega_1$, $\omega_2 \in \mathbb{R}_+$\footnote{It is possible to define a good quantum mechanical problem also for $\omega_1$, $\omega_2$ $\in \mathbb{C}$, $\omega_1 = \overline{\omega}_2$ as considered for example in \cite{Kashaev:2017zmv}; in this case it is often chosen for definiteness 
$\text{Im}(\omega_1/\omega_2) > 0$, i.e. $\vert q \vert <1$. Taking the limit 
$\text{Im}(\omega_1/\omega_2) \rightarrow 0$ from this regime leads us to the relativistic Toda chain under consideration\label{fnrel}.}; these are related to the Planck constant and an additional parameter that can naively be thought as the ``speed of light'' of the relativistic system, and we defined 
\begin{equation}
q = e^{2\pi i \omega_1/\omega_2}, \;\;\;\; \vert q \vert = 1.
\end{equation}
The ``non-relativistic'' Toda Hamiltonians $\widehat{H}_m$ \eqref{hams} are recovered from \eqref{Htodarel} by taking an appropriate $\omega_1 \rightarrow 0$ limit. 
As for the Toda chain, we impose the boundary condition
\begin{equation}
x_{N+1} = x_1 - \frac{\omega_2}{2\pi} \ln Q_{\text{rel}} \,,\;\;\;\;\; Q_{\text{rel}} = e^{-\frac{2 \pi m_0}{\omega_2}} \in \mathbb{R}_+
\end{equation}
to better distinguish between the open and closed chain:
\begin{itemize}
\item $Q_{\text{rel}} = 0$: open relativistic Toda chain;
\item $Q_{\text{rel}} > 0$: closed relativistic Toda chain.
\end{itemize}
Similarly to the Toda case, relativistic open chains have a continuous spectrum while relativistic closed chains admit a discrete spectrum.
Relativistic Toda chains are also integrable systems, their $N$ commuting Hamiltonians being
\begin{eqnarray}
\widehat{\mathcal{H}}_1 &=& \sum_{m=1}^N \left[ 1 + q^{-1/2} 
e^{\frac{2\pi}{\omega_2}(x_m - x_{m+1})} \right] e^{\omega_1 p_m}, \nonumber \\
&\vdots & \nonumber \\
\widehat{\mathcal{H}}_{N-1} &=& \sum_{m=1}^N \left[ 1 + q^{-1/2} 
e^{\frac{2\pi}{\omega_2}(x_m - x_{m+1})} \right] e^{-\omega_1 p_{m+1}} \nonumber \\ 
\widehat{\mathcal{H}}_N &=& \prod_{m=1}^N e^{\omega_1 p_m}, \label{hamrel}
\end{eqnarray}
with $p_{N+1} = p_1$. Decoupling the center of mass in this case is equivalent to impose $\widehat{\mathcal{H}}_N = 1$. 
Despite the many similarities with the non-relativistic Toda chain,
a peculiar and very important property of the relativistic Toda chain is the existence of its \textit{modular dual} version (see \cite{Kharchev:2001rs}, based on \cite{1995LMaPh..34..249F,Faddeev:1999fe}); this is defined by the set of $N$ commuting Hamiltonians
\begin{eqnarray}
\widehat{\widetilde{\mathcal{H}}}_1 &=& \sum_{m=1}^N \left[ 1 + \widetilde{q}^{-1/2} \omega_2^2 e^{\frac{2\pi}{\omega_1}(x_m - x_{m+1})} \right] e^{\omega_2 p_m}, \nonumber \\
&\vdots & \nonumber \\
\widehat{\widetilde{\mathcal{H}}}_{N-1} &=& \sum_{m=1}^N \left[ 1 + \widetilde{q}^{-1/2} \omega_2^2 e^{\frac{2\pi}{\omega_1}(x_m - x_{m+1})} \right] e^{-\omega_2 p_{m+1}} \nonumber \\ 
\widehat{\widetilde{\mathcal{H}}}_N &=& \prod_{m=1}^N e^{\omega_2 p_m}, \label{hamreldual}
\end{eqnarray}
with boundary condition
\begin{equation}
x_{N+1} = x_1 - \frac{\omega_1}{2\pi} \ln \widetilde{Q}_{\text{rel}} \,,\;\;\;\;\; \widetilde{Q}_{\text{rel}} = e^{-\frac{2 \pi m_0}{\omega_1}} \in \mathbb{R}_+,
\end{equation}
and where we introduced\footnote{More in general, tilded variables will always denote quantities in the modular dual system; these are related to the analogous quantities in the original model by the exchange $\omega_1 \leftrightarrow \omega_2$. When considering $\omega_1$, $\omega_2$ $\in \mathbb{C}$, $\omega_1 = \overline{\omega}_2$ the canonical choice 
$\text{Im}(\omega_1/\omega_2) > 0 $ implies $\vert \widetilde{q}^{-1} \vert <1$.}
\begin{equation}
\widetilde{q} = e^{2\pi i \omega_2/\omega_1}, \;\;\;\; \vert \widetilde{q} \vert = 1.
\end{equation}
The dual Hamiltonians $\widehat{\widetilde{\mathcal{H}}}_{m}$ are not really independent from the original ones since the two sets are related by the exchange $\omega_1 \leftrightarrow \omega_2$; however because of this relation the original and dual set of Hamiltonians commute with each other by construction:
\begin{equation}
[\widehat{\mathcal{H}}_{m}, \widehat{\widetilde{\mathcal{H}}}_{n}] = 0.
\end{equation}
This means that the eigenfunctions of the relativistic Toda chain will also be eigenfunctions of the dual system; we therefore expect them to be symmetric under $\omega_1 \leftrightarrow \omega_2$. 
Among other things, the existence of the modular dual system plays a key role in eliminating some of the ambiguities of the relativistic Toda chain solution:
in fact differently from ordinary quantum mechanics in which eigenfunctions are solutions to a differential equation and as such are only defined modulo an overall constant, eigenfunctions of finite-difference operators like the relativistic Toda one \eqref{hamrel} are only defined modulo an $i \omega_1$-periodic function (much like the solution to the non-relativistic Toda chain Baxter equation \eqref{baxclosed}); this ambiguity can however be reduced to the usual overall constant normalization by requiring them to be eigenfunctions of the dual relativistic Toda operators \eqref{hamreldual} as well.\footnote{Of course there may still be ambiguities given by doubly-periodic functions in $i\omega_1$, $i \omega_2$.}

To sum up, our spectral problem in this case consists of constructing common eigenfunctions $\Psi_{\vec{\mathcal{E}},\vec{\widetilde{\mathcal{E}}}}^{\text{rel}}(\vec{x}, \omega_1, \omega_2)$ of the relativistic Toda Hamiltonians $\widehat{\mathcal{H}}_m$ and dual Hamiltonians $\widehat{\widetilde{\mathcal{H}}}_m$, that is
\begin{equation}
\begin{split}
& \widehat{\mathcal{H}}_m \Psi_{\vec{\mathcal{E}},\vec{\widetilde{\mathcal{E}}}}^{\text{rel}}(x_1, \ldots, x_N, \omega_1, \omega_2) = \mathcal{E}_m \Psi_{\vec{\mathcal{E}},\vec{\widetilde{\mathcal{E}}}}^{\text{rel}}(x_1, \ldots, x_N, \omega_1, \omega_2) \,,\;\;\;\; m = 1, \ldots, N, \\[4 pt]
\label{sptodarel}
& \widehat{\widetilde{\mathcal{H}}}_m \Psi_{\vec{\mathcal{E}},\vec{\widetilde{\mathcal{E}}}}^{\text{rel}}(x_1, \ldots, x_N, \omega_1, \omega_2) = \widetilde{\mathcal{E}}_m \Psi_{\vec{\mathcal{E}},\vec{\widetilde{\mathcal{E}}}}^{\text{rel}}(x_1, \ldots, x_N, \omega_1, \omega_2) \,,\;\;\;\; m = 1, \ldots, N, 
\end{split}
\end{equation}
where $\vec{\mathcal{E}} = (\mathcal{E}_1, \ldots, \mathcal{E}_N)$, $\vec{\widetilde{\mathcal{E}}} = (\widetilde{\mathcal{E}}_1, \ldots, \widetilde{\mathcal{E}}_N)$ are the corresponding eigenvalues (with $\mathcal{E}_N = \widetilde{\mathcal{E}}_N = 1$ after decoupling the center of mass), satisfying the appropriate boundary conditions: 
\begin{itemize}
\item for the relativistic open Toda chain, $\Psi_{\vec{\mathcal{E}},\vec{\widetilde{\mathcal{E}}}}^{\text{rel}}(\vec{x}, \omega_1, \omega_2)$ should vanish fast enough as $x_k - x_{k+1} \rightarrow \infty$;
\item for the relativistic closed Toda chain, normalizability requires $\Psi_{\vec{\mathcal{E}},\vec{\widetilde{\mathcal{E}}}}^{\text{rel}}(\vec{x}, \omega_1, \omega_2) \in L^2(\mathbb{R}^{N-1})$. 
\end{itemize}
More precise statements about the relativistic Toda spectral problem can be found in \cite{Kharchev:2001rs}. 
Similarly to what we did in Section \ref{sectoda}, we can rewrite this spectral problem in a more compact notation: if we define 
\begin{equation}
w = e^{-\frac{2\pi \sigma}{\omega_2}}, \;\;\;\; \widetilde{w} = e^{-\frac{2\pi \sigma}{\omega_1}}
\end{equation}
and introduce the generating functions of the relativistic Hamiltonians
\begin{equation}
\begin{split}
& \widehat{t}_{\text{rel}}(\sigma) = \sum_{m=0}^N (-1)^m w^{m -\frac{N}{2}}\widehat{\mathcal{H}}_m, \;\;\;\;\;\; \widehat{\mathcal{H}}_0 = 1, \\
& \widehat{\widetilde{t}}_{\text{rel}}(\sigma) = \sum_{m=0}^N (-1)^m \widetilde{w}^{m-\frac{N}{2}}\widehat{\widetilde{\mathcal{H}}}_m, \;\;\;\;\;\; \widehat{\widetilde{\mathcal{H}}}_0 = 1,
\end{split}
\end{equation}
satisfying 
\begin{equation}
\begin{split}
& [\widehat{t}_{\text{rel}}(\sigma), \widehat{t}_{\text{rel}}(\sigma')] = 0, \\
& [\widehat{\widetilde{t}}_{\text{rel}}(\sigma), \widehat{\widetilde{t}}_{\text{rel}}(\sigma')] = 0, \\
& [\widehat{t}_{\text{rel}}(\sigma), \widehat{\widetilde{t}}_{\text{rel}}(\sigma')] = 0, 
\end{split}
\end{equation}
the spectral problem \eqref{sptodarel} becomes
\begin{equation}
\begin{split}
& \widehat{t}_{\text{rel}}(\sigma)\Psi^{\text{rel}}_{\vec{\mathcal{E}},\vec{\widetilde{\mathcal{E}}}}(\vec{x}, \omega_1, \omega_2) = t_{\text{rel}}(\sigma, \vec{\mathcal{E}})\Psi^{\text{rel}}_{\vec{\mathcal{E}},\vec{\widetilde{\mathcal{E}}}}(\vec{x}, \omega_1, \omega_2), \\[4 pt]
& \widehat{\widetilde{t}}_{\text{rel}}(\sigma)\Psi^{\text{rel}}_{\vec{\mathcal{E}},\vec{\widetilde{\mathcal{E}}}}(\vec{x}, \omega_1, \omega_2) = \widetilde{t}_{\text{rel}}(\sigma, \vec{\widetilde{\mathcal{E}}})\Psi^{\text{rel}}_{\vec{\mathcal{E}},\vec{\widetilde{\mathcal{E}}}}(\vec{x}, \omega_1, \omega_2), 
\end{split}
\end{equation}
where 
\begin{equation}
\begin{split}
& t_{\text{rel}}(\sigma, \vec{\mathcal{E}}) = \sum_{m=0}^N (-1)^{m} w^{m-\frac{N}{2}} \mathcal{E}_m, \;\;\;\;\;\; \mathcal{E}_0 = 1, \\
& \widetilde{t}_{\text{rel}}(\sigma, \vec{\widetilde{\mathcal{E}}}) = \sum_{m=0}^N (-1)^{m} \widetilde{w}^{m-\frac{N}{2}} \widetilde{\mathcal{E}}_m, \;\;\;\;\;\; \widetilde{\mathcal{E}}_0 = 1
\end{split}
\end{equation}
are the generating functions of the relativistic Toda and dual Toda eigenvalues. These generating functions also enter in the definition of the spectral curves of the classical relativistic Toda chain and its dual: these are Riemann surfaces embedded in $(w,y)$ or $(\widetilde{w}, \widetilde{y}) \in \mathbb{C}^{\times} \times \mathbb{C}^{\times}$ defined by the equations
\begin{equation}
\begin{split}
& y + Q_{\text{rel}}\, y^{-1} = 
\mathcal{E}_N^{-\frac{1}{2}}\,
t_{\text{rel}}(\sigma,\vec{\mathcal{E}}), \\
& \widetilde{y} + \widetilde{Q}_{\text{rel}}\, \widetilde{y}^{-1} = 
\widetilde{\mathcal{E}}_N^{-\frac{1}{2}}\,
\widetilde{t}_{\text{rel}}(\sigma,\vec{\widetilde{\mathcal{E}}}). \label{curverel}
\end{split}
\end{equation}
Similarly to what we saw in Section \ref{sec2}, quantization of the spectral curves \eqref{curverel} will provide the Baxter equations associated to the relativistic Toda chain and its modular dual, whose solution will be necessary in order to construct the relativistic Toda eigenfunctions in the context of Separation of Variables.
We may also re-express $t_{\text{rel}}(\sigma,\vec{\mathcal{E}}) $, $\widetilde{t}_{\text{rel}}(\sigma,\vec{\widetilde{\mathcal{E}}})$, which are polynomials in $w$, $\widetilde{w}$, in terms of auxiliary sets of variables $(a_1, \ldots, a_N)$ or $(\tau_1, \ldots, \tau_N)$ for the open and closed case respectively as
\renewcommand\arraystretch{1.8}
\begin{eqnarray}
\mathcal{E}_N^{-\frac{1}{2}}\,
t_{\text{rel}}(\sigma,\vec{\mathcal{E}})  &=& \left\{ \begin{array}{ll}
\prod_{m=1}^N 2\sinh \left[ \dfrac{\pi}{\omega_2}(\sigma - a_m) \right] \;\;\;\;\Longrightarrow & \;\;\;\; \mathcal{E}_m = \mathcal{E}_m(\vec{a}) \;\;\;\; \text{(open chain)},\\
\prod_{m=1}^N 2\sinh \left[ \dfrac{\pi}{\omega_2}(\sigma - \tau_m) \right]\;\;\;\;\Longrightarrow & \;\;\;\; \mathcal{E}_m = \mathcal{E}_m(\vec{\tau}) \;\;\;\;\; \text{(closed chain)},
\end{array}
\right.  \label{trel} \\ 
\widetilde{\mathcal{E}}_N^{-\frac{1}{2}}\,
\widetilde{t}_{\text{rel}}(\sigma,\vec{\widetilde{\mathcal{E}}})  &=& \left\{ \begin{array}{ll}
\prod_{m=1}^N 2\sinh \left[ \dfrac{\pi}{\omega_1}(\sigma - a_m) \right]\;\;\;\;\Longrightarrow & \;\;\;\; \widetilde{\mathcal{E}}_m = \widetilde{\mathcal{E}}_m(\vec{a}) \;\;\;\; \text{(dual open chain)},\\
\prod_{m=1}^N 2\sinh \left[ \dfrac{\pi}{\omega_1}(\sigma - \tau_m) \right]\;\;\;\;\Longrightarrow & \;\;\;\; \widetilde{\mathcal{E}}_m = \widetilde{\mathcal{E}}_m(\vec{\tau}) \;\;\;\;\; \text{(dual closed chain)}.
\end{array}
\right.  \label{treldual}
\end{eqnarray}
\renewcommand\arraystretch{1}
\hspace{-0.29 cm} The auxiliary variables $a_m$, $\tau_m$ are the same ones we used in the four-dimensional case \eqref{tpol} and are such that $a_1 + \ldots + a_N = \tau_1 + \ldots + \tau_N$; sometimes it will be useful to write them via the combinations
\begin{equation}
 \mu_m = e^{\frac{2\pi a_m}{\omega_2}},\;\;\; \widetilde{\mu}_m = e^{\frac{2\pi a_m}{\omega_1}}, \;\;\;  \delta_m = e^{\frac{2\pi \tau_m}{\omega_2}},\;\;\; \widetilde{\delta}_m = e^{\frac{2\pi \tau_m}{\omega_1}}. \label{egalite}
 \end{equation} 
As in the four-dimensional case, the $\tau_m = \tau_m(\vec{a})$ variables can be thought as functions of the $a_m$ ones and reduce to these when $Q_{\text{rel}}$, $\widetilde{Q}_{\text{rel}}$ are set to zero (i.e. $m_0 \rightarrow \infty$); therefore also the relativistic closed Toda and dual Toda eigenvalues $\mathcal{E}_m = \mathcal{E}_m(\vec{\tau}(\vec{a}))$ and $\widetilde{\mathcal{E}}_m = \widetilde{\mathcal{E}}_m(\vec{\widetilde{\tau}}(\vec{\widetilde{a}}))$ can be thought as functions of the $a_m$'s, and in fact this is the most natural parametrization of the spectrum if we look for a solution constructed via gauge theory.


\subsection{Numerical study of spectrum and eigenfunctions} \label{sec3.2}


Before moving to discuss how the relativistic Toda chain can be solved in the framework of gauge theory, let us pause a moment to perform a numerical study of this system;
numerical eigenvalues and eigenfunctions computed in this Section will be later used to provide some check of the validity of our proposed gauge theory solution.
As explained in Section \ref{2.3.2}, numerical analysis is performed by diagonalizing the Hamiltonian $\widehat{\mathcal{H}}$ of interest in terms of some orthonormal basis in the appropriate Hilbert space; for definiteness we will choose the basis given by the eigenfunctions $\psi_k(x)$ of a harmonic oscillator of mass $m$, frequency $\omega$ and Hamiltonian $\widehat{H} = \frac{p^2}{2m} + \frac{1}{2}m \omega^2 x^2$ (with $[p, x] = - i \hbar$): 
\begin{equation}
\psi_k(x) = \dfrac{1}{\sqrt{2^k k!}} \left( \dfrac{m \omega}{\pi \hbar} \right)^{\frac{1}{4}} e^{-\frac{m \omega x^2}{2\hbar}} H_k \left( \sqrt{\frac{m \omega}{\hbar}} x \right) \,, \;\;\;\; k \geqslant 0. \label{harmonicbasisbis}
\end{equation}
We then compute the matrix elements $\langle \psi_{k_1} \vert \widehat{\mathcal{H}}  \vert \psi_{k_2} \rangle$ up to a certain $k_1$, $k_2$ and evaluate numerically the eigenvalues of this finite-dimensional matrix; these eigenvalues should approach the ones of our Hamiltonian $\widehat{\mathcal{H}}$ when increasing the size of the matrix. Numerical (normalized) eigenfunctions are then obtained by looking at the eigenvectors of this matrix. In actual computations we usually consider $300 \times 300$ matrices and fix $m$ and $\omega$ by expanding $\widehat{\mathcal{H}}$.

\begin{figure}[]
  \centering
  \subfloat[Original Baxter operator]{\includegraphics[width=0.9\textwidth]{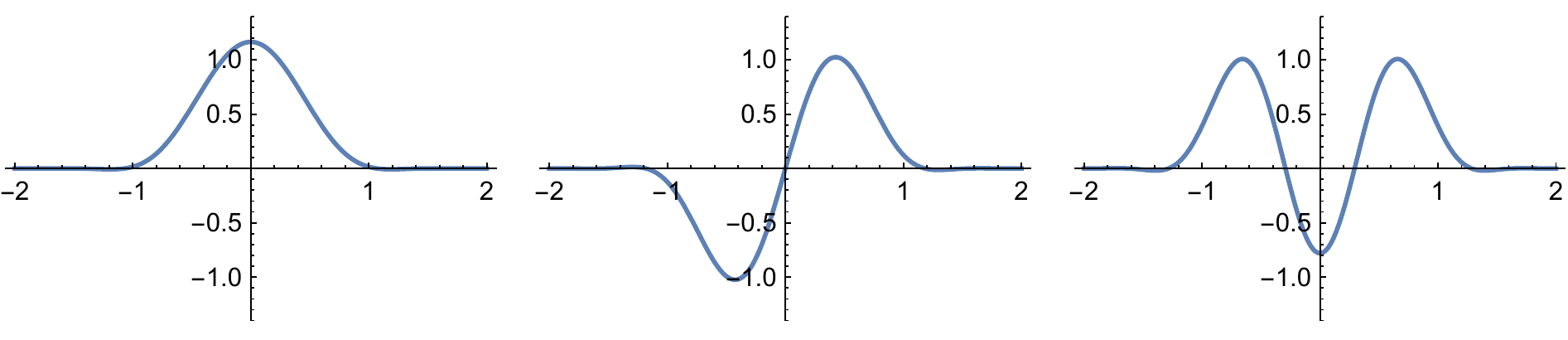}\label{bax5dexnum}}
\newline
  \subfloat[Dual Baxter operator]{\includegraphics[width=0.9\textwidth]{bax5dexnum.pdf}\label{bax5dexnumdual}}
\newline
  \subfloat[Difference]{\includegraphics[width=0.9\textwidth]{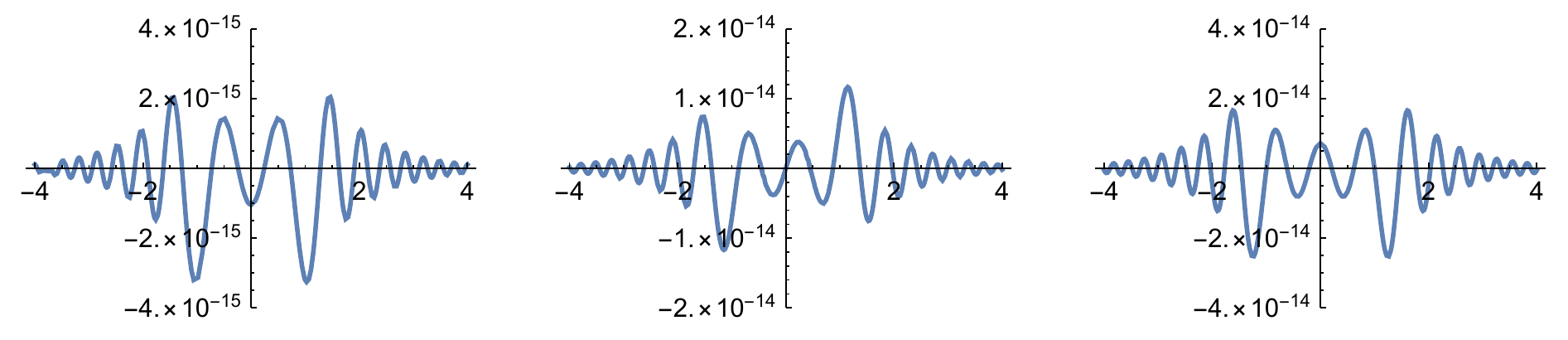}\label{bax5dexnumdiff}}
 \caption{
 Numerical Baxter solution $\overline{q}(\sigma)$ at $\omega_1 = 2^{-1/4}$, $\omega_2 = 2^{1/4}$, $m_0 = 0$.}
 \label{plotexnum}
\end{figure}

\begin{figure}[]
  \centering
  \subfloat[Original Baxter operator]{\includegraphics[width=0.9\textwidth]{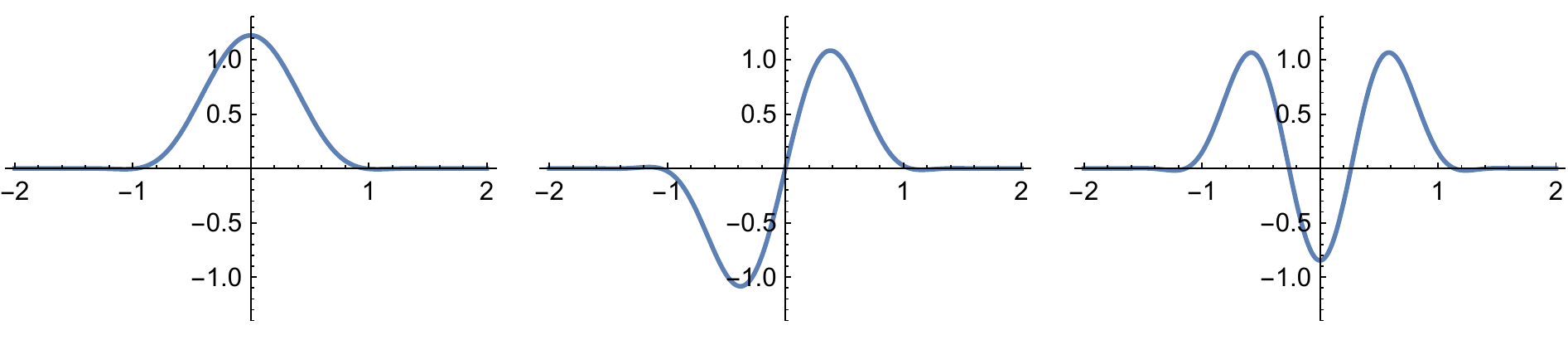}\label{bax5dmodnum}}
\newline
  \subfloat[Dual Baxter operator]{\includegraphics[width=0.9\textwidth]{bax5dmodnum.pdf}\label{bax5dmodnumdual}}
\newline
  \subfloat[Difference]{\includegraphics[width=0.9\textwidth]{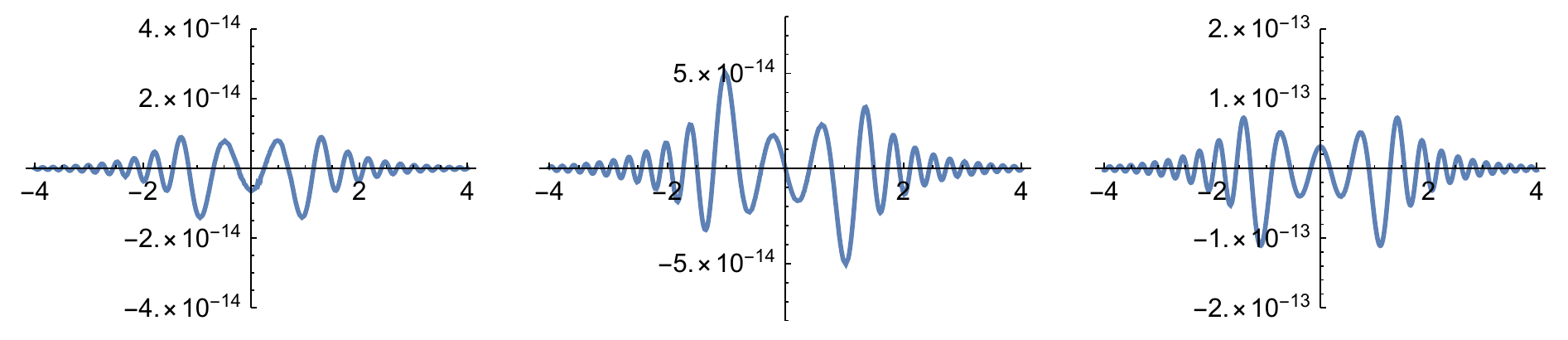}\label{bax5dmodnumdiff}}
 \caption{
 Numerical Baxter solution $\overline{q}(\sigma)$ at $\omega_1 = \frac{1}{\sqrt{2}}$, $\omega_2 = 1$, $m_0 = -\frac{\ln 3}{2 \pi}$.}
 \label{plotmodnum}
\end{figure}

Let us show how this works in the case of a relativistic 2-particle Toda chain with center of mass decoupled (that is $a_1 = - a_2 = \frac{a}{2}$ or $\mu_1 = \mu_2^{-1} = \mu^{1/2}$); as for the non-relativistic case, we will be interested in two different operators:
\begin{itemize}
\item The first one is the operator associated to the Baxter equation for the system and its dual version (see \eqref{baxclosedrel} in Section \ref{sec3.4}), which can be obtained by quantizing the spectral curves \eqref{curverel}; after a slight redefinition of variables these reduce to
\begin{subequations} \label{Bax5d2Toda}
\begin{align}
& \left[ e^{i \omega_1 \partial_{\sigma}} + Q_{\text{rel}} e^{-i \omega_1 \partial_{\sigma}} + e^{\frac{2\pi \sigma}{\omega_2}} + e^{-\frac{2\pi \sigma}{\omega_2}} - \mathcal{E}_1 \right] q(\sigma, a, \omega_1, \omega_2, m_0) = 0, \label{Bax5d2Todaa} \\
& \left[ e^{i \omega_2 \partial_{\sigma}} + \widetilde{Q}_{\text{rel}} e^{-i \omega_2 \partial_{\sigma}} + e^{\frac{2\pi \sigma}{\omega_1}} + e^{-\frac{2\pi \sigma}{\omega_1}} - \widetilde{\mathcal{E}}_1\right] q(\sigma, a, \omega_1, \omega_2, m_0) = 0, \label{Bax5d2Todab}
\end{align}
\end{subequations}
where the energies $\mathcal{E}_1$, $\widetilde{\mathcal{E}}_1$ depend on $\omega_1$, $\omega_2$, $m_0$ and $a$ ($a$ will be discretized because of the quantization conditions). The two Baxter operators are related by the exchange $\omega_1 \leftrightarrow \omega_2$ and commute with each other; the solution $q(\sigma) \in L^2(\mathbb{R})$ should then be their simultaneous eigenfunction and as such should be symmetric under $\omega_1 \leftrightarrow \omega_2$, and is also expected to be an entire function based on the analogy with the non-relativistic chain. For numerical computations it is actually more convenient to consider the function 
\begin{equation}
\overline{q}(\sigma, a, \omega_1, \omega_2, m_0) = e^{-\frac{i \pi m_0 \sigma}{\omega_1 \omega_2}} q(\sigma, a, \omega_1, \omega_2, m_0),
\end{equation}
which is a simultaneous eigenfunction of the more symmetric equations
\begin{subequations} \label{Bax5d2Todamod}
\begin{align}
& \left[ Q_{\text{rel}}^{\frac{1}{2}} \left(e^{i \omega_1 \partial_{\sigma}} + e^{-i \omega_1 \partial_{\sigma}}\right) + e^{\frac{2\pi \sigma}{\omega_2}} + e^{-\frac{2\pi \sigma}{\omega_2}} - \mathcal{E}_1 \right] \overline{q}(\sigma, a, \omega_1, \omega_2, m_0) = 0, \label{Bax5d2Todamoda} \\
& \left[ \widetilde{Q}_{\text{rel}}^{\frac{1}{2}} \left(e^{i \omega_2 \partial_{\sigma}} + e^{-i \omega_2 \partial_{\sigma}}\right) + e^{\frac{2\pi \sigma}{\omega_1}} + e^{-\frac{2\pi \sigma}{\omega_1}} - \widetilde{\mathcal{E}}_1 \right] \overline{q}(\sigma, a, \omega_1, \omega_2, m_0) = 0. \label{Bax5d2Todamodb}
\end{align}
\end{subequations}
Expanding \eqref{Bax5d2Todamod} and comparing with the harmonic oscillator Hamiltonian we can fix the values of $m$, $\omega$, $\hbar$ to be used in numerical computations:
\begin{equation}
\begin{split}
& -\omega_1^2 Q_{\text{rel}}^{\frac{1}{2}} \partial_{\sigma}^2  + \frac{4\pi^2}{\omega_2^2} \sigma^2 = \mathcal{E}_1 \;\;\;\Longrightarrow \;\;\; m = \frac{1}{2} Q_{\text{rel}}^{-\frac{1}{2}},\;\; \omega = \frac{4 \pi}{\omega_2} Q_{\text{rel}}^{\frac{1}{4}},\;\; \hbar = \omega_1, 
\\
& -\omega_2^2 \widetilde{Q}_{\text{rel}}^{\frac{1}{2}} \partial_{\sigma}^2  + \frac{4\pi^2}{\omega_1^2} \sigma^2 =  \widetilde{\mathcal{E}}_1\;\;\;\Longrightarrow \;\;\; m = \frac{1}{2} \widetilde{Q}_{\text{rel}}^{-\frac{1}{2}},\;\; \omega = \frac{4 \pi}{\omega_1} \widetilde{Q}_{\text{rel}}^{\frac{1}{4}},\;\; \hbar = \omega_2.
\end{split}
\end{equation}
Examples of numerical results for the energy and dual energy of the ground state and the first two excited states of equations \eqref{Bax5d2Todamod} at various values of $\omega_1$, $\omega_2$, $m_0$ can be found in Table \ref{numenergy5d}, while Figure 
\ref{plotexnum}, \ref{plotmodnum} 
show the corresponding eigenfunctions. As we can see from the plots, already by considering $300 \times 300$ matrices there is a very small difference between the numerical eigenfunction obtained from diagonalizing \eqref{Bax5d2Todamoda} and the one obtained by diagonalizing \eqref{Bax5d2Todamodb}, supporting the expectation that there should be only one common eigenfunction $\overline{q}(\sigma)$ symmetric under the exchange $\omega_1 \leftrightarrow \omega_2$.\footnote{The self-dual point $\omega_1 = \omega_2 = 1$, $m_0 = 0$ is somewhat special since in this case the Baxter equation and its dual coincide.}

\renewcommand\arraystretch{1.5}
\begin{table}[]
\begin{center}
\begin{tabular}{ccc}
\hline 
\rule[-1ex]{0pt}{2.5ex} &  
$\renewcommand\arraystretch{1.1}\begin{array}{c} \omega_1 = 2^{-1/4},\, \omega_2 = 2^{1/4} \\ m_0 = 0 \end{array}\renewcommand\arraystretch{1.5}$ & 
$\renewcommand\arraystretch{1.1}\begin{array}{c} \omega_1 = \frac{1}{\sqrt{2}},\, \omega_2 = 1 \\ m_0 = -\frac{\ln 3}{2\pi} \end{array}\renewcommand\arraystretch{1.5}$ \\ 
\hline 
\rule[-1ex]{0pt}{2.5ex} $\ln \mathcal{E}_1^{(0)}$& 2.4605242719\ldots & 2.7528481019\ldots \\ 
\rule[-1ex]{0pt}{2.5ex} $\ln \widetilde{\mathcal{E}}_1^{(0)}$ & 3.4605592909\ldots & 3.8720036669\ldots \\ 
\hline 
\rule[-1ex]{0pt}{2.5ex} $\ln \mathcal{E}_1^{(1)}$& 3.5984708772\ldots & 3.8838346782\ldots \\ 
\rule[-1ex]{0pt}{2.5ex} $\ln \widetilde{\mathcal{E}}_1^{(1)}$ & 5.0869593531\ldots & 5.4902688202\ldots \\ 
\hline 
\rule[-1ex]{0pt}{2.5ex} $\ln \mathcal{E}_1^{(2)}$ & 4.4628893132\ldots & 4.7460288538\ldots \\ 
\rule[-1ex]{0pt}{2.5ex} $\ln \widetilde{\mathcal{E}}_1^{(2)}$ & 6.3111090796\ldots & 6.7114798487\ldots \\ 
\hline 
\end{tabular} \caption{Numerical eigenvalues $\mathcal{E}_1^{(n)}$, $\widetilde{\mathcal{E}}_1^{(n)}$ of \eqref{Bax5d2Todamod} at level $n = 0,1,2$.
} \label{numenergy5d}
\end{center} 
\end{table}
\renewcommand\arraystretch{1}


\item The second one is the operator corresponding to the relativistic 2-particle Hamiltonian $\widehat{\mathcal{H}}_1$ itself \eqref{hamrel}, together with its dual $\widehat{\widetilde{\mathcal{H}}}_1$ \eqref{hamreldual}; after a redefinition of variables the spectral problem \eqref{sptodarel} reads
\begin{subequations} \label{5d2Toda}
\begin{align}
& \left[ e^{i \omega_1 \partial_{x}} + e^{-i \omega_1 \partial_{x}} + e^{\frac{2\pi x}{\omega_2}} + Q_{\text{rel}} e^{-\frac{2\pi x}{\omega_2}} - \mathcal{E}_1 \right] \Psi^{\text{rel}}_a(x, \omega_1, \omega_2, m_0) = 0, \label{5d2Todaa} \\
& \left[ e^{i \omega_2 \partial_{x}} + e^{-i \omega_2 \partial_{x}} + e^{\frac{2\pi x}{\omega_1}} + \widetilde{Q}_{\text{rel}} e^{-\frac{2\pi x}{\omega_1}} - \widetilde{\mathcal{E}}_1\right] \Psi^{\text{rel}}_a(x, \omega_1, \omega_2, m_0) = 0. \label{5d2Todab}
\end{align}
\end{subequations}
Clearly, problems \eqref{Bax5d2Toda} and \eqref{5d2Toda} coincide when $m_0 = 0$; moreover, since in the special $N = 2$ case \eqref{5d2Toda} is simply the Fourier-transformed version of \eqref{Bax5d2Toda}, the energies $\mathcal{E}_1$, $\widetilde{\mathcal{E}}_1$ are the same as for the Baxter problem even for $m_0 \neq 0$. Our $\Psi^{\text{rel}}_a(x, \omega_1, \omega_2, m_0) \in L^2(\mathbb{R})$ should be a simultaneous eigenfunction of \eqref{5d2Todaa}, \eqref{5d2Todab} since these operators commute, and as such is expected to be symmetric under $\omega_1 \leftrightarrow \omega_2$. For numerical computations it may be more convenient to work in terms of the variable $\overline{x} = x + \frac{m_0}{2}$ which makes the problem more symmetric. The numerical spectrum obtained by diagonalizing \eqref{5d2Todaa}, \eqref{5d2Todab} in the harmonic oscillator basis coincides with the one in Table \ref{numenergy5d} as expected; numerical  eigenfunctions (symmetric with respect to $\overline{x}$) for the ground state and the first two excited states are as in Figure 
\ref{plotexnum} in the example where $m_0 = 0$, while in the example with $m_0 = -\frac{\ln 3}{2 \pi}$ they are plotted in Figure \ref{plotTodamodnum}. 
\end{itemize}
The $N$-particle problem can be approached numerically in a similar way (see \cite{2015arXiv151102860H} for the $N = 3$ case and \cite{Franco:2015rnr} for a related computation concerning other finite-difference systems). 

\begin{figure}[]
  \centering
  \subfloat[Original Toda operator]{\includegraphics[width=0.9\textwidth]{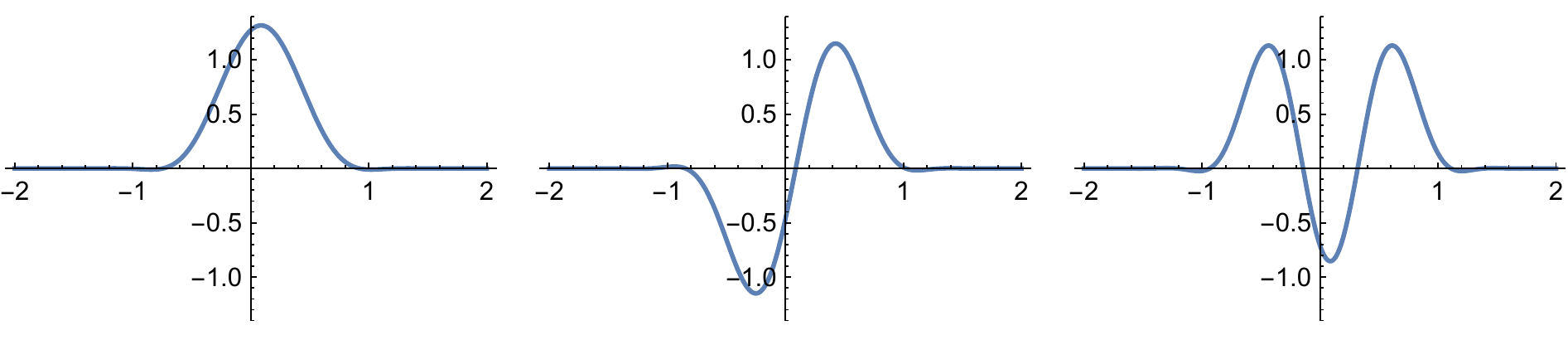}\label{Toda5dmodnum}}
\newline
  \subfloat[Dual Toda operator]{\includegraphics[width=0.9\textwidth]{Toda5dmodnum.pdf}\label{Toda5dmodnumdual}}
\newline
  \subfloat[Difference]{\includegraphics[width=0.9\textwidth]{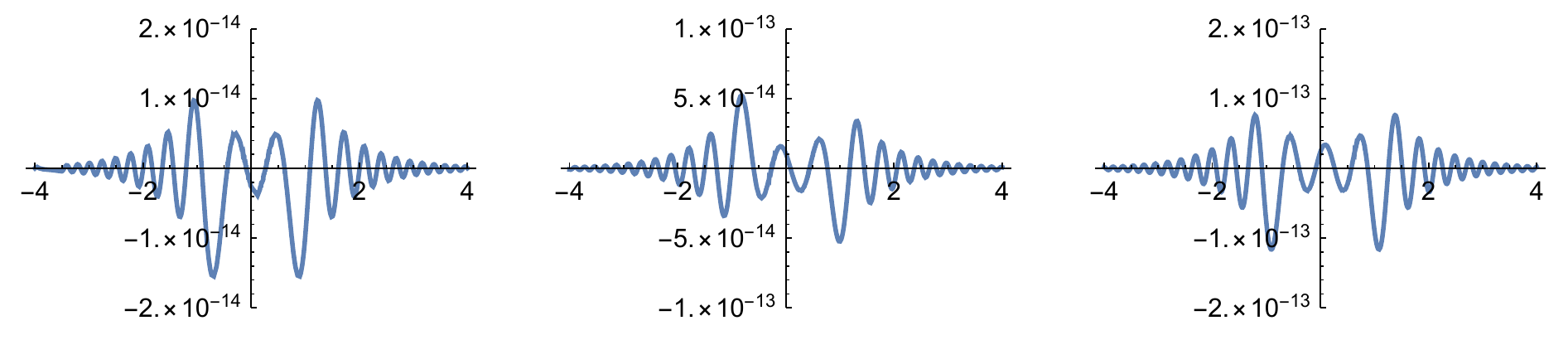}\label{Toda5dmodnumdiff}}
  \caption{
  Numerical Toda solution $\Psi_a(x)$ at $\omega_1 = \frac{1}{\sqrt{2}}$, $\omega_2 = 1$, $m_0 = -\frac{\ln 3}{2 \pi}$. }
 \label{plotTodamodnum}
\end{figure}


\subsection{Solution via gauge theory: quantization conditions and energy spectrum} \label{sec3.3}

As quickly reviewed in Section \ref{sec2.3}, the $N$-particle (non-relativistic) closed Toda chain can be completely solved in terms of various observables of the four-dimensional $\mathcal{N} = 2$ $SU(N)$ Yang-Mills theory on $\mathbb{R}^4_{\epsilon_1, \epsilon_2}$ (or $D_{\epsilon_1} \times \mathbb{R}^2_{\epsilon_2}$) in the NS limit $\epsilon_1 = i \hbar$, $\epsilon_2 = 0$ (with $\hbar \in \mathbb{R}_+$). 
Schematically, the dictionary between Toda chain quantities and four-dimensional gauge theory observables is as follows: 

\renewcommand\arraystretch{1.5}
\begin{table}[h]
\begin{center}
\begin{tabular}{ccc}
Quantization conditions & $\Longleftrightarrow$ & Extremization of  $\mathcal{W}_{\text{4d}}(\vec{a}, i\hbar, Q_{\text{4d}})$ (NS limit) \\
Energy spectrum $E_m$ & $\Longleftrightarrow$ &  Codimension 4 defects $\langle \text{Tr}\, \phi^m \rangle_{\text{NS}}$ (NS limit) \\ 
Baxter eigenfunction $q(\sigma)$ & $\Longleftrightarrow$ &  Codimension 2 defects (type II, NS limit) \\
Toda eigenfunction $\Psi_{\vec{a}}(\vec{x})$ & $\Longleftrightarrow$ &  Codimension 2 defects (type I, NS limit) \\  
\end{tabular} 
\end{center} 
\end{table}
\renewcommand\arraystretch{1}
\vspace{-0.5 cm}
\noindent It is then a natural question to ask if it is possible to find a solution for the $N$-particle relativistic closed Toda chain via similar gauge theory arguments. In this Section we will discuss what gauge theory can say about quantization conditions and energy spectrum, while leaving the study of the eigenfunctions to Section \ref{sec3.4}. 

\subsubsection{The naive proposal: five-dimensional gauge theory on flat space} \label{3.3.1}

A first attempt to answer this question was done in \cite{2010maph.conf..265N}, where it was suggested (based on previous observations \cite{Nekrasov:1996cz}) that the relativistic version of the Toda chain may have something to do with the five-dimensional uplift of the pure $SU(N)$ theory: 
more precisely, it was proposed to consider the five-dimensional $\mathcal{N} = 1$ $SU(N)$ Yang-Mills theory on flat space $\mathbb{R}^4_{\epsilon_1, \epsilon_2} \times S^1_R$ in the NS limit $\epsilon_1 = i \omega_1$, $\epsilon_2 = 0$ with $\omega_1 \in \mathbb{R}_+$. The radius $R = \omega_2^{-1} \in \mathbb{R}_+$ of the extra circle will play the role of the (inverse) speed of light, so that when $R \rightarrow 0$ we recover the Toda chain discussed in Section \ref{sec2}. The details of the proposal are the same as in the four-dimensional case. Let us start by fixing notations: in the following we will denote
\begin{equation}
q_1 = e^{2 \pi R \epsilon_1}, \;\;\; q_2 = e^{2 \pi R \epsilon_2}, \;\;\; Q_{\text{5d}} = e^{-2\pi R m_0}, \;\;\; \mu_m = e^{2 \pi R a_m} \;\;(m = 1, \ldots, N),
\end{equation}
where $a_m$ are the vacuum expectation values of (the Cartan part of) the adjoint scalar field in the $\mathcal{N} = 1$ vector multiplet and $m_0 = 4\pi^2/g^2_{\text{5d}}$ is related to the five-dimensional Yang-Mills coupling constant. We can now consider the partition function on $\mathbb{R}^4_{\epsilon_1, \epsilon_2} \times S^1_R$:
\begin{equation}
Z_{\text{5d}}(\vec{\mu}, q_1, q_2, Q_{\text{5d}}) = 
Z_{\text{5d}}^{\text{pert}}(\vec{\mu}, q_1, q_2, Q_{\text{5d}})
Z_{\text{5d}}^{\text{inst}}(\vec{\mu}, q_1, q_2, Q_{\text{5d}}). \label{pf5d}
\end{equation}
Sometimes it will also be useful to rewrite this partition function as
\begin{equation}
Z_{\text{5d}}(\vec{\mu}, q_1, q_2, Q_{\text{5d}}) = \exp \big(- F_{\text{5d}}(\vec{\mu}, q_1, q_2, Q_{\text{5d}})\big),
\end{equation}
where the prepotential
\begin{equation}
F_{\text{5d}}(\vec{\mu}, q_1, q_2, Q_{\text{5d}}) = 
F^{0}_{\text{5d}}(\vec{a}, \epsilon_1, \epsilon_2, R, m_0) + 
F^{\text{BPS}}_{\text{5d}}(\vec{\mu}, q_1, q_2, Q_{\text{5d}}) \label{prepotential5d}
\end{equation}
coincides with the refined closed topological string prepotential resummed \`{a} la Gopakumar-Vafa 
once expanded around large Kahler parameters $t_m$, $m = 1, \ldots, N$ given by
\begin{equation}
t_i = a_i - a_{i+1} \;\;\;(i = 1, \ldots, N-1) \;\;\;\;\;\; \text{and} \;\;\;\;\;\; t_N = m_0 + \sum_{i = 1}^{N - 1} t_i \,;\label{kahler}
\end{equation}
the $F^{0}_{\text{5d}}$ term is a cubic polynomial in $\vec{a}$, $m_0$, while the $F^{\text{BPS}}_{\text{5d}}$ term only depends on the number of BPS states $N_{j_L, j_R}^{\mathbf{d}}$ of given left, right spins $j_L$, $j_R$ of our theory and can be expressed as 
\begin{equation}
\begin{split}
F_{\text{5d}}^{\text{BPS}} & = \sum_{n \geqslant 1} \sum_{\mathbf{d} > \mathbf{0}} \sum_{j_L, j_R} \sum_{s_L = -j_L}^{j_L} \sum_{s_R = -j_R}^{j_R} 
\dfrac{N_{j_L,j_R}^{\mathbf{d}} }{n}
\dfrac{q_1^{n(s_R + s_L +1/2)}q_2^{n(s_R - s_L + 1/2)}}{\left(1 - q_1^{n}\right)\left(1 - q_2^{n}\right)}
\mathbf{Q}^{n \mathbf{d}}, \label{BPS}
\end{split}
\end{equation}
where we used the notation $\mathbf{d} = (d_1, \ldots, d_N)$ and $\mathbf{Q} = (Q_1, \ldots, Q_N)$ with $Q_i = e^{-2\pi R t_i}$.
The NS limit of the partition function can be used to define the five-dimensional version of the twisted effective superpotential $\mathcal{W}_{\text{5d}}(\vec{\mu}, q_1, Q_{\text{5d}})$ as
\begin{equation}
\mathcal{W}_{\text{5d}}(\vec{\mu}, q_1, Q_{\text{5d}}) = \lim_{\epsilon_2 \rightarrow 0}\left[ -\epsilon_2 \log Z_{\text{5d}}(\vec{\mu}, q_1, q_2, Q_{\text{5d}})\right],  \label{aziza}
\end{equation}
which is also equal to the NS limit of the refined closed topological string prepotential:
\begin{equation}
\mathcal{W}_{\text{5d}}(\vec{\mu}, q_1, Q_{\text{5d}}) = F_{\text{5d}}^{\text{NS}}(\vec{\mu}, q_1, Q_{\text{5d}})  = \lim_{\epsilon_2 \rightarrow 0} \left[ \epsilon_2 F_{\text{5d}}(\vec{\mu}, q_1, q_2, Q_{\text{5d}})\right]. \label{aziza2}
\end{equation}
The twisted effective superpotential can be separated into its perturbative (classical + 1-loop) and instanton part
\begin{equation}
\mathcal{W}_{\text{5d}}(\vec{\mu}, q_1, Q_{\text{5d}}) = \mathcal{W}_{\text{5d}}^{\text{pert}}(\vec{\mu}, q_1, Q_{\text{5d}})  + \mathcal{W}_{\text{5d}}^{\text{inst}}(\vec{\mu}, q_1, Q_{\text{5d}}), \label{split}
\end{equation}
or can also be divided as
\begin{equation}
\begin{split}
\mathcal{W}_{\text{5d}}(\vec{\mu}, q_1, Q_{\text{5d}}) &= 
\mathcal{W}_{\text{5d}}^{0}(\vec{a}, \epsilon_1, R, m_0) + 
\mathcal{W}_{\text{5d}}^{\text{BPS}}(\vec{\mu}, q_1, Q_{\text{5d}}) \\[5 pt]
& = F_{\text{5d}}^{\text{NS}, 0}(\vec{a}, \epsilon_1, R, m_0) + 
F_{\text{5d}}^{\text{NS}, \text{BPS}}(\vec{\mu}, q_1, Q_{\text{5d}}), \label{split2}
\end{split}
\end{equation}
where (from \eqref{BPS})
\begin{equation}
\begin{split}
& \mathcal{W}_{\text{5d}}^{\text{BPS}}(\vec{\mu}, q_1, Q_{\text{5d}}) = F_{\text{5d}}^{\text{NS}, \text{BPS}}(\vec{\mu}, q_1, Q_{\text{5d}}) = \\[5 pt]
& = \dfrac{1}{4\pi R} \sum_{n \geqslant 1} \sum_{\mathbf{d} > \mathbf{0}} \sum_{j_L, j_R} \dfrac{N_{j_L, j_R}^{\mathbf{d}}}{n^2} \dfrac{\sinh \left[ (2j_L+1)\pi n R \epsilon_1 \right] \sinh \left[ (2j_R+1)\pi n R \epsilon_1 \right]}{\sinh^3 \left[\pi n R \epsilon_1 \right]}
\mathbf{Q}^{n \mathbf{d}}. \label{BPSNS}
\end{split}
\end{equation}
At this point if we identify 
\begin{equation}
\epsilon_1 = i \omega_1, \;\;\;R = \omega_2^{-1}, \;\;\;q_1 = q, \;\;\;\frac{4\pi^2}{g^2_{\text{5d}}} = m_0, \;\;\;Q_{\text{5d}} = Q_{\text{rel}}, \label{lelouch}
\end{equation}
and $\vec{\mu}$ with the parameters appearing in \eqref{egalite}, the quantization conditions for the relativistic Toda chain proposed by \cite{2010maph.conf..265N} would be equivalent to the set of supersymmetric vacua equations
\begin{equation}
\exp\left( - \dfrac{\partial}{\partial a_m} \mathcal{W}_{\text{5d}}(\vec{\mu}, q, Q_{\text{rel}}) \right) = 1 \;,\;\;\; m = 1, \ldots, N,
\end{equation}
or alternatively
\begin{equation}
\dfrac{\partial}{\partial a_m} \mathcal{W}_{\text{5d}}(\vec{\mu}, q, Q_{\text{rel}}) = 2\pi i n_m \;, \;\;\; n_m \in \mathbb{Z},\;\;\; m = 1, \ldots, N. \label{falseqc}
\end{equation}
Moreover, always according to \cite{2010maph.conf..265N} the energy spectrum $\mathcal{E}_m$ of the $m$-th Hamiltonian $\widehat{\mathcal{H}}_m$ at level $(\vec{n}) = (n_1, \ldots, n_N)$ should correspond to the NS limit of the vacuum expectation value of a Wilson loop $W^{SU(N)}_{\Lambda^m}$ in the $m$-th antisymmetric representation $\Lambda^m$ wrapping $S^1_R$:
\begin{equation}
\mathcal{E}_m^{(\vec{n})} = W_{\Lambda^m,\,\text{NS}}^{SU(N), (\vec{n})}. \label{wilNS}
\end{equation} 
Finally, although not discussed in \cite{2010maph.conf..265N} we could expect from what happens in the four-dimensional case that eigenfunctions of the relativistic Toda Hamiltonians or of the associated Baxter equation should be given by the NS limit of the five-dimensional partition function in the presence of codimension two defects of type I (Figure \ref{TSUN}) or II (Figure \ref{freechiral}) living on $D_{\epsilon_1} \times S^1_R$ (or $\mathbb{R}^2_{\epsilon_1} \times S^1_R$ if we sum over all vacua). 
For example, if we define
\begin{equation}
w = e^{-2\pi R \sigma},
\end{equation}
we may expect the type II defect partition function with chiral multiplets 
\begin{equation}
Z_{\text{3d/5d}}^{(\text{c})}(w, \vec{\mu}, q_1, q_2, Q_{\text{5d}}) = 
Z_{\text{3d/5d}}^{(\text{c}),\text{pert}}(w, \vec{\mu}, q_1, q_2, Q_{\text{5d}})
Z_{\text{3d/5d}}^{(\text{c}),\text{inst}}(w, \vec{\mu}, q_1, q_2, Q_{\text{5d}}) \label{coso}
\end{equation}
to satisfy the relativistic closed Toda chain Baxter equation (see \eqref{baxclosedrel} in Section \ref{sec3.4}).
It is sometimes useful to rewrite the type II defect partition function as
\begin{equation}
Z_{\text{3d/5d}}^{(\text{c})}(w, \vec{\mu}, q_1, q_2, Q_{\text{5d}}) = 
\exp\big(-F^{(\text{c})}_{\text{3d/5d}}(w, \vec{\mu}, q_1, q_2, Q_{\text{5d}})\big),
\end{equation}
where the prepotential
\begin{equation}
F^{(\text{c})}_{\text{3d/5d}}(w, \vec{\mu}, q_1, q_2, Q_{\text{5d}}) = 
F^{(\text{c}),0}_{\text{3d/5d}}(\sigma, \vec{a}, \epsilon_1, \epsilon_2, R, m_0) + 
F^{(\text{c}),\text{BPS}}_{\text{3d/5d}}(w, \vec{\mu}, q_1, q_2, Q_{\text{5d}})
\end{equation}
corresponds to the refined open topological string prepotential resummed \`{a} la Gopakumar-Vafa once expanded in the appropriate closed ($t_i$) and open ($\sigma$) 
Kahler moduli; the $F^{(\text{c}),0}_{\text{3d/5d}}$ term is a quadratic polynomial in $\vec{a}$, $m_0$, $\sigma$ while $F^{(\text{c}),\text{BPS}}_{\text{3d/5d}}$ only depends on the number of open BPS states $D^{s_1, s_2}_{m,\mathbf{d}}$ and reads \cite{Aganagic:2011sg,Kashani-Poor:2016edc}\footnote{The standard definition of the open topological string partition function does not contain the term $m = 0$ in the sum, which is a constant in $\sigma$ (or $w$); we will however include this constant term since it naturally appears from gauge theory formulae.}
\begin{equation}
F^{(\text{c}),\text{BPS}}_{\text{3d/5d}} = \sum_{n = 1}^{\infty} \sum_{\mathbf{d}} \sum_{m \in \mathbb{Z}} \sum_{s_1, s_2} D^{s_1, s_2}_{m, \mathbf{d}} \dfrac{q_1^{n s_1} q_2^{n s_2}}{n(1-q_1^n)} \mathbf{Q}^{n \mathbf{d}} w^{m n},
\end{equation}
which in the NS limit $\epsilon_2 \rightarrow 0$ reduces to
\begin{equation}
F^{(\text{c}),\text{BPS}}_{\text{3d/5d}, \text{NS}} = \sum_{n = 1}^{\infty} \sum_{\mathbf{d}} \sum_{m \in \mathbb{Z}} \sum_{s_1} D^{s_1}_{m, \mathbf{d}} \dfrac{q_1^{n s_1}}{n(1-q_1^n)} \mathbf{Q}^{n \mathbf{d}} w^{m n}, \label{opentop}
\end{equation}
where $D^{s_1}_{m, \mathbf{d}} = \sum_{s_2} D^{s_1, s_2}_{m, \mathbf{d}}$.

As we can see, the proposal by \cite{2010maph.conf..265N} is nothing else than the naive five-dimensional uplift of the four-dimensional gauge theory solution of the (non-relativistic) Toda chain which we reviewed in Section \ref{sec2.3}, and reduces to it in the limit $R \rightarrow 0$\footnote{This limit involves some scaling for $Q_{\text{5d}}$.}. However, it turns out that this naive proposal is incorrect, or at least incomplete.
For example, as shown in \cite{Grassi:2014zfa} the ``naive'' quantization conditions \eqref{falseqc} cannot be the correct ones: in fact it is easy to see from \eqref{BPSNS} that these are divergent when $\omega_1$, $\omega_2$ $\in \mathbb{R}_+$ with $\omega_1/\omega_2 \in \mathbb{Q}_+$, and even if we use these quantization conditions for $\omega_1/\omega_2$ irrational the energy spectrum we obtain from \eqref{wilNS} does not match with numerical computation of the eigenvalues \cite{Grassi:2014zfa}.
Similarly the ``naive'' Baxter eigenfunction \eqref{coso}, although formally satisfying the Baxter equation \cite{Gaiotto:2014ina,Bullimore:2014awa}, cannot be the correct one since it diverges at $\omega_1/\omega_2 \in \mathbb{Q}_+$; finally, the proposal by \cite{2010maph.conf..265N} only focusses on the relativistic Toda operators $\widehat{\mathcal{H}}_m$ without taking into account the existence of the modular dual operators $\widehat{\widetilde{\mathcal{H}}}_m$ \cite{Kashani-Poor:2016edc,Sciarappa:2016ctj}. 
All these problems imply that if we want to solve the relativistic closed Toda system by means of gauge theory, we should consider something more elaborated than $\mathcal{N} = 1$ $SU(N)$ theories on $\mathbb{R}^4_{\epsilon_1, \epsilon_2} \times S^1_R$ in the NS limit. 

\subsubsection{Exact quantization conditions and spectrum} \label{3.3.2}

Exact quantization conditions for the 2-particle relativistic closed Toda chain have been proposed in \cite{Grassi:2014zfa}; these conditions arise from requiring a certain spectral determinant to vanish, and share some similarity with \eqref{tauzero} (in fact they reduce to \eqref{tauzero} in an appropriate four-dimensional limit \cite{Bonelli:2016idi}). This spectral determinant can be written in terms of a particular combination of both the unrefined ($\epsilon_1 = -\epsilon_2 = i\hbar$) and the NS limits of the five-dimensional partition function \eqref{pf5d}; the key property of this combination is that it involves \textit{non-perturbative} (in $\hbar$) contributions, i.e. terms of the form $e^{1/\hbar}$, chosen in such a way that the resulting quantization conditions are free from divergences at $\omega_1/\omega_2 \in \mathbb{Q}_+$. When substituted in \eqref{wilNS} (modulo subtleties which will be discussed later), the values of $a_m$ solving these exact quantization conditions provide an energy spectrum that matches with the numerical one. 

It has later been realized in \cite{Wang:2015wdy} (and proven in \cite{Grassi:2016nnt}) that the exact quantization conditions for the 2-particle relativistic Toda chain proposed by \cite{Grassi:2014zfa} can be equivalently expressed in a different way which only involves the NS limit of the partition function \eqref{pf5d}; this different expression can be written as
\begin{equation}
\dfrac{\partial}{\partial a_m} \left[ \mathcal{W}^{0}(\vec{a}, m_0, \omega_1, \omega_2) + \mathcal{W}^{\text{BPS}}(\vec{\mu}, q, Q_{\text{rel}}) + \mathcal{W}^{\text{BPS}}(\vec{\widetilde{\mu}}, \widetilde{q}, \widetilde{Q}_{\text{rel}}) \right] = 2 \pi i n_m. \label{sdual}
\end{equation}
For what the general $N$-particle relativistic Toda chain is concerned, exact quantization conditions in the form \eqref{sdual} and its spectrum were studied in \cite{2015arXiv151102860H} and were checked to match with numerical results for $N = 3$. 
Here the function $\mathcal{W}^{0}$ is a cubic polynomial in $\vec{a}$, $m_0$ which contains $\mathcal{W}_{5d}^{0}$ in \eqref{split2} as well as some additional piece, while the function $\mathcal{W}^{\text{BPS}}$ basically coincides with $\mathcal{W}_{5d}^{\text{BPS}}$ in \eqref{split2} modulo a correction due to the $B$-field, i.e. a vector $\mathbf{B}$ such that 
\begin{equation}
(-1)^{\mathbf{B} \cdot \mathbf{d}} = (-1)^{2j_L + 2j_R + 1} \label{magic}
\end{equation}
for all those $\mathbf{d}$, $j_L$, $j_R$ whose BPS invariant $N^{\mathbf{d}}_{j_L, j_R}$ is non-zero; more precisely, with such a modification we have 
\begin{equation}
\mathcal{W}^{\text{BPS}} =  
\dfrac{\omega_2}{4\pi i} \sum_{n \geqslant 1}\sum_{j_L, j_R} \sum_{\mathbf{d} > \mathbf{0}} \dfrac{(-1)^{(2j_L + 2j_R + 1)n}}{n^2} N_{j_L, j_R}^{\mathbf{d}} \dfrac{\sin \left[ (2j_L+1)\pi n \frac{\omega_1}{\omega_2} \right] \sin \left[ (2j_R+1)\pi n \frac{\omega_1}{\omega_2} \right]}{\sin^3 \left[\pi n \frac{\omega_1}{\omega_2} \right]}
\mathbf{Q}^{n \mathbf{d}}  \label{withB}
\end{equation}
instead of \eqref{BPSNS}. The $B$-field is crucial in order to ensure poles cancellation in the quantization conditions \eqref{sdual}, and for the five-dimensional $\mathcal{N} = 1$ $SU(N)$ theories of interest to us it can be chosen to be
\begin{equation}
\mathbf{B} = \left\{
\begin{array}{ll}
(0, 0, \ldots, 0, 0), & \;\;N \;\,\text{even},\\
(0, 0, \ldots, 0, 1), & \;\;N \;\,\text{odd}.
\end{array} \right.
\end{equation}
The $B$-field is therefore irrelevant for $N$ even, while when $N$ is odd it has the effect of changing sign $Q_{\text{5d}} \rightarrow -Q_{\text{5d}}$, $\widetilde{Q}_{\text{5d}} \rightarrow -\widetilde{Q}_{\text{5d}}$ in \eqref{BPSNS}; alternatively, because of \eqref{magic} one can think of the effect of the $B$-field not as changing the sign of $Q_{\text{5d}}$, $\widetilde{Q}_{\text{5d}}$ but as identifying $\epsilon_1 = i(\omega_1 + \omega_2)$ instead of $\epsilon_1 = i\omega_1$ in \eqref{lelouch}.
We can roughly think of \eqref{sdual} as the naive quantization conditions \eqref{falseqc} corrected by a non-perturbative contribution which depends on tilded variables. The remarkable property of \eqref{sdual} is its \textit{$S$-duality} symmetry, that is its invariance under the exchange $\omega_1 \leftrightarrow \omega_2$: in fact it turns out that $\mathcal{W}^{0}$ is invariant under this exchange, while the second and third terms in the left hand side of \eqref{sdual} are mapped into each other. $S$-duality tells us that perturbative (or WKB) and non-perturbative sectors contribute in the same way to the quantization conditions.
$S$-duality also fits naturally with the existence of the modular dual relativistic Toda system, since this is related to the original relativistic Toda chain by the exchange $\omega_1 \leftrightarrow \omega_2$ and the two systems should admit a common solution. 

To sum up, as we already saw happening in four dimensions (Section \ref{sec2.4}), also for the $2$-particle relativistic Toda chain (and its dual) the quantization conditions admit two different representations:
\begin{itemize}
\item[A)] The one of \cite{Wang:2015wdy},\cite{2015arXiv151102860H} given by \eqref{sdual}, manifestly invariant under $S$-duality and written in terms of the \textit{NS limit} of the five-dimensional partition function (corrected by the $B$-field when necessary) + non-perturbative (in $\hbar$) corrections; this is the five-dimensional analogue of \eqref{museum};
\item[B)] The one of \cite{Grassi:2014zfa} (which we only mentioned), equivalent to requiring a certain spectral determinant to vanish, where this spectral determinant can be written in terms of the \textit{unrefined limit} of the five-dimensional partition function + non-perturbative (in $\hbar$) corrections; this is the five-dimensional analogue of \eqref{tauzero}.
\end{itemize} 
A similar statement can be formulated for general $N$, although the situation in this case is slightly more involved.\footnote{The approach of \cite{Grassi:2014zfa} was extended to quantum spectral curves of higher genus in \cite{Codesido:2015dia,Gu:2015pda}. These works treat the quantum spectral curve as a quantum mechanical operator and not as the Baxter equation of an integrable system: for the case at hand, this implies they only provide 1 quantization condition instead of $N-1$. It is however possible to recover the remaining quantization conditions, as explained in \cite{Sun:2016obh,Grassi:2016nnt}.}
As shown in \cite{Grassi:2016nnt} (see also \cite{Sun:2016obh}), these two representations are not independent but can be related via the five-dimensional blow-up equations of \cite{Nakajima:2005fg,Gottsche:2006bm}. 
It might however be possible to give an alternative explanation for the compatibility of these two representations by considering hyperKahler moduli spaces of double-periodic monopoles. In fact, while four-dimensional $\mathcal{N} = 2$ $SU(N)$ theories are theories of class $\mathcal{S}$ and as such they can be related to hyperKahler moduli spaces of Hitchin systems \cite{Gaiotto:2009hg}, five-dimensional $\mathcal{N} = 1$ $SU(N)$ theories are instead associated to hyperKahler moduli spaces of double-periodic monopoles \cite{Cherkis:2012qs,Cherkis:2014vfa}. It is important to remark that, differently from the Hitchin system moduli spaces, double-periodic monopoles are 
self-dual under Nahm transform.
Being hyperKahler, these moduli spaces admit many possible complex structures usually labelled by a parameter $\zeta \in \mathbb{CP}^1$, and we can denote a basis of complex structures as $I$, $J$, $K$. As suggested in \cite{2014arXiv1411.3692W}, relativistic closed Toda chains may appear differently according to the complex structure under consideration; based on what happens in the four-dimensional case \cite{Bonelli:2016idi,Bonelli:2016qwg}, one would be led to associate quantization conditions in the representation A) to complex structure $I$ and quantization conditions in the form B) to complex structure $J$. It would certainly be interesting to understand this point in more detail. 


\subsubsection{A more refined proposal: five-dimensional gauge theory on curved space} \label{3.3.3}

As we have just discussed in Section \ref{3.3.3}, there are two main differences between the naive quantization conditions \eqref{falseqc} proposed by \cite{2010maph.conf..265N} and the exact quantization conditions \eqref{sdual} in the form B) (related to the ones in form A)) proposed by \cite{Wang:2015wdy},\cite{2015arXiv151102860H}:
\begin{itemize}
\item Naive quantization conditions are missing the non-perturbative contributions in $\hbar$;
\item Naive quantization conditions are missing the correction due to the $B$-field.
\end{itemize}
Is there a way to solve these two problems in a gauge theory framework? \\

A possibility would be to consider five-dimensional gauge theories on \textit{curved} spaces, rather than on the trivial flat background $\mathbb{R}^4_{\epsilon_1, \epsilon_2} \times S^1_R$.
In fact as noticed in \cite{2015arXiv150704799H} and further elaborated in \cite{Sun:2016obh}, exact quantization conditions in the A)-form \eqref{sdual} seem to be compatible with a proposal put forward in \cite{2012arXiv1210.5909L}: this proposal states that the non-perturbative completion of the refined closed topological string partition function (that is, the five-dimensional partition function \eqref{pf5d} for the cases of interest to us) is given by the \textit{integrand} of the partition function of our gauge theory on a squashed $S^5_{\omega_1, \omega_2, \omega_3}$ background. In an ``NS-like'' limit $\omega_3 \rightarrow 0$ this geometry will reduce to $S^3_{\omega_1, \omega_2} \times \mathbb{R}^2$, and the supersymmetric vacua equations derived from the (integrand of the) $S^5_{\omega_1, \omega_2, \omega_3}$ partition function seem to reproduce \eqref{sdual}. This interpretation naturally incorporates the symmetry $\omega_1 \leftrightarrow \omega_2$ of the exact quantization conditions in the form A), as well as of the relativistic Toda and dual Toda chain; let us however remark that there might be many five-dimensional geometries which reduce to $S^3_{\omega_1, \omega_2} \times \mathbb{R}^2$ in some limit, and since it feels a bit unnatural to consider the integrand of a partition function instead of the partition function itself one might prefer to consider some non-compact five-dimensional geometry which could eliminate the problem of integration. Although this will be important if one wants to discuss non-perturbative completions of refined topological strings at generic values of Omega background parameters, we will not be concerned with this problem since we expect all possible different five-dimensional geometries to give the same answer when reduced to $S^3_{\omega_1, \omega_2} \times \mathbb{R}^2$; therefore in the following we will always start from the integrand of the partition function on $S^5_{\omega_1, \omega_2, \omega_3}$, keeping in mind that this might not be the most appropriate starting point.

Apart from this caveat, let us try to see in more detail if the proposal by \cite{2012arXiv1210.5909L} effectively leads to \eqref{sdual} in the limit $\omega_3 \rightarrow 0$ or not; the discussion here mostly follows Appendix E of \cite{Sciarappa:2016ctj}. We start by considering the squashed $S^5_{\omega_1, \omega_2, \omega_3}$ geometry; this geometry can be parametrized as 
\begin{equation}
\omega_1^2 \vert z_1 \vert^2 + \omega_2^2 \vert z_2 \vert^2 + \omega_3^2 \vert z_3 \vert^2 = 1, \;\;\;\; z_1, z_2, z_3 \in \mathbb{C}, \;\;\;\; \omega_1, \omega_2, \omega_3 \in \mathbb{R}_+,
\end{equation}
and admits a $U(1)^3$ isometry group. Alternatively, we can think of $S^5_{\omega_1, \omega_2, \omega_3}$ as an $(S^1)^3$ fibration over a triangle: the three vertices of the triangle correspond to the three circles $S^1_{(i)}$ of radii $\omega_i^{-1}$, $i = 1,2,3$ fixed under the action of the $U(1)^3$ isometry group, while the three edges correspond to three squashed three-spheres $S^3_{\omega_i, \omega_j}$, $i \neq j$. The partition function of $\mathcal{N} = 1$ gauge theories on $S^5_{\omega_1, \omega_2, \omega_3}$ has been discussed in \cite{Kim:2012ava,Kim:2012qf,Kallen:2012cs,Hosomichi:2012ek,Kallen:2012va,Imamura:2012xg,Imamura:2012bm}; in the case of a pure $\mathcal{N}=1$ $SU(N)$ theory without Chern-Simons term, it can schematically be written as
\begin{equation}
Z_{S^5} = \int [d a] \,Z^{\text{cl}}_{S^5}(\vec{a}, \omega_1, \omega_2, \omega_3, m_0) Z^{\text{1l}}_{S^5}(\vec{a}, \omega_1, \omega_2, \omega_3) Z^{\text{inst}}_{S^5}(\vec{a}, \omega_1, \omega_2, \omega_3, m_0), \label{integral}
\end{equation}
with $m_0 = 4\pi^2/g^2_{\text{5d}}$ as defined in \eqref{lelouch}. The classical term of the integrand is given by
\begin{equation}
Z^{\text{cl}}_{S^5}(\vec{a}, \omega_1, \omega_2, \omega_3, m_0) = \prod_{m=1}^N e^{- \frac{\pi m_0 a_m^2}{\omega_1 \omega_2 \omega_3}}, \label{classical}
\end{equation}
while the one-loop part reads ($\vert \omega \vert = \omega_1 + \omega_2 + \omega_3$)
\begin{equation}
\begin{split}
Z^{\text{1l}}_{S^5}(\vec{a}, \omega_1, \omega_2, \omega_3) &\;=\; \prod_{\alpha \in \Delta_+} S_3\left(i \alpha(\vec{a})\vert \omega_1, \omega_2, \omega_3\right)
S_3\left(- i \alpha(\vec{a}) \vert \omega_1, \omega_2, \omega_3\right) \\
&\;=\; \prod_{\alpha \in \Delta_+} S_3\left(i \alpha(\vec{a})\vert \omega_1, \omega_2, \omega_3\right)
S_3\left( i \alpha(\vec{a}) + \vert \omega \vert \vert \omega_1, \omega_2, \omega_3\right),
 \label{oneloop}
\end{split}
\end{equation}
where $\alpha \in \Delta_+$ are the positive roots of $SU(N)$ and $S_3\left( x \vert \omega_1, \omega_2, \omega_3\right)$ is the triple sine function defined in Appendix \ref{appdoublesine}. 
By making use of the properties of the triple sine function it is possible to show that the $S^5_{\omega_1, \omega_2, \omega_3}$ one-loop term  \eqref{oneloop} factorizes into the product of three $\mathbb{R}^4_{\epsilon_1, \epsilon_2} \times S^1_R$ one-loop terms: therefore at least for what the 1-loop terms are concerned, the (integrand of the) $S^5_{\omega_1, \omega_2, \omega_3}$ partition function can be thought as the product of three copies of the $\mathbb{R}^4_{\epsilon_1, \epsilon_2} \times S^1_R$ partition function \eqref{pf5d}, where each copy is associated to one of the three fixed circles of the $U(1)^3$ action and $\epsilon_1$, $\epsilon_2$, $R$ are identified with $\omega_1$, $\omega_2$, $\omega_3$ differently for each of the three copies according to Table \ref{tabo}\footnote{The unusual identification used in this Table follows from the discussion in \cite{2012arXiv1210.5909L}.}.
\renewcommand\arraystretch{1.4}
\begin{table}
\begin{center}
\begin{tabular}{|c|c|c|c|}
\hline           & $-i\epsilon_1$ & $-i\epsilon_2$ & $1/R$ \\
\hline $S^1_{(1)}$ &   $\omega_3$   & $\omega_2 + \omega_1$     & $\omega_1$ \\ 
       $S^1_{(2)}$ &   $\omega_1 + \omega_2$   & $\omega_3$     & $\omega_2$ \\ 
       $S^1_{(3)}$ &   $\omega_1$   & $\omega_2 + \omega_3$     & $\omega_3$ \\ 
\hline
\end{tabular} \caption{Identification $S^5_{\omega_1, \omega_2, \omega_3}$ squashing parameters - gauge theory parameters.} \label{tabo}
\end{center}
\end{table}
\renewcommand\arraystretch{1} 
\hspace{-0.3 cm}For what the instanton term $Z^{\text{inst}}_{S^5}$ is concerned, the situation is less clear: based on what happens with the one-loop term it is expected that $Z^{\text{inst}}_{S^5}$ factorizes into the product of three copies of the instanton partition function $Z^{\text{inst}}_{\text{5d}}(\vec{a},\epsilon_1, \epsilon_2, R, m_0)$ on flat space $\mathbb{R}^4_{\epsilon_1, \epsilon_2} \times S^1_R$, that is
\begin{equation}
\begin{split}
& Z^{\text{inst}}_{S^5}(\vec{a}, \omega_1, \omega_2, \omega_3, m_0) \;= \\ 
&\;\;\;\;\;\; = Z^{\text{inst}}_{\text{5d},(1)}(\vec{a}, i \omega_3, i (\omega_2 + \omega_1), \omega_1^{-1}, m_0) \,
Z^{\text{inst}}_{\text{5d},(2)}(\vec{a}, i (\omega_1 + \omega_2), i \omega_3, \omega_2^{-1}, m_0) \, \\
&\;\;\;\;\;\;\;\;\;\;\; Z^{\text{inst}}_{\text{5d},(3)}(\vec{a}, i \omega_1, i (\omega_2 + \omega_3), \omega_3^{-1}, m_0), \label{instanton}
\end{split}
\end{equation}
where again $\epsilon_1, \epsilon_2$, $R$ are identified with $\omega_1, \omega_2, \omega_3$ differently for each of the three copies as in Table \ref{tabo}; however this expectation is still conjectural.\footnote{We prefer to indicate the dependence on the gauge coupling via $m_0$ rather than via the five-dimensional instanton counting parameter $Q_{\text{5d}} = e^{-2\pi R m_0}$ since $Q_{\text{5d}}$ depends on $R$ and will therefore be different for each of the three copies.}. Now, according to the proposal of \cite{2012arXiv1210.5909L} we should  consider the integrand of the partition function \eqref{integral}, that is
\begin{equation}
Z_{S^5}^{\text{int}}(\vec{a}, \omega_1, \omega_2, \omega_3, m_0) = Z^{\text{cl}}_{S^5}(\vec{a}, \omega_1, \omega_2, \omega_3, m_0) Z^{\text{1l}}_{S^5}(\vec{a}, \omega_1, \omega_2, \omega_3) Z^{\text{inst}}_{S^5}(\vec{a}, \omega_1, \omega_2, \omega_3, m_0).
\label{integrand}
\end{equation}
It is actually more convenient to rewrite \eqref{integrand} in a slightly different form, by separating the BPS states contribution from the contribution due to classical terms; in practice we rearrange the various terms as
\begin{equation}
Z_{S^5}^{\text{int}} = Z^{0}_{S^5} Z^{\text{BPS}}_{S^5} = e^{- F_{S^5}} = e^{- F^{0}_{S^5} - F^{\text{BPS}}_{S^5}}. \label{pf5}
\end{equation}
The $Z^{0}_{S^5}$ part is simply
\begin{equation}
Z^{0}_{S^5} = e^{- F^{0}_{S^5}} = \prod_{m=1}^N e^{- \frac{4\pi^3}{\omega_1 \omega_2 \omega_3} \frac{a_m^2}{g^2_{5d}}} 
\prod_{m<n}^N e^{-\frac{\pi}{3\omega_1 \omega_2 \omega_3}\left(a_m - a_n\right)^3 + 
\frac{\pi}{6}\frac{\omega_1^2 + \omega_2^2 + \omega_3^2}{\omega_1 \omega_2 \omega_3}(a_m - a_n) +
\frac{\pi}{2}\frac{\omega_1 \omega_2 + \omega_1 \omega_3 + \omega_2 \omega_3}{\omega_1 \omega_2 \omega_3}(a_m-a_n)
} \label{nonBPS5d}
\end{equation}
and contains \eqref{classical} as well as the exponential terms in \eqref{oneloop} coming from the $B_{3,3}$ polynomials in the definition of triple sine functions \eqref{pppp}, while the BPS part
\begin{equation}
Z^{\text{BPS}}_{S^5} = e^{- F^{\text{BPS}}_{S^5}} = Z_{\text{5d},(1)}^{\text{inst}} \,
Z_{\text{5d},(2)}^{\text{inst}} \, Z_{\text{5d},(3)}^{\text{inst}} \prod_{j<k}^N \mathcal{S}_{\omega_1, \omega_2, \omega_3}\Big( i(a_j - a_k) \Big) \mathcal{S}_{\omega_1, \omega_2, \omega_3}\Big(i(a_j - a_k) + \vert \omega \vert \Big) \label{BPSS5}
\end{equation}
contains the $\mathcal{S}_{\omega_1, \omega_2, \omega_3}$ functions (defined in Appendix \ref{appdoublesine}) coming from \eqref{oneloop} together with the instanton terms \eqref{instanton}. The BPS part can be more compactly written as
\begin{equation}
e^{- F^{\text{BPS}}_{S^5}} = e^{- F^{\text{BPS}}_{\text{5d},(1)} - F^{\text{BPS}}_{\text{5d},(2)} - F^{\text{BPS}}_{\text{5d},(3)}}
\end{equation}
in terms of three copies of the function \eqref{BPS}, where as usual $\epsilon_1, \epsilon_2$, $R$ are identified with $\omega_1, \omega_2, \omega_3$ for each of the three copies as in Table \ref{tabo}. The $t_i$ parameters are the flat K\"{a}hler moduli and mass parameters used in \cite{Grassi:2014zfa,2015arXiv151102860H} and correspond to linear combinations of the $a_i$ parameters and Yang-Mills coupling $m_0 = 4\pi^2/g^2_{5d}$ as in \eqref{kahler}, that is
\begin{equation}
t_i = a_i - a_{i+1} \;\;\;(i = 1, \ldots, N-1) \;\;\;\;\;\; \text{and} \;\;\;\;\;\; t_N = m_0 + \sum_{i = 1}^{N - 1} t_i.
\end{equation}


We now want to take the ``NS limit'' of \eqref{pf5}; what we mean by this is the following. As we saw, $Z_{S^5}^{\text{int}}$ roughly factorizes into three copies of the partition function on flat space $\mathbb{R}^4_{\epsilon_1, \epsilon_2} \times S^1_R$. Let us focus for the moment on the second copy and take the NS limit here: since $\epsilon_2 = i \omega_3$ in this copy, this means that we have to take the limit $i \omega_3 \rightarrow 0$, and as a consequence the squashed $S^5_{\omega_1, \omega_2, \omega_3}$ geometry reduces to $S^3_{\omega_1, \omega_2} \times \mathbb{R}^2$. What happens to the other copies? For what the first copy is concerned, the limit we are considering corresponds again to an NS limit because of the symmetry $\epsilon_1 \leftrightarrow \epsilon_2$ of the partition function on flat space; on the third copy instead we are sending $R \rightarrow \infty$, which implies that $F^{\text{BPS}}_{\text{5d},(3)} \rightarrow 0$.\footnote{Large-volume formulae like \eqref{BPS} are only valid for $t_i \rightarrow \infty$.} We therefore expect the ``NS limit'' of \eqref{pf5} to roughly reduce to two copies of the flat space twisted effective superpotential \eqref{aziza}, related by the exchange $\omega_1 \leftrightarrow \omega_2$; it is however important to notice that these two copies will be slightly modified with respect to \eqref{BPSNS} because of the non-trivial identifications in Table \ref{tabo}, and the modification exactly coincides with the effect of the $B$-field thus giving \eqref{withB}. Being more precise, in complete analogy with the definition in \eqref{aziza2} we can define the twisted effective superpotential on $S^5_{\omega_1, \omega_2, \omega_3}$ as
\begin{equation}
\mathcal{W}_{S^5}(\vec{a}, \omega_1, \omega_2, m_0) = \lim_{\omega_3 \rightarrow 0} \left[ i \omega_3 F_{S^5}(\vec{a}, \omega_1, \omega_2, \omega_3, m_0) \right].
\end{equation}
This can be decomposed according to
\begin{equation}
\mathcal{W}_{S^5}(\vec{a}, \omega_1, \omega_2, m_0) =
\mathcal{W}_{S^5}^{0}(\vec{a}, \omega_1, \omega_2, m_0) + 
\mathcal{W}_{S^5}^{\text{BPS}}(\vec{a}, \omega_1, \omega_2, m_0), \label{ziopaperone}
\end{equation}
with
\begin{equation}
\mathcal{W}_{S^5}^{\text{BPS}}(\vec{a}, \omega_1, \omega_2, m_0)  = 
\mathcal{W}_{\text{5d}}^{\text{BPS}}(\vec{a}, i(\omega_1+ \omega_2), \omega_2^{-1}, m_0)  + 
\mathcal{W}_{\text{5d}}^{\text{BPS}}(\vec{a}, i(\omega_2 + \omega_1), \omega_1^{-1}, m_0) 
\end{equation}
given in terms of the flat-space function $\mathcal{W}_{\text{5d}}^{\text{BPS}}(\vec{a}, \epsilon_1, R, m_0)$  appearing in \eqref{split2}; let us however stress once more that the non-trivial identifications for the $\epsilon_i$ parameters lead to the $B$-field corrected formula \eqref{withB}. \\

To sum up, from the ``NS limit'' of the proposal by \cite{2012arXiv1210.5909L} we obtain the putative quantization conditions
\begin{equation}
\dfrac{\partial }{\partial a_m} \mathcal{W}_{S^5}(\vec{a}, \omega_1, \omega_2, m_0)  = 2\pi i n_m \;, \;\;\;\; n_m \in \mathbb{Z}, \;\;\;\; m = 1, \ldots, N. \label{procione}
\end{equation}
As we can see, the two problems that were affecting the naive quantization conditions \eqref{falseqc} are now solved in \eqref{procione}:
\begin{itemize}
\item The non-perturbative corrections (restoring $S$-duality) arise from a second copy of the flat space twisted effective superpotential, related to the first copy by $\omega_1 \leftrightarrow \omega_2$, naturally incorporated in the $S^3_{\omega_1, \omega_2} \times \mathbb{R}^2$ geometry;
\item The $B$-field is taken into account by the identification in Table \ref{tabo}. 
\end{itemize}
Working out the explicit expressions, in particular the non-BPS term $\mathcal{W}_{S^5}^{0}$ in \eqref{ziopaperone}, it is not hard to verify that \eqref{procione} precisely coincides with the known exact quantization conditions.
Let us consider for example the pure 5d $\mathcal{N} = 1$ $SU(2)$ theory; this will be associated to the 2-particle relativistic Toda chain and its dual, whose Hamiltonians can be written as (by redefining variables)
\begin{equation}
\begin{split}
\widehat{\mathcal{H}}_1 &= e^{-i \omega_1 \partial_x} + e^{i \omega_1 \partial_x} + e^{\frac{2 \pi x}{\omega_2}} + Q_{\text{rel}} e^{-\frac{2 \pi x}{\omega_2}}, \\
\widehat{\widetilde{\mathcal{H}}}_1 &= e^{-i \omega_2 \partial_x} + e^{i \omega_2 \partial_x} + e^{\frac{2 \pi x}{\omega_1}} + \widetilde{Q}_{\text{rel}} e^{-\frac{2 \pi x}{\omega_1}}. \label{H2Ht2}
\end{split}
\end{equation}
By setting $a_1 = - a_2 = \frac{a}{2}$ the parameter $a = a_1 - a_2$ gets identified with the Kahler modulus $t$, while $m_0$ is interpreted as a mass parameter in the language of topological strings. 
Formula \eqref{procione} for $a_1$ gives
\begin{equation}
\begin{split}
2 \pi n & = \dfrac{1}{i}\dfrac{\partial \mathcal{W}_{S^5}}{\partial a_1} = 
\dfrac{2}{i} \dfrac{\partial \mathcal{W}_{S^5}}{\partial a} = \\
& = \dfrac{2\pi a^2}{\omega_1 \omega_2} + \dfrac{2\pi m_0 a}{\omega_1 \omega_2} 
- \dfrac{\pi}{3}\dfrac{\omega_1^2 + \omega_2^2}{\omega_1 \omega_2} - \pi \\
& + \sum_{j_L,j_R} \sum_{n \geqslant 1} \sum_{d_1, d_2}
 \dfrac{(-1)^{(2j_L + 2j_R + 1)n}}{n} (d_1 + d_2) N_{j_L,j_R}^{d_1, d_2} \\ 
& \times \dfrac{\sin \left[ (2j_L+1)\pi n \frac{\omega_1}{\omega_2} \right] \sin \left[ (2j_R+1)\pi n \frac{\omega_1}{\omega_2} \right]}{\sin^3 \left[\pi n \frac{\omega_1}{\omega_2} \right]} e^{-2\pi n d_1a/\omega_2} e^{-2\pi n d_2 (m_0 + a)/\omega_2} \\
& + \sum_{j_L,j_R} \sum_{n \geqslant 1} \sum_{d_1, d_2}
\dfrac{(-1)^{(2j_L + 2j_R + 1)n}}{n} (d_1 + d_2) N_{j_L,j_R}^{d_1, d_2} \\
& \times \dfrac{\sin \left[ (2j_L+1)\pi n \frac{\omega_2}{\omega_1} \right] \sin \left[ (2j_R+1)\pi n \frac{\omega_2}{\omega_1} \right]}{\sin^3 \left[\pi n \frac{\omega_2}{\omega_1} \right]} e^{-2\pi n d_1a/\omega_1} e^{-2\pi n d_2 (m_0 + a)/\omega_1}.   \label{ex2toda5d}
\end{split}
\end{equation}
The same equation is obtained from taking the derivative with respect to $a_2$. 
Notice that the factor $(-1)^{(2j_L + 2j_R + 1)n}$ coming from the identification in Table \ref{tabo} exactly coincides with the effect of introducing the $B$-field.
It is easy to check that the quantization conditions \eqref{ex2toda5d} coincide with the exact ones given in \cite{Wang:2015wdy}; as such they are manifestly ``$S$-dual'' (i.e. invariant under the exchange $\omega_1 \leftrightarrow \omega_2$) and free of poles at $\omega_1/\omega_2 \in \mathbb{Q}_+$. 
\\

\renewcommand\arraystretch{1.7}
\begin{table}[]
\begin{center}
\begin{tabular}{ccc}
\hline 
\rule[-1ex]{0pt}{2.5ex} &  
$\renewcommand\arraystretch{1.1}\begin{array}{c} \omega_1 = 2^{-1/4},\, \omega_2 = 2^{1/4} \\ m_0 = 0 \end{array}\renewcommand\arraystretch{1.5}$ & 
$\renewcommand\arraystretch{1.1}\begin{array}{c} \omega_1 = \frac{1}{\sqrt{2}},\, \omega_2 = 1 \\ m_0 = -\frac{\ln 3}{2\pi} \end{array}\renewcommand\arraystretch{1.5}$ \\ 
\hline 
\rule[-1ex]{0pt}{2.5ex} $a^{(0)}$&   0.9257432179\ldots & 0.8709463196\ldots \\ 
\rule[-1ex]{0pt}{2.5ex} $\ln \left[ W_{\square ,\,,\text{NS}}^{SU(2),(0)} \right]$ & 2.4605242719\ldots & 2.7528481019\ldots \\ 
\rule[-1ex]{0pt}{2.5ex} $\ln \left[ \widetilde{W}_{\square ,\,,\text{NS}}^{SU(2),(0)} \right]$ & 3.4605592909\ldots & 3.8720036669\ldots \\ 
\hline 
\rule[-1ex]{0pt}{2.5ex} $a^{(1)}$& 1.3615838159\ldots & 1.2357226878\ldots \\ 
\rule[-1ex]{0pt}{2.5ex} $\ln \left[ W_{\square ,\,,\text{NS}}^{SU(2),(1)} \right]$ & 3.5984708772\ldots & 3.8838346782\ldots \\ 
\rule[-1ex]{0pt}{2.5ex} $\ln \left[ \widetilde{W}_{\square ,\,,\text{NS}}^{SU(2),(1)} \right]$ & 5.0869593531\ldots & 5.4902688202\ldots \\ 
\hline 
\rule[-1ex]{0pt}{2.5ex} $a^{(2)}$&  1.6892652995\ldots & 1.5106117940\ldots \\ 
\rule[-1ex]{0pt}{2.5ex} $\ln \left[ W_{\square ,\,,\text{NS}}^{SU(2),(2)} \right]$ & 4.4628893132\ldots & 4.7460288538\ldots \\ 
\rule[-1ex]{0pt}{2.5ex} $\ln \left[ \widetilde{W}_{\square ,\,,\text{NS}}^{SU(2),(2)} \right]$ & 6.3111090796\ldots & 6.7114798487\ldots \\ 
\hline 
\end{tabular} \caption{Energies $W_{\square ,\,,\text{NS}}^{SU(2),(n)}$, $ \widetilde{W}_{\square ,\,,\text{NS}}^{SU(2),(n)}$  at level $n = 0,1,2$.
}  \label{ii3}
\end{center} 
\end{table}
\renewcommand\arraystretch{1}

Having obtained exact quantization conditions for the $N$-particle relativistic Toda chain from the partition function of $\mathcal{N} = 1$ $SU(N)$ gauge theories on $S^3_{\omega_1, \omega_2} \times \mathbb{R}^2$, we can now move on to discuss how to compute its energy spectrum. As we said in \eqref{wilNS}, the expectation is that energies will correspond to the NS limit of Wilson loops in an appropriate representation wrapping a circle \cite{Gaiotto:2014ina,Bullimore:2014upa,Bullimore:2014awa}. The background $S^3_{\omega_1, \omega_2} \times \mathbb{R}^2$ naturally admits two such circles, of radii $\omega_1$ and $\omega_2$: this is because  $S^3_{\omega_1, \omega_2}$ can be thought of as a non-trivial $S^1$ fibration over $S^2$ which contains two special circles at the north and south pole of $S^2$. It is then natural to identify Wilson loops wrapping these two circles with the energies $\mathcal{E}_m$ and dual energies $\widetilde{\mathcal{E}}_m$ of the relativistic Toda chain and its dual. 

For example, in the 2-particle case we only have Wilson loops in the fundamental representation, so that\footnote{Since $N=2$ is even, the $B$-field plays no role and \eqref{W2f} coincides with the flat space result; for $N$ odd instead the $B$-field will induce a change of sign for $Q_{\text{5d}}$, $\widetilde{Q}_{\text{5d}}$ compared to to the flat space results.}

\begin{equation}
\begin{split}
\mathcal{E}_1 & = W_{\square ,\,,\text{NS}}^{SU(2)} = \mu^{1/2} + \mu^{-1/2} - Q_{5d\,} (\mu^{1/2} + \mu^{-1/2}) \dfrac{q}{(1 - q \mu)(1 - q \mu^{-1})} + o(Q^2_{5d}), \\
\widetilde{\mathcal{E}}_1 & = \widetilde{W}_{\square \,,\, \text{NS}}^{SU(2)} = \widetilde{\mu}^{1/2} + \widetilde{\mu}^{-1/2} - \widetilde{Q}_{5d\,} (\widetilde{\mu}^{1/2} + \widetilde{\mu}^{-1/2}) \dfrac{\widetilde{q}}{(1 - \widetilde{q} \widetilde{\mu})(1 - \widetilde{q} \widetilde{\mu}^{-1})} + o(\widetilde{Q}^2_{5d}). \label{W2f}
\end{split}
\end{equation}
These are invariant under the exchange $q \leftrightarrow q^{-1}$, $\widetilde{q} \leftrightarrow \widetilde{q}^{-1}$ and, differently from what happens with the quantization conditions, do not present any divergence at $\vert q \vert = 1$, $\vert \widetilde{q} \vert = 1$ since they are one-dimensional observables; moreover, they coincide with the (inverse) quantum mirror map of \cite{Grassi:2014zfa}.
If we now fix the value of $\omega_1$, $\omega_2$, $m_0$ and substitute the solution $a^{(n)}$ of \eqref{ex2toda5d} for $n \geqslant 0$ into the expressions for $W_{\square ,\,,\text{NS}}^{SU(2)}$, $\widetilde{W}_{\square ,\,,\text{NS}}^{SU(2)}$ in \eqref{W2f}, we should be obtaining the exact spectrum for the 2-particle quantum relativistic Toda chain and its modular dual. A few examples of energies for different values of $\omega_1$, $\omega_2$, $m_0$,  obtained via a 5-instanton computation, are listed in Table \ref{ii3}; already at 5-instantons these can be seen to coincide with the energies in Table \ref{numenergy5d} obtained by a numerical study of the operators \eqref{H2Ht2}. Using the values of $a^{(n)}$ generated by the naive quantization conditions \eqref{falseqc} would instead produce an energy spectrum which does not match with the numerical one. \\

We therefore conclude that it is possible to obtain exact quantization conditions and energy spectrum for the relativistic closed Toda chain and its modular dual while remaining in the framework of gauge theory (in the NS limit); the big difference with respect to the naive proposal \eqref{falseqc} is that differently from what happens in the four-dimensional case, in five dimensions we have to consider a curved background $S^3_{\omega_1, \omega_2} \times \mathbb{R}^2$ instead of the flat one 
$\mathbb{R}^2_{\epsilon_1}\times \mathbb{R}^2 \times S^1_R$, thing which naturally takes into account the existence of the modular dual relativistic Toda system. As a consequence, we expect that in order to obtain exact eigenfunctions we should be considering codimension-two defects living on $S^3_{\omega_1, \omega_2}$ instead of the naive $D_{\epsilon_1} \times S^1_R$; this is what we are going to verify next.




\subsection{Solution via gauge theory: eigenfunctions} \label{sec3.4}

As we mentioned at the beginning of Section \ref{sec3}, while eigenfunctions for the $N$-particle relativistic open Toda chain have been constructed in \cite{Kharchev:2001rs}, much less is known about the eigenfunctions of the relativistic closed Toda chain (or the associated Baxter equation) apart from the works \cite{Marino:2016rsq,Kashaev:2017zmv}. Here we will propose a solution to this problem via gauge theory. Based on the analogy with the ``non-relativistic'' (or four-dimensional) case reviewed in Section \ref{2.3.3}, and from what we have seen in Section \ref{3.3.3}, the eigenfunctions we are looking for should correspond to partition functions on $S^3_{\omega_1, \omega_2} \times \mathbb{R}^2$ (rather than flat space 
$D_{\epsilon_1}\times \mathbb{R}^2 \times S^1_R$) of codimension two defects of type II or I wrapping $S^3_{\omega_1, \omega_2}$ coupled to the five-dimensional $\mathcal{N} = 1$ $SU(N)$ theory. 
Here we mostly focus on studying the solution to the Baxter equation for the case $N = 2$ and compare it against the numerical results obtained in Section \ref{sec3.2}; in Section \ref{sec3.5} we will comment on the general $N$ case and provide some check for $N = 3$. \\


\noindent\textit{Open relativistic Toda chains} \\

\noindent Let us start by looking for simultaneous eigenfunctions $\Psi^{\text{rel}}_{\vec{a}}(\vec{x},\omega_1,\omega_2)$ of the $N$-particle relativistic open Toda chain and its dual. The solution to this problem is known from Separation of Variables \cite{Kharchev:2001rs} and can be expressed recursively as 
\begin{equation}
\begin{split}
& \Psi^{\text{rel}}_{a_1, \ldots, a_N}(x_1, \ldots, x_N,\omega_1,\omega_2) =  \\
& = \int_{\mathcal{C}}
\left[\prod_{j=1}^{N-1} d\sigma_j \right] \mu(\vec{\sigma}, \hbar)\, \mathcal{Q}(\vec{\sigma}, \vec{a}, \omega_1,\omega_2)
e^{\frac{i \pi}{\omega_1 \omega_2} \sum_{j=1}^{N-1} \sigma_j \left( \sum_{j=1}^{N-1} \sigma_j - \sum_{m=1}^N a_m \right)}
\\
&\;\;\;\;\;\; \Psi^{\text{rel}}_{\sigma_1, \ldots, \sigma_{N-1}}(x_1, \ldots, x_{N-1},\omega_1,\omega_2)  e^{\frac{2 \pi i x_N}{\omega_1 \omega_2}\left( \sum_{m=1}^N a_m - \sum_{j=1}^{N-1}\sigma_j \right)} \;\;\;\;\;\; \text{(open chain)}, \label{eigopenrel}
\end{split}
\end{equation}
starting from the eigenfunction of the 1-particle problem $\Psi^{\text{rel}}_{a_1}(x_1,\omega_1, \omega_2) = e^{i a_1 x_1/\omega_1 \omega_2}$. Here the integration contour $\mathcal{C}$ is as in \eqref{thisopen}, $\mu(\vec{\sigma}, \hbar)$ is an integration measure,
\begin{equation}
\mu(\vec{\sigma}, \omega_1, \omega_2) = \prod_{j < k}^{N-1} \sinh\left( \pi\frac{\sigma_j - \sigma_k}{\omega_1} \right) \sinh\left( \pi\frac{\sigma_j - \sigma_k}{\omega_2} \right),
\end{equation}
while
\begin{equation}
\mathcal{Q}(\vec{\sigma}, \vec{a}, \omega_1, \omega_2) = \prod_{j=1}^{N-1} q(\sigma_j, \vec{a}, \omega_1, \omega_2)= \prod_{j=1}^{N-1} \prod_{m=1}^{N} 
S_2^{-1}\big(-i(\sigma_j - a_m)\vert \omega_1, \omega_2\big) \label{Qtildeopenrel}
\end{equation}
is the wave-function in separated variables, where $S_2(\sigma \vert \omega_1, \omega_2)$ is defined in Appendix \ref{appdoublesine}. Exactly as it happens in the non-relativistic case, the function $q(\sigma, \vec{a}, \omega_1, \omega_2)$, although not entire, formally satisfies the one-dimensional Baxter equation for the relativistic open Toda chain and its dual:
\begin{equation}
\begin{split}
& (i)^N q(\sigma + i \omega_1, \vec{a}, \omega_1, \omega_2) = \mathcal{E}_N^{-\frac{1}{2}}\, t_{\text{rel}}(\sigma, \vec{\mathcal{E}}(\vec{a})) q(\sigma, \vec{a}, \omega_1, \omega_2), \\
& (i)^N q(\sigma + i \omega_2, \vec{a}, \omega_1, \omega_2) = \widetilde{\mathcal{E}}_N^{-\frac{1}{2}}\, \widetilde{t}_{\text{rel}}(\sigma, \vec{\widetilde{\mathcal{E}}}(\vec{a})) q(\sigma, \vec{a}, \omega_1, \omega_2), \label{baxopenrel}
\end{split}
\end{equation}
with $t(\sigma, \vec{\mathcal{E}}(\vec{a})))$, $\widetilde{t}(\sigma, \vec{\widetilde{\mathcal{E}}}(\vec{a})))$ as in the first line of \eqref{trel}, \eqref{treldual}.  

From the point of view of gauge theory, the functions $\Psi^{\text{rel}}_{\vec{a}}(\vec{x},\omega_1,\omega_2) $ and $q(\sigma, \vec{a}, \omega_1, \omega_2) $ can be interpreted respectively as partition functions of three-dimensional $\mathcal{N} = 2$ theories of type I (Figure \ref{TSUN}) and II (Figure \ref{freechiral}) living on a squashed three-sphere $S^3_{\omega_1,\omega_2}$ \cite{Hama:2011ea}; these three-dimensional theories should be considered as codimension two defects for the five-dimensional $\mathcal{N} = 1$ $SU(N)$ theory on $S^3_{\omega_1,\omega_2} \times \mathbb{R}^2$, in the limit in which the five-dimensional gauge interaction is decoupled (i.e. $m_0 = 4 \pi^2/g^2_{5d} \rightarrow \infty$). More precisely:
\begin{itemize}
\item The $S^3_{\omega_1,\omega_2}$ partition function for defects of type II, that is $N$ free chiral/antichiral multiplets, 
is given by
\begin{equation}
\begin{split}
q^{(\text{c})}(\sigma, \vec{a}, \omega_1, \omega_2) &= \prod_{m=1}^N S_2^{-1}\big( -i(\sigma - a_m) \vert \omega_1, \omega_2\big) \;\;\;\;\;\;\;\; \text{(chiral)},\\
q^{(\text{ac})}(\sigma, \vec{a}, \omega_1, \omega_2) &= \prod_{m=1}^N S_2^{-1}\big(i(\sigma - a_m) \vert \omega_1, \omega_2\big) \;\;\;\;\;\; \text{(antichiral)}; \label{papparel}
\end{split}
\end{equation}
these functions satisfy the equations
\begin{equation}
\begin{split}
& (i)^N q^{(\text{c})}(\sigma + i \omega_1, \vec{a}, \omega_1, \omega_2) = \mathcal{E}_N^{-\frac{1}{2}}\, t_{\text{rel}}(\sigma, \vec{\mathcal{E}}(\vec{a})) q^{(\text{c})}(\sigma, \vec{a}, \omega_1, \omega_2), \\
& (i)^N q^{(\text{c})}(\sigma + i \omega_2, \vec{a}, \omega_1, \omega_2) = \widetilde{\mathcal{E}}_N^{-\frac{1}{2}}\, \widetilde{t}_{\text{rel}}(\sigma, \vec{\widetilde{\mathcal{E}}}(\vec{a})) q^{(\text{c})}(\sigma, \vec{a}, \omega_1, \omega_2), 
\end{split}
\end{equation}
and
\begin{equation}
\begin{split}
& (i)^{-N} q^{(\text{ac})}(\sigma - i \omega_1, \vec{a}, \omega_1, \omega_2) = \mathcal{E}_N^{-\frac{1}{2}}\, t_{\text{rel}}(\sigma, \vec{\mathcal{E}}(\vec{a})) q^{(\text{ac})}(\sigma, \vec{a}, \omega_1, \omega_2), \\
& (i)^{-N} q^{(\text{ac})}(\sigma - i \omega_2, \vec{a}, \omega_1, \omega_2) = \widetilde{\mathcal{E}}_N^{-\frac{1}{2}}\, \widetilde{t}_{\text{rel}}(\sigma, \vec{\widetilde{\mathcal{E}}}(\vec{a})) q^{(\text{ac})}(\sigma, \vec{a}, \omega_1, \omega_2), 
\end{split}
\end{equation}
respectively, with
\begin{equation}
\mathcal{E}_N = \prod_{m=1}^N e^{\frac{2\pi a_m}{\omega_2}}, \;\;\;\; 
\widetilde{\mathcal{E}}_N = \prod_{m=1}^N e^{\frac{2\pi a_m}{\omega_1}}.
\end{equation}
The chiral partition function $q^{(\text{c})}(\sigma, \vec{a}, \omega_1, \omega_2)$ exactly coincides with the function $q(\sigma, \vec{a}, \omega_1, \omega_2)$ appearing in \eqref{Qtildeopenrel}.
\item The $S^3_{\omega_1,\omega_2}$ partition function for defects of type I with chiral multiplets 
(plus Chern-Simons and mixed Chern-Simons terms)\footnote{Different Chern-Simons terms (last line of \eqref{thisopenrel}) correspond to redefinitions of $\vec{p}$, $\vec{x}$ variables in the relativistic open Toda Hamiltonians 
$\widehat{\mathcal{H}}_m$, $\widehat{\widetilde{\mathcal{H}}}_m$.} reads
\begin{eqnarray}
&& \Psi^{\text{rel},(\text{c})}_{a_1, \ldots, a_N}(x_1, \ldots, x_N, \omega_1, \omega_2) = \nonumber \\ 
&& = e^{\frac{2 \pi i x_N}{\omega_1 \omega_2}\sum_{m=1}^N a_m} \int_{\mathcal{C}} \prod_{s=1}^{N-1} \prod_{j=1}^s d\sigma^{(s)}_j
\prod_{s=1}^{N-1}\prod_{j<k}^s \sinh \left[ \pi \frac{\sigma_j^{(s)} - \sigma_k^{(s)}}{\omega_1} \right] \sinh \left[ \pi \frac{\sigma_j^{(s)} - \sigma_k^{(s)}}{\omega_2} \right] \nonumber \\ 
&& \;\;\;\;\;\;
\prod_{s=1}^{N-1}\prod_{j=1}^{s}\prod_{k=1}^{s+1} S^{-1}_2\big( -i (\sigma_j^{(s)} - \sigma_k^{(s+1)}) \big)
\prod_{s=1}^{N-1} e^{2 \pi i\frac{x_s - x_{s+1}}{\omega_1 \omega_2}\sum_{j=1}^s \sigma_j^{(s)}} \nonumber \\
&& \;\;\;\;\;\; \prod_{s=1}^{N-1} e^{\frac{i \pi}{\omega_1 \omega_2} \sum_{j=1}^s \sigma_j^{(s)} \left( \sum_{j=1}^s \sigma_j^{(s)}  - \sum_{k=1}^{s+1} \sigma_k^{(s + 1)} \right)}
, \label{thisopenrel}
\end{eqnarray}
where $\sigma^{(N)}_m = a_m$ and the integration contour $\mathcal{C}$ is as in \eqref{thisopen}; this function is a simultaneous eigenfunction of the relativistic open Toda and dual Toda Hamiltonians $\widehat{\mathcal{H}}_m$ \eqref{hamrel}, $\widehat{\widetilde{\mathcal{H}}}_m$ \eqref{hamreldual}. A similar expression can be obtained if we consider antichiral instead of chiral multiplets. It is easy to show that \eqref{thisopenrel} can be re-expressed in a recursive form equivalent to \eqref{eigopenrel}. 
\end{itemize}
As we can see, the defects we are considering are the same ones we used in Section \ref{2.3.3} for constructing the non-relativistic open Toda chain eigenfunctions; the only differences are the dimensionality of the defect theories (three dimensions versus two dimensions) and the background in which these theories live ($S^3_{\omega_1, \omega_2}$ versus the disc $D_{\epsilon_1}$). \\

\noindent\textit{Closed relativistic Toda chains} \\

\noindent What about the relativistic closed Toda chain? According to the analysis of \cite{Kharchev:2001rs}, Separation of Variables tells us that eigenfunctions of the $N$-particle relativistic \textit{closed} Toda chain can be obtained from the eigenfunction of the $(N-1)$-particle relativistic \textit{open} Toda chain via 
\begin{equation}
\begin{split}
& \Psi^{\text{rel}}_{a_1, \ldots, a_N}(x_1, \ldots, x_N, \omega_1, \omega_2, m_0) =  \\
& = \int_{\mathcal{C}} \left[\prod_{j=1}^{N-1} d\sigma_j \right] \mu(\vec{\sigma}, \omega_1, \omega_2) \, \mathcal{Q}(\vec{\sigma}, \vec{\tau}(\vec{a}), \omega_1, \omega_2, m_0) 
\, e^{\frac{i \pi}{\omega_1 \omega_2} \sum_{j=1}^{N-1} \sigma_j \left( \sum_{j=1}^{N-1} \sigma_j - \sum_{m=1}^N a_m \right)} \\
&\;\;\;\;\;\; \Psi^{\text{rel}}_{\sigma_1, \ldots, \sigma_{N-1}}(x_1, \ldots, x_{N-1}, \omega_1, \omega_2)  e^{\frac{2 \pi i x_N}{\omega_1 \omega_2} \left( \sum_{m=1}^N a_m - \sum_{j=1}^{N-1}\sigma_j \right)} \;\;\;\;\;\; \text{(closed chain)}. \label{eigclosedrel}
\end{split}
\end{equation}
Here
\begin{equation}
\mathcal{Q}(\vec{\sigma}, \vec{\tau}(\vec{a}), \omega_1, \omega_2, m_0) = \prod_{j=1}^{N-1} q(\sigma_j, \vec{\tau}(\vec{a}), \omega_1, \omega_2, m_0), \label{Qseprel}
\end{equation}
and $q(\sigma, \vec{\tau}(\vec{a}), \omega_1, \omega_2, m_0)$ satisfies the one-dimensional Baxter equation for the relativistic closed Toda chain and its dual
\begin{equation}
\begin{split}
& (i)^N q(\sigma + i \omega_1, \vec{\tau}(\vec{a}), \omega_1, \omega_2, m_0) + Q_{\text{rel}} (i)^{-N} q(\sigma - i \omega_1, \vec{\tau}(\vec{a}), \omega_1, \omega_2, m_0) \\
& = \mathcal{E}_N^{-\frac{1}{2}}\, t_{\text{rel}}(\sigma, \vec{\mathcal{E}}(\vec{\tau}(\vec{a}))) q(\sigma, \vec{\tau}(\vec{a}), \omega_1, \omega_2, m_0), \\[10 pt]
& (i)^N q(\sigma + i \omega_2, \vec{\tau}(\vec{a}), \omega_1, \omega_2, m_0) + \widetilde{Q}_{\text{rel}} (i)^{-N} q(\sigma - i \omega_2, \vec{\tau}(\vec{a}), \omega_1, \omega_2, m_0) \\
& = \widetilde{\mathcal{E}}_N^{-\frac{1}{2}}\, \widetilde{t}_{\text{rel}}(\sigma, \vec{\widetilde{\mathcal{E}}}(\vec{\tau}(\vec{a}))) q(\sigma, \vec{\tau}(\vec{a}), \omega_1, \omega_2, m_0),
 \label{baxclosedrel}
\end{split}
\end{equation}
with $t(\sigma, \vec{\mathcal{E}}(\vec{\tau}(\vec{a}))))$, $\widetilde{t}(\sigma, \vec{\widetilde{\mathcal{E}}}(\vec{\tau}(\vec{a}))))$ as in the second line of \eqref{trel}, \eqref{treldual}. As we already mentioned, these equations can be thought as the quantized versions of the classical spectral curves \eqref{curverel}. Always according to \cite{Kharchev:2001rs}, the function \eqref{eigclosedrel} is a solution to the closed Toda spectral problem (i.e. is an eigenfunction of the relativistic closed Toda and dual Toda Hamiltonians $\widehat{\mathcal{H}}_m$, $\widehat{\widetilde{\mathcal{H}}}_m$ with the appropriate boundary conditions) only if the solution $q(\sigma, \vec{\tau}(\vec{a}), \omega_1, \omega_2, m_0)$ to the Baxter and dual Baxter equations \eqref{baxclosedrel} is entire in $\sigma$ and goes to zero fast enough as $\vert \sigma \vert \rightarrow \infty$. From what we have seen in the non-relativistic case, we expect these conditions will be satisfied only for certain values of the $a_m$ (or $\tau_m$) auxiliary parameters, or equivalently only for certain quantized values of the energies $\mathcal{E}_m$, $\widetilde{\mathcal{E}}_m$. Finding such a solution to \eqref{baxclosedrel} is still an open problem; what we are going to do next is to propose a solution to this problem via gauge theory, keeping in mind what we learnt in Section \ref{sec2.3} and Section \ref{sec3.3}. \\

Let us focus here on the $N = 2$ case. Let us decouple the center of mass and set the auxiliary variables as $a_1 = -a_2 = \frac{a}{2}$.
For what the relativistic open chain is concerned, we know from \eqref{papparel} that when the five-dimensional gauge theory is decoupled (that is $g^2_{\text{5d}} = 0$ or equivalently $m_0 = \infty$) the $S^3_{\omega_1, \omega_2}$ partition function of a type II defect with chiral multiplets
\begin{equation}
q^{(\text{c})}(\sigma, a, \omega_1, \omega_2) = S_2^{-1}\big( -i(\sigma - a/2) \vert \omega_1, \omega_2 \big) S_2^{-1}\big( -i(\sigma + a/2) \vert \omega_1, \omega_2 \big) \label{qcrelopen}
\end{equation}
formally satisfies
\begin{equation}
\begin{split}
\left[ e^{i \omega_1 \partial_{\sigma}} + e^{\frac{2\pi \sigma}{\omega_2}} + e^{-\frac{2\pi \sigma}{\omega_2}} - \mathcal{E}_1 \right] q^{(\text{c})}(\sigma, a, \omega_1, \omega_2) &= 0, \\
\left[ e^{i \omega_2 \partial_{\sigma}} + e^{\frac{2\pi \sigma}{\omega_1}} + e^{-\frac{2\pi \sigma}{\omega_1}} - \widetilde{\mathcal{E}}_1 \right] q^{(\text{c})}(\sigma, a, \omega_1, \omega_2) &= 0, \label{Baxcopenrel}
\end{split}
\end{equation}
while the $S^3_{\omega_1, \omega_2}$ partition function of a type II defect with antichiral multiplets
\begin{equation}
q^{(\text{ac})}(\sigma, a, \omega_1, \omega_2) = S_2^{-1}\big( i(\sigma - a/2) \vert \omega_1, \omega_2 \big) S_2^{-1}\big( i(\sigma + a/2) \vert \omega_1, \omega_2 \big) \label{qacrelopen}
\end{equation}
formally satisfies
\begin{equation}
\begin{split}
\left[ e^{-i \omega_1 \partial_{\sigma}} + e^{\frac{2\pi \sigma}{\omega_2}} + e^{-\frac{2\pi \sigma}{\omega_2}} - \mathcal{E}_1 \right] q^{(\text{ac})}(\sigma, a, \omega_1, \omega_2) &= 0, \\
\left[ e^{-i \omega_2 \partial_{\sigma}} + e^{\frac{2\pi \sigma}{\omega_1}} + e^{-\frac{2\pi \sigma}{\omega_1}} - \widetilde{\mathcal{E}}_1 \right] q^{(\text{ac})}(\sigma, a, \omega_1, \omega_2) &= 0, \label{Baxacopenrel}
\end{split}
\end{equation}
where
\begin{equation}
\mathcal{E}_1 = e^{\frac{\pi a}{\omega_2}} + e^{-\frac{\pi a}{\omega_2}}, \;\;\;\; \widetilde{\mathcal{E}}_1 = e^{\frac{\pi a}{\omega_1}} + e^{-\frac{\pi a}{\omega_1}}
\end{equation}
are nothing but the $m_0 \rightarrow \infty$ limit of \eqref{W2f}.\footnote{$SU(2)$ gauge Wilson loops \eqref{W2f} reduce to flavour Wilson loops as $m_0 \rightarrow \infty$.} Because of the properties of the double sine function, our expressions \eqref{qcrelopen}, \eqref{qacrelopen} are not entire but present poles. This is not a problem, and is not unexpected either: in fact we encountered a similar situation in Section \ref{2.3.3}, where instead of double sine functions we had Gamma functions. In that case we found that, by moving the Gamma functions from the numerator to the denominator, \eqref{pappa} can be rewritten as the product of an entire function times a function giving a lattice of poles at $\sigma = \pm a + i n \hbar$, $n \in \mathbb{Z}$.
Although not necessary for our purposes, let us just mention that something along the same lines can be done also for the case at hand: more precisely,
after introducing the $\theta$-function
\begin{equation}
\theta\left(\frac{i \sigma}{\omega_2}, \frac{\omega_1}{\omega_2}\right) = 
\left(e^{-\frac{2\pi \sigma}{\omega_2}}; e^{\frac{2\pi i \omega_1}{\omega_2}}\right)_{\infty}\left(e^{\frac{2\pi \sigma}{\omega_2}}e^{\frac{2\pi i \omega_1}{\omega_2}}; e^{\frac{2\pi i \omega_1}{\omega_2}}\right)_{\infty},
\end{equation}
which is periodic in $i \omega_2$ and quasi-periodic in $i \omega_1$ and is such that
\begin{equation}
\theta\left(\frac{i \sigma}{\omega_1}, -\frac{\omega_2}{\omega_1}\right) =
e^{-\frac{i \pi \sigma^2}{\omega_1 \omega_2} - \frac{\pi \sigma}{\omega_1} + \frac{\pi \sigma}{\omega_2} + i \pi\frac{\omega_1^2 + \omega_2^2}{6 \omega_1 \omega_2} -i\frac{\pi}{2}} 
\theta\left(\frac{i \sigma}{\omega_2}, \frac{\omega_1}{\omega_2}\right),
\end{equation}
we can use the properties of the double sine function to rewrite \eqref{qcrelopen}, \eqref{qacrelopen} as
\begin{equation}
\begin{split}
q^{(\text{c})}(\sigma, a, \omega_1, \omega_2) \;=\; &
e^{-\frac{i \pi}{\omega_1 \omega_2}\left(\sigma^2 + \frac{a^2}{4}\right) - \pi \frac{\omega_1 + \omega_2}{\omega_1 \omega_2}\sigma + i\pi\frac{\omega_1^2 + \omega_2^2}{6\omega_1 \omega_2} + i \frac{\pi}{2}} \\
& \dfrac{(q w \mu^{\frac{1}{2}};q)_{\infty}(q w \mu^{-\frac{1}{2}};q)_{\infty}
(\widetilde{q}^{-1} \widetilde{w}^{-1} \widetilde{\mu}^{\frac{1}{2}};\widetilde{q}^{-1})_{\infty}(\widetilde{q}^{-1} \widetilde{w}^{-1} \widetilde{\mu}^{-\frac{1}{2}};\widetilde{q}^{-1})_{\infty}}{\theta\left(i\frac{\sigma - a/2}{\omega_1}, -\frac{\omega_2}{\omega_1}\right)\theta\left(i\frac{\sigma + a/2}{\omega_1}, -\frac{\omega_2}{\omega_1}\right)},
\\
q^{(\text{ac})}(\sigma, a, \omega_1, \omega_2) \;=\; &
e^{\frac{i \pi}{\omega_1 \omega_2}\left(\sigma^2 + \frac{a^2}{4}\right) - \pi \frac{\omega_1 + \omega_2}{\omega_1 \omega_2}\sigma - i\pi\frac{\omega_1^2 + \omega_2^2}{6\omega_1 \omega_2} - i \frac{\pi}{2}} \\
& \dfrac{(q w^{-1} \mu^{\frac{1}{2}};q)_{\infty}(q w^{-1} \mu^{-\frac{1}{2}};q)_{\infty}
(\widetilde{q}^{-1} \widetilde{w} \widetilde{\mu}^{\frac{1}{2}};\widetilde{q}^{-1})_{\infty}(\widetilde{q}^{-1} \widetilde{w} \widetilde{\mu}^{-\frac{1}{2}};\widetilde{q}^{-1})_{\infty}}{\theta\left(i\frac{\sigma - a/2}{\omega_2},\frac{\omega_1}{\omega_2}\right)\theta\left(i\frac{\sigma + a/2}{\omega_2},\frac{\omega_1}{\omega_2}\right)}, \label{second}
\end{split}
\end{equation}
where we remind that 
\begin{equation}
q=e^{\frac{2\pi i \omega_1}{\omega_2}}, \;\; w=e^{-\frac{2\pi \sigma}{\omega_2}}, \;\; \mu=e^{\frac{2\pi a}{\omega_2}} \;\;\;\; \text{and} \;\;\;\; \widetilde{q}=e^{\frac{2\pi i \omega_2}{\omega_1}}, \;\; \widetilde{w}=e^{-\frac{2\pi \sigma}{\omega_1}}, \;\; \widetilde{\mu}=e^{\frac{2\pi a}{\omega_1}}. \label{curvedvar}
\end{equation}
Expressions \eqref{second} are actually only valid for Im$(\omega_1/\omega_2) > 0$, but can also be used when Im$(\omega_1/\omega_2) = 0$ with $\omega_1/\omega_2 \notin \mathbb{Q}_+$ (in the case $\omega_1/\omega_2 \in \mathbb{Q}_+$ instead we can only use the original formulae \eqref{qcrelopen}, \eqref{qacrelopen} written in terms of double sine functions). In this form we can easily see the similarity between \eqref{second} and \eqref{cacchio1}: in particular the terms at the numerator give an entire function, while the quasi-periodic $\theta$-functions at the denominator generate a lattice of simple poles at $\sigma = \pm \frac{a}{2} + i m \omega_1 + i n \omega_2$, $m, n \in \mathbb{Z}$ and play the same role of the $\sinh$ functions at the denominator of \eqref{cacchio1} in the non-relativistic Toda case. \\

Let us now move to the relativistic 2-particle closed Toda chain. Here we should look for an entire $L^2(\mathbb{R})$ solution $q(\sigma, a, \omega_1, \omega_2, m_0)$ to the relativistic closed chain Baxter and dual Baxter equations \eqref{baxclosedrel}, that is
\begin{subequations} \label{Bax5d2Todater}
\begin{align}
& \left[ e^{i \omega_1 \partial_{\sigma}} + Q_{\text{rel}} e^{-i \omega_1 \partial_{\sigma}} + e^{\frac{2\pi \sigma}{\omega_2}} + e^{-\frac{2\pi \sigma}{\omega_2}} - \mathcal{E}_1 \right] q(\sigma, a, \omega_1, \omega_2, m_0) = 0 ,
\label{Bax5d2Todaater} 
\\
& \left[ e^{i \omega_2 \partial_{\sigma}} + \widetilde{Q}_{\text{rel}} e^{-i \omega_2 \partial_{\sigma}} + e^{\frac{2\pi \sigma}{\omega_1}} + e^{-\frac{2\pi \sigma}{\omega_1}} - \widetilde{\mathcal{E}}_1\right] q(\sigma, a, \omega_1, \omega_2, m_0) = 0,
\label{Bax5d2Todabter}
\end{align}
\end{subequations}
where the energies $\mathcal{E}_1$, $\widetilde{\mathcal{E}}_1$ are given by \eqref{W2f}. Sometimes it will be more convenient to consider the function 
\begin{equation}
\overline{q}(\sigma, a, \omega_1, \omega_2, m_0) = e^{-\frac{i \pi m_0 \sigma}{\omega_1 \omega_2}} q(\sigma, a, \omega_1, \omega_2, m_0), \label{combobar}
\end{equation}
which should be an entire solution of the more symmetric equations
\begin{subequations} \label{sym5d}
\begin{align}
& \left[ Q_{\text{rel}}^{\frac{1}{2}} \left(e^{i \omega_1 \partial_{\sigma}} + e^{-i \omega_1 \partial_{\sigma}}\right) + e^{\frac{2\pi \sigma}{\omega_2}} + e^{-\frac{2\pi \sigma}{\omega_2}} - \mathcal{E}_1 \right] \overline{q}(\sigma, a, \omega_1, \omega_2, m_0) = 0, 
\\
& \left[ \widetilde{Q}_{\text{rel}}^{\frac{1}{2}} \left(e^{i \omega_2 \partial_{\sigma}} + e^{-i \omega_2 \partial_{\sigma}}\right) + e^{\frac{2\pi \sigma}{\omega_1}} + e^{-\frac{2\pi \sigma}{\omega_1}} - \widetilde{\mathcal{E}}_1 \right] \overline{q}(\sigma, a, \omega_1, \omega_2, m_0) = 0, 
\end{align}
\end{subequations}
studied numerically in Section \ref{sec3.2}. Again, based on the analogy with the non-relativistic case \eqref{tot} and the analysis of \cite{Kharchev:2001rs} we expect a linear combination of the form
\begin{equation}
q(\sigma, a, \omega_1, \omega_2, m_0) \; \propto \; q^{\text{(ac)}}(\sigma, a, \omega_1, \omega_2, m_0) - \xi q^{\text{(c)}}(\sigma, a, \omega_1, \omega_2, m_0) \label{combo}
\end{equation}
to be the correct eigenfunction of the Baxter and dual Baxter problem (modulo an overall $\sigma$-independent factor), where $\xi \in \mathbb{C}$ and, similarly to \eqref{pappaQ},
\begin{equation}
\begin{split}
q^{\text{(c)}}(\sigma, a, \omega_1, \omega_2, m_0) & = q^{\text{(c)}}(\sigma, a, \omega_1, \omega_2) q^{\text{(c)},\text{inst}}_{\text{NS}}(\sigma, a, \omega_1, \omega_2, m_0), \\
q^{\text{(ac)}}(\sigma, a, \omega_1, \omega_2, m_0) & = e^{\frac{2 i \pi m_0 \sigma}{\omega_1 \omega_2}} q^{\text{(ac)}}(\sigma, a, \omega_1, \omega_2) q^{\text{(ac)},\text{inst}}_{\text{NS}}(\sigma, a, \omega_1, \omega_2, m_0), \label{mireia}
\end{split}
\end{equation}
should be the ``NS limit'' of the partition function of the five-dimensional $\mathcal{N} = 1$ $SU(2)$ theory on $S^3_{\omega_1, \omega_2} \times \mathbb{R}^2$ in the presence of a type II codimension-two defect on $S^3_{\omega_1, \omega_2}$ with chiral/antichiral multiplets, in line with the discussion of Section \ref{3.3.3}. 
More precisely we expect \eqref{mireia} will have the correct asymptotic behaviour, but will not be entire functions if taken separately (because of the poles coming from the $\theta$-functions/double sine functions); nevertheless their linear combination should be an entire function, even if only for those values of $a$ satisfying the exact quantization conditions \eqref{ex2toda5d}. 

In order to check this expectation, we need to compute the functions \eqref{mireia}; 
since we already know the perturbative part from \eqref{qcrelopen}, \eqref{qacrelopen}, all it remains to understand is how to include instanton corrections. According to \cite{Bullimore:2014upa}, and similarly to what suggested for the $S^5_{\omega_1, \omega_2, \omega_3}$ partition function, the instanton contributions $q^{\text{(c)},\text{inst}}_{\text{NS}}$, $q^{\text{(ac)},\text{inst}}_{\text{NS}}$ should factorize into the product of two copies of the type II defect instanton partition function $Z^{(\text{c}), \text{inst}}_{\text{3d/5d}, \text{NS}}$ or $Z^{(\text{ac}), \text{inst}}_{\text{3d/5d}, \text{NS}}$ on flat space $\mathbb{R}^2_{\epsilon_1} \times \mathbb{R}^2 \times S^1_R$, with $\epsilon_1$, $R$ and $\omega_1$, $\omega_2$ identified as in the first two lines\footnote{The ``NS limit'' on $S^5_{\omega_1, \omega_2, \omega_3}$ we are considering corresponds to sending $\omega_3 \rightarrow 0$; in this limit only the first two fixed points remain, corresponding to north and south pole of $S^3_{\omega_1, \omega_2}$ in $S^3_{\omega_1, \omega_2} \times \mathbb{R}^2$.} of Table \ref{tabo}. The functions $Z^{(\text{c}), \text{inst}}_{\text{3d/5d}, \text{NS}}$, $Z^{(\text{ac}), \text{inst}}_{\text{3d/5d}, \text{NS}}$  can be computed as discussed in \cite{Gaiotto:2014ina,Bullimore:2014awa}; we collected the relevant formulae in Appendix \ref{appA}. In terms of the flat space variables
\begin{equation}
q_1=e^{2\pi R \epsilon_1}, \;\; w=e^{-2\pi R \sigma}, \;\; \mu=e^{2\pi R a}, \;\; 
Q_{\text{5d}} = e^{-2\pi R m_0} \label{flatvar}
\end{equation}
we find
\begin{equation}
Z^{(\text{c}), \text{inst}}_{\text{3d/5d}, \text{NS}}(w, \mu, q_1, Q_{\text{5d}}) = 
1 + \dfrac{Q_{\text{5d}} q_1^2 w(\mu^{1/2} + \mu^{-1/2} - q_1 w - q_1^2 w)}{(1-q_1)(1-q_1 \mu)(1-q_1 \mu^{-1})(1-q_1w\mu^{1/2})(1-q_1w\mu^{-1/2})} + o(Q^2_{\text{5d}}), \label{lockhart}
\end{equation}
while
\begin{equation}
Z^{(\text{ac}), \text{inst}}_{\text{3d/5d}, \text{NS}}(w, \mu, q_1, Q_{\text{5d}}) = 
Z^{(\text{c}), \text{inst}}_{\text{3d/5d}, \text{NS}}(w, \mu, q_1^{-1}, Q_{\text{5d}}).
\label{relation}
\end{equation}
A few properties of the function \eqref{lockhart} should be mentioned:
\begin{itemize}
\item The flat space defect instanton partition function \eqref{lockhart} presents poles at $-i R \epsilon_1 \in \mathbb{Q}_+$ (or $\omega_1 / \omega_2 \in \mathbb{Q}_+$): these are the same poles that were affecting the ``naive'' quantization conditions \eqref{falseqc} one would obtain from considering gauge theories in flat space. Moving to curved background $S^3_{\omega_1, \omega_2} \times \mathbb{R}^2$ we get two copies of this function, related by the exchange $\omega_1 \leftrightarrow \omega_2$, and the expectation is that their poles will cancel each other (similarly to what happens for the exact quantization conditions \eqref{ex2toda5d}).
\item The function \eqref{lockhart} (and similarly \eqref{relation}) can also be written in the form
\begin{equation}
Z^{(\text{c}), \text{inst}}_{\text{3d/5d}, \text{NS}}(w, \mu, q_1, Q_{\text{5d}}) =
\exp\left(- F^{(\text{c}), \text{inst}}_{\text{3d/5d}, \text{NS}}(w, \mu, q_1, Q_{\text{5d}}) \right), \label{lo}
\end{equation}
where 
\begin{equation}
F^{(\text{c}), \text{inst}}_{\text{3d/5d}, \text{NS}}(w, \mu, q_1, Q_{\text{rel}}) = 
\sum_{n = 1}^{\infty} \sum_{s_1} \sum_{d_1, d_2 > 0} \sum_{m \in \mathbb{Z}}
D^{s_1}_{m,d_1,d_2}\dfrac{q_1^{n s_1}}{n(1-q_1^n)} (\mu^{-1})^{n d_1} (\mu^{-1} Q_{\text{rel}})^{n d_2} w^{mn}. \label{rivellinoflat}
\end{equation}
Here $d_2$ is taken to run over $\mathbb{N}$, while $d_1$ is allowed to take values in $\mathbb{N}/2$.\footnote{This is in contrast to the usual conventions in the literature, see for example \cite{Kashani-Poor:2016edc}, in which $\sigma$ is shifted by $a/2$ in such a way to have $d_1$ integer; as a consequence, our BPS invariants $D^{s_1}_{m,d_1,d_2}$ will be different from the ones in the literature.} This expression is almost the same as the BPS part of the open topological string partition function \eqref{opentop}, except for the fact that the BPS states associated to the 1-loop factor in \eqref{coso} (which in the present setting would correspond to $(q w \mu^{\frac{1}{2}};q)_{\infty}(q w \mu^{-\frac{1}{2}};q)_{\infty}$ in the numerator of \eqref{second}) are missing in \eqref{rivellinoflat}.
\item The function \eqref{lockhart} is symmetric under exchange $\mu \leftrightarrow \mu^{-1}$, i.e.
\begin{equation}
Z^{(\text{c}), \text{inst}}_{\text{3d/5d}, \text{NS}}(w, \mu, q_1, Q_{\text{5d}}) 
\; = \;
Z^{(\text{c}), \text{inst}}_{\text{3d/5d}, \text{NS}}(w, \mu^{-1}, q_1, Q_{\text{5d}}),
\end{equation}
and also satisfies
\begin{equation}
Z^{(\text{c}), \text{inst}}_{\text{3d/5d}, \text{NS}}(w, \mu, q_1, Q_{\text{5d}}) 
\; \propto \;
Z^{(\text{c}), \text{inst}}_{\text{3d/5d}, \text{NS}}(w^{-1}, \mu, q_1^{-1}, Q_{\text{5d}}) , \label{catilina}
\end{equation}
where the proportionality factor is $\sigma$-independent.
\end{itemize}
At this point, after moving from flat space variables \eqref{flatvar} to curved space ones \eqref{curvedvar} and using the identifications in Table \ref{tabo}, according to \cite{Bullimore:2014upa} we should have 
\begin{equation}
\begin{split}
q^{\text{(c)},\text{inst}}_{\text{NS}}(\sigma, a, \omega_1, \omega_2, m_0) 
&\, \propto \, Z^{(\text{c}), \text{inst}}_{\text{3d/5d}, \text{NS}}(w, \mu, e^{2\pi i}q, Q_{\text{rel}})
Z^{(\text{c}), \text{inst}}_{\text{3d/5d}, \text{NS}}(\widetilde{w}^{-1}, \widetilde{\mu}, e^{-2\pi i}\widetilde{q}^{-1}, \widetilde{Q}_{\text{rel}}), \\[5 pt]
q^{\text{(ac)},\text{inst}}_{\text{NS}}(\sigma, a, \omega_1, \omega_2, m_0) 
&\, \propto \, Z^{(\text{ac}), \text{inst}}_{\text{3d/5d}, \text{NS}}(w^{-1}, \mu, e^{-2\pi i}q^{-1}, Q_{\text{rel}})
Z^{(\text{ac}), \text{inst}}_{\text{3d/5d}, \text{NS}}(\widetilde{w}, \widetilde{\mu}, e^{2\pi i}\widetilde{q}, \widetilde{Q}_{\text{rel}}) \\[5 pt]
&\, =\, Z^{(\text{c}), \text{inst}}_{\text{3d/5d}, \text{NS}}(w^{-1}, \mu, e^{2\pi i}q, Q_{\text{rel}})
Z^{(\text{c}), \text{inst}}_{\text{3d/5d}, \text{NS}}(\widetilde{w}, \widetilde{\mu}, e^{-2\pi i}\widetilde{q}^{-1}, \widetilde{Q}_{\text{rel}}), \label{falseinst}
\end{split}
\end{equation}
modulo $\sigma$-independent terms.
Here powers of $w^{\pm 1}$, $q^{\pm 1}$ in the various factors are chosen in such a way that, when combined with the numerators in \eqref{second}, the functions $Z^{(\text{c}), \text{inst}}_{\text{3d/5d}, \text{NS}}$ will be completed to full open topological string partition functions, but this is not the only possibility: in fact by using  property \eqref{catilina} other choices are possible, all of them only differing by $\sigma$-independent terms. 
In fact a more convenient choice would be to remove these $\sigma$-independent terms completely, in the following sense: remembering property \eqref{lo}, 
we can separate the $m \in \mathbb{Z}$ sum in  
\begin{equation}
F^{(\text{c}), \text{inst}}_{\text{3d/5d}, \text{NS}}(w, \mu, e^{2\pi i} q, Q_{\text{rel}}) = 
\sum_{n = 1}^{\infty} \sum_{s_1} \sum_{d_1, d_2 > 0} \sum_{m \in \mathbb{Z}}
D^{s_1}_{m,d_1,d_2}\dfrac{(-1)^{2n s_1}q^{n s_1}}{n(1-q_1^n)} (\mu^{-1})^{n d_1} (\mu^{-1} Q_{\text{rel}})^{n d_2} w^{mn} \label{rivellino}
\end{equation}
into three sums over $m > 0$, $m = 0$ and $m < 0$:
\begin{equation}
\begin{split}
& F^{(\text{c}), \text{inst}}_{\text{3d/5d}, \text{NS}}(w, \mu, e^{2\pi i}q, Q_{\text{rel}}) = \\[5 pt]
& F^{(\text{c}), \text{inst}, (+)}_{\text{3d/5d}, \text{NS}}(w, \mu, e^{2\pi i}q, Q_{\text{rel}}) +
F^{(\text{c}), \text{inst}, (0)}_{\text{3d/5d}, \text{NS}}(\mu, e^{2\pi i}q, Q_{\text{rel}}) +
F^{(\text{c}), \text{inst}, (-)}_{\text{3d/5d}, \text{NS}}(w, \mu, e^{2\pi i}q, Q_{\text{rel}}) ;
\end{split}
\end{equation}
the $m = 0$ part is independent of $\sigma$, while the $m > 0$ and $m < 0$ parts turn out to be related by
\begin{equation}
F^{(\text{c}), \text{inst}, (+)}_{\text{3d/5d}, \text{NS}}(w, \mu, e^{2\pi i}q, Q_{\text{rel}}) =
F^{(\text{c}), \text{inst}, (-)}_{\text{3d/5d}, \text{NS}}(w^{-1}, \mu, e^{-2\pi i}q^{-1}, Q_{\text{rel}}). \label{emile} 
\end{equation}
Because of this property, if we remove the ($\sigma$-independent) $m = 0$ contribution the modified prepotential
\begin{equation}
\begin{split}
\widehat{F}^{(\text{c}), \text{inst}}_{\text{3d/5d}, \text{NS}}(w, \mu, e^{2\pi i}q, Q_{\text{rel}}) & =
F^{(\text{c}), \text{inst}}_{\text{3d/5d}, \text{NS}}(w, \mu, e^{2\pi i}q, Q_{\text{rel}}) - F^{(\text{c}), \text{inst}, (0)}_{\text{3d/5d}, \text{NS}}(\mu, e^{2\pi i}q, Q_{\text{rel}}) \\[5 pt]
& = F^{(\text{c}), \text{inst}, (+)}_{\text{3d/5d}, \text{NS}}(w, \mu, e^{2\pi i}q, Q_{\text{rel}}) + F^{(\text{c}), \text{inst},(-)}_{\text{3d/5d}, \text{NS}}(w, \mu, e^{2\pi i}q, Q_{\text{rel}}) \\[5 pt]
\end{split}
\end{equation}
and the modified type II defect instanton partition function
\begin{equation}
\widehat{Z}^{(\text{c}), \text{inst}}_{\text{3d/5d}, \text{NS}}(w, \mu, e^{2\pi i}q, Q_{\text{rel}}) =
\exp\left(- \widehat{F}^{(\text{c}), \text{inst}}_{\text{3d/5d}, \text{NS}}(w, \mu, e^{2\pi i}q, Q_{\text{rel}}) \right)
\end{equation}
satisfy the additional property
\begin{equation}
\begin{split}
\widehat{F}^{(\text{c}), \text{inst}}_{\text{3d/5d}, \text{NS}}(w, \mu, e^{2\pi i}q, Q_{\text{rel}}) & = \widehat{F}^{(\text{c}), \text{inst}}_{\text{3d/5d}, \text{NS}}(w^{-1}, \mu, e^{-2\pi i}q^{-1}, Q_{\text{rel}}), \\[5 pt]
\widehat{Z}^{(\text{c}), \text{inst}}_{\text{3d/5d}, \text{NS}}(w, \mu, e^{2\pi i}q, Q_{\text{rel}}) & = \widehat{Z}^{(\text{c}), \text{inst}}_{\text{3d/5d}, \text{NS}}(w^{-1}, \mu, e^{-2\pi i}q^{-1}, Q_{\text{rel}}), \label{propertybis}
\end{split}
\end{equation}
instead of \eqref{catilina} which involves a proportionality factor. 
If we now choose to fix the $\sigma$-independent term in \eqref{falseinst} according to
\begin{equation}
\begin{split}
q^{\text{(c)},\text{inst}}_{\text{NS}}(\sigma, a, \omega_1, \omega_2, m_0) 
& = \widehat{Z}^{(\text{c}), \text{inst}}_{\text{3d/5d}, \text{NS}}(w, \mu, e^{2\pi i}q, Q_{\text{rel}})
\widehat{Z}^{(\text{c}), \text{inst}}_{\text{3d/5d}, \text{NS}}(\widetilde{w}^{-1}, \widetilde{\mu}, e^{-2\pi i}\widetilde{q}^{-1}, \widetilde{Q}_{\text{rel}}), \\[5 pt]
q^{\text{(ac)},\text{inst}}_{\text{NS}}(\sigma, a, \omega_1, \omega_2, m_0) 
& = \widehat{Z}^{(\text{ac}), \text{inst}}_{\text{3d/5d}, \text{NS}}(w^{-1}, \mu, e^{-2\pi i}q^{-1}, Q_{\text{rel}})
\widehat{Z}^{(\text{ac}), \text{inst}}_{\text{3d/5d}, \text{NS}}(\widetilde{w}, \widetilde{\mu}, e^{2\pi i}\widetilde{q}, \widetilde{Q}_{\text{rel}}) \\[5 pt]
& = \widehat{Z}^{(\text{c}), \text{inst}}_{\text{3d/5d}, \text{NS}}(w^{-1}, \mu, e^{2\pi i}q, Q_{\text{rel}})
\widehat{Z}^{(\text{c}), \text{inst}}_{\text{3d/5d}, \text{NS}}(\widetilde{w}, \widetilde{\mu}, e^{-2\pi i}\widetilde{q}^{-1}, \widetilde{Q}_{\text{rel}}), \label{trueinst}
\end{split}
\end{equation}
then $q^{\text{(c)},\text{inst}}_{\text{NS}}$, $q^{\text{(ac)},\text{inst}}_{\text{NS}}$ are mapped into each other under complex conjugation\footnote{This is true both at $\omega_1$, $\omega_2 \in \mathbb{R}$ and $\omega_1 = \overline{\omega}_2 \in \mathbb{C}$ for $\sigma$, $a$, $m_0 \in \mathbb{R}$.} as well as under the exchange $\sigma \leftrightarrow -\sigma$; these two properties, together with property \eqref{complconj} for the double sine function, imply that our proposed solution $\overline{q}(\sigma, a ,\omega_1, \omega_2, m_0)$ (modulo an overall normalization)
\begin{equation}
\begin{split}
& \overline{q}(\sigma, a ,\omega_1, \omega_2, m_0) \; \propto \; \\
& e^{\frac{i \pi m_0 \sigma}{\omega_1 \omega_2}} 
S_2^{-1}\big( i(\sigma - a/2) \vert \omega_1, \omega_2 \big) S_2^{-1}\big( i(\sigma + a/2) \vert \omega_1, \omega_2 \big) q^{\text{(ac)},\text{inst}}_{\text{NS}}(\sigma, a, \omega_1, \omega_2, m_0) \\
& - \xi \,e^{-\frac{i \pi m_0 \sigma}{\omega_1 \omega_2}} S_2^{-1}\big( -i(\sigma - a/2) \vert \omega_1, \omega_2 \big) S_2^{-1}\big( -i(\sigma + a/2) \vert \omega_1, \omega_2 \big) q^{\text{(c)},\text{inst}}_{\text{NS}}(\sigma, a, \omega_1, \omega_2, m_0) 
\label{proposal}
\end{split}
\end{equation}
with instanton contributions as in \eqref{trueinst} will be either purely real or purely imaginary and with definite $\sigma \leftrightarrow -\sigma$ parity if we choose $\xi = -(-1)^n$ for the level $n$ eigenfunction: this is as expected, due to the symmetry $\sigma \leftrightarrow - \sigma$ of the problem \eqref{sym5d}. 

We now have all the ingredients entering our proposed solution \eqref{proposal} to the Baxter and dual Baxter equations associated to the 2-particle relativistic Toda chain.
It is easy to check that the two functions entering in the linear combination \eqref{proposal} separately are at least formal solutions to \eqref{sym5d} perturbatively in $Q_{\text{rel}}$, $\widetilde{Q}_{\text{rel}}$; however, there are a few problems we should face: 
\begin{itemize}
\item  First, we have to check that poles at $\omega_1/\omega_2 \in \mathbb{Q}_+$ of the function \eqref{lockhart} will cancel if we consider two copies of it in the combinations \eqref{trueinst};
\item Second, it is not clear that our proposed solution \eqref{proposal} will be entire and $L^2(\mathbb{R})$ for the values of $a = a^{(n)}$ satisfying the exact quantization conditions and for some appropriate value of $\xi$ (which we fixed to $\xi = -(-1)^{n}$ due to the symmetry $\sigma \leftrightarrow -\sigma$ of the problem \eqref{sym5d}).
\end{itemize}
The first point can be addressed easily, since it has already been shown in \cite{Kashani-Poor:2016edc} that the combinations 
\begin{equation}
\widehat{F}^{(\text{c}), \text{inst}}_{\text{3d/5d}, \text{NS}}(w^{\pm}, \mu, e^{2\pi i}q, Q_{\text{rel}}) + \widehat{F}^{(\text{c}), \text{inst}}_{\text{3d/5d}, \text{NS}}(\widetilde{w}^{\pm 1}, \widetilde{\mu}, e^{2\pi i}\widetilde{q}, \widetilde{Q}_{\text{rel}})
\end{equation}
appearing in \eqref{trueinst} are free of poles at $\omega_1/\omega_2 \in \mathbb{Q}_+$; a key ingredient for the poles cancellation argument of \cite{Kashani-Poor:2016edc} is the term $(-1)^{2 n s_1}$ in \eqref{rivellino}, which arises due to the identification of Table \ref{tabo} (or alternatively to the $B$-field).\footnote{As we already mentioned the $B$-field is trivial in the $N = 2$ case or more in general for $N$ even, but it has an important effect for odd $N$.} Since the perturbative contributions \eqref{qcrelopen}, \eqref{qacrelopen} do not present poles at $\omega_1/\omega_2 \in \mathbb{Q}_+$ because of the definition of double sine function, we can conclude that \eqref{proposal} will be free of divergences at $\omega_1/\omega_2 \in \mathbb{Q}_+$.

\begin{figure}[]
  \centering
  \subfloat[Numerical results]{\includegraphics[width=1\textwidth]{bax5dmodnum.pdf}\label{bax5dmodnumbis}}
\newline
  \subfloat[Gauge theory results]{\includegraphics[width=1\textwidth]{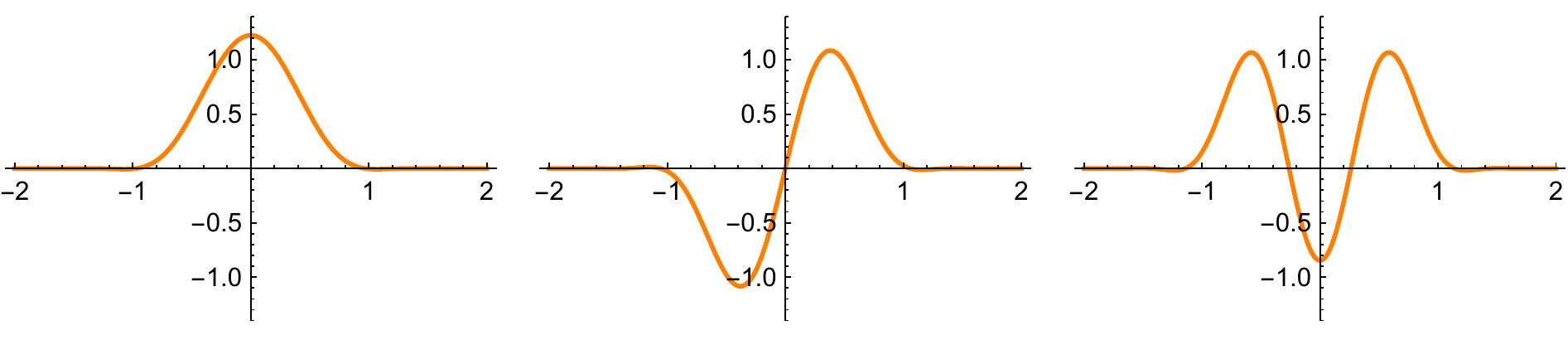}\label{bax5dmodgauge}}
\newline
  \subfloat[Difference]{\includegraphics[width=1\textwidth]{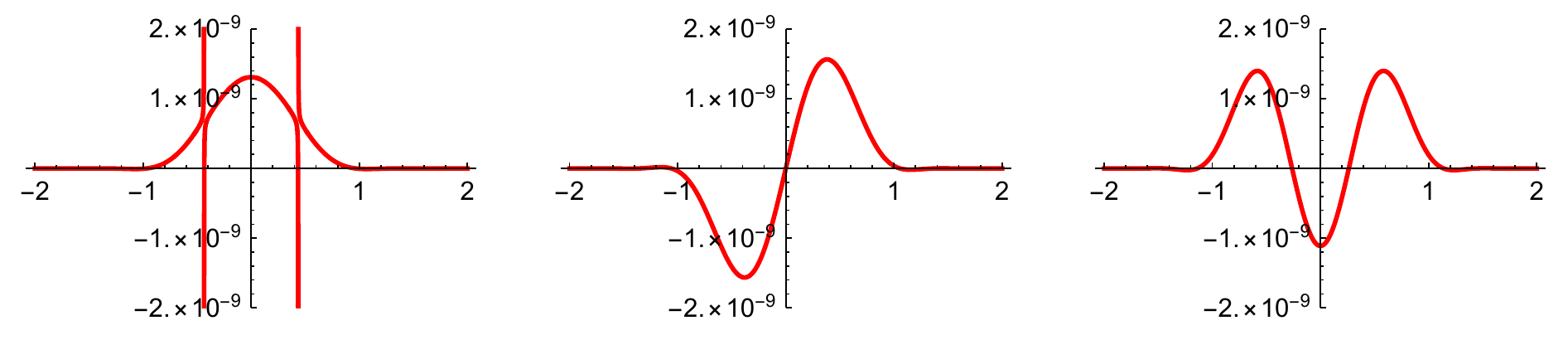}\label{bax5dmodcomp}}
  \caption{Level $n = 0, 1, 2$ gauge theory 2-particle Baxter solution $\overline{q}(\sigma)$ for \\$\omega_1 = \frac{1}{\sqrt{2}}$, $\omega_2 = 1$, $m_0 = -\frac{\ln 3}{2 \pi}$. }
 \label{plobax5dmod}
\end{figure}

\begin{figure}[]
  \centering
  \subfloat[Numerical results]{\includegraphics[width=0.9\textwidth]{bax5dexnum.pdf}\label{bax5dexnumbis}}
\newline
  \subfloat[Gauge theory results]{\includegraphics[width=0.9\textwidth]{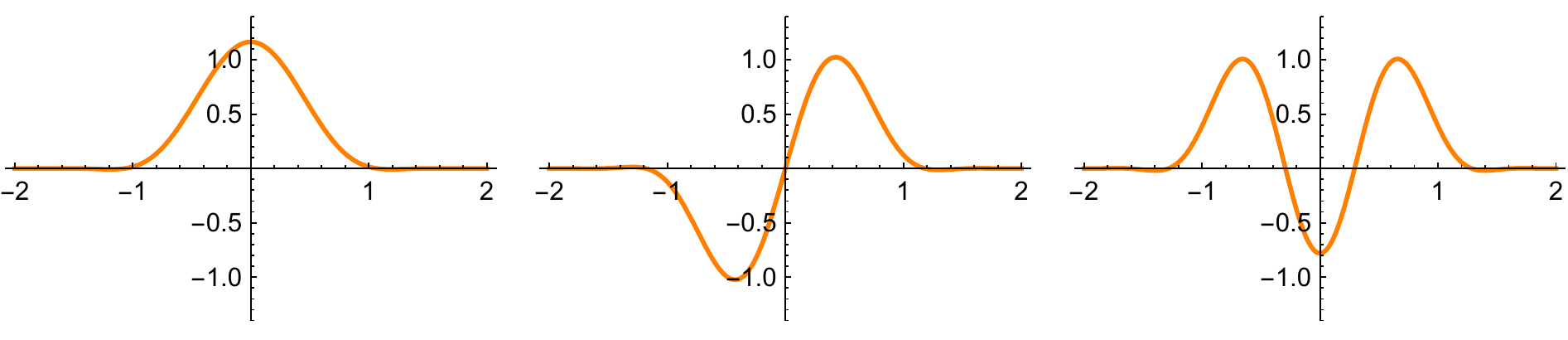}\label{bax5dexgauge}}
\newline
\subfloat[Difference]{\includegraphics[width=0.9\textwidth]{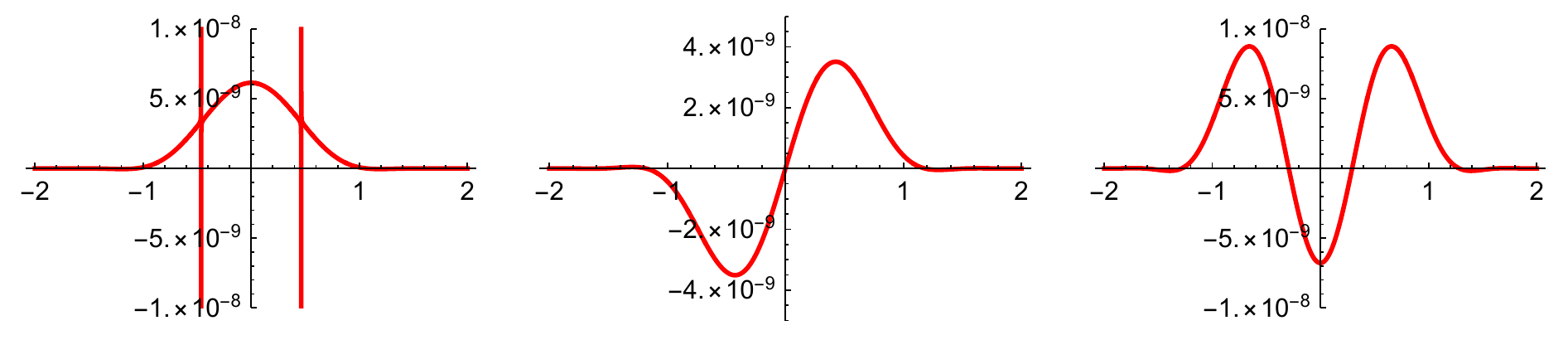}\label{bax5dexcomp}}
  \caption{Level $n = 0, 1, 2$ gauge theory 2-particle Baxter solution $\overline{q}(\sigma)$ for \\ $\omega_1 = 2^{-\frac{1}{4}}$, $\omega_2 = 2^{\frac{1}{4}}$, $m_0 =0$. }
 \label{plobax5dex}
\end{figure}

However, we should still understand if our proposal \eqref{proposal} is an entire and $L^2(\mathbb{R})$ function when the quantization conditions are satisfied.
One observation is that the poles in $\sigma$ coming from the instanton contributions \eqref{trueinst} seem to be cancelled by the zeroes of the numerators in \eqref{second}, thus leaving only the poles coming from the $\theta$-functions in \eqref{second} as possible singularities of $\overline{q}(\sigma, a ,\omega_1, \omega_2, m_0)$. It seems however hard to show analytically that these poles actually disappear at the particular values $a = a^{(n)}$ determined by the exact quantization conditions \eqref{ex2toda5d} since we do not have a closed form or TBA-like expression for the instanton contribution \eqref{trueinst}; for the same reason it would be hard to prove $L^2(\mathbb{R})$-normalizability. We will therefore content ourselves with performing some numerical check of our proposal: more precisely, as we did in Section \ref{2.3.3}, we plot our proposed solution $\overline{q}(\sigma, a ,\omega_1, \omega_2, m_0)$ (computed up to 4 instantons) for $\sigma \in \mathbb{R}$ and compare the resulting plot against the Baxter eigenfunctions obtained via numerical methods in Section \ref{sec3.2}.\footnote{We thank Ivan Chi-Ho Ip for explanations on \cite{2011arXiv1108.5376C} about the problem of plotting the double sine function.} 
For example, in Figure \ref{bax5dmodgauge} we fixed $\omega_1 = \frac{1}{\sqrt{2}}$, $\omega_2 = 1$, $m_0 = -\frac{\ln 3}{2\pi}$ and plotted our proposed gauge theory Baxter solution Re$[\overline{q}(\sigma)]$, Im$[\overline{q}(\sigma)]$, Re$[\overline{q}(\sigma)]$ for energy levels $n = 0, 1, 2$ respectively (the other real or imaginary component being zero), evaluated at those $a^{(n)}$ solving the exact quantization conditions (see Table \ref{ii3}) and for $\xi = -(-1)^n$;
Figure \ref{bax5dmodcomp} instead represents the difference between gauge theory and numerical solutions (Figure \ref{bax5dmodnumbis}).\footnote{As for the non-relativistic Toda chain studied in Section \ref{sec2.3}, the Baxter solution obtained from gauge theory \eqref{proposal} is not normalized to 1; we therefore divide it by its norm in order to make comparison against numerical results.} 
\linebreak

\begin{figure}[h]
  \centering
\includegraphics[width=0.5\textwidth]{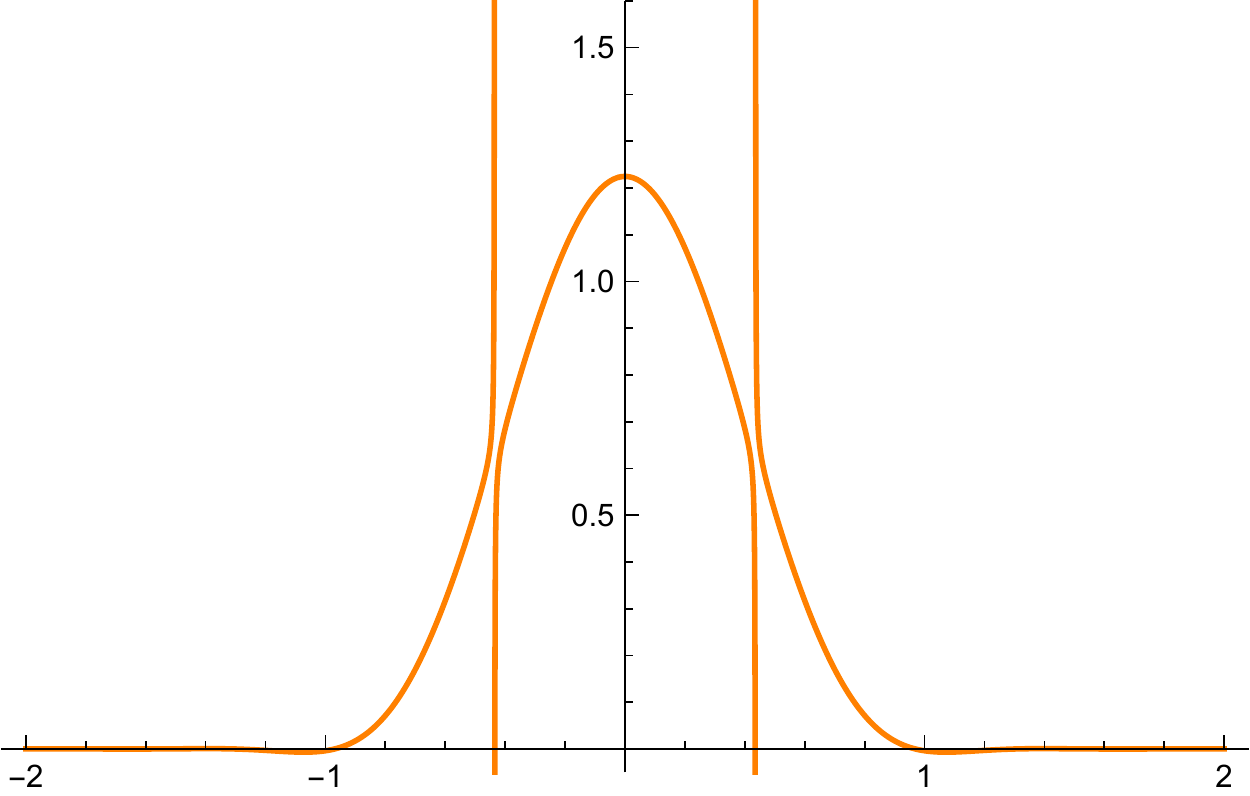}
  \caption{Level $n = 0$ gauge theory 2-particle Baxter solution $\overline{q}(\sigma)$ for $\omega_1 = \frac{1}{\sqrt{2}}$, $\omega_2 = 1$, $m_0 = -\frac{\ln 3}{2 \pi}$ without instanton corrections. }
  \label{bax5dmodgaugenoinst}
\end{figure}

\vspace{0.5 cm}

\begin{figure}[h]
  \centering
\includegraphics[width=0.5\textwidth]{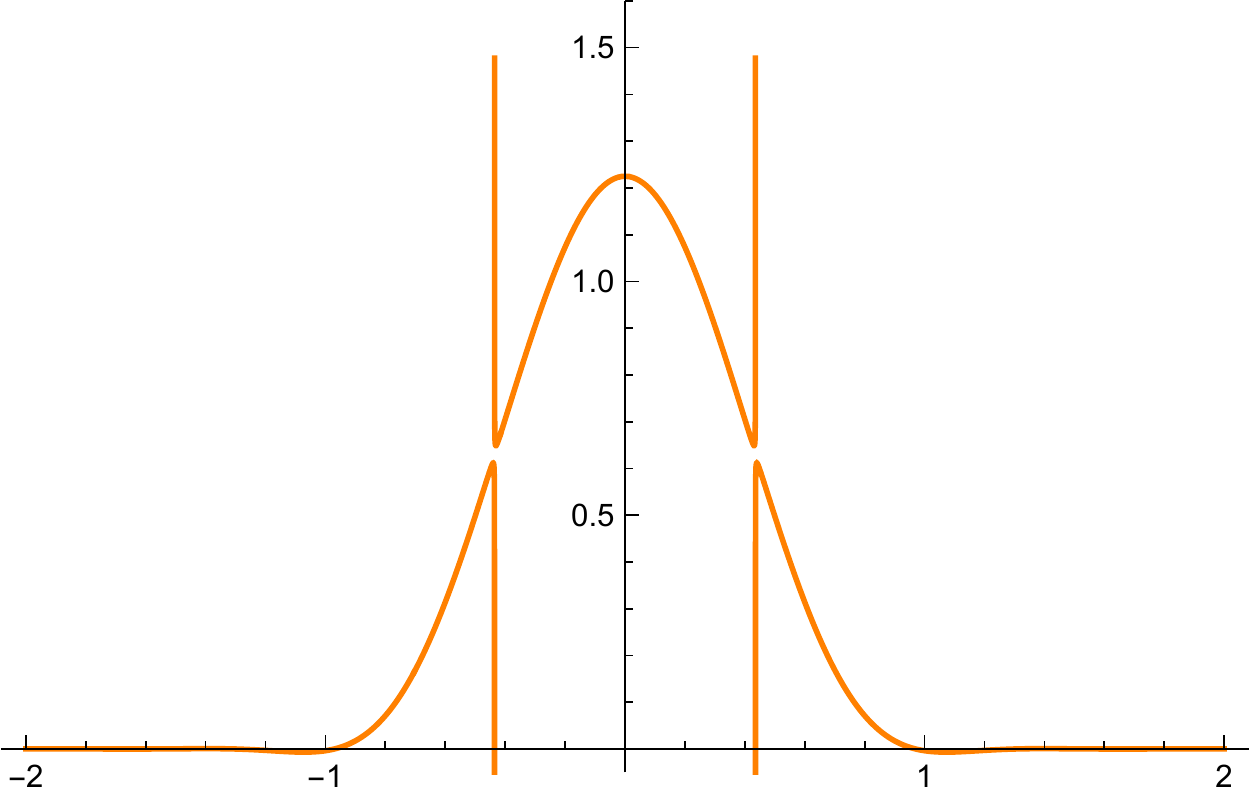}
  \caption{Level $n = 0$ gauge theory 2-particle Baxter solution $\overline{q}(\sigma)$ for $\omega_1 = \frac{1}{\sqrt{2}}$, $\omega_2 = 1$, $m_0 = -\frac{\ln 3}{2 \pi}$ with only one copy of the instanton corrections. }
  \label{gaugenosecondcopy}
\end{figure}

\vspace{0.5 cm}

\begin{figure}[h]
  \centering
\includegraphics[width=0.5\textwidth]{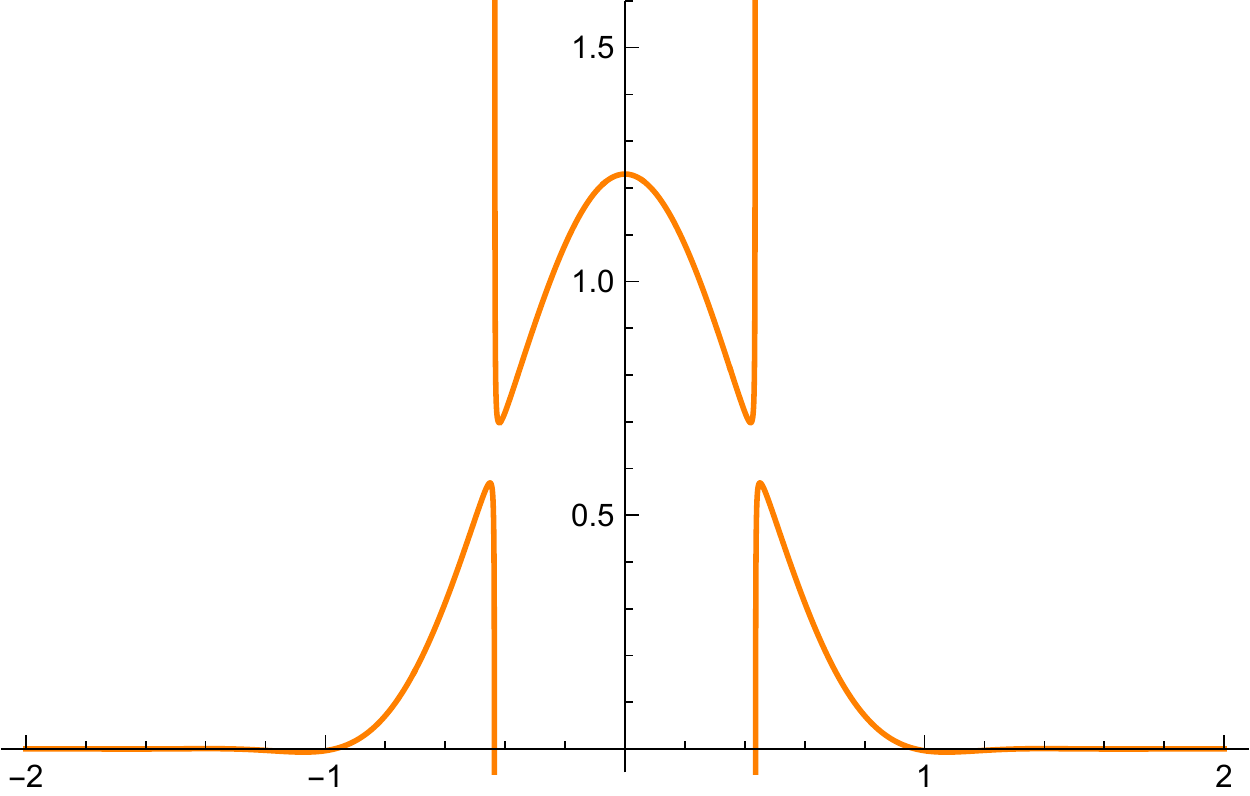}
  \caption{Gauge theory 2-particle Baxter solution $\overline{q}(\sigma)$ for $\omega_1 = \frac{1}{\sqrt{2}}$, $\omega_2 = 1$, $m_0 = -\frac{\ln 3}{2 \pi}$ at $a = 0.87$ (value not satisfying the quantization conditions). }
  \label{bax5dmodgaugewrong}
\end{figure}

\begin{figure}[]
  \centering
\includegraphics[width=1\textwidth]{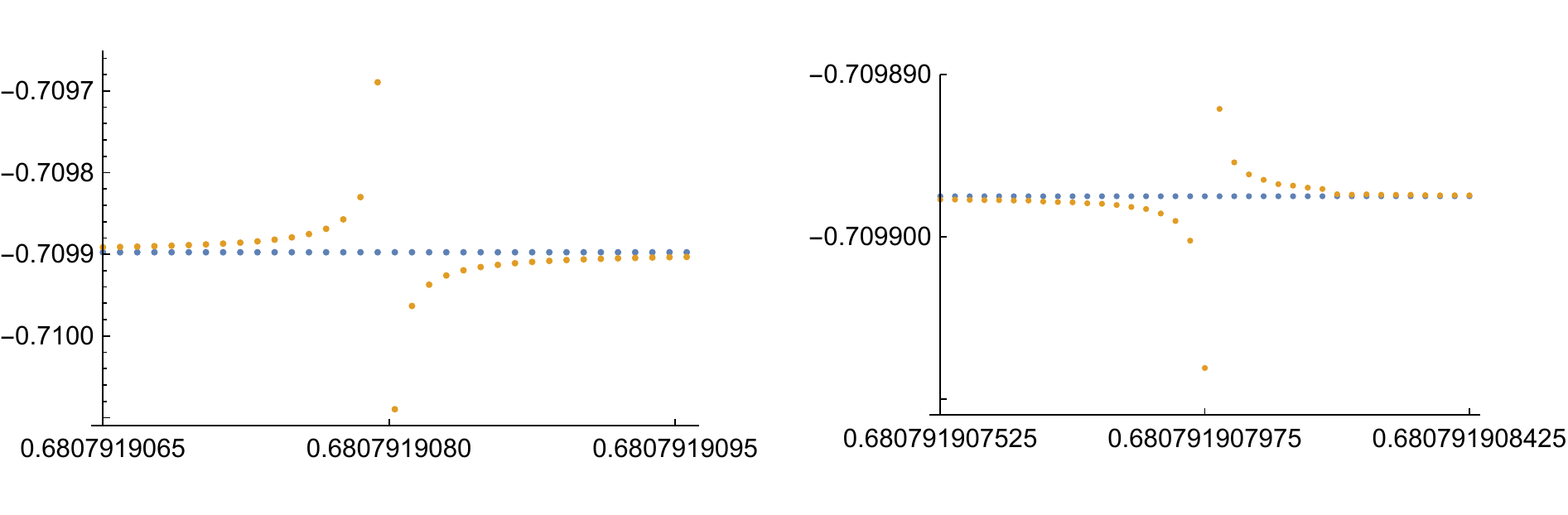}
 \caption{Level $n = 1$ numerical (blue dots) and gauge theory (orange dots) \\ 2-particle Baxter solution $\overline{q}(\sigma)$ for $\omega_1 = 2^{-\frac{1}{4}}$, $\omega_2 = 2^{\frac{1}{4}}$, $m_0 =0$ near $\sigma = a^{(1)}/2$. \\
 Left: 3-instanton results; right: 4-instanton results.}
  \label{evolution5d}
\end{figure}

Another example for $\omega_1 = 2^{-\frac{1}{4}}$, $\omega_2 = 2^{\frac{1}{4}}$, $m_0 =0$ is presented in Figure \ref{plobax5dex}. 
By the naked eye, numerical (blue) and gauge theory (orange) Baxter solutions seem to agree: 
in particular, the singularities at $\sigma = \pm a^{(n)}/2$ of our proposed gauge theory solution seem to disappear\footnote{Clearly one of the two poles can always be eliminated by appropriately choosing $\xi$; however, the expectation is that both poles disappear only when $a = a^{(n)}$ satisfies the exact quantization conditions.} already with a 4-instantons computation, in contrast to what we would get without considering instanton corrections (Figure \ref{bax5dmodgaugenoinst}), or without considering both copies of the defect instanton partition function in \eqref{trueinst} (Figure \ref{gaugenosecondcopy}, in which we did not consider the contribution coming from the second copy with tilded variables), or by using a value for $a$ not satisfying the exact quantization conditions (Figure \ref{bax5dmodgaugewrong}). 
In reality the singularities are still there, as they should disappear only when considering \textit{all} instanton contributions, but we may need a bigger resolution in our plots to see them. For example these singularities can be seen in the plot of the difference (red) between level $n = 0$ numerical and gauge theory solutions in Figure \ref{bax5dmodcomp}, but we need to consider lengths of order $\sim 10^{-9}$ to notice them; the same plot for levels $n = 1,2$ seems not to have singularities, but this simply means we need an even bigger resolution (so we should more appropriately think of these red plots as indicating the difference between numerical and gauge theory solution away from singular points). 
More precisely what we should expect is that, by adding more and more instanton corrections, the gauge theory solution approaches the exact one better and better, the singularities are smoothed away and the discontinuities get smaller and smaller.
An example of this behaviour is shown in Figure \ref{evolution5d} for the level $n = 1$ solution in the case $\omega_1 = 2^{-\frac{1}{4}}$, $\omega_2 = 2^{\frac{1}{4}}$, $m_0 =0$: while the numerical (blue dots) solution is smooth around $\sigma = a^{(1)}/2$, the gauge theory one (orange dots) diverges, but this divergence gets milder and tends to disappear by increasing the number of instantons (from 3 to 4 in this example), and we need to ``zoom'' more in order to see it. This is completely analogous to what we saw happening in Section \ref{sec2.3} for the non-relativistic Toda chain, and provides a few numerical checks of the validity of our proposal.


\subsection{Additional remarks} \label{seccomments5d}

There are a number of points which deserve further comments.

\begin{itemize}

\item As we discussed, our proposed solution \eqref{proposal} is free of poles at $\omega_1/\omega_2 \in \mathbb{Q}_+$ thanks to the argument of \cite{Kashani-Poor:2016edc}; nevertheless, we were only able to plot it for $\omega_1/\omega_2 \notin \mathbb{Q}_+$. The problem is not in the double sine functions, which admit nice representations when $\omega_1/\omega_2 \in \mathbb{Q}_+$ \cite{Garoufalidis:2014ifa}, but it is due to the complicated expressions for the instanton corrections 
which make it hard to study these particular values of the $\omega_i$'s.
In fact, already in order to show the absence of poles it is more convenient to rewrite the instanton corrections in an open topological string form; this alternative form is however not very good for plots since it is not exact in $w$ but is a Laurent series. 
It would be important to be able to study at least the self-dual point $\omega_1 = \omega_2 = 1$, $m_0 = 0$ in order to compare our proposal to the eigenfunction given in \cite{Marino:2016rsq,Kashaev:2017zmv}.
In any case, we think that this issue can be solved by a better computer analysis of \eqref{proposal}.  \\

\item In addition to studying the 2-particle relativistic closed Toda eigenfunction at special values $\omega_1 = \omega_2 = 1$ and $m_0 = 0$ (case in which Toda equations \eqref{5d2Toda} and Baxter equations \eqref{Bax5d2Toda} coincide), the authors of \cite{Kashaev:2017zmv} also considered the case $\omega_1$, $\omega_2 \in \mathbb{C}$ with $\omega_1 = \overline{\omega}_2$ and $m_0 = 0$.\footnote{As we already mentioned (see footnote \ref{fnrel}) it is possible to define good quantum mechanical problems both at $\omega_1$, $\omega_2 \in \mathbb{R}$ and at $\omega_1 = \overline{\omega}_2 \in \mathbb{C}$ \cite{Kharchev:2001rs}.}
It is not difficult to show that the gauge theory solution to the Baxter equation we proposed in Section \ref{sec3.4} reduces to the one of \cite{Kashaev:2017zmv} for complex $\omega_i$'s if we re-adjust the constant factors. Let us choose for definiteness Im$(\omega_1/\omega_2) > 0$, which implies $\vert q \vert < 1$, $\vert \widetilde{q}^{-1} \vert < 1$, and consider the linear combination
\begin{equation}
\begin{split}
& \overline{q}'(\sigma, a, \omega_1, \omega_2, m_0) = \\[5pt]
& e^{\frac{i \pi m_0 \sigma}{\omega_1 \omega_2}} \dfrac{e^{\frac{i \pi}{\omega_1 \omega_2}\left(\sigma^2 + \frac{a^2}{4}\right) - \pi \frac{\omega_1 + \omega_2}{\omega_1 \omega_2}\sigma - i\pi\frac{\omega_1^2 + \omega_2^2}{6\omega_1 \omega_2} - i \frac{\pi}{2}}}{\theta\left(i\frac{\sigma - a/2}{\omega_2},\frac{\omega_1}{\omega_2}\right)\theta\left(i\frac{\sigma + a/2}{\omega_2},\frac{\omega_1}{\omega_2}\right)}  \\[5pt]
& \times (q w^{-1} \mu^{\frac{1}{2}};q)_{\infty}(q w^{-1} \mu^{-\frac{1}{2}};q)_{\infty}
Z^{(\text{c}), \text{inst}}_{\text{3d/5d}, \text{NS}}(w^{-1}, \mu, e^{2\pi i}q, Q_{\text{rel}}) \\[5pt]
& \times (\widetilde{q}^{-1} \widetilde{w} \widetilde{\mu}^{\frac{1}{2}};\widetilde{q}^{-1})_{\infty}(\widetilde{q}^{-1} \widetilde{w} \widetilde{\mu}^{-\frac{1}{2}};\widetilde{q}^{-1})_{\infty}
Z^{(\text{c}), \text{inst}}_{\text{3d/5d}, \text{NS}}(\widetilde{w}, \widetilde{\mu}, e^{-2\pi i}\widetilde{q}^{-1}, \widetilde{Q}_{\text{rel}}) \\[5 pt]
& - \xi \, e^{-\frac{i \pi m_0 \sigma}{\omega_1 \omega_2}}\dfrac{e^{-\frac{i \pi}{\omega_1 \omega_2}\left(\sigma^2 + \frac{a^2}{4}\right) - \pi \frac{\omega_1 + \omega_2}{\omega_1 \omega_2}\sigma + i\pi\frac{\omega_1^2 + \omega_2^2}{6\omega_1 \omega_2} + i \frac{\pi}{2}}}{\theta\left(i\frac{\sigma - a/2}{\omega_1}, -\frac{\omega_2}{\omega_1}\right)\theta\left(i\frac{\sigma + a/2}{\omega_1}, -\frac{\omega_2}{\omega_1}\right)} 
\\[5pt]
& \times (q w \mu^{\frac{1}{2}};q)_{\infty}(q w \mu^{-\frac{1}{2}};q)_{\infty}
Z^{(\text{c}), \text{inst}}_{\text{3d/5d}, \text{NS}}(w, \mu, e^{2\pi i}q, Q_{\text{rel}}) \\[5pt]
& \times (\widetilde{q}^{-1} \widetilde{w}^{-1} \widetilde{\mu}^{\frac{1}{2}};\widetilde{q}^{-1})_{\infty}(\widetilde{q}^{-1} \widetilde{w}^{-1} \widetilde{\mu}^{-\frac{1}{2}};\widetilde{q}^{-1})_{\infty} Z^{(\text{c}), \text{inst}}_{\text{3d/5d}, \text{NS}}(\widetilde{w}^{-1}, \widetilde{\mu}, e^{-2\pi i}\widetilde{q}^{-1}, \widetilde{Q}_{\text{rel}}) \\[5pt]. \label{inter}
\end{split}
\end{equation}
This is very similar to our proposed solution \eqref{proposal}, but instead of using the function $\widehat{Z}^{(\text{c}), \text{inst}}_{\text{3d/5d}, \text{NS}}$ appearing in \eqref{trueinst} as instanton correction we are using $Z^{(\text{c}), \text{inst}}_{\text{3d/5d}, \text{NS}}$ (which only differ by a constant factor). Because of the symmetry $\sigma \leftrightarrow -\sigma$ we can choose $\xi = \pm 1$; then \eqref{inter} will be either purely real or purely imaginary since under complex conjugation $\omega_1$ and $\omega_2$ get exchanged.\footnote{This would no longer be true if we were to consider $\omega_1$, $\omega_2 \in \mathbb{R}$. In that case, in order to get a purely real or purely imaginary function we should remove an appropriate constant term from the instanton corrections; this is what we did in \eqref{proposal} where we used  $\widehat{Z}^{(\text{c}), \text{inst}}_{\text{3d/5d}, \text{NS}}$ instead of $Z^{(\text{c}), \text{inst}}_{\text{3d/5d}, \text{NS}}$.} At this point we notice that the coefficients $c_n(\mathcal{E}_1,q, Q_{\text{rel}})$ of the $w \sim 0$ expansion of 
\begin{equation}
\begin{split}
& (q w \mu^{\frac{1}{2}};q)_{\infty}(q w \mu^{-\frac{1}{2}};q)_{\infty}
Z^{(\text{c}), \text{inst}}_{\text{3d/5d}, \text{NS}}(w, \mu, e^{2\pi i}q, Q_{\text{rel}})
 = \\
& \sum_{n \geqslant 0}^{\infty} c_n(\mathcal{E}_1,q, Q_{\text{rel}}) \dfrac{w^n}{(q^{-1};q^{-1})_n} = \sum_{n \geqslant 0}^{\infty} (-1)^n q^{\frac{n(n+1)}{2}} c_n(\mathcal{E}_1,q, Q_{\text{rel}}) \dfrac{w^n}{(q;q)_n}
\end{split}
\end{equation}
can be written directly in terms of the energy $\mathcal{E}_1$ (and $q$, $Q_{\text{rel}}$) rather than the auxiliary variable $\mu$; for example
\begin{equation}
\begin{split}
& c_0 = 1, \;\;\;\;\; c_1 = \mathcal{E}_1, \;\;\;\;\; 
c_2 = \mathcal{E}_1^2 + q^{-1} (1-q)(1-q Q_{\text{rel}}), \\[5 pt]
& c_3 = \mathcal{E}_1 \Big[\mathcal{E}_1^2 + q^{-1} (1-q)(1-q Q_{\text{rel}}) + q^{-2} (1-q^2)(1-q^2 Q_{\text{rel}})\Big], \;\;\; \ldots.
\end{split}
\end{equation}
This series was shown in \cite{Kashaev:2017zmv} to be convergent for $\vert q \vert < 1$ (and $m_0 = 0$) and was a key ingredient in their construction of the 2-particle Toda eigenfunction at $\omega_1 = \overline{\omega}_2$. In fact it is easy to realize, after this observation, that \eqref{inter} coincides with the solution of \cite{Kashaev:2017zmv} (modulo constant factors), where the $\theta$-functions in \eqref{inter} due to the double sine functions coincide with the contribution of the Wronskian of two special solutions to the main functional equation analysed in \cite{Kashaev:2017zmv}. We therefore expect our proposed solution \eqref{proposal} to be an eigenfunction of both the quantum mechanical problem at $\omega_1$, $\omega_2 \in \mathbb{R}$ and the quantum mechanical problem at $\omega_1 = \overline{\omega}_2 \in \mathbb{C}$, after the appropriate normalization factors (or values of $\xi$) are taken into account.

\item Cancellation of poles at $\sigma = \pm \frac{a^{(n)}}{2}$ for our proposed solution \eqref{proposal} requires (the relativistic analogue of) conditions \eqref{quantcond} to be satisfied or, equivalently, that
\begin{equation}
\text{Res}\, \Big[\overline{q}(\sigma, a, \omega_1, \omega_2, m_0)\Big]\Big\vert_{\sigma = \pm \frac{a^{(n)}}{2}} = 0.
\end{equation}
Since in our case $\xi = -(-1)^n$, focussing on $\sigma = - \frac{a^{(n)}}{2}$ this condition reduces to
\begin{equation}
(-1)^n = e^{-i \pi n} = e^{-\frac{i \pi m_0 a^{(n)}}{\omega_1 \omega_2}} 
\dfrac{S_2^{-1}\left(-i a^{(n)} \vert \omega_1, \omega_2\right)}{S_2^{-1}\left(i a^{(n)} \vert \omega_1, \omega_2\right)}
\dfrac{q_{\text{NS}}^{\text{(ac),inst}}(-\frac{a^{(n)}}{2}, a^{(n)}, \omega_1, \omega_2, m_0)}{q_{\text{NS}}^{\text{(c),inst}}(-\frac{a^{(n)}}{2}, a^{(n)}, \omega_1, \omega_2, m_0)}. \label{treno}
\end{equation}
These are quantization conditions for the parameter $a$: only when $a = a^{(n)}$ satisfies \eqref{treno} our solution \eqref{proposal} will be free of singularities in $\sigma$. By consistency, \eqref{treno} should then be equivalent to the exact quantization conditions \eqref{ex2toda5d}; let us try to see if this is the case. First of all, by making use of the properties of the double sine function we can rewrite \eqref{treno} as
\begin{equation}
\begin{split}
e^{-i \pi n} \;=\; & e^{-\frac{i \pi m_0 a}{\omega_1 \omega_2} - \frac{i \pi a^2}{\omega_1 \omega_2} + \frac{i \pi}{6} \frac{\omega_1^2 + \omega_2^2}{\omega_1 \omega_2} + \frac{i \pi}{2}} \dfrac{\left(q e^{-\frac{2\pi a}{\omega_2}}; q\right)_{\infty}\left(e^{-\frac{2\pi a}{\omega_2}}; q\right)_{\infty}}{\left(\widetilde{q}^{-1} e^{-\frac{2\pi a}{\omega_1}}; \widetilde{q}^{-1}\right)_{\infty}\left(e^{-\frac{2\pi a}{\omega_1}}; \widetilde{q}^{-1}\right)_{\infty}} \\
& \dfrac{q_{\text{NS}}^{\text{(ac),inst}}(-\frac{a}{2}, a, \omega_1, \omega_2, m_0)}{q_{\text{NS}}^{\text{(c),inst}}(-\frac{a}{2}, a, \omega_1, \omega_2, m_0)} \;\;\;\;\;\; \text{at} \;\; a = a^{(n)}.
\end{split}
\end{equation}
Moving to logarithms (and multiplying by $2i$) we obtain
\begin{equation}
\begin{split}
2 \pi n \;=\; & \frac{2 \pi m_0 a}{\omega_1 \omega_2} + \frac{2 \pi a^2}{\omega_1 \omega_2} - \frac{\pi}{3} \frac{\omega_1^2 + \omega_2^2}{\omega_1 \omega_2} - \pi - 2i \sum_{k \geqslant 1} \dfrac{1}{k} \dfrac{1 + q^k}{1 - q^k} \mu^{-k}
- 2i \sum_{k \geqslant 1} \dfrac{1}{k} \dfrac{1 + \widetilde{q}^k}{1 - \widetilde{q}^k} \widetilde{\mu}^{-k} \\[5 pt]
& -2i \widehat{F}^{\text{(ac),inst}}_{\text{3d/5d,NS}}(\mu^{1/2}, \mu, e^{2\pi i}q, Q_{\text{rel}}) 
-2i \widehat{F}^{\text{(ac),inst}}_{\text{3d/5d,NS}}(\widetilde{\mu}^{1/2},\widetilde{ \mu}, e^{2\pi i}\widetilde{q}, \widetilde{Q}_{\text{rel}})  \\[5 pt]
& + 2i \widehat{F}^{\text{(c),inst}}_{\text{3d/5d,NS}}(\mu^{1/2}, \mu, e^{2\pi i}q, Q_{\text{rel}})
+ 2i \widehat{F}^{\text{(c),inst}}_{\text{3d/5d,NS}}(\widetilde{\mu}^{1/2}, \widetilde{\mu}, e^{2\pi i}\widetilde{q}, \widetilde{Q}_{\text{rel}}) \label{questocoso}
\end{split}
\end{equation}
at $a = a^{(n)}$. It is easy to recognize that the first line of \eqref{questocoso} precisely reproduces the classical part of the exact quantization conditions \eqref{ex2toda5d} as well as the contribution coming from the BPS states with $d_2 = 0$, i.e. $(j_L, j_R)_{d_1,d_2} = (0, \frac{1}{2})_{1,0}$ with $N_{0,\frac{1}{2}}^{1,0} = 1$; a full matching between \eqref{ex2toda5d} and \eqref{questocoso} will therefore be obtained if 
\begin{equation}
\begin{split}
& \frac{2}{i} \partial_a \mathcal{W}_{\text{5d}}^{\text{inst}}(\mu, e^{2\pi i}q, Q_{\text{rel}}) = \\[5 pt]
&  -2i \widehat{F}^{\text{(ac),inst}}_{\text{3d/5d,NS}}(\mu^{1/2}, \mu, e^{2\pi i}q, Q_{\text{rel}}) 
 + 2i \widehat{F}^{\text{(c),inst}}_{\text{3d/5d,NS}}(\mu^{1/2}, \mu, e^{2\pi i}q, Q_{\text{rel}}), \\[5pt]
& \frac{2}{i} \partial_a \mathcal{W}_{\text{5d}}^{\text{inst}}(\widetilde{ \mu}, e^{2\pi i}\widetilde{q}, \widetilde{Q}_{\text{rel}}) =  \\[5 pt]
&  -2i \widehat{F}^{\text{(ac),inst}}_{\text{3d/5d,NS}}(\widetilde{\mu}^{1/2},\widetilde{ \mu}, e^{2\pi i}\widetilde{q}, \widetilde{Q}_{\text{rel}}) 
 + 2i \widehat{F}^{\text{(c),inst}}_{\text{3d/5d,NS}}(\widetilde{\mu}^{1/2}, \widetilde{\mu}, e^{2\pi i}\widetilde{q}, \widetilde{Q}_{\text{rel}}), \label{chissa}
\end{split}
\end{equation}
with $\mathcal{W}_{\text{5d}}^{\text{inst}}$ as in \eqref{split}. In order for \eqref{chissa} to be true, some non-trivial relation should exist between closed BPS invariants $N_{d_1, d_2}^{j_L,j_R}$ appearing in $\mathcal{W}_{\text{5d}}^{\text{inst}}$ and open BPS invariants $D^{s_1}_{m, d_1, d_2}$ appearing in $\widehat{F}^{\text{(c),inst}}_{\text{3d/5d,NS}}$, as already suggested in \cite{Kashani-Poor:2016edc} based on previous works \cite{Aganagic:2003qj,Mironov:2009uv,Aganagic:2011mi}. Although it may be difficult to prove \eqref{chissa}, we were able to check its validity up to 4-instantons; it is therefore plausible for \eqref{treno} to coincide with the exact quantization conditions \eqref{ex2toda5d} as expected.

\item Assuming our proposal \eqref{proposal} is the entire $L^2(\mathbb{R})$ solution to the Baxter and dual Baxter equations associated to the 2-particle relativistic closed Toda system, the simultaneous eigenfunction $\Psi^{\text{rel}}_a(x, \omega_1, \omega_2, m_0)$ of the relativistic Toda and dual Toda Hamiltonians $\widehat{\mathcal{H}}_1$, $\widehat{\widetilde{\mathcal{H}}}_1$ can be obtained via Separation of Variables as in \eqref{eigclosedrel}: in our case, this means that
\begin{equation}
\Psi^{\text{rel}}_a(x, \omega_1, \omega_2, m_0) =
\int_{\mathbb{R}} d\sigma \, q(\sigma, a, \omega_1, \omega_2, m_0) e^{\frac{2\pi i \sigma x}{\omega_1 \omega_2}} \label{paperina}
\end{equation}
will be the (not normalized) eigenfunction satisfying\footnote{These operators are obtained from \eqref{hamrel}, \eqref{hamreldual} after a redefinition of variables; the eigenfunction for the original operators is obtained from \eqref{eigclosedrel}, and requires adding Chern-Simons terms in the integrand.}
\begin{subequations} 
\begin{align}
& \left[ e^{i \omega_1 \partial_{x}} + e^{-i \omega_1 \partial_{x}} + e^{\frac{2\pi x}{\omega_2}} + Q_{\text{rel}} e^{-\frac{2\pi x}{\omega_2}} - \mathcal{E}_1 \right] \Psi^{\text{rel}}_a(x, \omega_1, \omega_2, m_0) = 0, \\
& \left[ e^{i \omega_2 \partial_{x}} + e^{-i \omega_2 \partial_{x}} + e^{\frac{2\pi x}{\omega_1}} + \widetilde{Q}_{\text{rel}} e^{-\frac{2\pi x}{\omega_1}} - \widetilde{\mathcal{E}}_1\right] \Psi^{\text{rel}}_a(x, \omega_1, \omega_2, m_0) = 0.
\end{align}
\end{subequations}
This is the analogue of \eqref{ven} for the non-relativistic Toda chain. In the non-relativistic chain, performing the integral leads to an expression like \eqref{rompi}; going to the relativistic system, the integral will produce a double infinite sum over $n, n' \in \mathbb{Z}$ involving $a + i n \omega_1 + i n' \omega_2$, which should reduce to a single sum over $n \in \mathbb{Z}$ when $\omega_1 = \omega_2 = 1$. This sum would be the same as the one appearing in the expression for the 2-particle relativistic Toda eigenfunction discussed in \cite{Marino:2016rsq}, although the summands we are considering are different: in our case we only have NS limits of partition functions in the presence of codimension 2 defects, while in \cite{Marino:2016rsq} both the NS and the unrefined limits are needed. It would be interesting to understand if the two expressions are actually the same, maybe because of some identity relating different limits of codimension two defect instanton partition functions.

In addition, as we already mentioned for the non-relativistic case, \eqref{paperina} is expected to coincide with (a linear combination of) type I defect partition functions when evaluated at those values $a = a^{(n)}$ satisfying the exact quantization conditions: it would be interesting to check this explicitly. The ideas in the work \cite{Gorsky:2017hro} may be helpful in this respect, although their focus is mostly on four-dimensional theories. \\

\item Finally, we would like to come back to a comment we made in Section \ref{sec2.4}. As we mentioned there, quantization conditions for the (non-relativistic) 2-particle closed Toda chain seem to be expressible in two different but equivalent ways: the first one is related to the NS limit of the four-dimensional $\mathcal{N} = 2$ $SU(2)$ partition function, while the second one is related to its unrefined limit. Independently on how we express them, these quantization conditions seem to determine the spectrum of two different quantum mechanical systems, that is the 2-particle closed Toda chain and a Fermi gas, whose ``off-shell'' energies are given by \eqref{E2} and \eqref{wilson} respectively. 

As hypothesized by \cite{2014arXiv1411.3692W}, the relation between these two systems may be clearer if we start from five dimensions / from the 2-particle relativistic closed Toda chain and take an appropriate limit. According to what we discussed until now, a practical way to realize this suggestion might be to take the four-dimensional limit of our five-dimensional $\mathcal{N} = 1$ $SU(2)$ theory on $S^3_{\omega_1, \omega_2} \times \mathbb{R}^2$ and see what happens. 
This limit will break the symmetry $\omega_1 \leftrightarrow \omega_2$; one possibility would be to send $\omega_2 \rightarrow \infty$ while keeping all other parameters fixed, and remembering that when going to four dimensions the gauge coupling scales according to
\begin{equation}
\dfrac{4\pi^2}{g^2_{\text{5d}}} = m_0 \; \longrightarrow  \; -\dfrac{\omega_2}{2\pi} \ln \left( \dfrac{2\pi}{\omega_2} Q_{\text{4d}}^{1/4} \right)^4. \label{4dlimit}
\end{equation}
If we further redefine $a \rightarrow 2a$, $\omega_1 \rightarrow \hbar$, we find that the ``off-shell'' relativistic Toda and dual Toda energies $\mathcal{E}_1$, $\widetilde{\mathcal{E}}_1$ \eqref{W2f} reduce to
\begin{equation}
\begin{split}
\mathcal{E}_1 & \; \longrightarrow \; 2 \,+\, \dfrac{4\pi^2}{\omega_2^2} E_2(a, \hbar, Q_{\text{4d}}) \,+\, o(\omega_2^{-3}), \\
\widetilde{\mathcal{E}}_1 & \; \longrightarrow \; e^{\frac{2\pi a}{\hbar}} + e^{-\frac{2\pi a}{\hbar}}.
\end{split}
\end{equation}
As we can see, the four-dimensional limit of $\mathcal{E}_1$ reproduces the (non-relativistic) \linebreak 2-particle closed Toda energy $E_2(a, \hbar, Q_{\text{4d}})$ \eqref{E2}, while the four-dimensional limit of $\widetilde{\mathcal{E}}_1$ reproduces the Fermi gas energy \eqref{wilson} modulo overall factors. In addition, relativistic Toda exact quantization conditions will reduce to the non-relativistic Toda ones (in the A)-representation according to the notation of Section \ref{sec2.4}); in this way we may be able to obtain the energy spectrum of both systems in an unified way.

It is important to stress that the interpretation of the four-dimensional limit of $\widetilde{\mathcal{E}}_1$ can be a bit tricky. The limit \eqref{4dlimit} basically sets $\widetilde{Q}_{\text{5d}} \rightarrow 0$ in $\widetilde{\mathcal{E}}_1$, but this also happens when we consider the dual relativistic \textit{open} Toda chain, which has the same ``off-shell'' energy of our Fermi gas although its energy spectrum is continuous. However, in our opinion it makes more sense to think of \eqref{wilson} as the energy of the Fermi gas when considering it as arising from the four-dimensional limit \eqref{4dlimit}, since in this case $\mathcal{E}_1$ does not reduce to the relativistic open Toda energy. On the other hand, \eqref{wilson} should be interpreted as the dual relativistic open Toda energy in the different limit $m_0 \rightarrow \infty$ with $\omega_2$ and all other parameters fixed, case in which also $\mathcal{E}_1$ reduces to the relativistic open Toda energy; this different limit is not related to dimensional reduction, but to turning off the five-dimensional gauge interaction. 
\end{itemize}

\subsection{A further example: the 3-particle case} \label{sec3.5}

Up to now a large part of our discussion has been valid for general $N$, but most of the details (especially about the solution to the Baxter equation) were only given for $N = 2$. Since this case might be a bit too special, we will consider $N = 3$ in this Section; we will thus be able to see the importance of the $B$-field, which is trivial for $N = 2$. However, increasing $N$ makes all gauge theory computations more cumbersome, and all we were able to do is considering instanton corrections up to order 3 for all the relevant gauge theory quantities; we therefore expect our results to be less precise than the ones we obtained for $N = 2$. \\

Let us start by considering quantization conditions and spectrum for $N = 3$, which were already studied in \cite{2015arXiv151102860H}. First of all, in this case we have three auxiliary parameters $a_1$, $a_2$, $a_3$ (the Coulomb branch parameters of our five-dimensional $SU(3)$ theory); decoupling the center of mass corresponds to imposing $a_1 + a_2 + a_3 = 0$, and this is usually done by moving to the variables 
\begin{equation}
a_{12} = a_1 - a_2 , \;\;\; a_{23} = a_2 - a_3,
\end{equation}
which are nothing else but the Kahler parameters $t_1$, $t_2$ in \eqref{kahler} (the remaining parameter being $t_3 = m_0 + a_{12} + a_{23}$). In terms of these variables, we have
\begin{equation}
a_{1} = \dfrac{2}{3}a_{12} + \dfrac{1}{3} a_{23}, \;\;\;
a_{2} = -\dfrac{1}{3}a_{12} + \dfrac{1}{3} a_{23}, \;\;\;
a_{3} = - \dfrac{1}{3}a_{12} - \dfrac{2}{3} a_{23}.
\end{equation}
We will also often use the notation
\begin{equation}
\mu_{12} = e^{\frac{2\pi a_{12}}{\omega_2}}, \;\;\; 
\mu_{23} = e^{\frac{2\pi a_{23}}{\omega_2}} \;\;\;\; \text{and} \;\;\;\;
\widetilde{\mu}_{12} = e^{\frac{2\pi a_{12}}{\omega_1}}, \;\;\; 
\widetilde{\mu}_{23} = e^{\frac{2\pi a_{23}}{\omega_1}}.
\end{equation}
As we discussed in Section \ref{3.3.3}, exact quantization conditions are obtained by extremizing the effective twisted superpotential \eqref{ziopaperone} which is computed from the ``NS limit'' of the (integrand of the) partition function on $S^5_{\omega_1, \omega_2, \omega_3}$; the non-BPS part in \eqref{ziopaperone} is easily extracted from \eqref{nonBPS5d}, while the BPS part is written in terms of the function \eqref{withB} corrected by the $B$-field (arising in this setting from the identifications in Table \ref{tabo}). We then obtain 
\begin{eqnarray}
\dfrac{1}{i}\mathcal{W}_{S^5}^{SU(3)} \;= \;& & \dfrac{\pi}{3\omega_1 \omega_2}\left( a_{12}^3 + a_{23}^3 + (a_{12} + a_{23})^3 \right) + \dfrac{2\pi m_0}{3\omega_1 \omega_2}\left( a_{12}^2 + a_{12}a_{23} + a_{23}^2 \right) \nonumber \\
& & - \dfrac{\pi}{3} \dfrac{\omega_1^2 + \omega_2^2}{\omega_1 \omega_2}(a_{12} + a_{23}) - \pi (a_{12} + a_{23}) \nonumber \\
& & - \dfrac{\omega_2}{4\pi} \sum_{n \geqslant 1} \sum_{d_1, d_2, d_3} \sum_{j_L,j_R}
 \dfrac{(-1)^{(2j_L + 2j_R + 1)n}}{n^2} N_{j_L,j_R}^{d_1, d_2, d_3} \label{SU3qc} \\
& & \times \dfrac{\sin \left[ (2j_L+1)\pi n \frac{\omega_1}{\omega_2} \right] \sin \left[ (2j_R+1)\pi n \frac{\omega_1}{\omega_2} \right]}{\sin^3 \left[\pi n \frac{\omega_1}{\omega_2} \right]} 
e^{-\frac{2\pi n d_1 a_{12}}{\omega_2}} 
e^{-\frac{2\pi n d_2 a_{23}}{\omega_2}} 
e^{-\frac{2\pi n d_3 (a_{12} + a_{23} + m_0)}{\omega_2}}  \nonumber \\
& & - \dfrac{\omega_1}{4\pi} \sum_{n \geqslant 1} \sum_{d_1, d_2, d_3} \sum_{j_L,j_R}
 \dfrac{(-1)^{(2j_L + 2j_R + 1)n}}{n^2} N_{j_L,j_R}^{d_1, d_2, d_3} \nonumber \\ 
& & \times \dfrac{\sin \left[ (2j_L+1)\pi n \frac{\omega_2}{\omega_1} \right] \sin \left[ (2j_R+1)\pi n \frac{\omega_2}{\omega_1} \right]}{\sin^3 \left[\pi n \frac{\omega_2}{\omega_1} \right]} 
e^{-\frac{2\pi n d_1 a_{12}}{\omega_1}} 
e^{-\frac{2\pi n d_2 a_{23}}{\omega_1}} 
e^{-\frac{2\pi n d_3 (a_{12} + a_{23} + m_0)}{\omega_1}},  \nonumber 
\end{eqnarray}
manifestly symmetric under exchange $\omega_1 \leftrightarrow \omega_2$ (the $N_{j_L,j_R}^{d_1, d_2, d_3}$ numbers can be found for example in \cite{Sun:2016obh}); exact quantization conditions \eqref{procione} will therefore reduce to solving the equations
\begin{equation}
\left\{
\begin{array}{l}
\dfrac{1}{i} \left( 2\partial_{a_{12}} - \partial_{a_{23}} \right) \mathcal{W}_{S^5}^{SU(3)} = 2\pi n_{12}, \\[15 pt]
\dfrac{1}{i} \left( -\partial_{a_{12}} + 2 \partial_{a_{23}} \right) \mathcal{W}_{S^5}^{SU(3)} = 2\pi n_{23},
\end{array} \right. \label{nonsopiu}
\end{equation}
in terms of the $a_{12}$, $a_{23}$ variables at fixed $\omega_1$, $\omega_2$, $m_0$ and for chosen levels $n_{12}$, $n_{23} \in \mathbb{N}$. These quantization conditions precisely match with the ones given in \cite{2015arXiv151102860H}.

Energies $\mathcal{E}_1^{(n_{12}, n_{23})}$, $\mathcal{E}_2^{(n_{12}, n_{23})}$ and dual energies $\widetilde{\mathcal{E}}_1^{(n_{12}, n_{23})}$, $\widetilde{\mathcal{E}}_2^{(n_{12}, n_{23})}$ at level $(n_{12}, n_{23})$ will instead be given by the NS limit of vacuum expectation values of Wilson loops in fundamental $(\tiny\yng(1))$ and antisymmetric $(\tiny\yng(1,1))$ representations of $SU(3)$ wrapping the two special circles in $S^3_{\omega_1, \omega_2} \times \mathbb{R}^2$, evaluated at those values $a_{12} = a_{12}^{(n_{12}, n_{23})}$, $a_{23} = a_{23}^{(n_{12}, n_{23})}$ satisfying the quantization conditions \eqref{SU3qc}. These Wilson loops are basically the same as the ones in flat space, modulo taking into account the effect of the $B$-field: more precisely, if we denote the (NS limit of the) fundamental and antisymmetric Wilson loops in flat space $\mathbb{R}^2_{\epsilon_1} \times \mathbb{R}^2 \times S^1_R$ wrapping $S^1_R$ as
\begin{equation}
W_{\tiny\yng(1) \,,\, \text{NS}}^{SU(3)} (a_{12}, a_{23}, \epsilon_1, R, m_0) , \;\;\;\; W_{\tiny\yng(1,1) \,,\, \text{NS}}^{SU(3)} (a_{12}, a_{23}, \epsilon_1, R, m_0),
\end{equation}
then the energies of the 3-particle relativistic closed Toda chain are\footnote{As we already mentioned, for $N$ odd identifying $\epsilon_1 = i(\omega_1 + \omega_2)$ is equivalent to identify $\epsilon_1 = i \omega_1$ and change sign to the instanton counting parameter ($Q_{\text{5d}} \rightarrow - Q_{\text{5d}}$, $\widetilde{Q}_{\text{5d}} \rightarrow - \widetilde{Q}_{\text{5d}}$).}
\begin{equation}
\begin{split}
\mathcal{E}_1(a_{12}, a_{23}, \omega_1, \omega_2, m_0) & = W_{\tiny\yng(1) \,,\, \text{NS}}^{SU(3)} (a_{12}, a_{23}, i(\omega_1 + \omega_2), \omega_2^{-1}, m_0), \\[5pt]
\mathcal{E}_2(a_{12}, a_{23}, \omega_1, \omega_2, m_0) & = W_{\tiny\yng(1,1) \,,\, \text{NS}}^{SU(3)} (a_{12}, a_{23}, i(\omega_1 + \omega_2), \omega_2^{-1}, m_0), 
\end{split}
\end{equation}
while the energies of the dual system are 
\begin{equation}
\begin{split}
\widetilde{\mathcal{E}}_1(a_{12}, a_{23}, \omega_1, \omega_2, m_0) & = 
W_{\tiny\yng(1) \,,\, \text{NS}}^{SU(3)} (a_{12}, a_{23}, i(\omega_2 + \omega_1), \omega_1^{-1}, m_0), \\[5pt]
\widetilde{\mathcal{E}}_2(a_{12}, a_{23}, \omega_1, \omega_2, m_0) & = 
W_{\tiny\yng(1,1) \,,\, \text{NS}}^{SU(3)} (a_{12}, a_{23}, i(\omega_2 + \omega_1), \omega_1^{-1}, m_0).
\end{split}
\end{equation}

\renewcommand\arraystretch{1.7}
\begin{table}[]
\begin{center}
\begin{tabular}{c|cc|cc}
& \multicolumn{2}{|c|}{$\renewcommand\arraystretch{1.3}\begin{array}{c} \text{Case} \;\, m_0 = 0: \\ a_{12}^{(0,0)} = a_{23}^{(0,0)} = 0.4927628697\ldots \end{array}\renewcommand\arraystretch{1.5}$} 
& \multicolumn{2}{|c}{$\renewcommand\arraystretch{1.3}\begin{array}{c} \text{Case} \;\, m_0 = 1: \\ a_{12}^{(0,0)} = a_{23}^{(0,0)} = 0.3342262919\ldots \end{array}\renewcommand\arraystretch{1.7}$} \\
\hline 
\rule[-1ex]{0pt}{2.5ex} & 
$\mathcal{E}_1^{(0,0)} = \mathcal{E}_2^{(0,0)}$ & $\widetilde{\mathcal{E}}_1^{(0,0)} = \widetilde{\mathcal{E}}_2^{(0,0)}$ & 
$\mathcal{E}_1^{(0,0)} = \mathcal{E}_2^{(0,0)}$ & $\widetilde{\mathcal{E}}_1^{(0,0)} = \widetilde{\mathcal{E}}_2^{(0,0)}$  \\ 
\hline 
\rule[-1ex]{0pt}{2.5ex} 
$\renewcommand\arraystretch{1}\begin{array}{c} \text{gauge theory} \\ \text{(3-instantons)} \end{array}\renewcommand\arraystretch{1.7}$ & 23.1546976231\ldots & 80.7368446681\ldots  &  9.2886369712\ldots & 20.5404347303\ldots \\
\hline 
\rule[-1ex]{0pt}{2.5ex} numerics & 23.1546976231\ldots & 80.7368446681\ldots 
&  9.2886369712\ldots & 20.5404347303\ldots\\
\end{tabular} \caption{3-particle relativistic Toda energies and dual energies for the ground state $(n_{12}, n_{23}) = (0,0)$ at $\omega_1 = \frac{1}{\sqrt{2}}$, $\omega_2 = 1$ and $m_0 = 0$ or $m_0 = 1$.}  \label{enerSU3}
\end{center} 
\end{table}
\renewcommand\arraystretch{1}

\noindent Explicit expressions can be obtained for example by computing the NS limit of the $SU(3)$ fundamental qq-character with the formulae of \cite{Kim:2016qqs}, but already at 1-instanton the results are too long to be written down here; let us however mention that our energies will be given by instanton series expansions starting as 
\begin{equation}
\begin{split}
& \mathcal{E}_1 = \mu_{12}^{\frac{2}{3}}\mu_{23}^{\frac{1}{3}} + \mu_{12}^{-\frac{1}{3}}\mu_{23}^{\frac{1}{3}} + \mu_{12}^{-\frac{1}{3}}\mu_{23}^{-\frac{2}{3}} + o(Q_{\text{5d}}), \\[5pt]
& \mathcal{E}_2 = \mu_{12}^{\frac{1}{3}}\mu_{23}^{\frac{2}{3}} + \mu_{12}^{\frac{1}{3}}\mu_{23}^{-\frac{1}{3}} + \mu_{12}^{-\frac{2}{3}}\mu_{23}^{-\frac{1}{3}} + o(Q_{\text{5d}}), \\[5pt]
& \widetilde{\mathcal{E}}_1 = \widetilde{\mu}_{12}^{\frac{2}{3}}\widetilde{\mu}_{23}^{\frac{1}{3}} + \widetilde{\mu}_{12}^{-\frac{1}{3}}\widetilde{\mu}_{23}^{\frac{1}{3}} + \widetilde{\mu}_{12}^{-\frac{1}{3}}\widetilde{\mu}_{23}^{-\frac{2}{3}} + o(\widetilde{Q}_{\text{5d}}), \\[5pt]
& \widetilde{\mathcal{E}}_2 = \widetilde{\mu}_{12}^{\frac{1}{3}}\widetilde{\mu}_{23}^{\frac{2}{3}} + \widetilde{\mu}_{12}^{\frac{1}{3}}\widetilde{\mu}_{23}^{-\frac{1}{3}} + \widetilde{\mu}_{12}^{-\frac{2}{3}}\widetilde{\mu}_{23}^{-\frac{1}{3}} + o(\widetilde{Q}_{\text{5d}}),
\end{split}
\end{equation}
and moreover
\begin{equation}
\mathcal{E}_1 \leftrightarrow \mathcal{E}_2 , \;\;\;\; \widetilde{\mathcal{E}}_1 \leftrightarrow \widetilde{\mathcal{E}}_2 \;\;\;\; \text{under} \;\;\;\; a_{12} \leftrightarrow a_{23}.
\end{equation}
An example of ground state energies computed via these formulae is shown in Table \ref{enerSU3}; as we can see, already at 3-instantons gauge theory results are in good agreement with numerical ones, and the agreement gets better and better by considering more instanton corrections.\footnote{Numerical results are obtained by considering $350 \times 350$ matrices as in Section \ref{sec3.2}; see \cite{2015arXiv151102860H,Franco:2015rnr} for more details on numerical computations.}
\\

Let us now move to discuss the solution $q(\sigma, a_{12}, a_{23}, \omega_1, \omega_2, m_0)$ to the Baxter and dual Baxter equations \eqref{baxclosedrel} (with center of mass decoupled), which in the $N = 3$ case read

\begin{equation}
\begin{split}
& \left[ (i)^3 e^{i \omega_1 \partial_{\sigma}} + Q_{\text{rel}} (i)^{-3} e^{-i \omega_1 \partial_{\sigma}} \right] q(\sigma, a_{12}, a_{23}, \omega_1, \omega_2, m_0) = \\
& \left[ e^{\frac{2\pi}{\omega_2} \frac{3\sigma}{2}} - \mathcal{E}_1 e^{\frac{2\pi}{\omega_2} \frac{\sigma}{2}} + \mathcal{E}_2 e^{-\frac{2\pi}{\omega_2} \frac{\sigma}{2}} -  e^{-\frac{2\pi}{\omega_2} \frac{3\sigma}{2}}\right] q(\sigma, a_{12}, a_{23}, \omega_1, \omega_2, m_0),  \\[5 pt]
& \left[ (i)^3 e^{i \omega_2 \partial_{\sigma}} + \widetilde{Q}_{\text{rel}} (i)^{-3} e^{-i \omega_2 \partial_{\sigma}} \right] q(\sigma, a_{12}, a_{23}, \omega_1, \omega_2, m_0) = \\
& \left[ e^{\frac{2\pi}{\omega_1} \frac{3\sigma}{2}} - \widetilde{\mathcal{E}}_1 e^{\frac{2\pi}{\omega_1} \frac{\sigma}{2}} + \widetilde{\mathcal{E}}_2 e^{-\frac{2\pi}{\omega_1} \frac{\sigma}{2}} -  e^{-\frac{2\pi}{\omega_1} \frac{3\sigma}{2}}\right] q(\sigma, a_{12}, a_{23}, \omega_1, \omega_2, m_0). \label{baxSU3}
\end{split}
\end{equation}
According to the discussion in Section \ref{sec3.4}, this solution should involve the NS limit of the partition function of codimension-two defects of type II wrapping $S^3_{\omega_1, \omega_2}$.
As usual, we first consider the $m_0 \rightarrow \infty$ limit (relativistic open Toda chain); the $S^3_{\omega_1, \omega_2}$ partition function for a type II chiral/antichiral defect in this limit is given for general $N$ in \eqref{papparel}, which for $N = 3$ and decoupling the center of mass reduces to
\begin{eqnarray}
q^{\text{(c)}}(\sigma, a_{12}, a_{23}, \omega_1, \omega_2) \, = \, 
&& S_2^{-1}\big( -i(\sigma - \frac{2 a_{12}}{3} - \frac{a_{23}}{3}) \vert \omega_1, \omega_2\big)
S_2^{-1}\big( -i(\sigma + \frac{a_{12}}{3} - \frac{a_{23}}{3}) \vert \omega_1, \omega_2\big) \nonumber \\[3 pt]
&& S_2^{-1}\big( -i(\sigma + \frac{a_{12}}{3} + \frac{2 a_{23}}{3}) \vert \omega_1, \omega_2\big), \label{stacosa} \\[3 pt]
q^{\text{(ac)}}(\sigma, a_{12}, a_{23}, \omega_1, \omega_2) \, = \, 
&& S_2^{-1}\big( i(\sigma - \frac{2 a_{12}}{3} - \frac{a_{23}}{3}) \vert \omega_1, \omega_2\big)
S_2^{-1}\big( i(\sigma + \frac{a_{12}}{3} - \frac{a_{23}}{3}) \vert \omega_1, \omega_2\big) \nonumber  \\[3 pt]
&& S_2^{-1}\big( i(\sigma + \frac{a_{12}}{3} + \frac{2 a_{23}}{3}) \vert \omega_1, \omega_2\big). \nonumber 
\end{eqnarray}
Moving to the closed chain, we expect the solution to \eqref{baxSU3} to be the linear combination
\begin{equation}
q(\sigma, a_{12}, a_{23}, \omega_1, \omega_2, m_0) \; \propto \; 
q^{(\text{ac})}(\sigma, a_{12}, a_{23}, \omega_1, \omega_2, m_0) -
\xi q^{(\text{c})}(\sigma, a_{12}, a_{23}, \omega_1, \omega_2, m_0) \label{linear3}
\end{equation}
modulo an overall (complex) $\sigma$-independent term and for some value of $\xi \in \mathbb{C}$, where
\begin{equation}
\begin{split}
q^{\text{(c)}}(\sigma, a_{12}, a_{23}, \omega_1, \omega_2, m_0) & = q^{\text{(c)}}(\sigma, a_{12}, a_{23}, \omega_1, \omega_2) q^{\text{(c)},\text{inst}}_{\text{NS}}(\sigma, a_{12}, a_{23}, \omega_1, \omega_2, m_0), \\
q^{\text{(ac)}}(\sigma, a_{12}, a_{23}, \omega_1, \omega_2, m_0) & = e^{\frac{2 i \pi m_0 \sigma}{\omega_1 \omega_2}} q^{\text{(ac)}}(\sigma, a_{12}, a_{23}, \omega_1, \omega_2) q^{\text{(ac)},\text{inst}}_{\text{NS}}(\sigma, a_{12}, a_{23}, \omega_1, \omega_2, m_0), 
\end{split}
\end{equation}
are the instanton-corrected version of the open chain ones \eqref{stacosa}. Again, the instanton contributions are expected to be given by two copies of the NS limit of the type II defect instanton partition function $Z^{(\text{c}), \text{inst}}_{\text{3d/5d}, \text{NS}}$ or $Z^{(\text{ac}), \text{inst}}_{\text{3d/5d}, \text{NS}}$ on flat space $\mathbb{R}^2_{\epsilon_1} \times \mathbb{R}^2 \times S^1_R$, which can be computed with the formulae in Appendix \ref{appA}. The results are too involved to be shown here: let us just mention that, in terms of the flat space variables
\begin{equation}
q_1 = e^{2\pi R \epsilon_1}, \;\;\; w = e^{-2\pi R \sigma}, \;\;\; \mu_{12} = e^{2 \pi R a_{12}}, \;\;\; \mu_{23} = e^{2\pi R a_{23}}, \;\;\; Q_{\text{5d}} = e^{-2\pi R m_0}, \label{lavander}
\end{equation}
these function are related as
\begin{equation}
Z^{(\text{ac}), \text{inst}}_{\text{3d/5d}, \text{NS}}(w,\mu_{12}, \mu_{23}, q_1, Q_{\text{5d}}) = 
Z^{(\text{c}), \text{inst}}_{\text{3d/5d}, \text{NS}}(w,\mu_{12}, \mu_{23}, q_1^{-1}, Q_{\text{5d}});
\end{equation} 
moreover
\begin{equation}
Z^{(\text{c}), \text{inst}}_{\text{3d/5d}, \text{NS}}(w,\mu_{12}, \mu_{23}, q_1, Q_{\text{5d}}) \;\propto \;
Z^{(\text{c}), \text{inst}}_{\text{3d/5d}, \text{NS}}(w^{-1},\mu_{12}^{-1}, \mu_{23}^{-1}, q_1^{-1}, Q_{\text{5d}}), \label{important}
\end{equation}
where the proportionality factor is independent of $\sigma$. In addition, the defect instanton partition function admits an open topological string like representation as
\begin{equation}
Z^{(\text{c}), \text{inst}}_{\text{3d/5d}, \text{NS}}(w,\mu_{12}, \mu_{23}, q_1, Q_{\text{5d}}) = \exp \left( - F^{(\text{c}), \text{inst}}_{\text{3d/5d}, \text{NS}}(w,\mu_{12}, \mu_{23}, q_1, Q_{\text{5d}}) \right), \label{opstring}
\end{equation}
with\footnote{As in \eqref{rivellinoflat}, not all BPS states are contained in the instanton contributions: some of them arise from the double sine functions in \eqref{stacosa}.}
\begin{equation}
\begin{split}
& F^{(\text{c}), \text{inst}}_{\text{3d/5d}, \text{NS}}(w,\mu_{12}, \mu_{23}, q_1, Q_{\text{5d}}) = \\
& \sum_{n=1}^{\infty} \sum_{s_1} \sum_{d_1, d_2, d_3 > 0} \sum_{m \in \mathbb{Z}} 
D^{s_1}_{m, d_1, d_2, d_3} \dfrac{q_1^{n s_1}}{n(1-q_1^n)} \mu_{12}^{-n d_1} \mu_{23}^{-n d_2} \left( Q_{\text{5d}} \mu_{12}^{-1} \mu_{23}^{-1} \right)^{n d_3} w^{m n}.
\end{split}
\end{equation}
Because of the property \eqref{important}, the instanton contributions can be written in many different ways, all equivalent modulo $\sigma$-independent terms and different values of $\xi$; here we will choose to write them as
\begin{equation}
\begin{split}
& q^{\text{(c)},\text{inst}}_{\text{NS}}(\sigma, a_{12}, a_{23}, \omega_1, \omega_2, m_0) = \\[4 pt]
& Z^{(\text{c}), \text{inst}}_{\text{3d/5d}, \text{NS}}(w, \mu_{12},  \mu_{23}, e^{2\pi i}q, Q_{\text{rel}})
Z^{(\text{c}), \text{inst}}_{\text{3d/5d}, \text{NS}}(\widetilde{w}, \widetilde{\mu}_{12}, \widetilde{\mu}_{23}, e^{2\pi i}\widetilde{q}, \widetilde{Q}_{\text{rel}}), \\[8 pt]
& q^{\text{(ac)},\text{inst}}_{\text{NS}}(\sigma, a_{12}, a_{23}, \omega_1, \omega_2, m_0) = \\[4 pt]
& Z^{(\text{c}), \text{inst}}_{\text{3d/5d}, \text{NS}}(w, \mu_{12}, \mu_{23}, e^{-2\pi i}q^{-1}, Q_{\text{rel}})
Z^{(\text{c}), \text{inst}}_{\text{3d/5d}, \text{NS}}(\widetilde{w}, \widetilde{\mu}_{12}, \widetilde{\mu}_{23}, e^{-2\pi i}\widetilde{q}^{-1}, \widetilde{Q}_{\text{rel}}), \label{trueinstSU3}
\end{split}
\end{equation}
where we moved from flat space variables \eqref{lavander} to curved space ones following Table \ref{tabo} (here $q = e^{2\pi i \omega_1 / \omega_2}$, $\widetilde{q} = e^{2\pi i \omega_2 / \omega_1}$). By using the representation \eqref{opstring}, it is immediate to apply the argument of \cite{Kashani-Poor:2016edc} and show that the combinations appearing in \eqref{trueinstSU3} are free of divergences at $\omega_1 / \omega_2 \in \mathbb{Q}_+$.

\begin{figure}[]
  \centering
\includegraphics[width=1\textwidth]{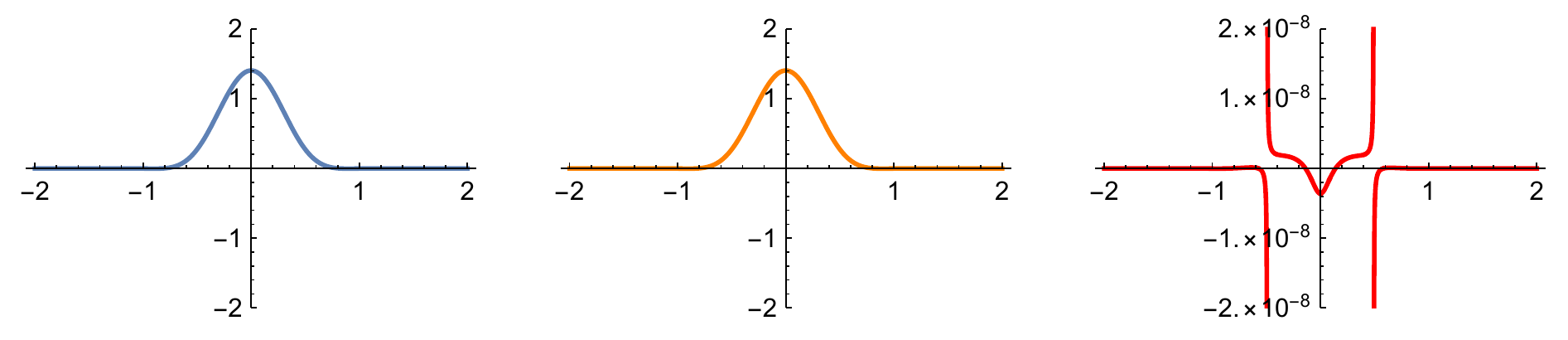}
  \caption{$N = 3$ Baxter solution $\overline{q}(\sigma)$ for the ground state $(0,0)$ at $\omega_1 = \frac{1}{\sqrt{2}}$, $\omega_2 = 1$, $m_0 = 0$. Left: numerical result; center: gauge theory result; right: difference.}
  \label{Bax5dSU3}
\end{figure}

\begin{figure}[]
  \centering
\includegraphics[width=1\textwidth]{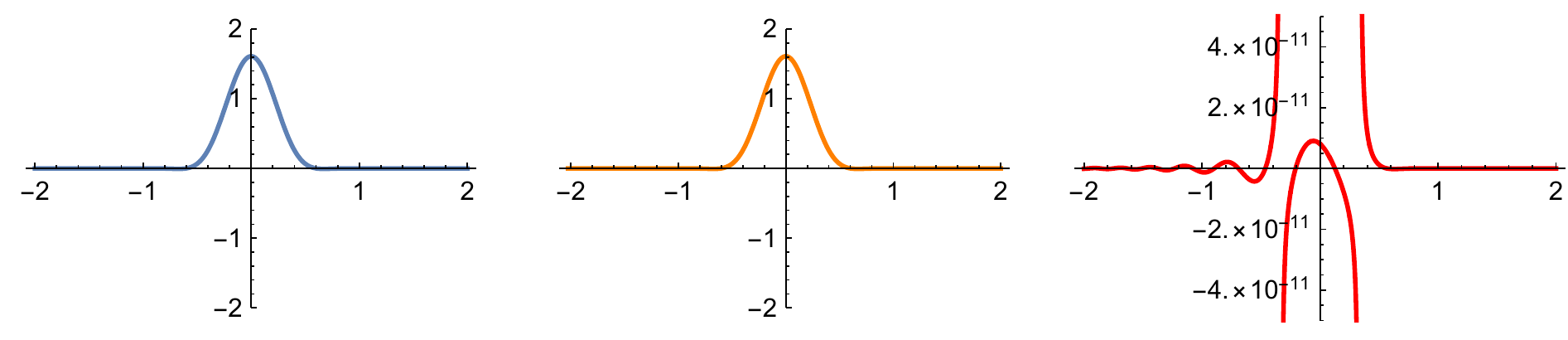}
  \caption{$N = 3$ Baxter solution $\overline{q}(\sigma)$ for the ground state $(0,0)$ at $\omega_1 = \frac{1}{\sqrt{2}}$, $\omega_2 = 1$, $m_0 = 1$. Left: numerical result; center: gauge theory result; right: difference.}
  \label{plotSU3fin}
\end{figure}

To sum up, our proposed solution to the Baxter and dual Baxter equations \eqref{baxSU3} associated to the 3-particle relativistic closed Toda chain is given by the linear combination \linebreak \eqref{linear3}, which involves the perturbative part \eqref{stacosa} and the instanton contributions \eqref{trueinstSU3}, modulo an overall complex $\sigma$-independent term and with $\xi \in \mathbb{C}$ determined as in \eqref{quantcond}; this is expected to be entire only at the values $a_{12} = a_{12}^{(n_{12}, n_{23})}$, $a_{23} = a_{23}^{(n_{12}, n_{23})}$ satisfying the exact quantization conditions \eqref{nonsopiu}. As usual, in order to check this proposal we plot it and compare against numerical results; for this purpose it is actually more convenient to consider the related function
\begin{equation}
\overline{q}(\sigma, a_{12}, a_{23}, \omega_1, \omega_2, m_0) = e^{-\frac{i \pi m_0 \sigma}{\omega_1 \omega_2}} q(\sigma, a_{12}, a_{23}, \omega_1, \omega_2, m_0), 
\end{equation}
since it satisfies a more symmetric problem. Figures \ref{Bax5dSU3} and \ref{plotSU3fin} show the ground state $(n_{12}, n_{23}) = (0,0)$ eigenfunction obtained via numerical methods (blue) and via our gauge theory proposal \eqref{linear3} (orange)\footnote{The orange plot shows the gauge theory function Re$\,[\overline{q}(\sigma)]$ divided by its norm; the overall complex $\sigma$-independent term has been fixed numerically in such a way to have Im$\,[\overline{q}(\sigma)] = 0$.} for $\omega_1 = \frac{1}{\sqrt{2}}$, $\omega_2 = 1$ and $m_0 = 0$ or $m_0 = 1$ respectively; as we can see from the left-most plots (red) the difference between numerical and gauge theory result away from singular points is quite small already considering 3-instanton expressions,\footnote{Here we fixed $\xi$ in such a way to cancel the singularity at $\sigma = 0$, so that only the other two singular points are visible. Moreover, the asymmetry in the red plots is due to numerical results, which are obtained not from diagonalizing \eqref{baxSU3} but from considering a closely related operator: see \cite{2015arXiv151102860H,Franco:2015rnr} for more details.} while divergences tend to close by considering more and more instanton corrections. Eigenfunctions for excited states can also be easily obtained from \eqref{linear3}, but comparison against numerical results becomes harder since excited stated are more problematic to treat numerically. \\

For general $N$, we expect a similar story to be true: the simultaneous solution to the Baxter and dual Baxter equations for the $N$-particle relativistic closed Toda chain will be given by a linear combination
\begin{equation}
q(\sigma, \vec{a}, \omega_1, \omega_2, m_0) \; \propto \;
q^{(\text{ac})}(\sigma, \vec{a}, \omega_1, \omega_2, m_0) - 
\xi q^{(\text{c})}(\sigma, \vec{a}, \omega_1, \omega_2, m_0), 
\end{equation}
modulo an overall $\sigma$-independent term and with $\xi \in \mathbb{C}$ determined by \eqref{quantcond}.\footnote{It would be desirable to have an explicit gauge theory expression for this $\sigma$-independent term and for $\xi$ in such a way that the resulting eigenfunction will be normalized to 1, but we weren't able to find it yet.} Here we have $\vec{a} = (a_1, \ldots, a_{N})$ with $a_1 + \ldots + a_N = 0$ and
\begin{equation}
\begin{split}
q^{\text{(c)}}(\sigma, \vec{a}, \omega_1, \omega_2, m_0) & = q^{\text{(c)}}(\sigma, \vec{a}, \omega_1, \omega_2) q^{\text{(c)},\text{inst}}_{\text{NS}}(\sigma, \vec{a}, \omega_1, \omega_2, m_0), \\
q^{\text{(ac)}}(\sigma, \vec{a}, \omega_1, \omega_2, m_0) & = e^{\frac{2 i \pi m_0 \sigma}{\omega_1 \omega_2}} q^{\text{(ac)}}(\sigma, \vec{a}, \omega_1, \omega_2) q^{\text{(ac)},\text{inst}}_{\text{NS}}(\sigma, \vec{a}, \omega_1, \omega_2, m_0), 
\end{split}
\end{equation}
where
\begin{equation}
\begin{split}
q^{(\text{c})}(\sigma, \vec{a}, \omega_1, \omega_2) &= \prod_{m=1}^N S_2^{-1}\big( -i(\sigma - a_m) \vert \omega_1, \omega_2\big), \\
q^{(\text{ac})}(\sigma, \vec{a}, \omega_1, \omega_2) &= \prod_{m=1}^N S_2^{-1}\big(i(\sigma - a_m) \vert \omega_1, \omega_2\big),
\end{split}
\end{equation}
while the instanton contributions can be written in terms of the NS limit of the flat space instanton partition function $Z^{(\text{c}), \text{inst}}_{\text{3d/5d}, \text{NS}}$ for pure five-dimensional $\mathcal{N} = 1$ $SU(N)$ Yang-Mills in the presence of a codimension-two defect of type II and can be chosen to be
\begin{equation}
\begin{split}
& q^{\text{(c)},\text{inst}}_{\text{NS}}(\sigma, \vec{a}, \omega_1, \omega_2, m_0) = Z^{(\text{c}), \text{inst}}_{\text{3d/5d}, \text{NS}}(w, \vec{\mu}, e^{2\pi i}q, Q_{\text{rel}})
Z^{(\text{c}), \text{inst}}_{\text{3d/5d}, \text{NS}}(\widetilde{w}, \vec{\mu}, e^{2\pi i}\widetilde{q}, \widetilde{Q}_{\text{rel}}), \\[8 pt]
& q^{\text{(ac)},\text{inst}}_{\text{NS}}(\sigma, \vec{a}, \omega_1, \omega_2, m_0) =  Z^{(\text{c}), \text{inst}}_{\text{3d/5d}, \text{NS}}(w, \vec{\mu}, e^{-2\pi i}q^{-1}, Q_{\text{rel}})
Z^{(\text{c}), \text{inst}}_{\text{3d/5d}, \text{NS}}(\widetilde{w}, \vec{\mu}, e^{-2\pi i}\widetilde{q}^{-1}, \widetilde{Q}_{\text{rel}}),
\end{split}
\end{equation}
assuming that for generic $N$ it still holds
\begin{equation}
Z^{(\text{ac}), \text{inst}}_{\text{3d/5d}, \text{NS}}(w,\vec{\mu}, q_1, Q_{\text{5d}}) = Z^{(\text{c}), \text{inst}}_{\text{3d/5d}, \text{NS}}(w,\vec{\mu}, q_1^{-1}, Q_{\text{5d}}).
\end{equation}
This solution will only be entire for those values of $\vec{a}$ satisfying the exact quantization conditions; energies for the Toda and dual Toda systems will then be given by vacuum expectation values of Wilson loop at the north and south poles of $S^3_{\omega_1, \omega_2}$ in appropriate representations of $SU(N)$.

\section{Conclusions} \label{sec4}

In this paper we proposed a gauge theory approach to solve 
the problem of constructing an entire, simultaneous solution to the Baxter and dual Baxter equations associated to the $N$-particle relativistic closed Toda chain, out of which relativistic Toda  and dual Toda eigenfunctions can be obtained via Separation of Variables techniques, and we tested our proposal against numerical results finding good agreement. Our proposal basically consists of considering various defects for the five-dimensional $\mathcal{N} = 1$ $SU(N)$ theory on the squashed $S^5_{\omega_1, \omega_2, \omega_3}$ background in the ``NS limit'' $\omega_3 \rightarrow 0$, rather than the flat $\mathbb{R}^2_{\epsilon_1} \times \mathbb{R}^2_{\epsilon_2} \times S^1_R$ background in the NS limit $\epsilon_2 \rightarrow 0$ usually considered in the literature. From this set-up, which seems to automatically incorporate the effect of the $B$-field and ``$S$-duality'' $\omega_1 \leftrightarrow \omega_2$ (related to the existence of the modular dual system), we were able to extract exact quantization conditions, exact spectrum and exact Baxter eigenfunctions in an unified setting; it is however not clear if the $S^5_{\omega_1, \omega_2, \omega_3}$ background is the most appropriate one to start with, since any other background that reduces to $S^3_{\omega_1, \omega_2} \times \mathbb{R}^2$ in some limit may be equivalently good. In any case it is important to notice that, although usually stated otherwise, the type II defect partition function on flat space $\mathbb{R}^2_{\epsilon_1} \times \mathbb{R}^2 \times S^1_R$ turns out not to be the correct solution to the Baxter equation by itself, but a second copy of it (containing non-perturbative contributions) is needed.

We should stress that, although it seems to match numerical results for the cases we checked, at this stage our proposal can only be regarded as a proposal, and we do not know of any way to demonstrate it. 
What is more, even assuming this proposal turns out to be correct, at the moment it is not clear to us why considering gauge theories on curved backgrounds should work. Our motivation to do so came from realizing that the known eigenfunctions of the relativistic open Toda chain given in \citep{Kharchev:2001rs} coincide with the $S^3_{\omega_1, \omega_2}$ partition function of some particular three-dimensional theory: based on this fact, on what happens in the non-relativistic (or four-dimensional) case, and on the existence of the modular dual structure of the relativistic Toda chain, we were naturally led to consider gauge theories on $S^3_{\omega_1, \omega_2} \times \mathbb{R}^2$; nevertheless, a good explanation for this has yet to be found. One could say that starting from the (integrand of the) partition function on $S^5_{\omega_1, \omega_2, \omega_3}$ may be meaningful because this was suggested in \cite{2012arXiv1210.5909L} to provide a non-perturbative completion of the topological string partition function, but this claim itself is simply another proposal. A deeper understanding of this point is surely desirable.


Another important remark is that, as we mentioned at various points in this paper, there seem to be two different gauge theory/topological string approaches to solve the 2-particle relativistic Toda chain (and, with the appropriate modifications, also the $N$-particle one): in both of them non-perturbative contributions are essential to get correct results, but these contributions appear in different ways. The first approach is the one followed in this work and, similarly, in \cite{Wang:2015wdy},\cite{2015arXiv151102860H} among others: this is based on considering formulas which are manifestly $S$-duality invariant, in agreement with the modular double structure of our problem, and only involves the NS limit of the refined topological string partition function (closed or open). The second approach instead, followed by \cite{Grassi:2014zfa,Marino:2016rsq} among others, is based on the study of spectral (Fredholm) determinants and  minors associated to an operator which basically is the inverse of the Baxter equation; these determinants and minors can be expressed in terms of (closed or open) refined topological string partition function, but these expressions involve both the unrefined and the NS limit. At least for what quantization conditions and spectrum are concerned, the two approaches are related \cite{Sun:2016obh,Grassi:2016nnt}; it would be interesting to understand if this continues to be true at the level of eigenfunctions. 

It is also worth noticing that these two approaches seem to have an analogue in four dimensions, although in this case they will be related to different Quantum Mechanical systems: the first approach simply reduces to the original gauge theory proposal of \citep{2010maph.conf..265N} for solving the non-relativistic Toda chain, which only involves the NS limit of four-dimensional gauge theory observables; the second approach instead is more related to the work \cite{Bonelli:2016idi}, where the spectral determinant for a Fermi gas associated to the $O(2)$ matrix model is expressed solely in terms of the unrefined limit of the four-dimensional partition function. It is possible that these two systems descend from the relativistic Toda and dual Toda chains respectively as we discussed in Section \ref{seccomments5d}, but further investigation is needed. \\

Finally, it would be very interesting to understand if our proposal for the solution to the Baxter and dual Baxter equation can be extended to include purely topological string cases like local $\mathbb{P}^2$, which are not associated to any gauge theory. This seems to be the case since our proposal can be written in terms of the NS limit of the refined open topological string partition function, which can be computed even for cases without a gauge theory interpretation; we hope to be able to discuss this point in more detail in the near future.

\section*{Acknowledgements}

We would like to thank Jean-Emile Bourgine, Yasuyuki Hatsuda, Ivan Chi-Ho Ip, Saebyeok Jeong, Joonho Kim, Marcos Marino, Tomoki Nosaka, Peng Zhao for useful discussions, comments and correspondence.


\appendix

\section{Instanton partition functions with defects} \label{appA}

In this Appendix we collect the relevant contour integral formulae computing the instanton contributions to the partition functions of 4d and 5d theories in the presence of various codimension two defects.

\subsection{Four-dimensional theories} \label{formula4d}

The perturbative part of the partition function for a four-dimensional $\mathcal{N} = 2$ $SU(N)$ theory living on $\mathbb{R}^2_{\epsilon_1} \times \mathbb{R}^2_{\epsilon_2}$ is given by
\begin{equation}
Z_{\text{4d}}^{\text{pert}}(\vec{a}, \epsilon_1, \epsilon_2, Q_{\text{4d}}) =
\exp \left(-\sum_{l\neq m}^N \gamma_{\epsilon_1, \epsilon_2}(a_l - a_m; Q_{\text{4d}})\right)
\end{equation}
in terms of the function
\begin{equation}
\gamma_{\epsilon_1, \epsilon_2}(x; Q_{\text{4d}}) = \dfrac{d}{ds} \left[\dfrac{Q_{\text{4d}}^{s/4}}{\Gamma(s)} \int_0^{\infty} \dfrac{dt}{t} t^s \dfrac{e^{-t x}}{(e^{\epsilon_1 t} - 1)(e^{\epsilon_2 t} - 1)}\right]_{s=0}.
\end{equation}
The instanton part instead reads
\begin{equation}
Z^{\text{inst}}_{\text{4d}}(\vec{a}, \epsilon_1, \epsilon_2, Q_{\text{4d}}) = \sum_{k = 0}^{\infty} Q_{\text{4d}}^k Z^{(k)}_{\text{4d}}(\vec{a}, \epsilon_1, \epsilon_2),
\end{equation}
where $\vec{a} = (a_1, \ldots, a_N)$ and
\begin{equation}
Z^{(k)}_{\text{4d}}(\vec{a}, \epsilon_1, \epsilon_2) = \dfrac{1}{k!} \oint \left[\prod_{s=1}^k \dfrac{d \phi_s}{2\pi i}\right]
Z^{(k), \text{int}}_{\text{4d}}(\vec{\phi}, \vec{a}, \epsilon_1, \epsilon_2),
\end{equation}
\begin{equation}
Z^{(k), \text{int}}_{\text{4d}}(\vec{\phi}, \vec{a}, \epsilon_1, \epsilon_2) = \left( -\dfrac{\epsilon_1 + \epsilon_2}{\epsilon_1 \epsilon_2} \right)^k
\prod_{s>t}^k \dfrac{\phi_{st}^2(\phi_{st}^2 - \epsilon_+^2)}{(\phi_{st}^2 - \epsilon_1^2)(\phi_{st}^2 - \epsilon_2^2)}
\prod_{m=1}^N \prod_{s=1}^k \dfrac{1}{(\phi_s - a_m - \frac{\epsilon_+}{2})(\phi_s - a_m + \frac{\epsilon_+}{2})}.
\end{equation}
Here we are using the short-hand notation $\epsilon_+ = \epsilon_1 + \epsilon_2$ and $\phi_{st} = \phi_s - \phi_t$. The poles to be considered are labelled by $N$-tuples of Young tableaux $\vec{Y} = (Y_1, \ldots, Y_N)$ and are located at
\begin{equation}
\phi_s = a_m + \frac{\epsilon_1 + \epsilon_2}{2} + (i - 1)\epsilon_1 + (j - 1)\epsilon_2, \label{poles}
\end{equation}
with $(i,j)$ box in the $m$-th Young tableau $Y_m$. \\

Similarly, the instanton part of the partition function of our 4d $\mathcal{N} = 2$ $SU(N)$ theory in the presence of a codimension two defect of type II with chiral multiplets (Figure \ref{freechiral}) is computed by 
\begin{equation}
Z^{\text{(c),inst}}_{\text{2d/4d}}(\sigma, \vec{a}, \epsilon_1, \epsilon_2, Q_{\text{4d}}) = \sum_{k = 0}^{\infty} Q_{\text{4d}}^kZ^{\text{(c)},(k)}_{\text{2d/4d}}(\sigma, \vec{a}, \epsilon_1, \epsilon_2), \label{baxfirst}
\end{equation}
with
\begin{equation}
Z^{\text{(c)},(k)}_{\text{2d/4d}}(\sigma, \vec{a}, \epsilon_1, \epsilon_2) = \dfrac{1}{k!} \oint \left[\prod_{s=1}^k \dfrac{d \phi_s}{2\pi i}\right]
Z^{(k), \text{int}}_{\text{4d}}(\vec{\phi}, \vec{a}, \epsilon_1, \epsilon_2)
Z^{\text{(c)}, (k), \text{int}}_{\text{2d/4d}}(\vec{\phi}, \sigma, \vec{a}, \epsilon_1, \epsilon_2), \label{baxsecond}
\end{equation}
\begin{equation}
Z^{\text{(c)}, (k), \text{int}}_{\text{2d/4d}}(\vec{\phi}, \sigma, \vec{a}, \epsilon_1, \epsilon_2)= \prod_{s=1}^k \dfrac{\phi_s - \sigma + \frac{\epsilon_1 + \epsilon_2}{2}}{\phi_s - \sigma + \frac{\epsilon_1 + \epsilon_2}{2} - \epsilon_2};
\end{equation}
the relevant poles to be considered are the same $N$-tuples of Young tableaux as in \eqref{poles}. 

If the type II defect consists of antichiral multiplets instead, the formulae are modified according to
\begin{equation}
Z^{\text{(ac),inst}}_{\text{2d/4d}}(\sigma, \vec{a}, \epsilon_1, \epsilon_2, Q_{\text{4d}}) = \sum_{k = 0}^{\infty} Q_{\text{4d}}^kZ^{\text{(ac)},(k)}_{\text{2d/4d}}(\sigma, \vec{a}, \epsilon_1, \epsilon_2), 
\end{equation}
\begin{equation}
Z^{\text{(ac)},(k)}_{\text{2d/4d}}(\sigma, \vec{a}, \epsilon_1, \epsilon_2) = \dfrac{1}{k!} \oint \left[\prod_{s=1}^k \dfrac{d \phi_s}{2\pi i}\right]
Z^{(k), \text{int}}_{\text{4d}}(\vec{\phi}, \vec{a}, \epsilon_1, \epsilon_2)
Z^{\text{(ac)}, (k), \text{int}}_{\text{2d/4d}}(\vec{\phi}, \sigma, \vec{a}, \epsilon_1, \epsilon_2), 
\end{equation}
\begin{equation}
Z^{\text{(ac)}, (k), \text{int}}_{\text{2d/4d}}(\vec{\phi}, \sigma, \vec{a}, \epsilon_1, \epsilon_2)= \prod_{s=1}^k \dfrac{\phi_s - \sigma - \frac{\epsilon_1 + \epsilon_2}{2}}{\phi_s - \sigma - \frac{\epsilon_1 + \epsilon_2}{2} + \epsilon_2};
\end{equation}
the poles contributing to the integral in this case are $(N+1)$-tuples of Young tableaux $\vec{Y} = (Y_1, \ldots, Y_N, Y_{N+1})$, where the first $N$ tableaux are as in \eqref{poles} while the tableau $Y_{N+1}$ contains poles at 
\begin{equation}
\phi_s = \sigma + \frac{\epsilon_1 + \epsilon_2}{2} - \epsilon_2 + (i - 1)\epsilon_1 + (j - 1)\epsilon_2,
\end{equation}
with $(i,j)$ box in $Y_{N+1}$. \\


Moving to defects of type I (Figure \ref{TSUN}), denoting $\vec{k} = (k_1, \ldots, k_N)$ and $k_{N+1} = k_1$ we have
\begin{equation}
Z^{\text{(I)}}_{\text{2d/4d}}(\vec{x}, \vec{a}, \epsilon_1, \epsilon_2, Q_{\text{4d}}) = \sum_{k_1, \ldots, k_N = 0}^{\infty} q_1^{k_1} \ldots q_N^{k_N} Z^{\text{(I)},(\vec{k})}_{\text{2d/4d}}(\vec{a}, \epsilon_1, \epsilon_2) \label{4d2dI}
\end{equation}
with $q_i = e^{x_i - x_{i+1}}$ for $i = 1, \ldots, N-1$ while $q_N = Q_{\text{4d}} e^{x_N - x_1}$ and
\begin{equation}
\begin{split}
Z^{\text{(I)},(\vec{k})}_{\text{2d/4d}}(\vec{a}, \epsilon_1, \epsilon_2)  = \dfrac{1}{k_1! \ldots k_N!} \oint \left[ \prod_{b=1}^N \prod_{s=1}^{k_b} \dfrac{d \phi^{(b)}_s}{2\pi i} \right] Z^{\text{(I)},(\vec{k}),\text{int}}_{\text{2d/4d}}(\{\vec{\phi} \}, \vec{a}, \epsilon_1, \epsilon_2), \label{4d2dIbis}
\end{split}
\end{equation}
\begin{equation}
\begin{split}
& Z^{\text{(I)},(\vec{k}),\text{int}}_{\text{2d/4d}}(\{\vec{\phi} \}, \vec{a}, \epsilon_1, \epsilon_2) = \epsilon_1^{-(k_1 + \ldots + k_N)} 
\prod_{b = 1}^N \prod_{s \neq t}^{k_b} \dfrac{\phi_s^{(b)} - \phi_t^{(b)}}{\phi_s^{(b)} - \phi_t^{(b)} + \epsilon_1}
\prod_{b = 1}^N \prod_{s = 1}^{k_b} \prod_{t = 1}^{k_{b+1}} 
\dfrac{\phi_s^{(b)} - \phi_t^{(b+1)} + \epsilon_1 + \frac{\epsilon_2}{N}}{\phi_s^{(b)} - \phi_t^{(b+1)} + \frac{\epsilon_2}{N}} \\
& \prod_{b = 1}^{N-1} \prod_{s = 1}^{k_b}\dfrac{1}{(\phi^{(b)}_s - a_b - \frac{b \,\epsilon_2}{N})(\phi^{(b)}_s - a_{b+1} + \epsilon_1 - \frac{b \,\epsilon_2}{N})}
\prod_{s = 1}^{k_N}\dfrac{1}{(\phi^{(N)}_s - a_N - \epsilon_2)(\phi^{(N)}_s - a_1 + \epsilon_1)},
\end{split}
\end{equation}
for $\{ \vec{\phi}\} = (\vec{\phi}^{(1)}, \ldots, \vec{\phi}^{(N)})$; more details on the computation of this partition function can be found for example in \cite{Kanno:2011fw}. 
\\

\noindent \textit{Thermodynamic Bethe Ansatz (TBA) formulae} \\

\noindent When considered in the NS limit $\epsilon_2 \rightarrow 0$, $\epsilon_1 = i \hbar$, some of these instanton partition functions admit a TBA expression, which is typically easier to work with. For example, by defining the function $\varphi(\mu)$ via
\begin{equation}
\varphi(\mu) = -\int_{-\infty}^{\infty} \dfrac{d \lambda}{2\pi} K(\mu - \lambda)
\ln\left(1 + Q_{\text{4d}} \Theta(\lambda) e^{-\varphi(\lambda)}\right),
\end{equation}
where
\begin{equation}
K(\mu) = \dfrac{2 \hbar}{\mu^2 + \hbar^2}, \;\;\;\;\;\; \Theta(\mu) = \prod_{m = 1}^N \dfrac{1}{(\mu - a_m - i \hbar/2)(\mu - a_m + i \hbar/2)},
\end{equation}
we have the following TBA formula for $\mathcal{W}_{\text{4d}}^{\text{inst}}$:
\begin{equation}
\mathcal{W}_{\text{4d}}^{\text{inst}} = -\dfrac{\hbar}{2\pi} \int_{-\infty}^{\infty} d\mu \left[ - \dfrac{1}{2} \varphi(\mu) \ln\left(1 + Q_{\text{4d}}\Theta(\mu) e^{-\varphi(\mu)}\right)
+ \text{Li}_2\left(-Q\, \Theta(\mu)e^{-\varphi(\mu)}\right) \right].
\end{equation}
There are also TBA formulae for $q^{\text{(c),NS}}_{\text{inst}}(\sigma)$ and $q^{\text{(ac),NS}}_{\text{inst}}(\sigma)$, introduced in \cite{Kozlowski:2010tv}:
\begin{equation}
\begin{split}
& \ln q^{\text{(c),NS}}_{\text{inst}}(\sigma)(\sigma) = \int_{-\infty}^{\infty} \dfrac{d\mu}{2\pi i}
\dfrac{1}{\sigma - \mu - i \hbar/2} \ln\left(1 + Q\, \Theta(\mu) e^{-\varphi(\mu)}\right), \\
& \ln q^{\text{(ac),NS}}_{\text{inst}}(\sigma) = - \int_{-\infty}^{\infty} \dfrac{d\mu}{2\pi i}
\dfrac{1}{\sigma - \mu + i \hbar/2} \ln\left(1 + Q\, \Theta(\mu) e^{-\varphi(\mu)}\right). \label{A22}
\end{split}
\end{equation}
These formulae are equivalent to the NS limit of the type II defect instanton partition function with chiral/antichiral multiplets respectively.

\subsection{Five-dimensional theories}

The instanton part of the partition function for a five-dimensional $\mathcal{N} = 1$ $SU(N)$ theory living on $\mathbb{R}^2_{\epsilon_1} \times \mathbb{R}^2_{\epsilon_2} \times S^1_R$ is given by the contour integral formula
\begin{equation}
Z^{\text{inst}}_{\text{5d}}(\vec{a}, \epsilon_1, \epsilon_2, R, Q_{\text{5d}}) = \sum_{k = 0}^{\infty} Q_{\text{5d}}^k Z^{(k)}_{\text{5d}}(\vec{a}, \epsilon_1, \epsilon_2, R),
\end{equation}
where 
\begin{equation}
Z^{(k)}_{\text{5d}}(\vec{a}, \epsilon_1, \epsilon_2, R) = \dfrac{1}{k!} \oint \left[\prod_{s=1}^k \dfrac{d (2 \pi R \phi_s)}{2\pi i}\right]
Z^{(k), \text{int}}_{\text{5d}}(\vec{\phi}, \vec{a}, \epsilon_1, \epsilon_2, R)
\end{equation}
and
\begin{equation}
\begin{split}
Z^{(k), \text{int}}_{\text{5d}}(\vec{\phi}, \vec{a}, \epsilon_1, \epsilon_2, R) = 
& \left(-\dfrac{2\sinh\left[\pi R(\epsilon_1 + \epsilon_2)\right]}{2\sinh\left[\pi R\epsilon_1\right] \cdot 2\sinh\left[\pi R\epsilon_2\right]} \right)^k \\
& \times \prod_{\substack{s,t=1 \\ s \neq t}}^k \dfrac{2\sinh\left[\pi R(\phi_s - \phi_t)\right] \cdot 2\sinh\left[\pi R(\phi_s - \phi_t + \epsilon_1 + \epsilon_2)\right]}{2\sinh\left[\pi R(\phi_s - \phi_t + \epsilon_1)\right] \cdot 2\sinh\left[\pi R(\phi_s - \phi_t + \epsilon_2)\right]} \\
& \times \prod_{s=1}^k \prod_{j=1}^N \dfrac{1}{2\sinh\left[\pi R(\phi_s - a_j + \frac{\epsilon_1 + \epsilon_2}{2})\right] \cdot 2\sinh\left[\pi R(- \phi_s + a_j + \frac{\epsilon_1 + \epsilon_2}{2})\right]}.
\end{split}
\end{equation}
Written in terms of $\sigma_s = e^{2\pi R\phi_s}$, $q_1 = e^{2\pi R \epsilon_1}$, $q_2 = e^{2\pi R \epsilon_2}$, $\mu_j = e^{2\pi R a_j}$ this becomes
\begin{equation}
Z^{(k)}_{\text{5d}}(\vec{\mu}, q_1, q_2) = \dfrac{1}{k!} \oint \left[ \prod_{s=1}^k \dfrac{d \sigma_s}{2\pi i \sigma_s} \right] Z^{(k), \text{int}}_{\text{5d}}(\vec{\sigma}, \vec{\mu}, q_1, q_2),
\end{equation}
with
\begin{equation}
\begin{split}
Z^{(k), \text{int}}_{\text{5d}}(\vec{\sigma}, \vec{\mu}, q_1, q_2) = & \left( \dfrac{1-q_1 q_2}{(1-q_1)(1-q_2)} \right)^k 
\prod_{\substack{s,t=1 \\ s \neq t}}^k \dfrac{(1-\sigma_s \sigma_t^{-1})(1-\sigma_s \sigma_t^{-1} q_1 q_2)}{(1-\sigma_s \sigma_t^{-1} q_1)(1-\sigma_s \sigma_t^{-1} q_2)} \\
& \times \prod_{s=1}^k \prod_{j=1}^N \dfrac{\sqrt{q_1 q_2}}{(1-\sigma_s \mu_j^{-1}\sqrt{q_1 q_2})(1-\sigma_s^{-1}\mu_j \sqrt{q_1 q_2})}.
\end{split}
\end{equation}
The poles to be considered for the evaluation of the integral are labelled by $N$-tuples of Young tableaux $\vec{Y} = (Y_1, \ldots, Y_N)$ and are located at
\begin{equation}
\phi_s = a_m + \frac{\epsilon_1 + \epsilon_2}{2} + (i - 1)\epsilon_1 + (j - 1)\epsilon_2, \label{poles5d}
\end{equation}
with $(i,j)$ box in the $m$-th Young tableau $Y_m$. \\

Moving to defects, it was shown in \cite{Gaiotto:2014ina,Bullimore:2014awa} that the instanton part of the partition function for a 5d $\mathcal{N} = 1$ $U(N)$ theory in the presence of a type II codimension-two defect (Figure \ref{freechiral}) living on $\mathbb{R}^2_{\epsilon_1} \times S^1_R$ represented by $N$ 3d chiral fields ($S$-dual to the simple defect/type III defect of Figure \ref{chiralgauge}) admits the contour integral representation
\begin{equation}
Z_{\text{3d/5d}}^{\text{(c),inst}}(\sigma, \vec{a}, \epsilon_1, \epsilon_2, R, Q_{\text{5d}}) = \sum_{k = 0}^{\infty} Q_{\text{5d}}^k Z_{\text{3d/5d}}^{\text{(c),}(k)}(\sigma, \vec{a}, \epsilon_1, \epsilon_2, R),
\end{equation}
where
\begin{equation}
Z_{\text{3d/5d}}^{\text{(c),}(k)}(\sigma, \vec{a}, \epsilon_1, \epsilon_2, R) = 
\dfrac{1}{k!} \oint \left[ \prod_{s=1}^k \dfrac{d \left(2\pi R \phi_{s}\right)}{2\pi i} \right] Z^{(k), \text{int}}_{\text{5d}}(\vec{\phi}, \vec{a}, \epsilon_1, \epsilon_2, R) Z^{\text{(c)},(k), \text{int}}_{\text{3d/5d}}(\sigma, \vec{\phi}, \vec{a}, \epsilon_1, \epsilon_2, R)
\end{equation}
and
\begin{equation}
\begin{split}
Z^{\text{(c)},(k), \text{int}}_{\text{3d/5d}}(\sigma, \vec{\phi}, \vec{a}, \epsilon_1, \epsilon_2, R) = \prod_{s=1}^k e^{\pi R\epsilon_2} \dfrac{2\sinh\left[\pi R (\phi_s - \sigma + \frac{\epsilon_1 + \epsilon_2}{2})\right]}{2\sinh\left[\pi R(\phi_s - \sigma + \frac{\epsilon_1 + \epsilon_2}{2} - \epsilon_2)\right]}.
\end{split}
\end{equation}
Written in terms of $\sigma_s = e^{2\pi R\phi_s}$, $q_1 = e^{2\pi R \epsilon_1}$, $q_2 = e^{2\pi R \epsilon_2}$, $\mu_j = e^{2\pi R a_j}$, $w = e^{-2\pi R \sigma}$ this becomes
\begin{equation}
Z_{\text{3d/5d}}^{\text{(c),}(k)}(w, \vec{\mu}, q_1, q_2) = \dfrac{1}{k!} \oint \left[ \prod_{s=1}^k \dfrac{d \sigma_s}{2\pi i \sigma_s} \right] 
Z^{(k), \text{int}}_{\text{5d}}(\vec{\sigma}, \vec{\mu}, q_1, q_2) 
Z^{\text{(c)},(k), \text{int}}_{\text{3d/5d}}(w, \vec{\sigma}, \vec{\mu}, q_1, q_2),
 \label{fff}
\end{equation}
with
\begin{equation}
\begin{split}
Z^{\text{(c)},(k), \text{int}}_{\text{3d/5d}}(w, \vec{\sigma}, \vec{\mu}, q_1, q_2) = \prod_{s=1}^k \dfrac{1-\sigma_s w\sqrt{q_1 q_2}}{1-\sigma_s w\, q_2^{-1}\sqrt{q_1 q_2}}.
\end{split}
\end{equation}
The contributing poles are as in \eqref{poles5d} and correspond to the usual Young tableau coming only from the $Z_{k}^{5d}$ part.

We can also consider a type II defect with antichiral fields; in this case we have
\begin{equation}
Z_{\text{3d/5d}}^{\text{(ac),inst}}(\sigma, \vec{a}, \epsilon_1, \epsilon_2, R, Q_{\text{5d}}) = \sum_{k = 0}^{\infty} Q_{\text{5d}}^k Z_{\text{3d/5d}}^{\text{(ac),}(k)}(\sigma, \vec{a}, \epsilon_1, \epsilon_2, R),
\end{equation}
where
\begin{equation}
Z_{\text{3d/5d}}^{\text{(ac)},(k)}(\sigma, \vec{a}, \epsilon_1, \epsilon_2, R) = 
\dfrac{1}{k!} \oint \left[ \prod_{s=1}^k \dfrac{d \left(2\pi R \phi_{s}\right)}{2\pi i} \right] Z^{(k), \text{int}}_{\text{5d}}(\vec{\phi}, \vec{a}, \epsilon_1, \epsilon_2, R) Z^{\text{(ac)},(k), \text{int}}_{\text{3d/5d}}(\sigma, \vec{\phi}, \vec{a}, \epsilon_1, \epsilon_2, R)
\end{equation}
and
\begin{equation}
\begin{split}
Z^{\text{(ac)},(k), \text{int}}_{\text{3d/5d}}(\sigma, \vec{\phi}, \vec{a}, \epsilon_1, \epsilon_2, R) = \prod_{s=1}^k e^{-\pi R\epsilon_2} \dfrac{2\sinh\left[\pi R (\phi_s - \sigma - \frac{\epsilon_1 + \epsilon_2}{2})\right]}{2\sinh\left[\pi R(\phi_s - \sigma - \frac{\epsilon_1 + \epsilon_2}{2} + \epsilon_2)\right]}.
\end{split}
\end{equation}
Written in terms of $\sigma_s = e^{2\pi R\phi_s}$, $q_1 = e^{2\pi R \epsilon_1}$, $q_2 = e^{2\pi R \epsilon_2}$, $\mu_j = e^{2\pi R a_j}$, $w = e^{-2\pi R \sigma}$ this becomes
\begin{equation}
Z_{\text{3d/5d}}^{\text{(ac),}(k)}(w, \vec{\mu}, q_1, q_2) = \dfrac{1}{k!} \oint \left[ \prod_{s=1}^k \dfrac{d \sigma_s}{2\pi i \sigma_s} \right] 
Z^{(k), \text{int}}_{\text{5d}}(\vec{\sigma}, \vec{\mu}, q_1, q_2) 
Z^{\text{(ac)},(k), \text{int}}_{\text{3d/5d}}(w, \vec{\sigma}, \vec{\mu}, q_1, q_2),
 \label{fff}
\end{equation}
with
\begin{equation}
\begin{split}
Z^{\text{(ac)},(k), \text{int}}_{\text{3d/5d}}(w, \vec{\sigma}, \vec{\mu}, q_1, q_2) = \prod_{s=1}^k \dfrac{1-\sigma_s w/\sqrt{q_1 q_2}}{1-\sigma_s w\, q_2/\sqrt{q_1 q_2}}.
\end{split}
\end{equation}
In this case poles are labelled by $(N+1)$-tuples of Young tableaux $\vec{Y} = (Y_1, \ldots, Y_N, Y_{N+1})$; the first $N$ Young tableaux correspond to the poles \eqref{poles5d}, while in the tableau $Y_{N+1}$ we have poles of the form
\begin{equation}
\phi_s = \sigma + \frac{\epsilon_1 + \epsilon_2}{2} - \epsilon_2 + (i - 1)\epsilon_1 + (j - 1)\epsilon_2,
\end{equation}
with $(i,j)$ box in $Y_{N+1}$. \\

Finally, the vacuum expectation value of a Wilson loop in the fundamental representation of $SU(N)$ wrapping $S^1_R$ can be computed as
\begin{equation}
W_{\square}^{SU(N)}(\vec{a}, \epsilon_1, \epsilon_2, R, Q_{\text{5d}}) = 
\sum_{k=0}^{\infty} Q^k_{\text{5d}} W^{SU(N),(k)}_{\square}(\vec{a}, \epsilon_1, \epsilon_2, R),
\end{equation}
where 
\begin{equation}
\begin{split}
& W^{SU(N),(k)}_{\square}(\vec{a}, \epsilon_1, \epsilon_2, R) =  
\dfrac{1}{k!} \oint \left[ \prod_{s=1}^k \dfrac{d \left(2\pi R \phi_{s}\right)}{2\pi i} \right] Z^{(k), \text{int}}_{\text{5d}}(\vec{\phi}, \vec{a}, \epsilon_1, \epsilon_2, R) \\
\;\;\;\;\;\; & \times \left[ \sum_{m=1}^N e^{2 \pi R a_m} - (1-e^{2\pi R \epsilon_1})(1-e^{2\pi R \epsilon_2})e^{-\pi R (\epsilon_1 + \epsilon_2)} \sum_{s = 1}^k e^{2\pi R \phi_s} \right].
\end{split}
\end{equation}
The contributing poles are once again the ones in \eqref{poles5d}.

\section{Special functions} \label{appdoublesine}

In this Appendix we collect the definition and main properties of the double sine and triple sine functions, which naturally appear when considering partition functions of supersymmetric gauge theories on curved spaces.

\subsection{Double sine function}

The function $\mathcal{S}_{\omega_1, \omega_2}(x)$ is defined as
\begin{equation}
\mathcal{S}_{\omega_1, \omega_2}(x)  = \text{exp} \left( \int_{\mathbb{R} + i 0} \dfrac{e^{xz}}{(e^{\omega_1 z} - 1)(e^{\omega_2 z} - 1)} \dfrac{dz}{z} \right),
\end{equation}
in the strip $0 < \text{Re}\,z  < \text{Re} (\omega_1 + \omega_2)$ when all $\text{Re} (\omega_j) > 0$.  This function is related to the quantum dilogarithm $\Phi_{\omega_1, \omega_2}(x)$ intrdoduced in \cite{1995LMaPh..34..249F}
\begin{equation}
\Phi_{\omega_1, \omega_2}(x) = \text{exp} \left( \int_{\mathbb{R} + i 0} \dfrac{e^{-2ixz}}{2\sinh(\omega_1 z)\cdot 2\sinh(\omega_2 z)} \dfrac{dz}{z} \right),
\end{equation}
according to
\begin{equation}
\mathcal{S}_{\omega_1, \omega_2}\left(-ix + \frac{\omega_1 + \omega_2}{2}\right) = \Phi_{\omega_1, \omega_2}(x).
\end{equation}
When Im$\,(\omega_1/\omega_2) > 0$ or $\omega_1/\omega_2 \notin \mathbb{Q}$ it admits the infinite product representation
\begin{equation}
\mathcal{S}_{\omega_1, \omega_2}\left(ix + \omega_1 + \omega_2\right) = 
\Phi_{\omega_1, \omega_2}\left(-x + i \frac{\omega_1 + \omega_2}{2}\right) = 
\dfrac{(q e^{-2\pi x/\omega_2};q)_{\infty}}{(e^{-2\pi x/\omega_1};\widetilde{q}^{-1})_{\infty}}, \label{pizza}
\end{equation}
where we introduced the parameters
\begin{equation}
q = e^{2\pi i \omega_1 /\omega_2} \;\;\;,\;\;\; \widetilde{q} = e^{2\pi i \omega_2 /\omega_1},
\end{equation}
and the $q$-Pochhammer symbol
\begin{equation}
(w;q)_{\infty} = \prod_{k=0}^{\infty} (1-q^k w).
\end{equation}
A useful formula related to the $q$-Pochhammer symbol is
\begin{equation}
\text{exp} \left( - \sum_{k \geqslant 1} \dfrac{q^k w^{k}}{k(1 - q^{k})} \right) =
\begin{cases}
\prod_{k \geqslant 0} (1-q^{k+1}w) \;\;\;, \;\;\; \vert q \vert < 1; \\
\prod_{k \geqslant 0} \dfrac{1}{(1-q^{-k}w)} \;\;\;, \;\;\; \vert q \vert > 1. 
\end{cases} 
\label{continuation}
\end{equation}
The $\mathcal{S}_{\omega_1, \omega_2}(x)$ function satisfies the identities
\begin{equation}
\begin{split}
& \mathcal{S}_{\omega_1, \omega_2}\left(ix + \omega_1 + \omega_2 - \omega_1 \right) 
= \left( 1 - e^{-2\pi x/\omega_2} \right) 
\mathcal{S}_{\omega_1, \omega_2}\left(ix + \omega_1 + \omega_2\right), \\
& \mathcal{S}_{\omega_1, \omega_2}\left(ix + \omega_1 + \omega_2 - \omega_2 \right) 
= \left( 1 - e^{-2\pi x/\omega_1} \right) 
\mathcal{S}_{\omega_1, \omega_2}\left(ix + \omega_1 + \omega_2\right), 
\end{split}
\end{equation}
and
\begin{equation}
\begin{split}
& \mathcal{S}_{\omega_1, \omega_2}\left(ix + \omega_1 + \omega_2 + \omega_1 \right) 
= \dfrac{1}{\left( 1 - q e^{-2\pi x/\omega_2} \right)}
\mathcal{S}_{\omega_1, \omega_2}\left(ix + \omega_1 + \omega_2\right), \\
& \mathcal{S}_{\omega_1, \omega_2}\left(ix + \omega_1 + \omega_2 +\omega_2 \right) 
= \dfrac{1}{\left( 1 - \widetilde{q}e^{-2\pi x/\omega_1} \right)}
\mathcal{S}_{\omega_1, \omega_2}\left(ix + \omega_1 + \omega_2\right), 
\end{split}
\end{equation}
as well as
\begin{equation}
\begin{split}
& \mathcal{S}_{\omega_1, \omega_2}\left(-ix + \frac{\omega_1 + \omega_2}{2}\right)\mathcal{S}_{\omega_1, \omega_2}\left(ix + \frac{\omega_1 + \omega_2}{2}\right) = \\
& = \; \Phi_{\omega_1, \omega_2}(x) \Phi_{\omega_1, \omega_2}(-x)
\; = \; e^{i \pi \frac{x^2}{\omega_1 \omega_2} + i \pi \frac{\omega_1^2 + \omega_2^2}{12\, \omega_1 \omega_2}}.
\end{split}
\end{equation}
We also introduce the double sine function $S_2(x \vert \omega_1, \omega_2)$ defined as
\begin{equation}
S_2(x \vert \omega_1, \omega_2) = e^{\frac{i \pi}{2}B_{2,2}(x\vert \omega_1, \omega_2)} \mathcal{S}_{\omega_1, \omega_2}(x),
\end{equation}
where
\begin{equation}
B_{2,2}(x\vert \omega_1, \omega_2) = \dfrac{x^2}{\omega_1 \omega_2} - \dfrac{\omega_1 + \omega_2}{\omega_1 \omega_2}x + \dfrac{\omega_1^2 + 3 \omega_1 \omega_2 + \omega_2^2}{6\omega_1 \omega_2}. 
\end{equation}
The double sine function can be thought as the regularization of the infinite product
\begin{equation}
S_2(x \vert \omega_1, \omega_2)  = \prod_{m,n \geqslant 0}\dfrac{m \omega_1 + n \omega_2 + x}{m \omega_1 + n \omega_2 + \omega_1 + \omega_2 - x}.
\end{equation}
This function satisfies 
\begin{equation}
S_2(x \vert \omega_1, \omega_2) S_2(-x +\omega_1 + \omega_2 \vert \omega_1, \omega_2) = 1,
\end{equation}
as well as
\begin{equation}
\overline{S_2(x \vert \omega_1, \omega_2)} = S_2(\overline{x} \vert \overline{\omega_1}, \overline{\omega_2}) \label{complconj}
\end{equation}
and
\begin{equation}
\begin{split}
& S_2^{-1}(-i x + \omega_1 \vert \omega_1, \omega_2)
= - i \cdot 2 \sinh\left[\frac{\pi x}{\omega_2}\right] 
S_2^{-1}(-i x \vert \omega_1, \omega_2), \\
& S_2^{-1}(-i x + \omega_2 \vert \omega_1, \omega_2)
= - i \cdot 2 \sinh\left[\frac{\pi x}{\omega_1}\right] 
S_2^{-1}(-i x \vert \omega_1, \omega_2), \\
& S_2^{-1}(-i x - \omega_1 \vert \omega_1, \omega_2)
= \dfrac{1}{- i \cdot 2 \sinh\left[\frac{\pi x}{\omega_2} - i \pi \frac{\omega_1}{\omega_2}\right]} S_2^{-1}(-i x \vert \omega_1, \omega_2),  \\
& S_2^{-1}(-i x - \omega_2 \vert \omega_1, \omega_2)
= \dfrac{1}{- i \cdot 2 \sinh\left[\frac{\pi x}{\omega_1} - i \pi \frac{\omega_2}{\omega_1}\right]} S_2^{-1}(-i x \vert \omega_1, \omega_2), \\
& S_2^{-1}(i x - \omega_1 \vert \omega_1, \omega_2)
= \dfrac{1}{i \cdot 2 \sinh\left[\frac{\pi x}{\omega_2} + i \pi \frac{\omega_1}{\omega_2}\right]} S_2^{-1}(i x \vert \omega_1, \omega_2), \\
& S_2^{-1}(i x - \omega_2 \vert \omega_1, \omega_2)
= \dfrac{1}{i \cdot 2 \sinh\left[\frac{\pi x}{\omega_1} + i \pi \frac{\omega_2}{\omega_1}\right]} S_2^{-1}(i x \vert \omega_1, \omega_2). \label{property}
\end{split}
\end{equation}

\subsection{Triple sine function}

Let us also introduce the function $\mathcal{S}_{\omega_1, \omega_2, \omega_3}(x)$, defined as
\begin{equation}
\mathcal{S}_{\omega_1, \omega_2, \omega_3}(x)  = 
\text{exp} \left( - \int_{\mathbb{R} + i 0} \dfrac{e^{xz}}{(e^{\omega_1 z} - 1)(e^{\omega_2 z} - 1)(e^{\omega_3 z} - 1)} \dfrac{dz}{z} \right)
\end{equation}
in the strip $0 < \text{Re}\,z  < \text{Re} (\omega_1 + \omega_2 + \omega_3)$ when all $\text{Re} (\omega_j) > 0$. When Im$\,(\omega_1/\omega_2) > 0$, Im$\,(\omega_1/\omega_3) > 0$, Im$\,(\omega_3/\omega_2) > 0$ or when $\omega_1/\omega_2, \omega_1/\omega_3, \omega_3/\omega_2 \notin \mathbb{Q}$ this admits the infinite product representation
\begin{equation}
\begin{split}
& \mathcal{S}_{\omega_1, \omega_2, \omega_3}\left(x \right) = \\
& = \dfrac{(e^{2\pi i x/\omega_2};e^{2\pi i \omega_1/\omega_2}; e^{2\pi i \omega_3/\omega_2})_{\infty} 
(e^{-2\pi i \omega_3/\omega_1} e^{-2\pi i \omega_2/\omega_1}e^{2\pi i x/\omega_1};e^{-2\pi i \omega_3/\omega_1}; e^{-2\pi i \omega_2/\omega_1})_{\infty}}{(e^{-2\pi i \omega_2/\omega_3} e^{2\pi i x/\omega_3};e^{2\pi i \omega_1/\omega_3}; e^{-2\pi i \omega_2/\omega_3})_{\infty}}, 
\end{split}
\end{equation}
where we introduced the notation
\begin{equation}
\begin{split}
& \left(e^{2\pi i x/\omega_2};e^{2\pi i \omega_1/\omega_2}; e^{2\pi i \omega_3/\omega_2} \right)_{\infty} = \prod_{j,k\geqslant 0}\left(1 - e^{2\pi i x/\omega_2}e^{2\pi i \omega_1 j/\omega_2} e^{2\pi i \omega_3 k/\omega_2} \right) = \\
& = \text{exp}\left( - \sum_{n \geqslant 1} \dfrac{e^{2\pi n i x/\omega_2}}{n(1 - e^{2\pi i \omega_1 n /\omega_2})(1 - e^{2\pi i \omega_3 n/\omega_2})} \right).
\end{split}
\end{equation}
The triple sine function is defined as
\begin{equation}
S_3(x \vert \omega_1, \omega_2, \omega_3) = e^{-\frac{i \pi}{6}B_{3,3}(x\vert \omega_1, \omega_2, \omega_3)} \mathcal{S}_{\omega_1, \omega_2, \omega_3}(x), \label{pppp}
\end{equation}
where
\begin{equation}
\begin{split}
B_{3,3}(x\vert \omega_1, \omega_2, \omega_3) & \;=\;  
\dfrac{x^3}{\omega_1 \omega_2 \omega_3} - \dfrac{3}{2}\dfrac{\omega_1 + \omega_2 + \omega_3}{\omega_1, \omega_2, \omega_3} x^2 \\
& \;+\; \dfrac{\omega_1^2 + \omega_2^2 + \omega_3^2 + 3 (\omega_1 \omega_2 + \omega_1 \omega_3 + \omega_2 \omega_3)}{2\omega_1 \omega_2 \omega_3} x \\
& \; - \; \dfrac{(\omega_1 + \omega_2 + \omega_3)(\omega_1 \omega_2 + \omega_1 \omega_3 + \omega_2 \omega_3)}{4\omega_1 \omega_2 \omega_3},
\end{split}
\end{equation}
and satisfies
\begin{equation}
S_3(x \vert \omega_1, \omega_2, \omega_3) = S_3(- x + \omega_1 + \omega_2 + \omega_3 \vert \omega_1, \omega_2, \omega_3). 
\end{equation}


\bibliography{bibl}
\bibliographystyle{JHEP}

\end{document}